\documentclass[ twoside, bibliography=totoc, openright,titlepage,numbers=noenddot,headinclude,
                11pt,a4paper,BCOR5mm,footinclude=true]{scrreprt}
\pdfoutput=1

\usepackage[latin1]{inputenc}
\usepackage[ngerman,american]{babel}
\usepackage[square,numbers]{natbib}
\usepackage[]{amsmath} 
\usepackage{appendix}
\usepackage{subfig}
\usepackage{multirow} 

\usepackage{classicthesis-ldpkg} 
\usepackage[eulerchapternumbers,listings,
						subfig,parts,dottedtoc]{classicthesis}
\usepackage{wrapfig}
\usepackage{upgreek}

\makeatletter \@addtoreset{figure}{chapter} \makeatother 					
\makeatletter \@addtoreset{table}{chapter} \makeatother 					

\usepackage{geometry}  
\usepackage[titles]{tocloft}   
\usepackage{lipsum}

\hypersetup{%
    linktocpage=true, pdfstartpage=5, pdfstartview=FitV,
    breaklinks=true, pdfpagemode=UseNone, pageanchor=true, pdfpagemode=UseOutlines,%
    plainpages=false, bookmarksnumbered, bookmarksopen=true, bookmarksopenlevel=1,%
    hypertexnames=true, pdfhighlight=/O,
    urlcolor=webbrown, linkcolor=RoyalBlue, citecolor=webgreen, 
    pdfsubject={},%
    pdfkeywords={},%
    pdfcreator={pdfLaTeX},%
    pdfproducer={LaTeX with hyperref and classicthesis}%
}
\renewcommand{\figurename}{Abb.}
\renewcommand{\tablename}{Tab.}
\usepackage{rotating} 

\graphicspath{ {./pics/} }

\begin{document}

\renewcommand{\contentsname}{Inhaltsverzeichnis} 
\renewcommand{\bibname}{Literaturverzeichnis}

\title{\huge Einfluss der Gewebegeometrie auf die Transversal Relaxation}
\date{} 
\author{\vspace{1cm}
         \includegraphics[width=0.48\linewidth]{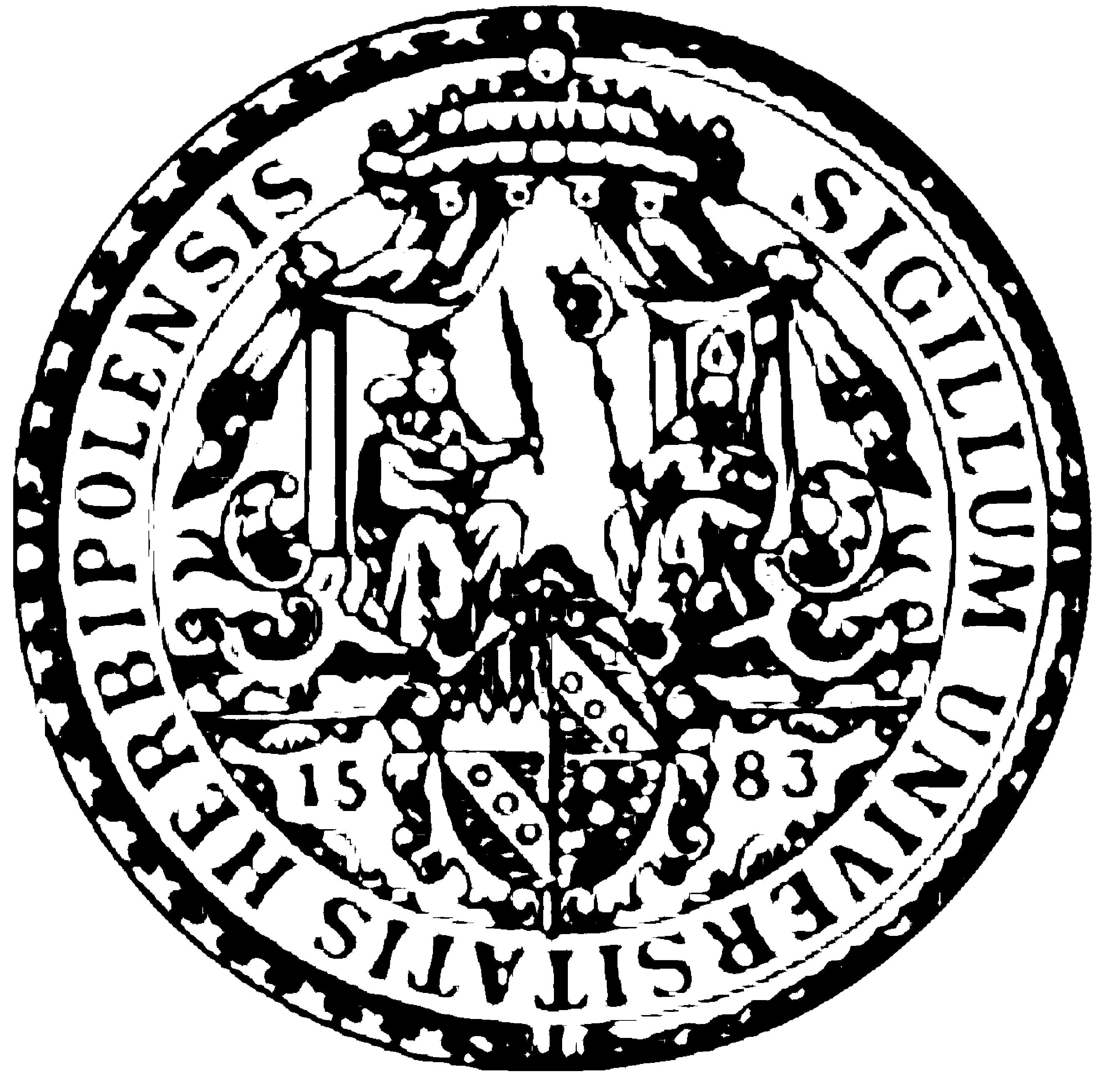} \\ 
         \vspace{1cm}
         Diplomarbeit im Fach Physik \\  01.11.2010 - 28.07.2011 \\  von Martin Rückl \\ \\
Fakultät für Physik und Astronomie der \\
Julius-Maximilians-Universität Würzburg \\
Lehrstuhl für Experimentelle Physik V \\
Prof. Dr. Peter Jakob
}
\maketitle
\thispagestyle{empty}
\begin{center} \textit{Besonderer Dank geht an alle meine Mitstreiter aus Zimmer E090\\ und meine zwei Betreuer Thomas Kampf und Christian Ziener.\\ Auch den vielen anderen Personen, die mir beim Erstellen dieser Arbeit geholfen haben, möchte ich danken.}     \end{center}    \thispagestyle{empty}   \clearpage

\begin{center} \textit{Hoffnung ist eben nicht Optimismus. Es ist nicht die Überzeugung, dass etwas gut ausgeht, sondern die Gewissheit, dass etwas Sinn hat - ohne Rücksicht darauf, wie es ausgeht.\newline
Václav Havel}
\end{center}

\newcommand{\sgn}{\mathsf{sgn}}
\newcommand{\Imag}{\mathsf{Im}}
\newcommand{\Real}{\mathsf{Re}}
\newcommand{\ICD}{\mathsf{ICD}}
\newcommand{\mum}{\upmu\mathsf{m}}
\newcommand{\ms}{\mathsf{ms}}
\newcommand{\dom}{\delta\omega}
\newcommand{\radps}{\,\mathsf{rad}/\mathsf{s}}
\newcommand{\mumsqpms}{\mum^2/\ms}
\newcommand{\ppm}{\mathsf{ppm}}

\pdfbookmark[1]{\contentsname}{toc} 
\tableofcontents 
\thispagestyle{empty}
\clearpage

~
\thispagestyle{empty}
\clearpage

\pagestyle{plain}
\pagenumbering{arabic}
\renewcommand{\baselinestretch}{1.1}\normalsize 
\chapter{Einführung}\label{kap:introduction}
\setcounter{page}{8}
\noindent In den letzten zwei Jahrzehnten hat sich in der medizinischen Praxis die Bildgebung mittels Magnetresonanztomographie oder kurz MRI (Magnetic Resonance Imaging) als Standard-Diagnoseverfahren immer mehr durchgesetzt. Der hohe Aufwand dieses Verfahrens wird dabei durch die vielen Vorteile gegenüber anderen Bildgebungsverfahren, die u.a. meist mit einer Stahlenbelastung einhergehen, gerechtfertigt. Vor allem die hohe Flexibilität bzgl. der Kontrastverteilung auf unterschiedliche Gewebearten nur durch kleine Variationen von Parametern im Messprotokoll macht MRI oft zum Mittel der Wahl.\newline
Ein Wermutstropfen besteht jedoch in dem hohen Aufwand, der für hochauflösende Bilder betrieben werden muss. Die dafür benötigten starken Magnetfelder und Gradienten setzen, vor allem in der medizinischen Praxis, eine Grenze der Voxelgröße bei ca. $(0.5\,\mathsf{mm})^3$. Dieser Optimal-wert wird allerdings in vielen Bereichen nicht erreicht. Für das schlagende Herz etwa liegt eine realistische Voxelgröße nur bei ca. $(1\,\mathsf{mm})^3$. In dieser Arbeit sollen nun die Möglichkeiten überprüft werden, auch unterhalb dieser Auflösungsgrenze noch Informationen über das Innere eines Voxels gewinnen zu können. Mit Informationen ist in diesem Zusammenhang speziell die räumliche Anordnung von Kapillaren im Muskelgewebe gemeint. Parameter die diese Anordnung beschreiben sind z.B. der mittlere Abstand der Kapillaren zueinander ($\ICD$) oder die Regelmäßigkeit der Anordnung, die sich wie in \cite{Karch2006} gezeigt, gut durch einen einzigen Parameter quantifizieren lässt. Diese Informationen sind z.B. in der Diagnostik von Arteriosklerose im Herzmuskelgewebe hilfreich, da eine Verengung der größeren Blutgefäße in der Regel durch Anpassung des Kapillar\-netzwerks teilweise kompensiert wird und so mit einer morphologischen Veränderung dieses Netzwerks einhergeht \cite{Karch2005f}.\newline
Um die Auswirkungen dieser morphologischen Veränderungen auf das Zeitverhalten der transversalen Magnetisierung zu verstehen und vorhersagen zu können, müssen der Einfluss der Feldinhomogenitäten und die Diffusion der signalgebenden Protonen auf den Dephasierungsprozess der Magnetisierung modelliert werden. Mit einem solchen Modell können dann verschiedene Kapillaranordnungen simuliert werden. Mit diesen Simulationen kann geprüft werden, ob und wie viel Einfluss die Gewebegeometrie, also z.B. der $\ICD$, auf den Magnetisierungsverlauf eines kompletten Voxels hat.\newline
Die kapillaren Blutgefäße im Muskelgewebe verlaufen über große Raumbereiche (bis zu mehreren $100\,\mum$) weitgehend parallel zueinander und zu den Muskelfasern. Die charakteristische Distanz, die die signalgebenden Protonen aufgrund des Diffusionsprozesses im Gewebe in einem für das Messsignal relevanten Zeitraum (ca. $2T_2^*$) zurücklegen, liegt nur in der Größenordnung der Kapillarabstände (wenige $\mum$) und ist somit weit kleiner als die durchschnittliche Länge einer Kapillare. Auf der für das Signal relevanten Zeitskala kann daher näherungsweise von einer Translationsinvarianz bzgl. der Kapillarachsen ausgegangen werden. Dies ermöglicht es in ein effizienteres zweidimensionales Modell zu wechseln.\newline
Der Unterschied der magnetischen Suszeptibilität von Kapillare und umgebendem Gewebe induziert mikroskopische Feldinhomogenitäten die direkt die geometrische Anordnung der Kapillaren wider spiegeln. Daher sollen zunächst diese sog. Offresonanzfelder einer ausführlichen Analyse unterzogen werden. Vernachlässigt man die Diffusion lassen sich die Auswirkungen der Feldinhomogenitäten auf das Signal mit den bereits im Jahr 1946 durch F. Bloch gefundenen Gleichungen beschreiben. Diese sind meist analytisch oder einfach numerisch lösbar \cite{Bloch46}.\newline
Als nächster Schritt soll die Diffusion der Protonen im Gewebe berücksichtigt werden. Dazu müssen, wie durch H. C. Torrey 1965 gezeigt, die Bloch-Gleichungen um einen Diffusions\-term erweitert werden \cite{Torrey56}. Die so entstehende Bloch-Torrey-Gleichung kann nur noch für sehr spezielle Randbedingungen und Offresonanzen analytisch gelöst werden. Für eine Anwendung auf realistischere Modellgeometrien muss daher auf numerische Methoden zurückgegriffen werden. Ein möglicher Ansatz, der hier weiter verfolgt werden soll, ist die so genannte Monte-Carlo-Methode. Dabei wird der Diffusionsprozess durch viele zufällige Random-Walk-Trajektorien modelliert. Mit Hilfe dieses Ansatzes sollen dann die untersuchten statischen Inhomogenitäten mit der Bloch-Torrey-Gleichung kombiniert werden.\newline
Die bisher am besten untersuchte zweidimensionale Gewebegeometrie stellt das Krogh-Modell dar. Seit kurzem ist für dieses sogar eine analytische Lösung der Bloch-Torrey-Gleichung bekannt \cite{ZienerPHDThesis}. Wie sich zeigen wird birgt jedoch die praktische Anwendung der analytischen Lösung einige bisher nicht untersuchte mathematische Probleme auf dem Bereich der Mathieu-Funktionen, denen daher ebenfalls ein Teil dieser Arbeit gewidmet werden soll.\newline
Für den direkten Anwendungsbezug sollen schließlich differenziertere und somit besser an reale Kapillarnetzwerke angepasste Geometrien untersucht werden. Ein erster Schritt geht dabei weg von der einzelnen Kapillare im Krogh-Modell hin zu einer regelmäßigen Struktur aus vielen Kapillaren. In einem zweiten Schritt soll die regelmäßige Struktur durch Störung aufgebrochen werden um eine bessere Modellierung realen Gewebes zu erreichen. Mit Hilfe der numerischen Implementierung kann dann der zeitliche Verlauf der transversalen Magnetisierung in Abhängigkeit der verschiedenen die Gewebegeo\-metrie beschreibenden Parameter simuliert werden. So kann eine Art Nachschlagewerk der Relaxationszeiten in Abhängigkeit der zu Grunde liegenden Geometrie erzeugt werden. Aus diesem Nachschlagewerk lässt sich dann ermitteln, ob zwischen den verschiedenen in der Realität vorkommenden Ausprägungen des Kapillarnetzwerks überhaupt messbare Unterschiede in der Relaxationszeit bestehen. Sollte dies der Fall sein, wäre prinzipiell eine Zuordnung gemessener Relaxationsraten zu den die Geometrie beschreibenden Parametern möglich.\newline

\chapter{Theorie}\label{kap:theory}
\renewcommand{\figurename}{Abb.}
\renewcommand{\tablename}{Tab.}

\section{Bloch- und Bloch-Torrey-Gleichungen für die transversale Magnetisierung}
	\begin{figure}
		\begin{center}\includegraphics[width=0.75\textwidth]{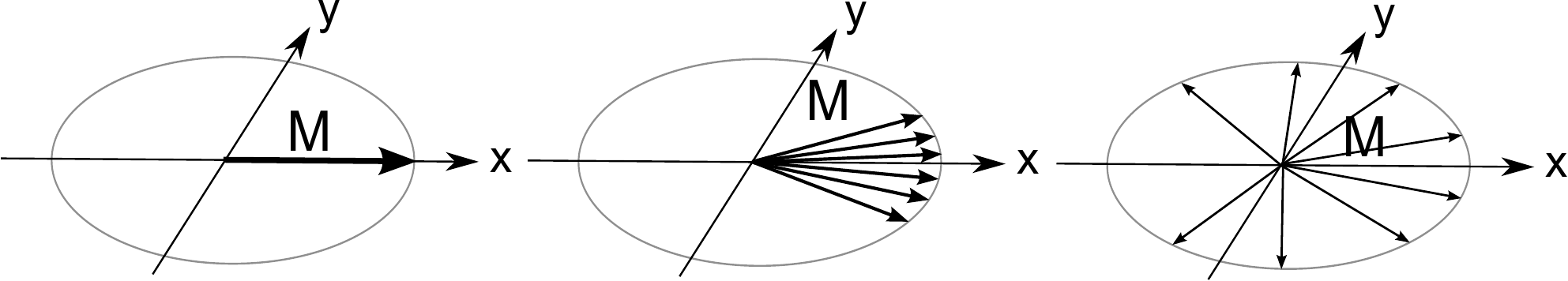}\end{center}
		\caption{In einem mit $\omega_0$ rotierenden Bezugssystem ohne Offresonanzen bliebe die Magnetisierungsdichte immer in Phase (links). Durch die unterschiedlichen Frequenzen der Präzession "fächert" sie jedoch mit der Zeit immer weiter auf (Mitte und rechts), die Gesamtmagnetisierung $\vec{M_T}$ (Vektorsumme) nimmt ab. Zurück im ruhenden Laborsystem lässt sich eine Abnahme der Amplitude der Magnetisierung messen.}
		\label{fig:transversal-dephasing}
	\end{figure}
Aus den Bloch-Gleichungen \cite{Bloch46} folgt, dass die transversale Magnetisierung in einem räumlich homogenen Magnetfeld $\vec{B_0}(\vec{r})=(0,0,B_0)$ überall gleichmäßig schnell mit der Lamor Frequenz $\omega_0 = \gamma B_0$ präzediert. $\gamma$ ist eine Atom spezifische Konstante und wird als gyromagnetisches Verhältnis bezeichnet. Für Wasserstoff gilt $\gamma=2.675\cdot10^8s^{-1}T^{-1}$.\newline
Auch für inhomogene Magnetfelder lassen sich die Bloch-Gleichungen nutzen um den Magnetisierungs-Zeit-Verlauf zu beschreiben. Besitzt das Magnetfeld $B_0$ räumliche Schwan\-kun\-gen, so beginnen räumlich getrennte Bereiche der Transversalmagnetisierung schneller oder langsamer zu präzedieren. Durch die unterschiedliche Präzession beginnt der Absolutbetrag der Gesamtmagnetisierung des Volumens $V$ abzunehmen, da die verschiedenen Magnetisierungsanteile dephasieren. Ersetzt man in den Bloch-Gleichungen $m_x$ und $m_y$ durch die komplexe Magnetisierungsdichte $m_T=m_x + \mbox{i} m_y$ so folgt für den transversalen Anteil der Gesamtmagnetisierung:
	\begin{align}
		M_T(t)=\int_{V}m_T(\vec{r},t)\mbox{d}^3r.
	\end{align}
Der zeitliche Verlauf der transversalen Magnetisierung wird auch als FID (Free Induction Decay) bezeichnet. Abb. \ref{fig:transversal-dephasing} veranschaulicht diesen Prozess in einem mit $\omega_0$ rotierenden Bezugssystem. In diesem "spürt" die Magnetisierung dann nur noch die auf dem Hintergrundfeld aufliegenden Inhomogenitäten. Man spricht von einem sog. "Offresonanz"-Feld $\delta\omega(\vec{r})$.\newline
Natürlich haben zusätzlich zum beschriebenen Dephasierungsprozess durch Feldinhomogenitäten auch Spin-Spin Wechselwirkung und andere Effekte Einfluss auf $M_T(t)$ (siehe z.B. \cite{Haacke} Kap. 4). Auf diese soll hier jedoch nicht weiter eingegangen werden.\newline
H. C. Torrey erweiterte zehn Jahre später die von Bloch $1946$ publizierten Gleichungen um den Diffusionseffekt \cite{Torrey56} zur sog. Bloch-Torrey-Gleichung
	\begin{align}
		\frac{\partial m_T(\vec{r}, t)}{\partial t} = \left[D \Delta +i \omega(\vec{r}) \right] m_T(\vec{r}, t).
		\label{eq:bloch-torrey}
	\end{align}
Auch hier wurde wieder die komplexe Schreibweise der Transversalmagnetisierung gewählt. Aufgrund der Rotationsinvarianz des Laplace-Operators in Gl. (\ref{eq:bloch-torrey}) lässt sich auch bei Berücksichtigung der Diffusion noch in das mit $\omega_0$ rotierende Koordinatensystem wechseln. Im Folgenden bezeichnet $\omega$ immer die Offresonanzen im rotierenden Koordinatensystem. Für eine genauere Herleitung der Bloch- bzw. Bloch-Torrey-Gleichungen sei auf die entsprechenden Facharktikel bzw. Literatur von Haacke \cite{Haacke} oder de Graaf \cite{deGraaf} verwiesen.
	
\section{Relaxationsraten und Diffusion}
In einem vollkommen homogenen Magnetfeld lässt sich der Zerfall der Transversalmagnetisierung nach den Bloch-Gleichungen als monoexponentiell beschreiben
	\begin{align}
		M_T(t)=M_0 \exp\left(-\frac{t}{T_2}\right).
	\end{align}
Durch Offresonanzen wird dieser Zerfall beschleunigt. Analog zu \cite{Haacke} (Kap. 4.3) folgt für den Zeitverlauf von $M_T$
	\begin{align}
		M_T(t)=M_0 \exp\left(-\frac{t}{T_2^*}\right) \ \ \mbox{mit} \ \ T_2^*=\left(\frac{1}{T_2}+\frac{1}{T_2'}\right)^{-1}\Rightarrow R_2^*=R_2+R_2'.
	\end{align}	
$T_2'$ charakterisiert die durch makroskopische Offresonanzen verursachte Beschleunigung des Zerfalls. Durch ein Spin-Echo-Experiment lässt sich dieser Zerfall rückgängig machen. $T_2'$ stellt also den reversiblen Anteil der durch die Offresonanzen verursachten Relaxation dar. Für einen monoexponentiellen Verlauf gilt außerdem die Mean-Relaxation-Time-Approximation $T_2^*\approx\int_0^\infty\mathsf{d}t M_T(t)$ exakt.\newline
Da ein Spin-Echo bei diffundierender Magnetisierung jedoch die Störung durch das Off\-resonanz\-feld nicht vollständig kompensieren kann, muss man die Zusammensetzung der Relaxationsraten noch genauer differenzieren:
	\begin{align}
		R_2^*&=R_2+R_2'\notag\\
				 &=R_2^D+R_2^{i}+R_2'
	\end{align}
\begin{itemize}
	\item $R_2^*$ ist die Abklingkonstante (Messgröße) des FID und beinhaltet alle Relaxationseffekte. Bei makroskopischen Offresonanzen hat $R_2^*$ für $D=0$ ein Maximum, da die Diffusion den Effekt der Offresonanzen durch Mittlung über die erfahrenen Felder abschwächt.
	\item $R_2$ ist die Abklingkonstante (Messgröße) in Spin-Echo-Experimenten. Für Echoabstände $T_E\rightarrow0$ gibt $R_2$ das rein intrinsische Abklingverhalten $R_2^i$ wieder, ist also unabhängig von makroskopischen Offresonanzen.
	\item $R_2'$ ist der reversible Anteil der Dephasierung durch Offresonanzen. $R_2'$ ist nicht direkt messbar, sondern bestimmt sich aus der Differenz   $R_2^*-R_2$. Für $D\rightarrow0$ hat $R_2'$ sein Maximum, der Echo-Puls erzeugt eine vollständige Refokussierung der Magnetisierung.
	\item	$R_2^D$ ist der irreversible Anteil der Dephasierung durch Offresonanzen. Unter Berücksichtigung der Diffusion kann auch ein SE die Magnetisierung nicht vollständig refokussieren. $R_2^D$ ist daher in $R_2$ enthalten. Für $D\rightarrow0$ oder $T_E\rightarrow0$ verschwindet der irreversible Anteil $R_2^D$.
	\item	$R_2^{i}$ ist der intrinsische Anteil der Relaxation. Er wird verursacht durch die mikro\-skopischen Wechselwirkungen der Spins und ist unabhängig von makroskopischen Off\-resonanzen. Ein SE mit üblichen $T_E$ hat auf diesen Anteil praktisch keinen Einfluss.
\end{itemize}
Während die letzten zwei Anteile messtechnisch nur schwer erfassbar sind, berücksichtigen die durchgeführten Simulationen überhaupt keine intrinsische Relaxation ($R_2^{i}=0$). Aus der Simulation erhaltene Relaxationsraten des FID
	\begin{align}
		R_{2,sim}^*=R_{2,sim}^D+R_{2,sim}'
	\end{align}
enthalten also nur die Dephasierung durch Offresonanzen. Eine realistische Abschätzung
	\begin{align}
		R_{2,real}^*=R_{2,sim}^*+R_2^{i}\approx R_{2,sim}^*+R_{2,real}
		\label{eq:r2s-relax-rate-approximation}
	\end{align}
aus Simulationsergebnissen ist daher nur über $R_2$ möglich. $R_{2,real}^*$ beinhaltet dann allerdings den $R_2^D$ Anteil doppelt, einmal aus der Simulation ($R_{2,sim}^D$), und einmal aus einer tatsächlich gemessenen Relaxationsrate $R_{2,real}$, und liefert daher eine systematisch zu schnelle Relaxation. Wie oben erwähnt kann jedoch in Messungen der Anteil $R_2^D$ in $R_2$ für sehr kurze Inter-Echo-Abstände klein gehalten werden, und die Näherung wird gut erfüllt. Da außerdem in der Simulation die Diffusion und die Echozeiten beliebig eingestellt werden können und keine intrinsische Relaxation stattfindet, kann $R_{2,sim}^D$ bzw. $R_{2,sim}'$ auch direkt berechnet werden. Ein gemessener Wert $R_{2,real}$ könnte so entsprechend korrigiert werden.

\section{Frequenzspektrum des FID}\label{kap:offresonance-distribution}
In den folgenden Kapiteln wird an Stelle des FID oft das normierte Frequenzspektrum des FID zur Darstellung von Simulationsergebnissen herangezogen.
	\begin{align}
		\rho(\omega)=\mathcal{F}\left[M_T(t)\right]
		\label{eq:fid-fourier}
	\end{align}
Wobei die kontinuierliche Fourier-Transformation $\mathcal{F}$ für die Simulationsdaten stets durch die diskrete Fast-Fourier-Transformation ersetzt wurde. Der Grund für den Wechsel in den Fourierraum liegt in der Tatsache, dass für das Static-Dephasing ($D=0$) die Frequenz\-verteilung des Offresonanzfeldes (bis auf eine Normierung) identisch mit der des FIDs ist \cite{Ziener07}. Durch den Wechsel in den Frequenzraum ist also immer eine Vergleichsbasis für Simulationen mit $D>0$ und $D=0$ vorhanden. Die Frequenzverteilung des Offresonanzfeldes ist definiert als
	\begin{align}
		\rho(\omega)=\frac{1}{V}\int_V\mathsf{d}^3r \delta\left(\omega-\omega(\vec{r})\right).
		\label{eq:offresonance-distribution}
	\end{align}
$\delta(\omega)$ bezeichnet die Dirac-Delta-Distribution. Wichtig dabei ist die genaue Definition des Integrationsvolumens $V$. In dieser Arbeit bezeichnet $V$ das Dephasierungsvolumen ohne das die Offresonanzen verursachende Objekt selbst, da dies auch das Verhalten der durchgeführten Simulationen wiedergibt. Möchte man auch den Signalbeitrag der Magnetisierung im Inneren des Objekts berücksichtigen, so muss man dort das Offresonanzfeld gesondert (meist konstant) definieren. Gl. (\ref{eq:offresonance-distribution}) lässt sich auch leicht auf die nach Kap. \ref{kap:feldberechnung} berechneten diskreten Interpolationsgitter anwenden.\newline
Eine lorentz-förmige Frequenzverteilung der Offresonanzen liefert ein besonderes Verhalten.
	\begin{align}
		\rho_{\mbox{\tiny Lorentz}}(\omega)=A \frac{\gamma }{\left(\omega-\mu\right)^2+\gamma ^2}
		\label{eq:lorentz-profile}
	\end{align}
Wenn der Mittelwert $\mu$ der Verteilung verschwindet, ergibt sich nach Gl. (\ref{eq:fid-fourier}) ein exakt monoexponentieller, um $t=0$ symmetrischer Zerfall des FID
	\begin{align}
		\mathcal{F}^{-1}\left[\rho_{\mbox{\tiny Lorentz}}(\omega)\right]=A' \exp\left(-\gamma|t|\right) = A' \exp\left(-R_2'|t|\right)\ \ \Rightarrow\ \ \gamma=R_2'.
		\label{eq:lorentz-fourier}
	\end{align}	
Die Halbwertsbreite des Peaks im Frequenzraum $\gamma$ ist identisch mit der Relaxationsrate $R_2'$ des Zerfalls. Wird Gl. (\ref{eq:fid-fourier}) auf ein nicht um $t=0$ symmetrisches Signal angewendet, so ergeben sich komplexe Amplituden im Frequenzraum. Eine Ergänzung des Signals für $t<0$ durch Spiegelung ($M_T(-t)=M_T(t)$) verhindert dies. Für eine kurze Auflistung weiterer Eigenschaften der Fourier-Transformation siehe z.B. \cite{HinrichsenCP}.\newline
Problematisch gestaltet sich eine Beschreibung der Transversalmagnetisierung für nicht lorentz-förmige Offresonanzverteilungen: Da hier der zusätzliche Relaxationseffekt nicht mehr zwangsläufig monoexponentiell verläuft, kann das Verhalten nicht auf einen einzigen Parameter $R_2'$ herunter gebrochen werden. Analog zur medizinischen Praxis, kann man den Signalverlauf trotzdem noch monoexponentiell anfitten um eine Vergleichsbasis zu schaffen. Man muss jedoch berücksichtigen, dass dadurch Informationen über einen eventuell für die Frequenzverteilung charakteristischen FID verloren gehen können.

\section{Kapillarfelder und Geometrien}\label{kap:theory-krogh-model}
	\begin{figure}
		\begin{center}\includegraphics[width=0.75\textwidth]{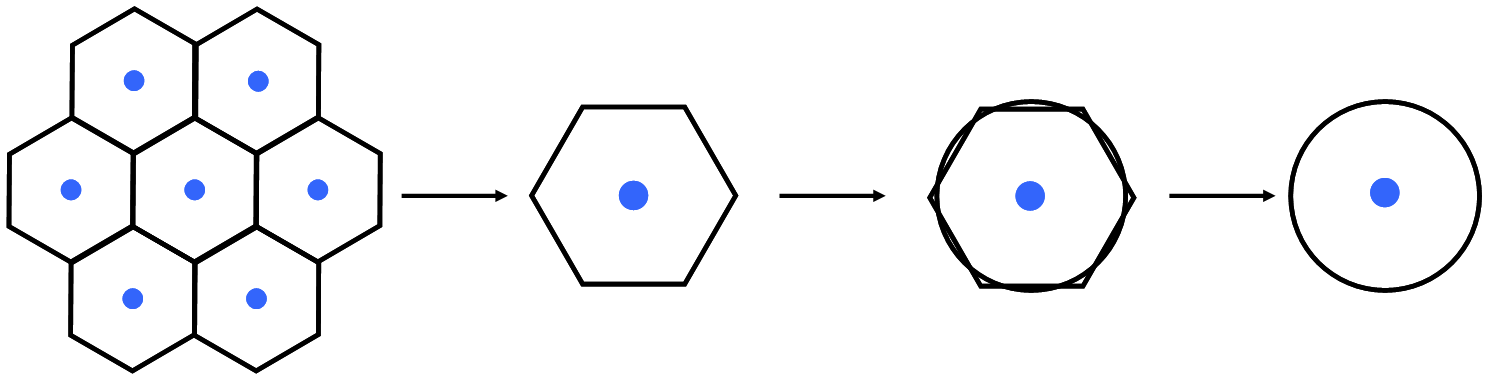}\end{center}
		\caption{Nach \cite{ZienerPHDThesis}: Im Krogh-Modell ersetzt der einfacher zu beschreibende Zylinder das komplexe Kapillargitter.}
		\label{fig:hex-krogh-transition}
	\end{figure}
\begin{figure}
		\begin{center}\includegraphics[width=0.45\textwidth]{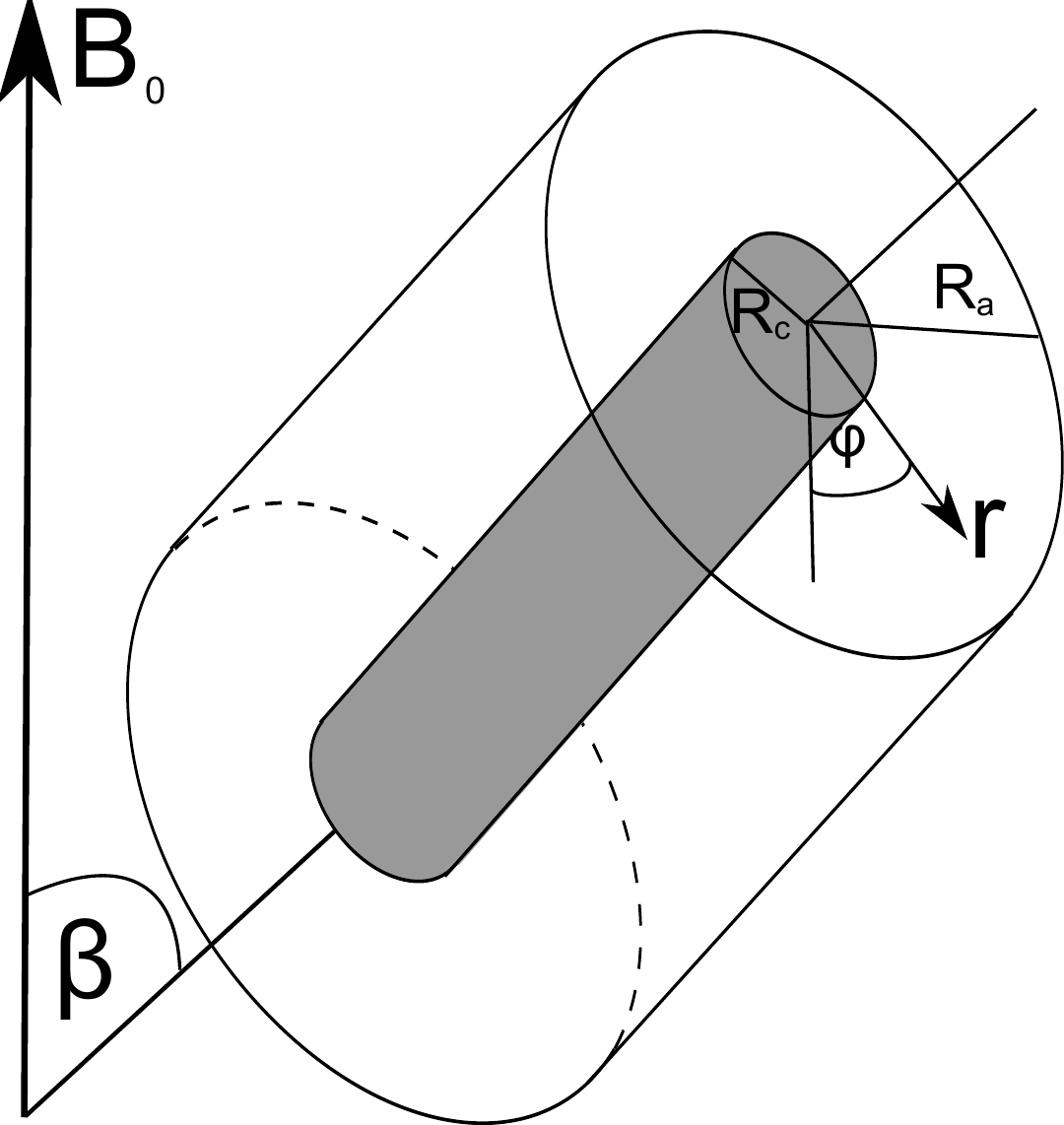}\end{center}
		\caption{Die Orientierung der Kapillare mit Radius $R_c$ und dem sie umgebenden zylinderförmigen Versorgungsgebiet (Radius $R_a$) gegenüber dem Magnetfeld wird durch den Winkel $\beta$ charakterisiert. Wird das Versorgungsgebiet nicht als Zylinder angenommen bzw. betrachtet man mehrere parallele Kapillaren, so liegt in der Schnittebene keine Rotationsinvarianz mehr vor, und der Winkel $\alpha$ aus Gl. (\ref{eq:single-capillary-field}) muss ebenfalls berücksichtigt werden.}
		\label{fig:coordinate-system}
	\end{figure}

\begin{figure}
		\begin{center}
		\subfloat[Offresonanzfeld im Krogh-Modell]{
			\includegraphics[width=0.35\textwidth]{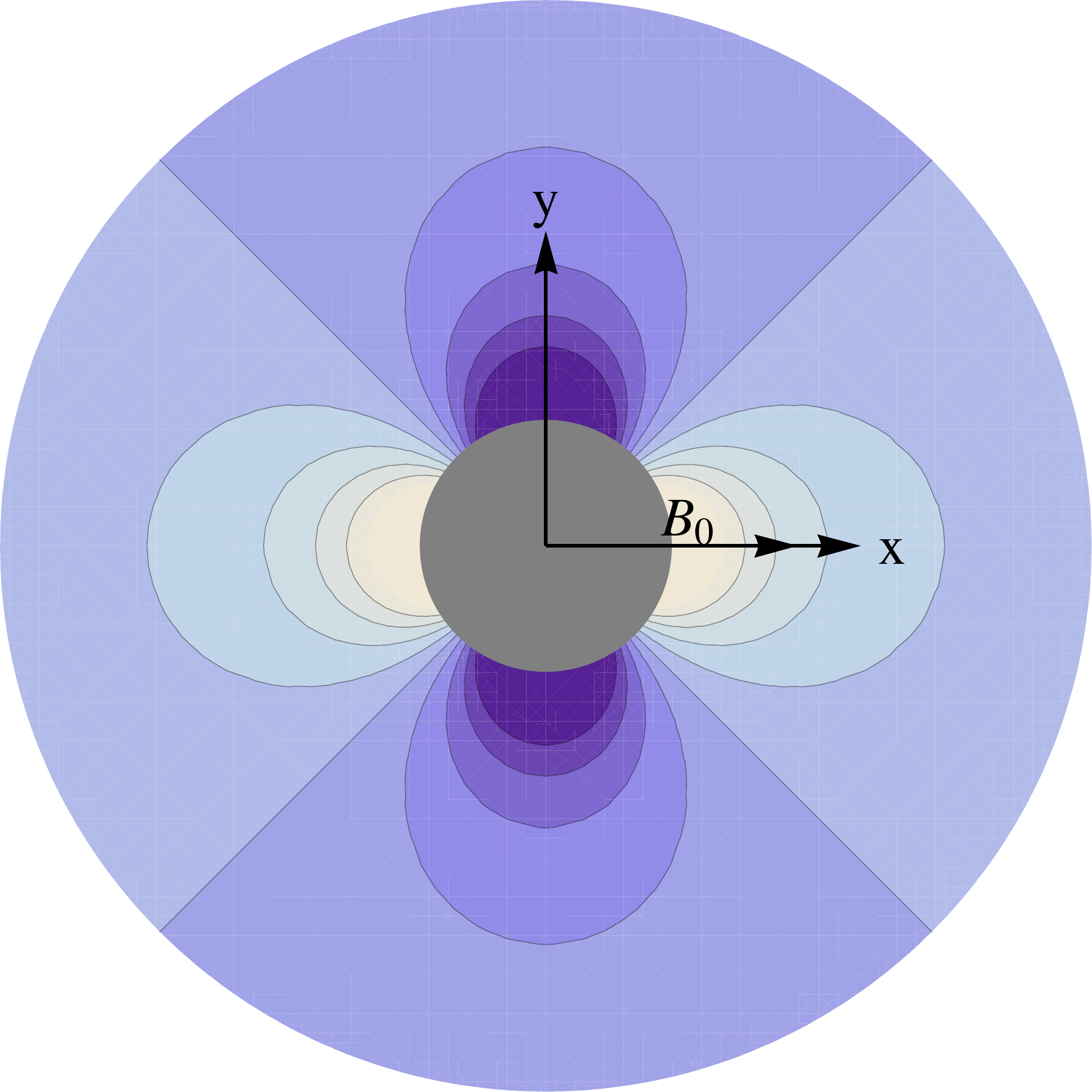}\label{fig:krogh-model-contour}
		}
		\subfloat[Frequenzverteilung im Krogh-Modell]{
			\includegraphics[width=0.53\textwidth]{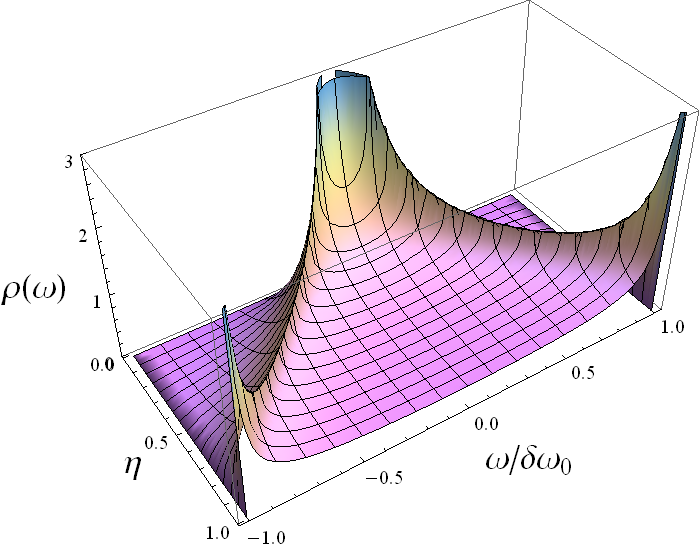}\label{fig:krogh-model-frequency-distribution}
		}
		\end{center}
		\caption{Für ein um die Kapillare konzentrisches Versorgungsgebiet kann wegen der Rotationsinvarianz das Koordinatensystem immer so gelegt werden, dass gilt $\alpha=0$.}
		\label{fig:krogh-model}
	\end{figure}

	
Das zweidimensionale Offresonanz-Feld eines einzelnen zylinderförmigen para\-magnetischen Objekts ergibt sich nach \cite{Reichenbach01} oder \cite{Haacke} (Kap. 25) zu
\begin{align}
	\begin{array}{rcl}
	\displaystyle	\omega\left(\vec{r}\right)&\displaystyle=&\displaystyle\delta\omega_0 R_c^2 \frac{\cos \left(2\varphi-2\alpha\right)}{\left|\vec{r}\right|^2}\\
	\displaystyle							&\displaystyle=&\displaystyle\delta\omega_0 R_c^2 \frac{(x-y)(x+y)\cos\left(2\alpha\right)-2 x y\sin\left(2\alpha\right)}{\left(x^2+y^2\right)^2}\\
			\mbox{mit}& & \delta\omega_0 = \frac{1}{2} \Delta \chi \gamma B_0\sin^2(\beta).
	\end{array}	\label{eq:single-capillary-field}
\end{align}
Dabei bezeichnet $R_c$ den Radius des Objekts, $\beta$ den Winkel zwischen $\vec{B_0}$ und der Kreisebene (bzw. einem senkrechten Schnitt durch den Zylinder, siehe Abb. \ref{fig:coordinate-system}) und $\alpha$ die Orientierung des Magnetfeldes in der Kreisebene (Abb. \ref{fig:hex-contour-10deg} und Abb. \ref{fig:square-contour-30deg}). $\Delta\chi$ gibt an wie stark sich die magnetische Suszeptibilität des Kapillarinneren von der des umgebenden Materials unterscheidet.\newline
In der Medizin wurde als Modell die konzentrische Anordnung von Kapillare und Versorgungszylinder aus Abb. \ref{fig:coordinate-system} und \ref{fig:krogh-model-contour} bereits 1919 durch Krogh eingeführt um die Versorgung des Gewebes mit Sauerstoff und Nährstoffen zu beschreiben. Aufgrund seiner hohen Symmetrie und einfachen Beschreibbarkeit durch die wenigen Parameter $R_c$, $R_a$ und $\delta\omega_0$ erfreut sich das Krogh-Modell nach wie vor großer Beliebtheit.\newline
Abb. \ref{fig:hex-krogh-transition} gibt die ursprüngliche Motivation zum Übergang in das Krogh-Modell wieder: Aufgrund der mehr oder weniger regelmäßigen parallelen Anordnung der Kapillaren im Gewebe wird angenommen, dass die Prozesse im Gewebe primär durch die am nächsten liegende Kapillare bestimmt sind. Der diese Kapillare umgebende Zylinder bestimmt das mittlere Volumen, welches von einer Kapillare versorgt wird. Die Problematik wird aber offen\-sichtlich, wenn man das Offresonanzfeld und dessen Frequenzverteilung einer einzelnen Kapillare (Abb. \ref{fig:krogh-model-contour}) mit der eines regelmäßigen Kapillargitters (Abb. \ref{fig:hex-contours}) vergleicht. Aus diesem Grund werden in den folgenden Kapiteln die Unterschiede im Relaxationsverhalten zwischen dem Krogh-Modell und der regelmäßigen Gitteranordnung bzw. eines dem realen Gewebe noch weiter angepassten Modell vorgestellt. Eine Liste der dabei meist verwendeten geometrischen Parameter findet sich in Tab. \ref{tab:model-parameters}.\newline
Für das Krogh-Modell kann die Frequenzverteilung der Offresonanzen analytisch berechnet werden. Der Verlauf von $\rho(\omega,\eta)$ nach Gl. (\ref{eq:krogh-offres-distrib}) (siehe auch \cite{Ziener07}) ist in Abb. \ref{fig:krogh-model-frequency-distribution} gezeigt.\newline
	\begin{align}
		\rho(\omega,\eta)= \begin{cases}
			\displaystyle\frac{1+\eta}{2\eta\dom_0\pi}	&\omega=0\\
			\displaystyle\frac{\eta}{1-\eta}\frac{\dom_0}{\pi\omega^2}\sqrt{1-\left(\frac{\omega}{\dom_0}\right)^2}& 		\eta\dom_0\leq|\omega|\leq\dom_0\\
\displaystyle\frac{\eta}{1-\eta}\frac{\dom_0}{\pi\omega^2}\left(\sqrt{1-\left(\frac{\omega}{\dom_0}\right)^2}-\sqrt{1-\left(\frac{\omega}{\eta\dom_0}\right)^2}\right) & |\omega|<\eta\dom_0\\
 0 & \mathsf{sonst}\\ 
\end{cases}
		\label{eq:krogh-offres-distrib}
	\end{align}

	\begin{table}[H]
		\renewcommand\arraystretch{1.7}
		\begin{center}
		\begin{tabular}[ht]{p{0.4\textwidth} >{$\displaystyle}p{.2\textwidth}<{$} >{$\displaystyle}p{.2\textwidth}<{$}} 
		   																		& \eta															& \mbox{CD}							\\
		  \hline
		  \hline 
		  Krogh-Modell ($R_a=\ICD/2$)															& \frac{R_c^2\pi}		{R_a^2\pi}				& \frac{1}{R_a^2\pi}			\\
		 	Hexagonales Gitter																			& \frac{2R_c^2\pi}	{\sqrt{3} \ICD^2}	& \frac{2}{\sqrt{3}\ICD^2}	\\
		  Quadratisches Gitter																		&	\frac{2R_c^2\pi}	{\ICD^2}					& \frac{1}{b^2}					\\
		  Plasma (Kap. \ref{kap:2D1CP-Theory}, $N\in\mathbb{N}$)	&	\frac{NR_c^2\pi}	{A}								&	\frac{N}{A}\\
		\end{tabular}
		\end{center}
		\caption{Für die verschiedenen Geometrien berechnen sich auch die charakteristischen Eigenschaften $\ICD$ (Intercapillary Distance), $\mbox{CD}$  (Capillary Density) und $\eta$ (auch $\mbox{RBV}$ für Regional Blood Volume) leicht unterschiedlich. In den Plasma-Konfigurationen ist $N$ die Anzahl der in der Simulationsbox mit Fläche $A$ enthaltenen Kapillaren.}
		\label{tab:model-parameters}
	\end{table}
	
\section{Offresonanzstärke und Kontrastmittel}\label{kap:offres-strength}
Möchte man aus dem Relaxationsverhalten der Transversalmagnetisierung Rückschlüsse auf die Gewebeanordnung ziehen, so ist eine genaue Kenntnis der Stärke der Offresonanzen $\dom_0$ nötig. $\Delta\chi$ aus Gl. (\ref{eq:single-capillary-field}) hängt für eine mit Blut gefüllte Kapillare von verschiedenen physiologischen Parametern ab \cite{Reichenbach01}:
	\begin{align}
		\Delta\chi=4 \pi \Delta\chi_{\mbox{\scriptsize do}} (1-Y) \mbox{Hct} .
		\label{eq:delta-chi}
	\end{align}
$Y$ bezeichnet den Oxygenierungsgrad ($0\!<\!Y\!<\!1$) des Blutes, der Hämatocrit-Wert $\mbox{Hct}$ für Kapillarblut des Menschen liegt nach \cite{Reichenbach01}  bei ca. $0.45\pm0.02$. $\Delta\chi_{\mbox{\scriptsize do}}$ bezeichnet den Suszeptibilitätsunterschied zwischen völlig oxygeniertem ($Y\!=\!1$) dia\-mag\-ne\-ti\-schen bzw. desoxygeniertem ($Y\!=\!0$) paramagnetischen Hämoglobin. Nach \cite{Weisskoff92} gilt $\Delta\chi_{\mbox{\scriptsize do}}=(0.183\pm0.014)\ppm$. Bei der Betrachtung des Oxygenierungsgrads tritt nun folgendes Problem auf: Während des Durchfließens der Kapillare gibt das Blut einen Großteil seines Sauerstoffs an das umgebende Gewebe ab. Für das zweidimensionale Modell hängt also $Y$ stark davon ab, ob der entsprechende Querschnitt am arteriellen oder venösen Ende einer Kapillare liegt. Messungen für den Oxygenierungsgrad für venöses Blut im Gehirn aus \cite{Sedlacik07} ergeben $Y=0.53\pm0.03$. Unter den Annahmen, dies gelte auch für venöse Blutgefäße am Herzen und beim Eintritt in die Kapillare sei das Blut vollständig oxygeniert ($Y\approx1$), so folgt für den Oxygenierungsgrad in Kapillaren $0.5\widetilde{<}Y\widetilde{<}1$ bzw. $Y\approx0.75\pm0.25$. Die Abhängigkeit der Offresonanzen vom Oxygenierungsgrad wird auch in der Auswertung (Kap. \ref{kap:2D-to-3D}) berücksichtigt. Außerdem hängen Oxygenierungsgrad und Hämatocrit-Wert von der momentanen Kreislaufbelastung, dem Versuchskandidaten (z.B. Mensch/Tier oder männlich/weiblich) und dem untersuchten Gewebetyp ab. Die Werte für $Y$ und $\mbox{Hct}$ stellen also nur bessere Schätz\-werte dar, und müssen bei echten Messungen individuell eingegrenzt werden.\newline
Aus der Orientierung des Herzens bzgl. des Magnetfeldes und der Position eines Voxels lässt sich auch der Winkel $\beta$ aus Gl. (\ref{eq:single-capillary-field}) abschätzen. Angenommen, für bestimmte Regionen des Herzens gilt $\beta=(90\pm20)^\circ$, dann ergibt sich nach Gl. (\ref{eq:single-capillary-field}) ein weiterer relativer Fehler in $\dom_0$ von ca. $12\%$.\newline
Berücksichtigt man alle dieser Fehlerquellen, führt dies zu typischen Werten von $\Delta\chi\approx(0.8\pm0.4)\ppm$ und Offresonanzen von $\dom_0\approx(160\pm90)\radps$ bei $1.5\mbox{T}$ bzw. $\dom_0\approx(730\pm420)\radps$ bei $7\mbox{T}$. In Tabelle \ref{tab:offres-influence} sind die in $\dom_0$ eingehenden Faktoren aufgelistet.\newline
Unter Anwendung von Kontrastmittel lässt sich $\Delta\chi$ nach Tabelle 2 in \cite{Kennan94} noch erhöhen. Eine Konzentration von $1\mathsf{mM}$ des Kontrastmittels Gd-DTPA im Blut führt zu einer Erhöhung von $\Delta\chi$ und damit auch von $\dom_0$ um ca $40\%$. Mit superparamagnetischen Eisenpartikeln lässt sich bei gleicher Konzentration ein Faktor von ca. $1500\%$ erreichen. Für die superparamagnetischen Partikel gilt dies allerdings nur bis zu einer gewissen Sättigungsfeldstärke.\newline
Durch die Verwendung intravasaler Kontrastmittel ist ein zusätzlicher Einfluss der Kontrastmittel auf die intrinsische $T_2$-Relaxation auf das Kapillarinnere beschränkt. Der Hauptteil der signalgebenden Protonen aus dem umgebenden Gewebe und damit auch die gemittelte $T_2$-Relaxation bleibt weitgehend unbeeinflusst.
	\begin{table}[H]
		\renewcommand\arraystretch{1.3}
		\begin{center}
		\begin{tabular}[ht]{l|cc} 
		   																			& Wert																	& rel. Fehler in $\dom_0$		\\
		  \hline
		  \hline 
		  $\Delta\chi_{\mbox{\scriptsize do}}$	&	$(0.183\pm0.014)\ppm$									& $ 8\%$										\\
		  $Y$																		& $0.75\pm0.25$													&	$33\%$										\\
		  Hct																		& $0.45\pm0.02$													&	$ 4\%$										\\
		  $\beta$																& $(90\pm20(10))^\circ$									&	$12\%(3\%)$								\\
		 	$\dom_0$(1.5T)												& $\dom_0\approx(160\pm90(75))\radps$		& $\approx57\%(48\%)$				\\
		 	$\dom_0$(7T)													& $\dom_0\approx(730\pm420(350))\radps$	& $\approx57\%(48\%)$				\\
		\end{tabular}
		\end{center}
		\caption{Die Stärke der Offresonanzen $\dom_0$ hängt nach Gl. (\ref{eq:single-capillary-field}) und Gl. (\ref{eq:delta-chi}) stark von den aufgelisteten Parametern ab. Während Hämatocrit-Wert und $\Delta\chi_{\mbox{\scriptsize do}}$ relativ gut bekannt sind, ergibt sich vor allem durch den praktisch unbekannten Oxygenierungsgrad $Y$ der große Fehler in $\dom_0$. Ist bekannt, ob die Kapillaren im untersuchten Voxel eher ateriell oder venös sind, ließe sich $Y$ eventuell noch weiter eingrenzen, was den relativen Fehler deutlich senken würde. Auch eine bessere Eingrenzung des Oxygenierungsgrades am arteriellen Ende der Kapillaren führt zu einer höheren Genauigkeit der Offresonanzen. Ist der Winkel $\beta$ messtechnisch genauer bestimmbar ($\Delta\beta=10^\circ$), so ergeben sich die in Klammern aufgeführten Werte. Die hier aufgelisteten Werte sind auf den Menschen bezogen. Bei Versuchstieren müssen sie voraussichtlich angepasst werden.}
		\label{tab:offres-influence}
	\end{table}

\renewcommand{\figurename}{Abb.}
\renewcommand{\tablename}{Tab.}

\section{Random Walk, Mastergleichung und Kontinuumslimes}
Bei einem Random Walk sind einzelne Trajektorien per Definition nicht vorhersagbar. Die Wahrscheinlichkeitsdichte $p(r,t)$ ein Teilchen zur Zeit $t$ am Ort $r$ zu finden ist jedoch deterministisch und durch die Eigenschaften des Random Walks bestimmt. Für einen Random Walk auf einem eindimensionalen Gitter lautet die so genannte Mastergleichung (siehe auch \cite{HinrichsenCP} oder Kap. 5.2 in \cite{Kinzel96})
	\begin{align}
		p(r,t+\Delta t) = \frac{1}{2}\left[ p(r-\Delta r,t) + p(r+\Delta r,t) \right].
		\label{eq:master-equation}
	\end{align}	
Dabei wird von einer Wahrscheinlichkeit von $50\%$ für einen Sprung nach links bzw. rechts ausgegangen. Durch eine Subtraktion von $p(r,t)$ auf beiden Seiten von Gl. (\ref{eq:master-equation}) und Divison durch $\Delta t$ und $\Delta r^2$ wird die Äquivalenz zur Diffusionsgleichung deutlich:
	\begin{align}
		\renewcommand\arraystretch{1.8}
		\begin{array}{ccc}
		\displaystyle\frac{p(r,t+\Delta t)-p(r,t)}{\Delta t} & = & \displaystyle D \frac{1}{2}\frac{p(r-\Delta r,t)-2 p(r,t) + p(r+\Delta r,t)}{\Delta r^2}\\
		\displaystyle\frac{\partial p(r,t)}{\partial t} 	   & = & \displaystyle D \nabla^2 p(r,t) 
		\end{array}	
 		\label{eq:master-equation-diffusion}
	\end{align}
Die zweite Gleichung folgt aus der Grenzwertbildung $\Delta t\rightarrow 0$, wobei gleichzeitig gelten muss $\Delta r\propto\sqrt{\Delta t}$. Die Diffusionskonstante $D=\Delta r^2/\Delta t$ ist dann genau die Proportionalitäts\-konstante zwischen $\Delta r$ und $\sqrt{\Delta t}$.\newline 
Prinzipiell ließe sich bei hinreichend hoher räumlicher und zeitlicher Auflösung bereits mit Gl. (\ref{eq:master-equation}) der Diffusionseffekt in der Bloch-Torrey-Gleichung (\ref{eq:bloch-torrey}) nachbilden. Hinsichtlich des benötigten Rechenaufwands ist es jedoch zweckmäßig zu einem räumlich kontinuierlichen Random Walk zu wechseln. Dazu wird im Folgenden kurz erläutert, wie die räumliche Diskretisierung zu Gunsten einer kontinuierlichen Schrittweite verworfen werden kann, und dass diese neue Schrittweitenverteilung auch die Diffusionsgleichung löst.\newline
Betrachtet man die Verteilung $s(S)$ der Summe $S=\sum{X_n}$ von $n$ gleichverteilten und unabhängigen Zufallszahlen $X_n$, so ergibt sich diese nach  \cite{HinrichsenCP} als Faltung der Verteilungen $p_n(X_n)=1$ für $X_n \in \left[-0.5,0.5\right]$
	\begin{align}
		s(S) = \int \ldots \int dX_1 \ldots dX_{n-1} \prod_{i=1}^{n-1}p_i(X_i) p_n(S-\sum_{i=1}^{n-1}X_i) .
 		\label{eq:randomnumber-addition-convolution}
	\end{align}
Die Verteilung $s(S)$ wird dabei mit jeder weiteren Faltung (also jedem weiteren Zeitschritt $\Delta t$) breiter und glatter. Die Mastergleichung führt, abgesehen von der Normierung, zum gleichen Schema wie das Pascallsche Dreieck bzw. des Galtonbretts. Für $n\rightarrow\infty$ schließlich ergibt sich mit Hilfe des Zentralen Grenzwertsatzes und des Faltungsatzes
	\begin{align}
		s(S) = \frac{1}{\sqrt{2\pi}\sigma}\exp{\left(\frac{-S^2}{2\sigma^2}\right)}.
		\label{eq:normal-distribution}
	\end{align}	
Die Standardabweichung $\sigma$ ist dabei abhängig von der Breite der $p_n$. Der Mittelwert $\left\langle S \right\rangle$ verschwindet wegen der Symmetrie der $p_n(X)$.\newline
Für $\sigma=\sqrt{2Dt}$ erfüllt Gl. (\ref{eq:normal-distribution}) die Diffusionsgleichung (\ref{eq:master-equation-diffusion}) für ein bei $r=0$ startendes Teilchen. Mit Hilfe von Gl. (\ref{eq:normal-distribution}) kann also eine sehr feine räumliche und zeitliche Diskretisierung in Gl. (\ref{eq:master-equation}) durch kontinuierliche normalverteilte Schritte mit $\sigma=\sqrt{2D\Delta t'}$ bei deutlich gröberer Zeitschrittweite $\Delta t'$ ersetzt werden.\newline
Allgemein gilt für die Schrittwahl in $d$ Dimensionen wegen der Entkopplung der einzelnen Richtungen
	\begin{align}
		p_{\vec{r}}(\vec{r'}) = p(\vec{r}\rightarrow \vec{r'}) = \frac{1}{\left(\sqrt{2\pi} \sigma\right)^d}\exp\left(\frac{-(\vec{r}'-\vec{r})^2}{2 \sigma^2}\right).
		\label{eq:steplength-distribution}
	\end{align}
Im Mittel wird dabei die Distanz $\left\langle \left|\vec{r'}-\vec{r} \right|\right\rangle = \sqrt{d\sigma^2}$ zurückgelegt.

\section{Zyklische und reflektive Randbedingungen}\label{kap:boundary-conditions}
Gleichung (\ref{eq:normal-distribution}) löst Gl. (\ref{eq:master-equation-diffusion}) in guter Näherung in räumlichen Bereichen mit großem Abstand zu möglichen Randbedingungen. Ist der Abstand vom Rand jedoch klein gegen die mittlere zurückgelegte Distanz, so muss eine Abbildung gefunden werden, welche die Sprünge über den Rand hinaus zurück ins Innere projiziert um die Normierung der Wahrscheinlichkeitsdichte zu bewahren. Prinzipiell kann zwischen zyklischen und reflektiven Randbedingungen ähnlich wie in \cite{Sommerfeld78} (Kap. 15 und 16) unterschieden werden:	
	\begin{align}
		\renewcommand\arraystretch{1.6}
		\begin{array}{lcccr}
		\mbox{zyklisch:}  & 	\displaystyle \left.\frac{\displaystyle \partial^n p}{\displaystyle \partial r^n}\right|_{r'} & = &	\displaystyle\left.\frac{\partial^n p}{\partial r^n}\right|_{r''}  & \forall \ r',r'' \in \ \mbox{Rand}\\
		\mbox{reflektiv:} & 	\displaystyle \left.\frac{\displaystyle \partial^n p}{\displaystyle \partial r^n}\right|_{r'}     & = & \displaystyle 0 & \forall \ r' \in \ \mbox{Rand}\\
		\end{array}
		\label{eq:boundary-conditions}
	\end{align}
Für die zyklischen Randbedingungen ist $r''$ der zu $r'$ gehörende gegenüberliegende Randpunkt und umgekehrt. Im Zweidimensionalen gelten dabei die entsprechenden Richtungsableitungen senkrecht auf den Rand. Die zyklischen Randbedingungen bieten meist den Vorteil, dass sogenannte Finite-Size-Effekte unterdrückt werden. Da die Ränder keine räumliche Sonderstellung einnehmen, kommt das phy\-si\-ka\-li\-sche Verhalten dem eines unendlich ausgedehnten Gebietes sehr nahe. Die zyklischen Randbedingungen stellen damit einen starken Gegensatz zum Krogh-Modell dar. Da in den durchgeführten Simulationen sowohl reflektive (Kollision mit Kapillare, Kollision mit Rand beim Krogh-Modell) als auch zyklische (äußerer Rand bei rechteckiger Simulationsbox) Randbedingungen verwendet wurden, soll im Folgenden kurz auf sie eingegangen werden. 

\section{Gerade Ränder}\label{kap:diffusion-gerade-raender}
\begin{figure}
		\begin{center}
		\subfloat[Reflektiv. Das System ist abgeschlossen]{
			\includegraphics[width=0.75\textwidth]{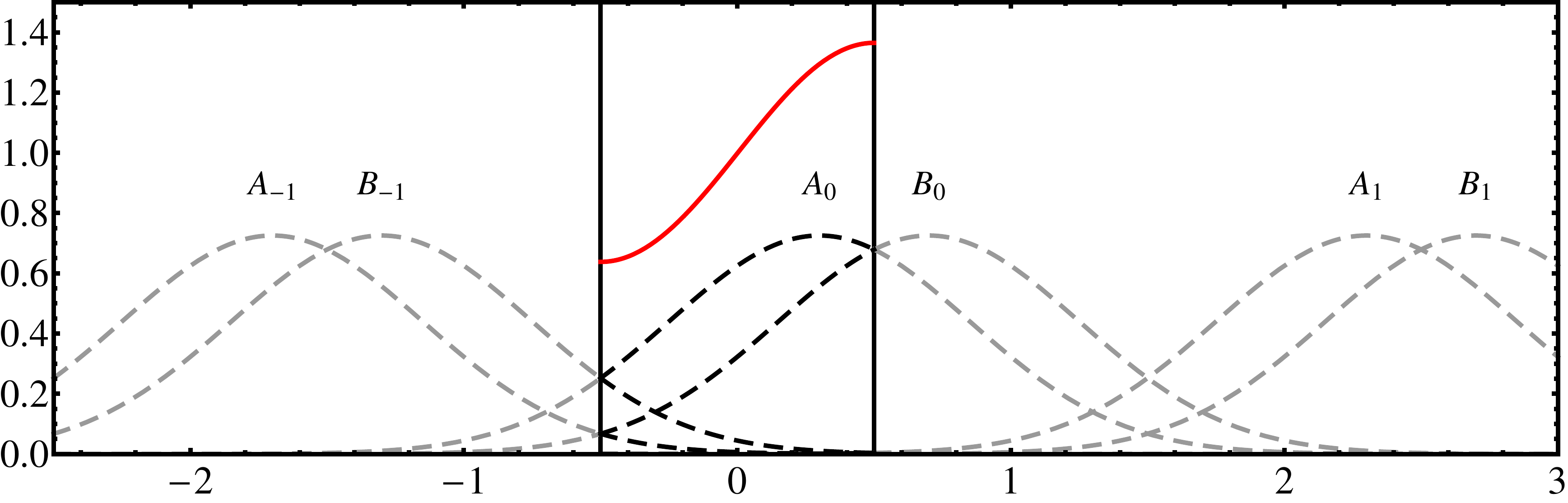}\label{fig:mirrored-density-1D-reflektive}
		}\\
		\subfloat[Zyklisch. Da die Ränder keine räumliche Sonderstellung einnehmen, kommt das physikalische Verhalten dem eines unendlich ausgedehnten Gebietes sehr nahe.]{
			\includegraphics[width=0.75\textwidth]{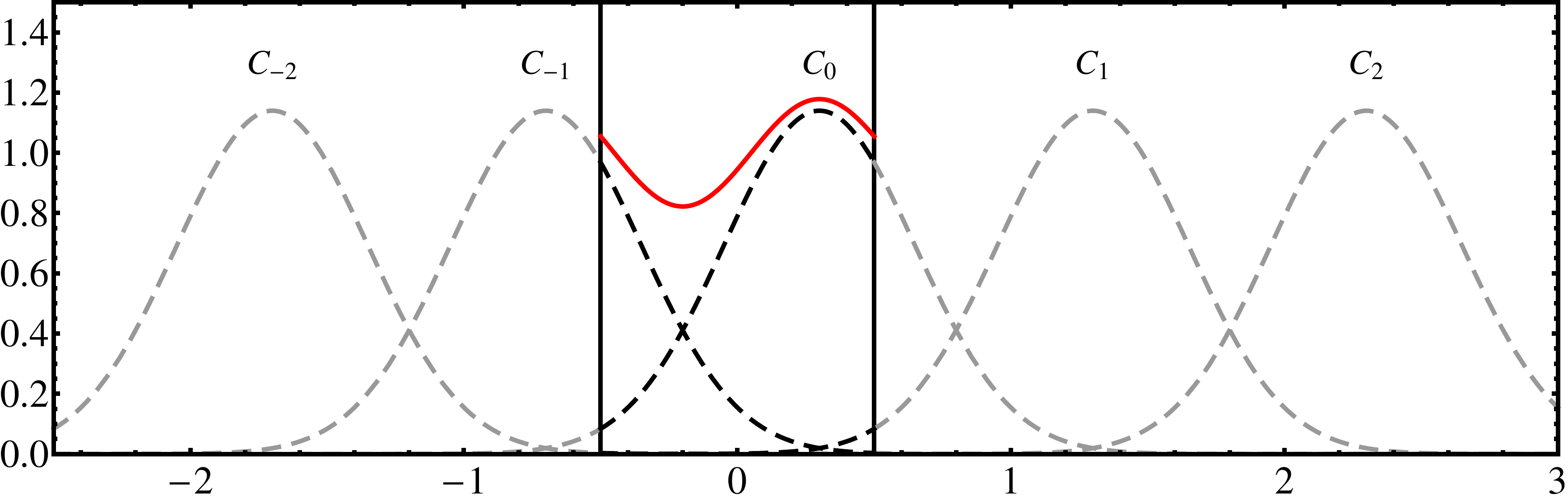}\label{fig:mirrored-density-1D-cyclic}
		}
		\end{center}
		\caption{Bei reflektiven und periodischen Randbedingungen müssen unterschiedliche Periodizitäten für die Spiegelverteilungen realisiert werden. Die $A_n$, $B_n$ und $C_n$ sind in Gl. (\ref{eq:mirrored-density-closed}) definiert.}
		\label{fig:mirrored-density-1D}
	\end{figure}

\begin{figure}
		\begin{center}
		\subfloat[]{
			\includegraphics[width=0.69\textwidth]{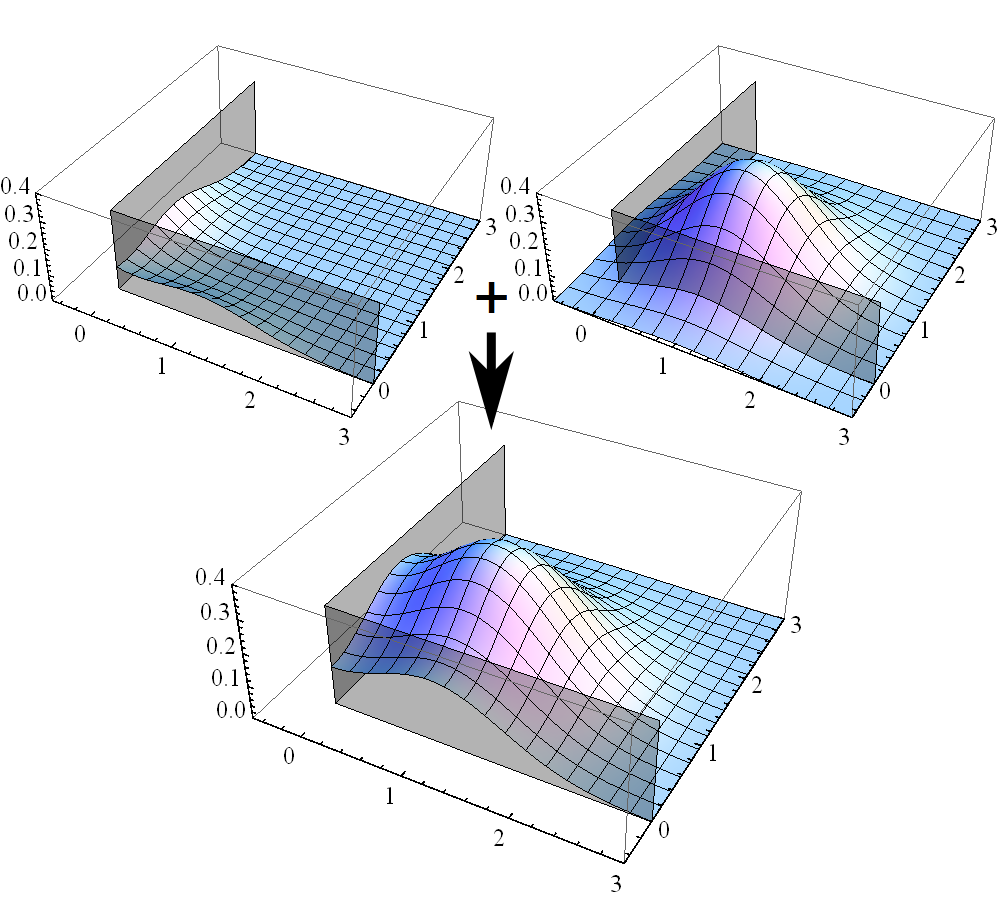}
		}\\
		\subfloat[]{
			\includegraphics[width=0.44\textwidth]{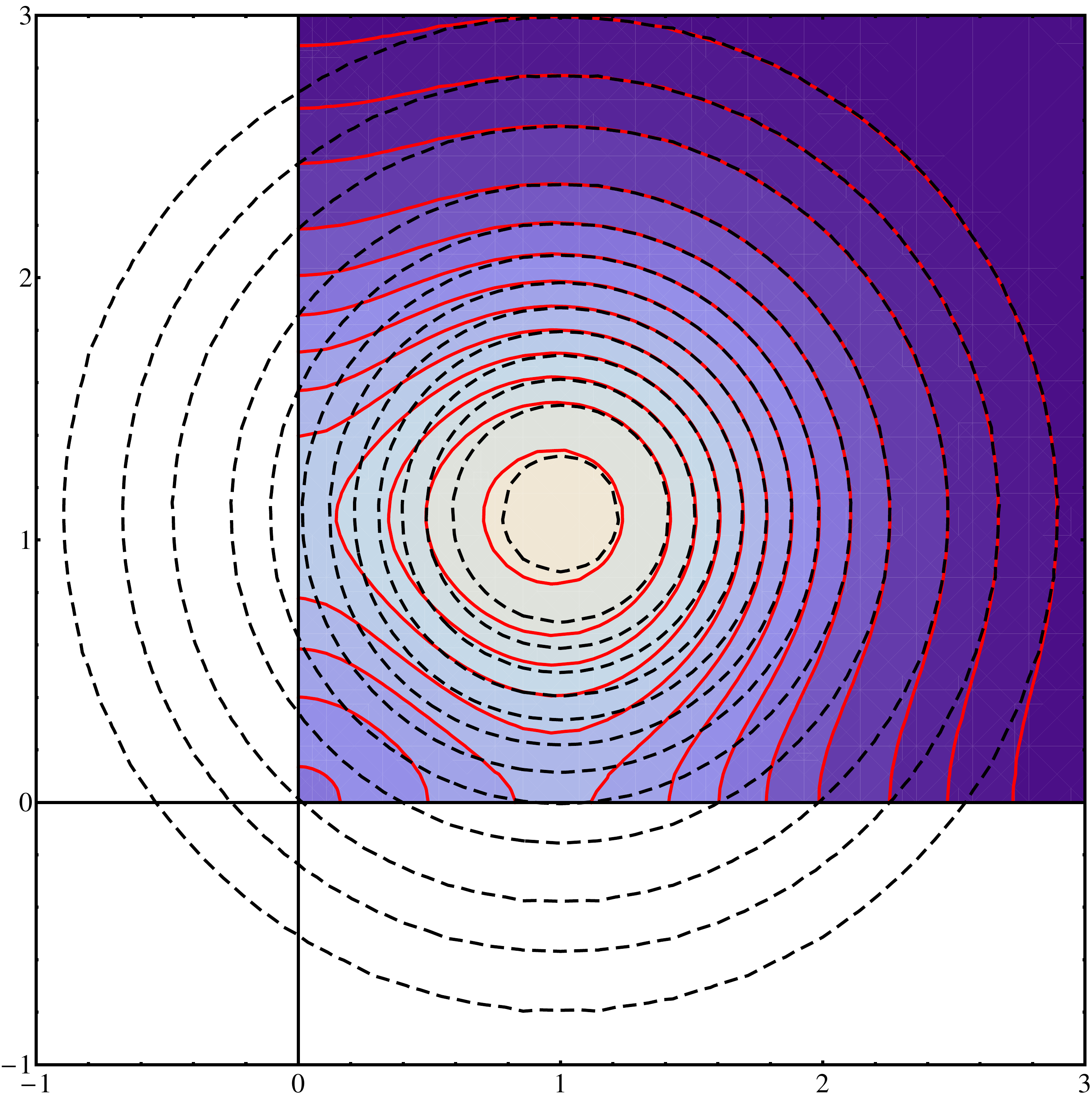}
		}
		\end{center}
		\caption{Im zweidimensionalen Fall muss für ein abgeschlossenes System ein regelmäßiges Gitter an Spiegelverteilungen berücksichtigt werden. Für einen "halboffenen" (d.h keine weitere Begrenzung nach oben oder rechts) oder sehr weit entfernten Rand können die entsprechenden Summen jedoch nach dem ersten Glied abgebrochen werden. Die Wahrscheinlichkeitsverteilung ergibt sich dann aus der Superposition der Gaußglocke im Inneren der Box mit den drei Spiegelverteilungen hinter dem unteren Rand, dem rechten Rand und "hinter" der Ecke. Im Kontur-Plot kann man die Gültigkeit von Gl. (\ref{eq:boundary-conditions}) prüfen: Die Äquipotentiallinien laufen alle senkrecht in die Ränder.}
		\label{fig:mirrored-density-2D-corner}
	\end{figure}

Bei geraden Rändern kann die Wahrscheinlichkeitsdichte mit Hilfe entsprechend positionierter Spiegelverteilungen hinter dem Rand dargestellt werden. Die Periodizität der Spiegelverteilungen ist davon abhängig ob, reflektive oder zyklische Randbedingungen vorliegen. Abb. \ref{fig:mirrored-density-1D-reflektive} und Abb. \ref{fig:mirrored-density-1D-cyclic} zeigen diese Periodizität für ein Intervall von $[-L/2,L/2]$.
Da es sich in beiden Fällen nur um Superpositionen von Gl. (\ref{eq:normal-distribution}) handelt, ist auch die Differentialgleichung (\ref{eq:master-equation-diffusion}) weiter erfüllt. Dass außerdem die Randbedingungen gelten folgt aus den Ableitungen von
	\begin{align}
		\begin{array}{lcccc}
			p_{\mbox{\small refl.}}(r') &=& \displaystyle\sum_{n=-\infty} ^{\infty} p_{A_n}(r') &+& \displaystyle\sum_{n=-\infty}^{\infty}p_{B_n}(r')\\
			p_{\mbox{\small cycl.}}(r') &=& \displaystyle\sum_{n=-\infty} ^{\infty} p_{C_n}(r') & & \\
		\end{array}	\label{eq:mirrored-density-closed}		\\
		\mbox{mit:}\ A_n = 2nL + r_0\ \mbox{,}\ B_n = (2n+1)L - r_0\ \mbox{und}\ C_n = nL + r_0.\notag
	\end{align}
Bei den zyklischen Randbedingungen sind die Ableitungen von $p_{C_n}(r')$ paarweise gleich $p_{C_{-n-1}}(-r')$. Für die reflektiven Ränder gilt
	\begin{align*}
		\begin{array}{rcccl}
			\left.\frac{\partial p_{A_n}(r')}{\partial r'}\right|_{L/2}  & + &\left.\frac{\partial p_{B_{-n}}(r')}{\partial r'}\right|_{L/2}    &=& 0\\
			\left.\frac{\partial p_{A_n}(r')}{\partial r'}\right|_{-L/2} & + &\left.\frac{\partial p_{B_{-n-1}}(r')}{\partial r'}\right|_{-L/2} &=& 0.\\
			\end{array}
	\end{align*}
Im Zweidimensionalen gilt dies ganz analog, nur dass hier ein periodisches Gitter aus Spiegelverteilungen betrachtet werden muss. Abb. \ref{fig:mirrored-density-2D-corner} zeigt die leichte Deformation der Wahrscheinlichkeitsdichte in einer durch zwei reflektive Ränder gebildeten Ecke.

\renewcommand{\figurename}{Abb.}
\renewcommand{\tablename}{Tab.}

\section{Zweidimensionales einkomponentiges Plasma}\label{kap:2D1CP-Theory}
In den letzten Jahren wurden starke Anstrengungen unternommen um die räumliche Anordnung von Kapillaren im Herzmuskelgewebe klassifizieren zu können. Bereits 2005 wurde dabei ein Ansatz über Voronoi-Zerlegungen des Versorgungsgebietes bzw. die Zweipunkt-Korrelationsfunktion der Abstände der Kapillaren untereinander verwendet \cite{Karch2005f}. Einen weiteren Fortschritt stellt die ein Jahr später präsentierte Beschreibung der Kapillar\-verteilung durch ein so genanntes zweidimensionales einkomponentiges Plasma dar ("2D1CP"). Obwohl mit diesem Modell die Verteilungsstatistik der Kapillaren durch einen einzigen Parameter $\Gamma$ festgelegt ist, erlaubt dieser Parameter trotzdem noch eine Klassifizierung des zugrunde liegenden Gewebes \cite{Karch2006}.\newline
Weiter wird der Versuch die Kapillaranordnung mittels eines 2D1CP zu beschreiben durch die Eigenschaften des 2D1CP motiviert: Die beste Nährstoffversorgung des Gewebes bei niedrigstem Aufwand (d.h. möglichst wenigen Kapillaren) würde eine vollkommen regelmäßige hexagonale Anordnung erzielen. Dieser Zustand entspricht der Kristallisation des Plasmas in ein festes Gitter. Gleichzeitig wird der Wachstumsprozess der Kapillaren jedoch durch andere Prozesse gestört, was das Ausbilden der regelmäßigen Struktur verhindert. Da in den folgenden Kapiteln viele der Simulationen in nach diesem Modell generierten Verteilungen stattgefunden haben, soll hier kurz auf die Theorie eingegangen werden die zur Erzeugung einer zu einem bestimmten $\Gamma$ gehörenden Konfiguration nötig ist.\newline
Die potentielle Energie $\Phi(r)$ zweier Punktladungen (mit Ladung $q$) im Abstand $r_{ij}$ sei durch $\Phi(r_{ij}) = - q^2 \log(r_{ij}/L)$ gegeben. $L$ ist eine beliebige Normierungslänge. Möchte man die Gesamtenergie $U$ von $N$ Partikeln in einer Elementarzelle (Basisvektoren $\vec{a_1}$ und $\vec{a_2}$) berechnen, so müssen auch die Wechselwirkungen mit Ladungen aus entsprechend verschobenen Elementarzellen berücksichtigt werden. Dies lässt sich am besten mit der Ewald-Methode \cite{Ewald21} bewerkstelligen.\newline
Bei der Ewald-Methode wird das Potential $\Phi(r)$ als Summe eines langreichweitigen und einen kurzreichweitigen Anteils definiert. Die Beiträge der langreichweitigen Wechselwirkung können dann unter Anwendung der Poisson-Summation im reziproken Raum addiert werden, die der kurzreichweitigen im Realraum. Nach \cite{Karch2006} bzw. \cite{Leeuw1982} folgt
	\begin{align}
		U = \frac{q^2}{4}\sum_{\vec{n}}\sum_{i,j}^{N}E_1(\eta^2 (\vec{r}_{ij}+\vec{n})^2) + \frac{\pi}{A}\sum_{\vec{k}\neq0}\frac{\exp(-k^2/4\eta^2)}{k^2} \left|\sum_{j=1}^N \exp(\mathsf{i}\vec{k}\vec{r_j})\right|^2 + U_{\mbox{\tiny const.}}.
		\label{eq:plasma-energy}
	\end{align}
Die erste Summe behandelt den kurzreichweitigen Energiebeitrag im Realraum. Der Summationsindex $\vec{n}$ läuft über die verschiedenen Einheitszellen (d.h. $\vec{n}=n_1 \vec{a_1} + n_2 \vec{a_2}$ mit $n_1,n_2 \in \mathbb{N}$), für $\vec{n}=0$ muss der Summand mit $i=j$ daher ausgelassen werden.\newline
Die zweite Summe über die reziproken Gittervektoren $\vec{k}$ addiert die langreichweitigen Beiträge im Fourierraum. Die reziproken Gittervektoren sind definiert als $\vec{k}=m_1 \vec{b_1}+ m_2 \vec{b_2}$ mit $\vec{b_1}=\frac{2\pi}{a_1}\widehat{a_1}$, $\vec{b_2}=\frac{2\pi}{a_2}\widehat{a_2}$ und $m_1,m_2\in\mathbb{N}$ mit $\widehat{a_i}=\frac{\vec{a_i}}{a_i}$. $U_{\mbox{\tiny const.}}$ bezeichnet einen möglichen Wechselwirkungsbeitrag mit einer gleichförmigen entgegengesetzten Hintergrundladung. $E_1(z)$ bezeichnet das Exponentialintegral
	\begin{align}
		E_1(z)=\int_z^\infty{\frac{\exp(-t)}{t}}\mathsf{d}t.
	\end{align}
Für den die Konvergenz beeinflussenden Parameter $\eta$ und die Abbruchbedingungen der Summation wurden die Werte aus \cite{Karch2006} übernommen.\newline
Mittels des Metropolis-Algorithmus \cite{Metropolis1953} kann einem solchen Teilchenensemble eine Temperatur $T=1/\Gamma$ zugeordnet bzw. die Temperatur des Systems eingestellt werden. Dazu wird die Position eines einzelnen Teilchens variiert und die dadurch entstehende Energieänderung $\Delta U=U_{\mbox{\tiny vorher}}-U_{\mbox{\tiny nachher}}$ des Systems berechnet ($U_{\mbox{\tiny const.}}$ aus Gl. (\ref{eq:plasma-energy}) spielt daher keine Rolle). Mit $\Delta U$ und $T$ wird die Wahrscheinlichkeit
	\begin{align}
		p(\Delta U)=\mbox{min}\left(\exp\left(\frac{\Delta U}{k_B T}\right),1\right)\quad \mbox{mit}\quad k_B=1
		\label{eq:metropolis-acceptance-probability}
	\end{align}
definiert. Wurde durch die Modifikation des Systems eine Energieabsenkung erreicht ($\Delta U\geq0$), so gilt nach Gl. (\ref{eq:metropolis-acceptance-probability}) $p(\Delta U)=1$ und die am System durchgeführte Veränderung wird beibehalten. Wurde jedoch die Energie erhöht ($\Delta U<0$), so folgt $0\leq p(\Delta U)\leq 1$. Jetzt wird eine gleichverteilte Zufallszahl $x$ aus $\left[0,1\right]$ gezogen. Gilt $x\leq p$ so wird die Veränderung ebenfalls akzeptiert, gilt allerdings $x>p$ so wird das System zurück in den Zustand vor der Modifikation versetzt. Da $p$ immer kleiner wird, je höher die Energiezunahme ausfällt, sinkt somit die Chance in energetisch ungünstigere Konfigurationen zu springen exponentiell ab. Gleichermaßen bedingt eine höhere Temperatur $T$ entsprechend größere Wahrscheinlichkeiten zwischen verschiedenen Zuständen zu wechseln, auch wenn dabei die Gesamtenergie des Systems zwischenzeitlich steigt.\newline
Wiederholt man diesen Prozess ausreichend oft (Thermalisierung), dann wechselt das System nur noch zwischen verschiedenen sog. Mikrozuständen des zu $T$ gehörenden Makrozustandes. Der Makrozustand ist vollständig durch den Parameter $T$ bzw. $\Gamma$ charakterisiert,das System ist im thermischen Gleichgewicht.\newline
Lässt man ein Plasma sukzessiv in niedrigere Temperaturen thermalisieren, so kann man bei $\Gamma\approx140$ eine Kristallisation des Plasmas beobachten. Bei einer weiteren Absenkung der Temperatur bis $T=0$ bildet sich dann ein perfektes hexagonales Gitter aus. $\Gamma$ quantifiziert somit den Gleichgewichtszustand zwischen Regelmäßigkeit und Störung. Nach \cite{Karch2006} (Tab. 2) liegen die Werte von $\Gamma$ ungefähr im Bereich von $2$ bis $5$, wobei die höher geordneten Konfigurationen ($\Gamma\approx5$) einem gesunden Muskelgewebe zuzuordnen sind.
\begin{figure}
		\begin{center}
		\subfloat[Thermalisierung für $\Gamma\!=\!2$(rot), $\Gamma\!=\!10$(grün), $\Gamma\!=\!100$(blau) und $\Gamma\!=\!200$(schwarz)]{
			\label{fig:2D1CP-thermalization}\includegraphics[width=0.79\textwidth]{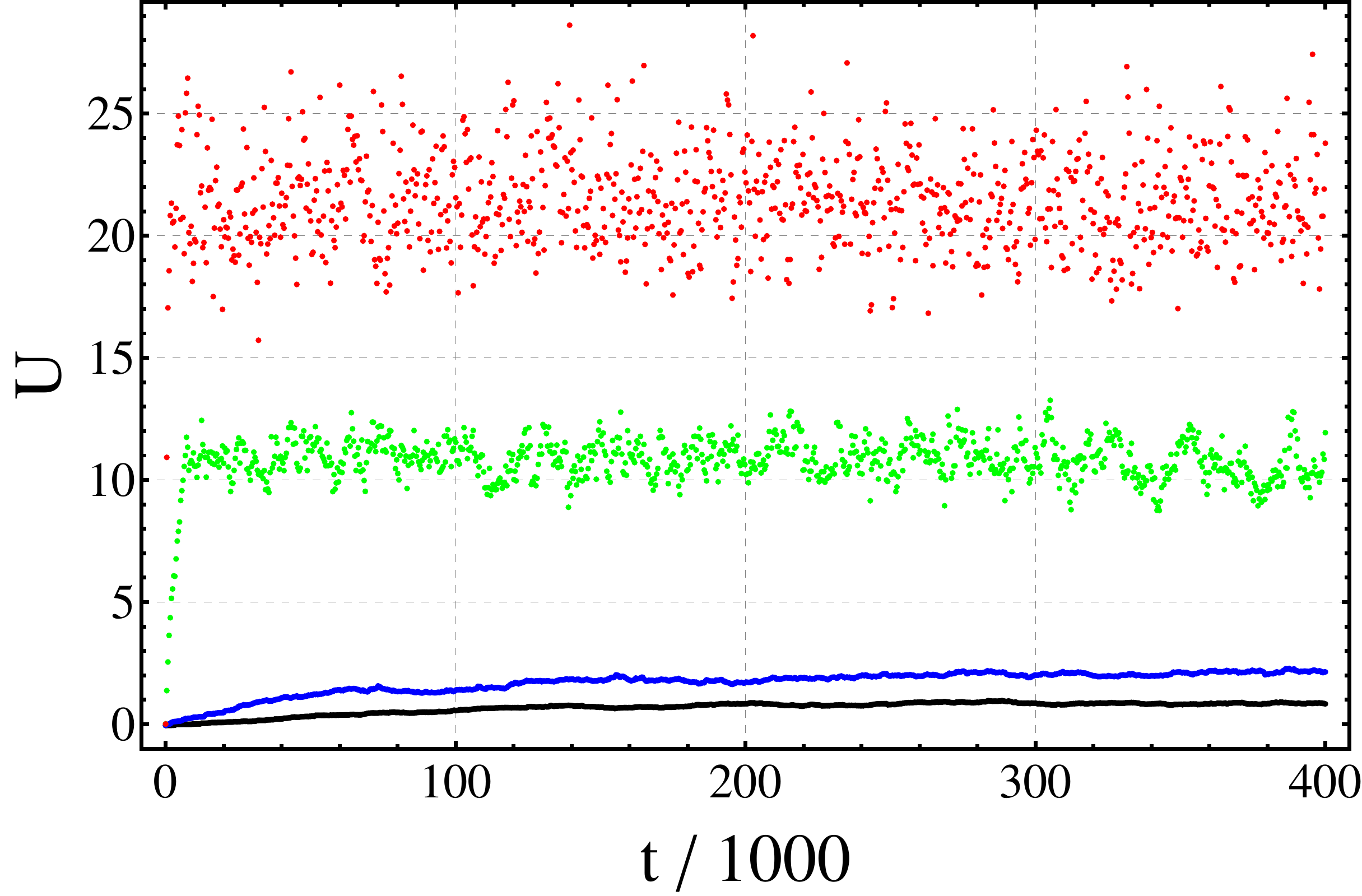}
		}\\
		\subfloat[Paar-Korrelation für $\Gamma\!=\!2$(rot), $\Gamma\!=\!10$(grün), $\Gamma\!=\!100$(blau) und $\Gamma\!=\!200$(schwarz)]{
			\label{fig:2D1CP-pair-correlation}\includegraphics[width=0.79\textwidth]{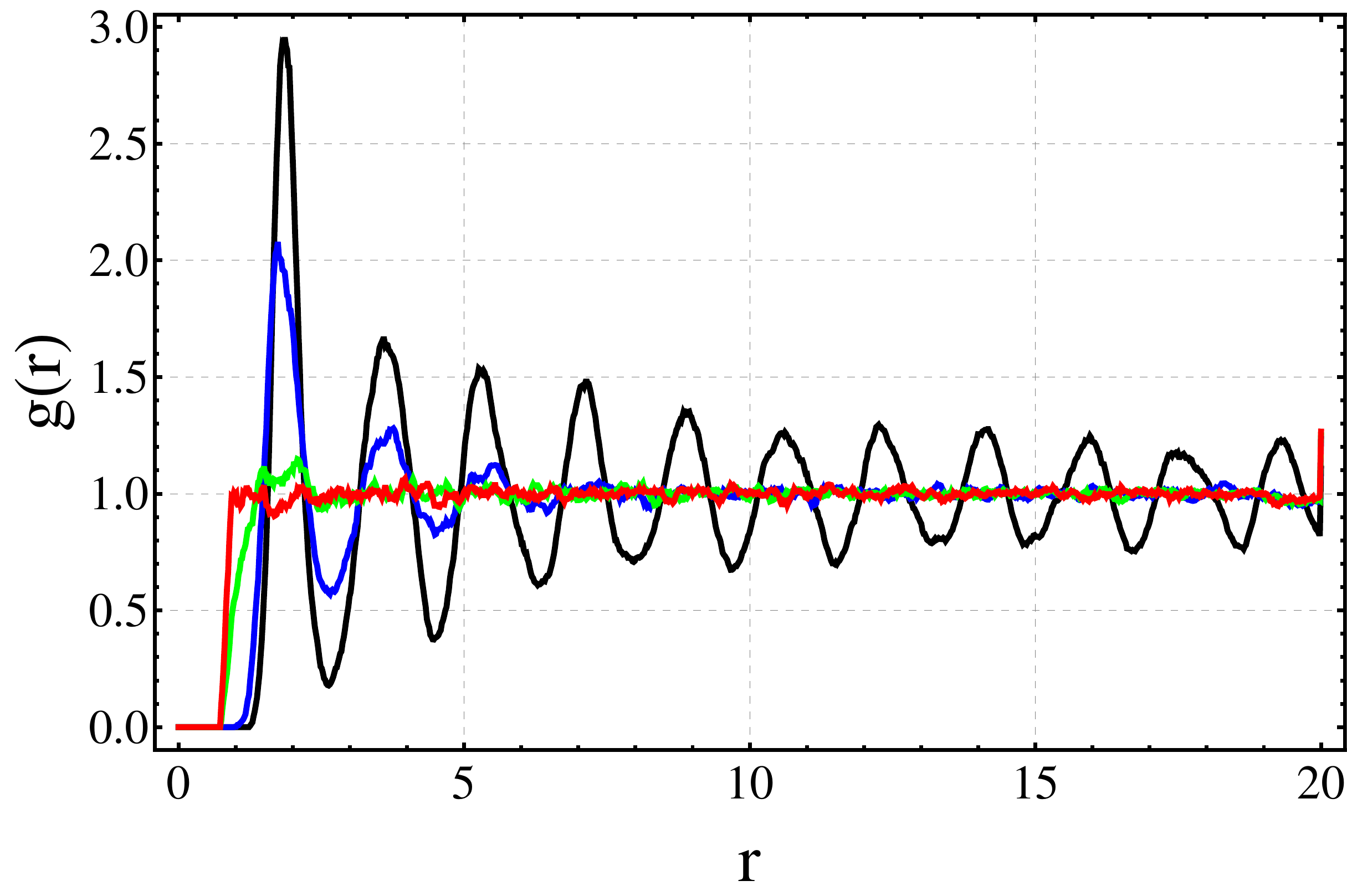}
		}
		\end{center}
		\caption{Bei den durchgeführten Thermalisierungen wurde immer ein hexagonales Gitter als Startverteilung der Ladungen gewählt. Die Energie der Konfiguration nimmt also während der Thermalisierung zu. Ein eventuell sehr langsamer Kristallisationsprozess wird dadurch vermieden.	Abb. \ref{fig:2D1CP-thermalization} zeigt den Energieverlauf der Mikrozustände gegen die Anzahl versuchter Monte-Carlo-Schritte $t$ für verschiedene  Thermalisierungen. Gut zu erkennen ist die schnellere Thermalisierung und die stärkeren Schwankungen um die zu $\Gamma$ gehörigen mittleren Energien für höhere Temperaturen. Abb. \ref{fig:2D1CP-pair-correlation} zeigt die über jeweils $10$ erzeugte Konfigurationen gemittelte Paar-Korrelations-Funktion. Gut zu erkennen ist der Übergang zu einer starken Fernordnung zwischen $\Gamma=100$ und $\Gamma=200$.}
		\label{fig:2D1CP-generation}
	\end{figure}

\chapter{Algorithmus}
\renewcommand{\figurename}{Abb.}
\renewcommand{\tablename}{Tab.}

\section{Parallelisierung und verwendete Bibliotheken}
\begin{figure}		
		\begin{center}\includegraphics[width=0.95\textwidth]{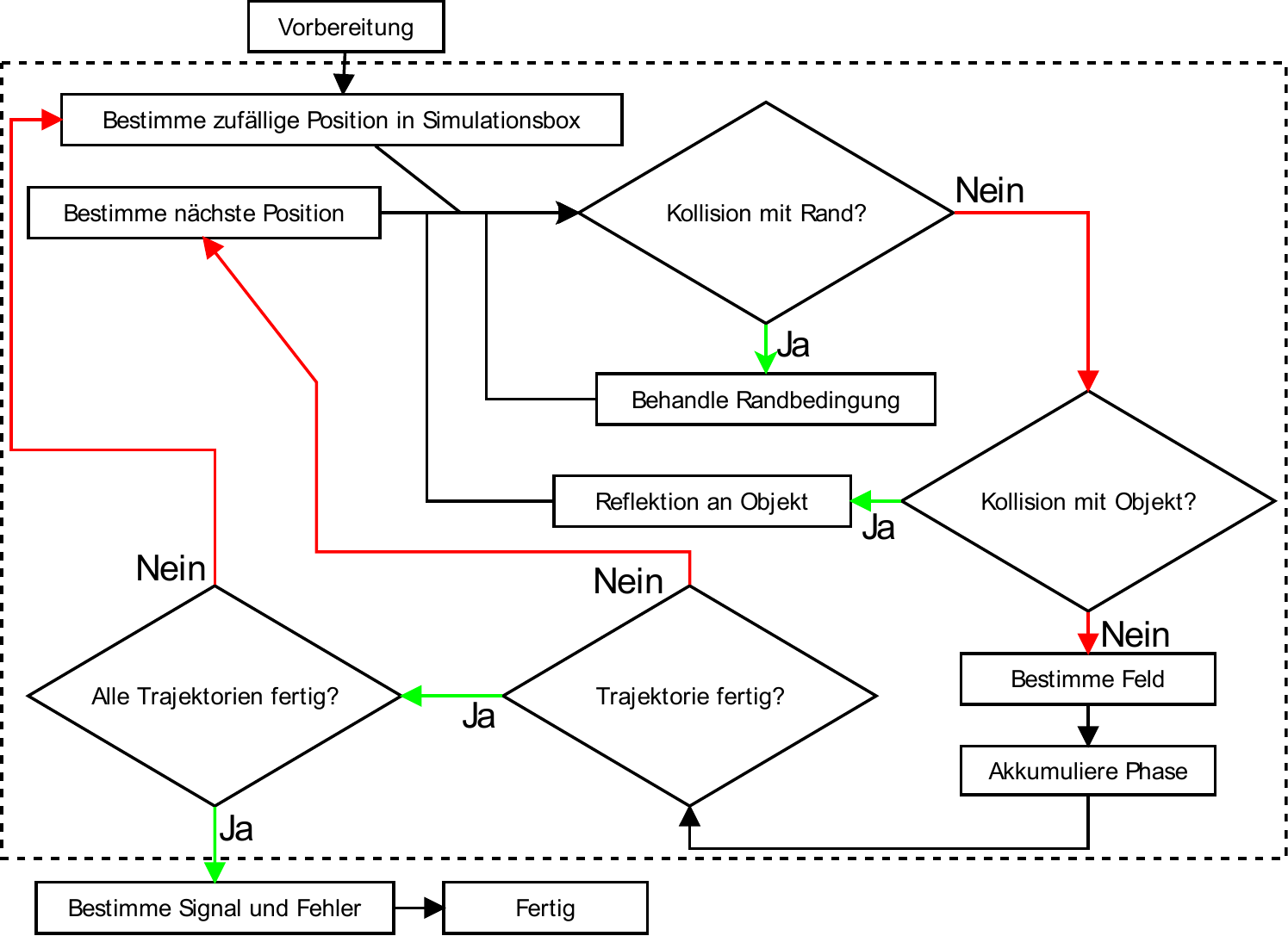}\end{center}
		\caption{Flussdiagramm des Simulationsprozesses. In der Vorbereitungsphase werden zunächst der in Kap. \ref{kap:recursive-collision-detection} erläuterte Kollisionsbaum erstellt und diverse Konsistenz Überprüfungen durchgeführt. Dann werden die parallel laufenden Arbeits-Threads gestartet. In dem abschließenden Mittelungsprozess wird dann aus den von den einzelnen Threads produzierten Daten das FID-Signal errechnet.}
		\label{fig:simulation-flowchart}
	\end{figure}

Um effektiv auf mehreren Prozessoren rechnen zu können wurde der komplette Programmfluss zwischen Vorbereitung und Fehler bzw. Signalbestimmung (eingerahmter Abschnitt in Abb. \ref{fig:simulation-flowchart}) mit Hilfe von \begin{itshape}OpenMP\end{itshape}\cite{OpenMP} parallelisiert. Für größere Mehrkernsysteme, bei welchen die verschiedenen CPUs nicht mehr über einen gemeinsamen Speicher (Shared Memory) verfügen, ist auch eine Parallelisierung mittels \begin{itshape}MPI\end{itshape} (Message Passing Interface) möglich \cite{MPI}.\newline
Die für den Random Walk nötigen Zufallszahlen wurden mit dem Random Paket der \begin{itshape}Boost\end{itshape}-Bibliothek (\cite{Boost}) erzeugt. Dabei wurden gleichverteilte durch den Pseudozufallsgenerator "rand48" erzeugte Zahlen mittels des Box-Muller-Algorithmus in normalverteilte Zufallszahlen transformiert.
	
\section{Diskrete Phasenakkumulation}
Im Folgenden bezeichnet $t_j=j\Delta t$ den Zeitpunkt nach $j$ Zeitschritten, $\vec{r}_n(t_j)=\vec{r}_{n,j}$ den Ort einer Trajektorie zur Zeit $t_j$ und entsprechend $\omega(\vec{r}_{n,j})=\omega_{n,j}$ das Offresonanzfeld welches die Trajektorie $n$ zum Zeitpunkt $t_j$ erfährt. Für die akkumulierte Phase $\Phi_{n}(t)$ einer Trajektorie $n$ ergibt sich dann
	\begin{equation}
		\Phi_{n}(t) = \sum_{j=0}^{m}{\omega(\vec{r}_n(t_j)) \Delta t} = \Delta t \sum_{j=0}^{m}{\omega_{n,j}}\quad \mbox{mit}\quad m=\left\lfloor \frac{t}{\Delta t}\right\rfloor.
		\label{eq:phase-accumulation}
	\end{equation}
Mit den Gaußklammern $\lfloor\,\rfloor$ wird der zum Zeitpunkt $t$ gehörende Zeitindex $m$ immer abgerundet. Der Frequenz- und Phasenverlauf einer Trajektorie ist schematisch in Abb. \ref{fig:phase-accumulation-large} dargestellt. Die normierte transversale Magnetisierung folgt aus der Mittelung über das Ensemble der $N$ Trajektorien:
	\begin{equation}
		M_T(t) = \frac{1}{N}\sum_{n=0}^{N-1}{\left[\cos(\Phi_n(t)) + i \sin(\Phi_n(t))\right]}.
		\label{eq:magnetisierung}
	\end{equation}
Zur Realisierung beliebiger Spin-Echo-Sequenzen kann nun einfach Gl. (\ref{eq:phase-accumulation}) angepasst werden
	\begin{equation}
		\Phi_{n}(t) = \Delta t \sum_{j=0}^{m}{w_j \omega_{n,j}}.
		\label{eq:phase-accumulation-sequence}
	\end{equation}
Wegen der Invertierung der Phase muss das an die Sequenz angepasstes Array $w_j=\pm1$ so gewählt werden, dass bei jedem $180^\circ$-Puls das Vorzeichen wechselt. Da die $\omega(\vec{r}_{n,j})$ nicht von der Sequenz abhängig sind, können eine beliebige Anzahl verschiedener Sequenzen in einer einzigen Simulation berechnet werden. Weil ein Großteil des Rechenaufwands auf den Random Walk und nicht die Auswertung der Phaseninkremente entfällt, stellt dies einen deutlichen Zeitgewinn dar.	

\section{Umsetzung der Randbedingungen}
\begin{figure}
		\begin{center}
		\subfloat[Elastischer Stoß]{
			\label{fig:reflection-elastic}\includegraphics[width=0.45\textwidth]{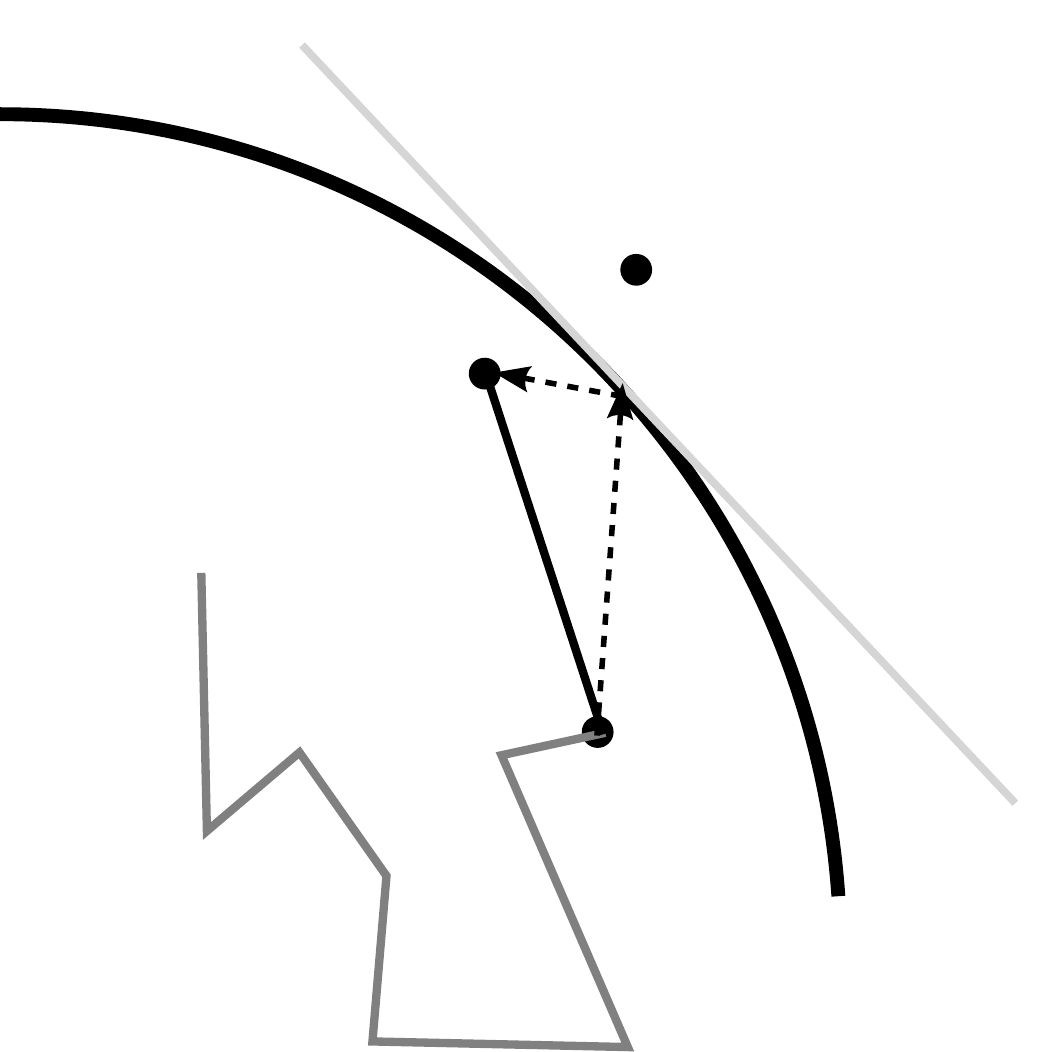}
		}
		\subfloat[Spiegelung an Tangente]{
			\label{fig:reflection-mirrored}\includegraphics[width=0.45\textwidth]{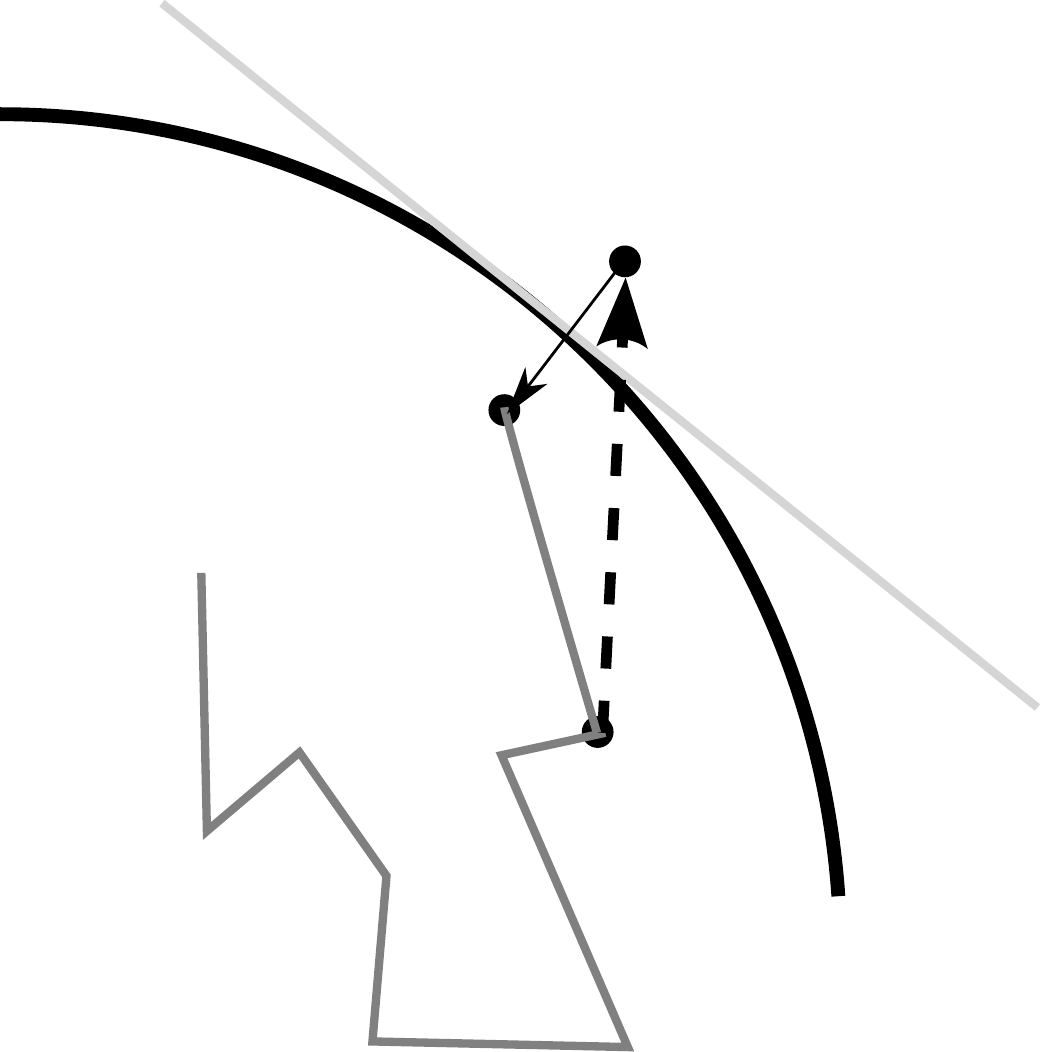}
		}
		\end{center}
		\caption{Möglichkeiten für die Abbildung eines Schrittes aus der Simulationsbox heraus zurück ins Innere. Als erste Möglichkeit erscheint ein elastischer Stoß sinnvoll, da dieser offensichtlich die reflektiven Randbedingungen aus Gl. (\ref{eq:boundary-conditions}) erfüllt. Der zweite Fall, die Spiegelung in radiale Richtung, entspricht einer Implementierung im Sinne eines geraden Randes. Wie bereits zu erkennen ist gehen für $\sigma \! \ll \! R_a$ die zwei Verfahren ineinander über.}
		\label{fig:reflection-outer-boundary}
	\end{figure}

\begin{figure}		
		\begin{center}\includegraphics[width=0.45\textwidth]{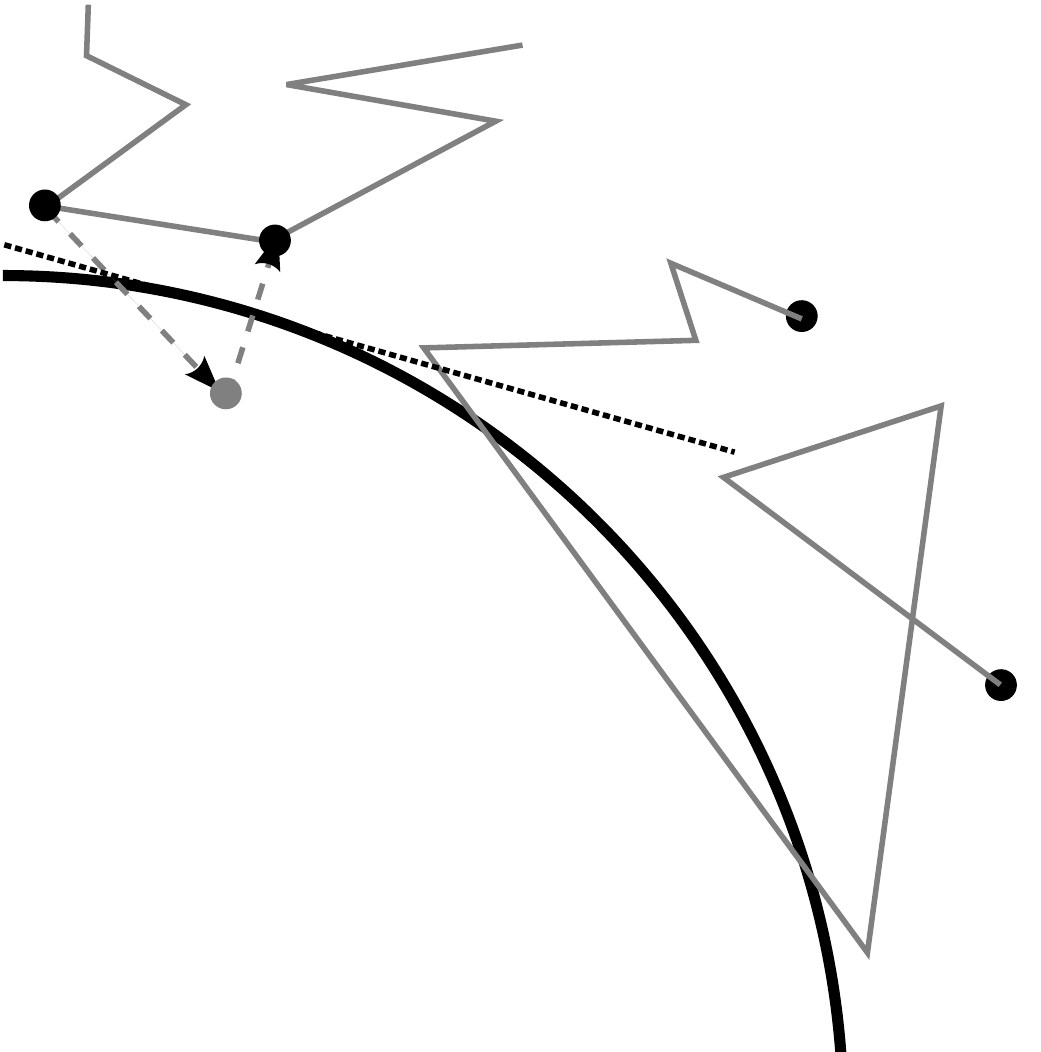}\end{center}
		\caption{Zwei mögliche Trajektorien nahe einer Kapillare. Bei der ersten Trajektorie wird ein Sprung ins Innere wie bei der Reflektion am Rand durch Spiegelung an der Tangente wieder nach außen abgebildet. Die zweite Trajektorie wird, da keine direkte Kollision mit der Kapillare besteht, ebenfalls als gültig gewertet. Aufgrund der maximalen mittleren Schrittweite von $1/4R_c$ sind solche Sprünge jedoch relativ selten. Liegt ein Startpunkt des Sprungs genau auf dem Kreis (ungünstigster Fall), so ist die Wahrscheinlichkeit einen solchen Sprung durchzuführen kleiner als $8\%$.}
		\label{fig:reflection-capillary}
	\end{figure}

\begin{figure}		
		\begin{center}\includegraphics[width=0.8\textwidth]{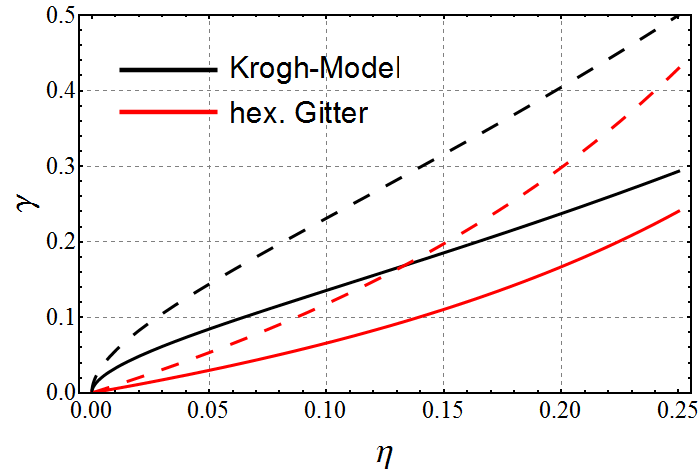}\end{center}
		\caption{Volumenbruchteil der randnahen Gebiete für $\sigma\!=\!R_c/4$ (gestrichelt) und $\sigma\!=\!0.44\mum$ (durchgezogen) für einen minimalen Kapillarradius $R_c\!=3\mum$. Die Schrittweite $\sigma\!=\!0.44\mum$ entspricht etwa $\Delta t\!\approx\!0.1\ms$ bei $D\!=\!1\mum^2/\ms$.}
		\label{fig:boundary-influenced-steps}
	\end{figure}

Bei der Implementierung der Ränder muss wie in Kap.\ref{kap:diffusion-gerade-raender} zwischen den verschiedenen Randtypen unterschieden werden. Für die periodischen Randbedingungen werden Schritte aus der Simulationsbox über die Modulo-Funktion wieder ins Innere abgebildet. D.h ein Teilchen welches z.B. um die Länge $l$ über den rechten Rand der Simulationsbox hinausläuft, wird mit Abstand $l$ vom gegenüber liegenden Rand wieder in die Box hinein gesetzt. Dieses Verfahren gibt exakt die in Kap. \ref{kap:diffusion-gerade-raender} berechnete Wahrscheinlichkeitsverteilung wieder. Da die einzelnen Trajektorien des Random Walks sich nicht gegenseitig beeinflussen, fällt ein Problem der zyklischen Randbedingungen weg, welches in vielen anderen Anwendungsbereichen auftritt: Wäre das Simulationsgebiet zu klein, so könnte es zu Wechselwirkungen eines Randes mit sich selbst kommen.\newline
Deutlich interessanter gestaltet sich die Implementierung der Reflexion an Kapillaren bzw. dem äußeren Rand im Krogh-Modell. Hierfür gibt es prinzipiell zwei Möglichkeiten (siehe Abb. \ref{fig:reflection-outer-boundary}). Für hinreichend kleine Schritte (gegenüber den Kapillarradien) führen beide Methoden zum gleichen Ergebnis.
Da eine einzelne zufällig gewählte Trajektorie nicht Impuls erhaltend sein muss und zudem das Impuls erhaltende Verfahren nach Abb. (\ref{fig:reflection-elastic}) weniger effizient implementiert werden kann, wurde das Verhalten aus Abb. \ref{fig:reflection-mirrored} umgesetzt. Bei der Kollision mit einer Kapillare tritt ein weiterer Effekt auf, welcher in Abb. \ref{fig:reflection-capillary} dargestellt ist.\newline
Betrachtet man den Anteil $\gamma$ der Schritte die nahe (d.h. Abstand kleiner als $\sigma$) an einem Rand oder einer Kapillare liegen, kommt man zu dem Schluss, dass für die im Rahmen dieser Arbeit untersuchten $\eta$ die Diffusionsstatistik weitestgehend randunabhängig ist. Für das Krogh-Modell lässt sich die Anzahl der durch Ränder beeinflussten Schritte wie folgt abschätzen
	\begin{align}
		\gamma=\frac{\left(R_c+\sigma\right)^2\pi -R_c^2\pi + R_a^2\pi -\left(R_a-\sigma\right)^2\pi}{R_a^2\pi -R_c^2\pi}=\frac{2\sigma}{R_a-R_c}=\frac{2\sigma\sqrt{\eta}}{R_c(1-\sqrt{\eta})}.
	\end{align}
Für ein hexagonales Gitter oder die in Kap. \ref{kap:2D1CP-Analysis} behandelten Plasma-Konfigurationen ergeben sich noch niedrigere Bruchteile $\gamma$, da hier die äußeren zyklischen Randbedingungen als Problem\-quelle wegfallen
	\begin{align}
		\gamma=\frac{2\left((R_c+\sigma)^2\pi -R_c^2\pi \right)}{\ICD^2\sqrt{3}-2R_c^2\pi }=\frac{2\pi \sigma(2 R_c+\sigma)}{ \sqrt{3} \ICD^2-2\pi R_c^2}=\frac{2\sigma R_c +\sigma^2}{R_c^2}\frac{\sqrt{3}\eta}{1-\sqrt{3}\eta}\approx\frac{2\sqrt{3}\sigma\eta}{R_c(1-\sqrt{3}\eta)}.
	\end{align}
In beiden Fällen führen also größere Schrittweiten und höheres $\eta$ zu mehr Randeinfluss und somit größeren möglichen systematischen Fehlern. Um einen Anstieg von $\gamma$ bei kleinen Kapillarradien möglichst gering zu halten, wurde bei allen Simulationen zu Beginn geprüft, ob die vorgegebene Zeitschrittweite $\Delta t$ und die damit verbundene mittlere räumliche Schritt\-weite $\sigma=\sqrt{2D\Delta t}$ kleiner ist als $1/4$ des kleinsten Kapillarradius. Bei einer Überschreitung dieses Schwellenwertes wurde die Zeitschrittweite dann soweit verringert, dass gilt $\sigma=R_c/4$. Sollten die Eingabewerte der Simulation bereits zu einem ausreichend kleinen $\sigma$ führen, so wurde keine Anpassung vorgenommen, da sonst durch die steigende Anzahl nötiger Zeitschritte auch die benötigte Rechenzeit mit ansteigt. Abb. \ref{fig:boundary-influenced-steps} zeigt $\gamma$ für Krogh-Modell und hexagonales Gitter. Erst ab sehr kleinem $\eta$ bzw. $R_c$ wird die automatische Anpassung der Schritt\-weite überhaupt nötig, da die normalerweise verwendete Zeitdiskretisierung $\Delta t=0.1\ms$ zu meist ausreichend kleinen Schrittweiten $\sigma=\sqrt{2\cdot1\mum^2/\ms\cdot0.1\ms} \approx0.45\mum<R_c/4$ führt. Wurde diese automatische Anpassung vorgenommen, so ist $\gamma$ wegen $\sigma=R_c/4$ nur noch von $\eta$ abhängig.
	
\section{Rekursive Collision Detection}\label{kap:recursive-collision-detection}
\begin{figure}
		\begin{center}
		\subfloat[Aufbau eines rekursiven Suchbaums. Liegt die zu prüfende Position in $Q_1$, $Q_3$ oder $Q_4$, so wird nach der Wurzel nur eine weitere Anfrage benötigt. Nur für Positionen in $Q_{2,2}$ muss der Suchbaum bis in die dritte Stufe durchlaufen werden. Nur die Knoten $Q_{2,2,3}$, $Q_{2,2,4}$, und $Q_{2,4}$ besitzen Informationen über Mittelpunkt und Radius von $k_1$, $Q_3$ kennt nur $k_2$, $k_3$ ist nur $Q_1$ und $Q_{2,3}$ bekannt.]{
			\includegraphics[width=0.75\textwidth]{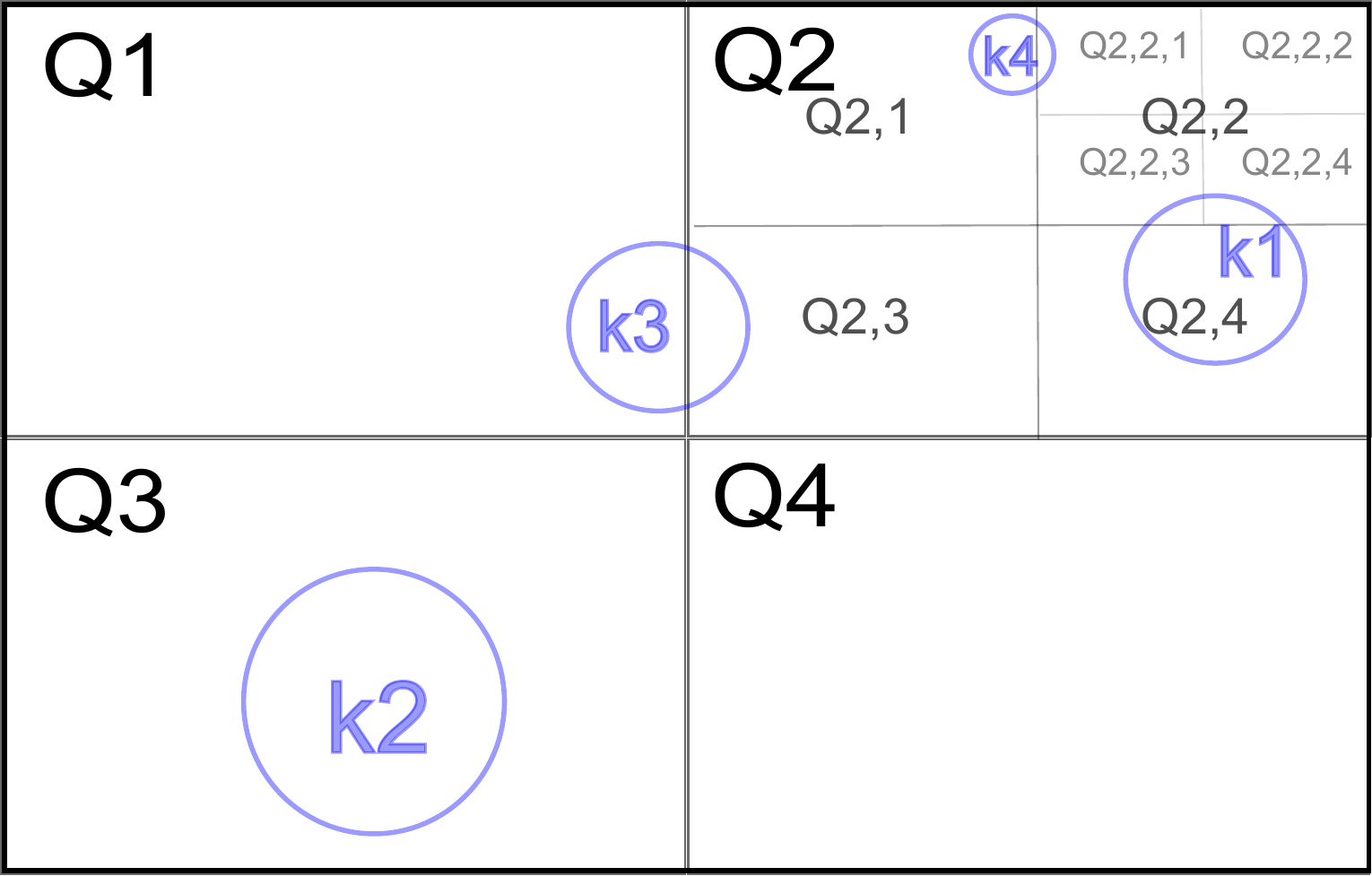}\label{fig:recursion-detailed}
		}\\
		\subfloat[Ausschnitt aus einem rekursiven Suchbaum. Beim Erstellen des Baumes wird die Tatsache genutzt, dass ein Überlapp der Kapillaren nicht möglich ist. Bei einer Überprüfung ob ein Punkt $\vec{r}$ im Inneren einer Kapillare liegt entsteht mit Hilfe des Suchbaums deutlich weniger Rechenaufwand. Die Schattierung der Quadranten gibt die Tiefe des Baums wieder. Läuft die Trajektorie durch eine hellere Region sind also deutlich weniger Rekursionen nötig.]{
			\includegraphics[width=0.75\textwidth]{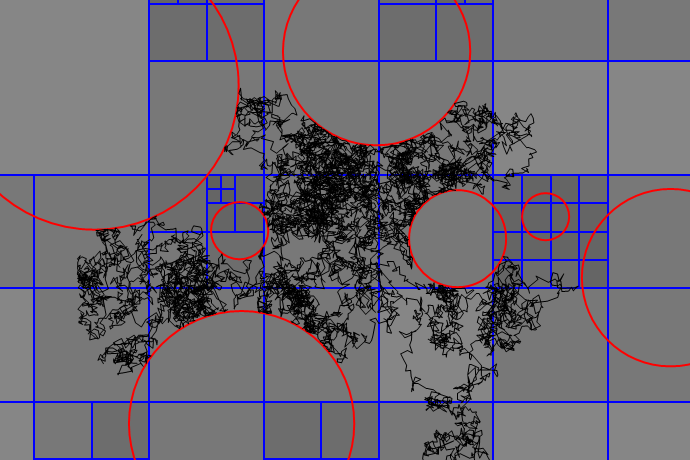}\label{fig:recursive-collision-detection}
		}
		\end{center}
		\caption{Aufbau und Funktionstest der verwendeten Suchbäume}
		\label{fig:recursion}
	\end{figure}

Um bei vielen Kapillaren die Überprüfung einer möglichen Kollision effizient zu gestalten wurde ein rekursiver Kollisionsbaum verwendet. Ein Suchbaum besteht aus einer Wurzel, welche Verweise auf Knoten besitzt. Diese Knoten können wieder Verweise auf weitere Knoten besitzen. Ist ein Knoten einem anderen untergeordnet, so spricht man von einem Kindknoten. Bis auf die Wurzel ist jeder Knoten immer Kindknoten von genau einem Elternknoten.\newline
In dem hier verwendeten Baum besitzt jeder Knoten entweder null oder vier Kindknoten, der Baum kann also auch nur aus der Wurzel bestehen. Jedem Knoten ist dabei ein rechteckiger Ausschnitt der Simulationsbox zugeordnet. Bei vorhandenen Kindknoten wird dieser Ausschnitt immer weiter gleichmäßig unterteilt (siehe Abb. \ref{fig:recursion}). Diese Unterteilung wird so lange fortgeführt, bis jeder Knoten höchstens eine einzige Kapillare schneidet oder beinhaltet. Um eine Position $\vec{r}$ auf mögliche Kollisionen zu testen wird beginnend mit dem Wurzelknoten geprüft, ob eine Unterteilung in Unterquadranten vorliegt. Falls dies der Fall ist wird die Anfrage an den entsprechenden Kindknoten weitergeleitet. Gibt es keine Kindknoten so wird auf eine Kollision mit einer evtl. verknüpften Kapillare geprüft.\newline
Abb. \ref{fig:recursive-collision-detection} zeigt einen Ausschnitt aus einem komplexeren Suchbaum mit Trajektorie. Wie man sich leicht überlegen kann sinkt der Rechenaufwand für die Kollisionsdetektion im besten Fall (möglichst gleichmäßig verteilte Kapillaren) von $O(n)=n$ auf $O(n)=\log n$ \cite{knuth2001art}. Da der Kollisionstest für jeden Zeitschritt jeder Trajektorie durchgeführt werden muss, macht er neben der Offresonanzberechnung (siehe Kap. \ref{kap:feldberechnung}) den Hauptteil der benötigten Rechenzeit aus. Die Beschleunigung die durch die rekursive Suche erreicht wird senkt den Gesamtrechenaufwand daher deutlich. Bei Tests liefert die rekursive Implementierung bei einer einzelnen Kapillare etwa die gleiche Leistung wie die klassische Überprüfung. Bereits bei ca. 50 Kapillaren und wenig dicht gepackter Anordnung lässt sich eine Beschleunigung der Simulation bis um den Faktor 10 erreichen, was hinsichtlich der erwarteten Simulationsdauern von mehreren Wochen (mit rekursiver Suche) eine notwendige Verbesserung darstellt.

\section{Feldberechnung}\label{kap:feldberechnung}
	\begin{figure}		
		\begin{center}\includegraphics[width=0.8\textwidth]{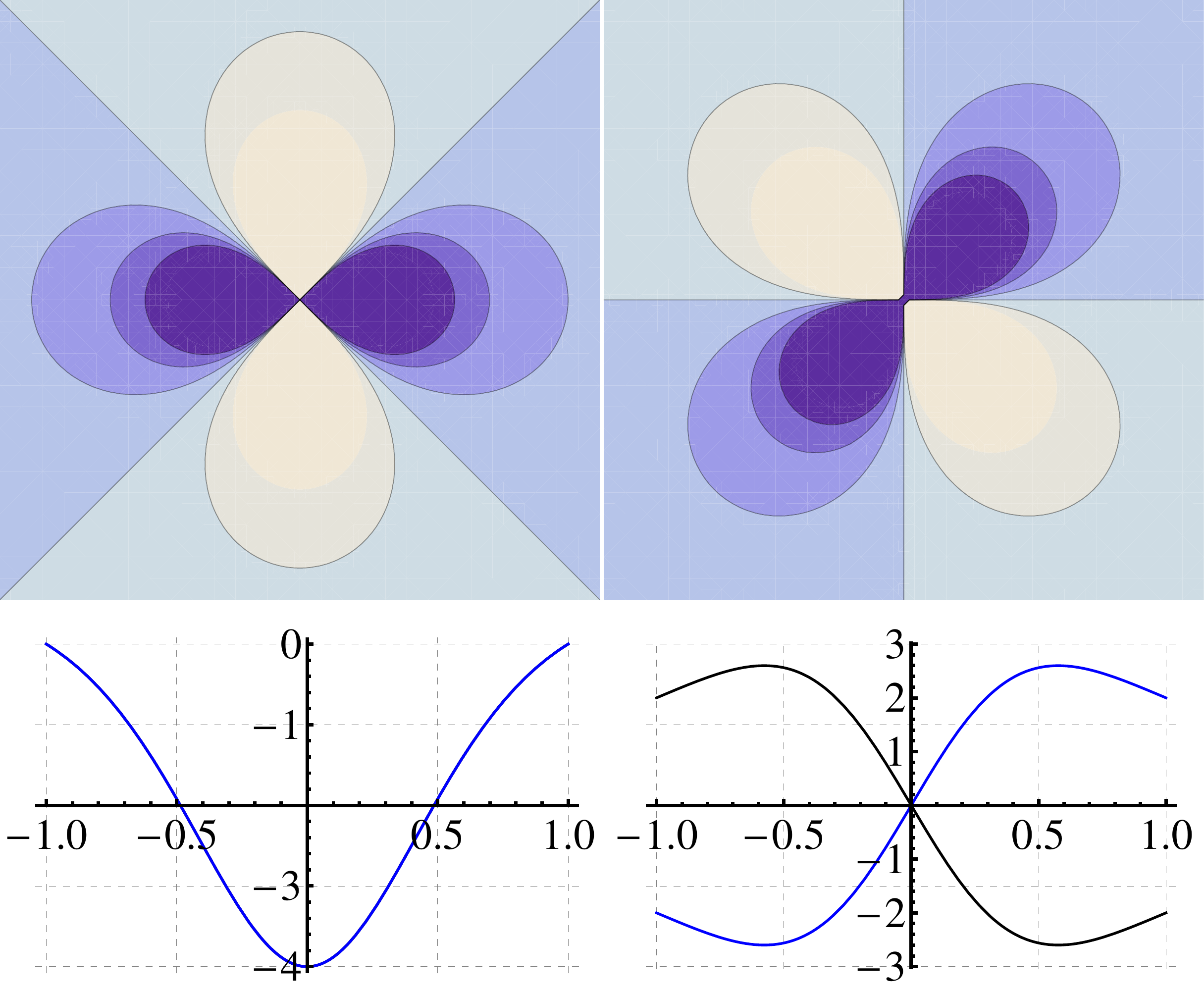}\end{center}
		\caption{Die obere Grafik zeigt die Offresonanzfelder für $\alpha=0$ (links) und $\alpha=45^\circ$ (rechts). In der unteren Grafik ist der Feldverlauf für den jeweils rechten und linken Rand aufgetragen. Bei $\alpha=0$ ist zwar $\omega$ über den Rand hinweg stetig, die erste Ableitung jedoch nicht. Für $\alpha=45^\circ$ weist bereits die Offresonanz selbst eine Unstetigkeit auf. Läuft eine Trajektorie durch die Randbedingungen, so wechselt das Vorzeichen von $\omega$.}
		\label{fig:feld-periodizitaet}
	\end{figure}
Für anfänglich durchgeführte Simulationen wurden die zu jedem Zeitschritt gehörenden Phaseninkremente mittels Gl. (\ref{eq:single-capillary-field}) zur Laufzeit berechnet. Dies ist jedoch nur für wenige Kapillaren praktikabel, da sonst die Summation über die einzelnen Feldbeiträge immer mehr Rechenzeit in Anspruch nimmt.\newline
Bei zyklischen Randbedingungen ist bei der Berechnung des Offresonanzfeldes besondere Vorsicht geboten. Da Trajektorien über den Rand der Simulationsbox hinaus laufen können und dann auf der anderen Seite fortgesetzt werden, muss auch das Offresonanzfeld selbst über den Rand hinweg den zyklischen Randbedingungen aus Gl. (\ref{eq:boundary-conditions}) gehorchen.
	\begin{align}
		\displaystyle \left.\frac{\displaystyle \partial^n \omega}{\displaystyle \partial r^n}\right|_{r'} = \displaystyle\left.\frac{\partial^n \omega}{\partial r^n}\right|_{r''}  \ \ \forall \ r',r'' \in \ \mbox{Rand und}\ \ n \in 0,1,\ldots N
		\label{eq:cyclic-offresonance-field}
	\end{align}
Auch hier ist $r'$ der zu $r''$ gehörige gegenüberliegende Punkt auf dem Rand der Simulationsbox. Für symmetrisch zum Rand angeordnete Kapillaren ist Gl. (\ref{eq:cyclic-offresonance-field}) zumindest für $n=0$ erfüllt, bereits die erste Ableitung ist jedoch unstetig. Für weniger symmetrische Anordnungen erfährt ein Teilchen bei einem Durchlauf des Randes sogar einen Sprung im Offresonanzfeld (Abb. \ref{fig:feld-periodizitaet}), was zu kritischen Fehlern in der Phasenakkumulation führen kann. Um die in Kap. \ref{kap:boundary-conditions} erläuterten Finite-Size-Effekte zu verhindern muss also die Periodizität von $\omega(\vec{r})$ wiederhergestellt werden. Dazu müssen für jeden Ort die Beiträge eines Gitters aus Feldquellen berücksichtigt werden. Analog zur Festkörperphysik entspricht die verwendete Simulationsbox der Elementarzelle des Gitters, die $L_k$ bezeichnen die Positionen der Kapillaren in der Elementarzelle. Es folgt
	\begin{align}
		\omega_{\mbox{\tiny gitter}}(\vec{r})=\sum_{\vec{n}}\sum_{k=0}^c \omega\!\left(\vec{r}-(\vec{L_k}+\vec{n})\right).
		\label{eq:lattice-field}
	\end{align}
Prinzipiell besteht eine starke Analogie zwischen der Feldberechnung und dem in Kap. \ref{kap:2D1CP-Theory} beschriebenen Ewald-Verfahren. Die Summe über $\vec{n}$ läuft analog zu Gl. (\ref{eq:plasma-energy}) über alle Gitterzellen. Eine direkte Anwendung der Ewald-Methode auf Gl. (\ref{eq:lattice-field}) ist jedoch nicht möglich, da $\omega$ im Gegensatz zur Energie $U$ zusätzlich vom Ort abhängt.\newline
Führt man einen Radius $R_{\mbox{\tiny max}}$ ein und begrenzt die Summation in Gl. (\ref{eq:lattice-field}) auf alle $\vec{n}$ mit $\left|\vec{n}\right|\leq R_{\mbox{\tiny max}}$ lässt sich Gl. \ref{eq:cyclic-offresonance-field} für sehr große $R_{\mbox{\tiny max}}$ zumindest in guter Näherung erfüllen. Falls nicht anders erwähnt, wurde in allen durchgeführten Simulationen $R_{\mbox{\tiny max}}=1650\mum$ verwendet.\newline
Das anfänglich verwendete Verfahren, die Berechnung von $\omega(\vec{r})$ zur Laufzeit, erweist sich daher für nahezu alle Geometrien (außer dem Krogh-Modell) als unpraktikabel und zu rechenintensiv, da die Summation aus Gl. (\ref{eq:lattice-field}) über mehrere zehntausend Summanden laufen kann. Eine deutlich schnellere Laufzeit erhält man, wenn das zur Geometrie gehörige Feld im Voraus berechnet wird. Über die Einheitszelle (bzw. Simulationsbox) wird dazu ein Gitter aus Stützpunkten gelegt für welche die Offresonanzen berechnet werden. Liegt die Auflösung dieses Gitters deutlich über der mittleren Schrittweite des Random Walks kann dann für jeden Zeitschritt durch bilineare Interpolation zwischen den jeweils vier nächsten Gitterpunkten das Feld in guter Näherung deutlich schneller berechnet werden.\newline
Die bilineare Interpolation nutzt die jeweils vier nächsten Stützpunkte des Gitters um den Funktionswert an einem Ort in dem so definierten Rechteck zu approximieren. Durch lineare Interpolation in $x$- und $y$-Richtung entsteht so eine Fläche zweiter Ordnung. Bei einer einfachen linearen Interpolation wären nur drei Stützpunkte nötig, welche eine Ebene definieren \cite{stoer2005numerische}.\newline
In vielen Simulationen wurde nur die Kapillardichte und das RBV (d.h. $\eta$, $R_c$ und $\ICD$) bzw. $\dom_0$ als Parameter variiert. Die relative Anordnung der Kapillaren zueinander bleibt jedoch häufig identisch. Dies ermöglicht eine Wiederverwertung des mittels Gl. (\ref{eq:lattice-field}) berechneten Interpolationsgitters für große Teilbereiche des Parameterraums.\newline	
Angenommen ein Interpolationsgitter wurde mit Auflösung $\Delta x$, Radien $R_c$, Abmessung $(a,b)$ und dem daraus folgenden $\eta$ berechnet, dann bestimmt man für eine Transformation nach $R_c'$ und $\eta'$ zunächst die neue nötige Abmessung $(a',b')$ und das daraus folgende $\Delta x'$ (die Anzahl der Stützpunkte bleibt konstant). Die Amplitude an jedem Stützpunkt muss dann um den Faktor $\frac{R_c'^2}{R_c^2}\frac{\Delta x^2}{\Delta x'^2}$ skaliert werden. Möchte man zusätzlich $\delta\omega_0$ skalieren, so kommt noch der Faktor $\frac{\delta\omega_0'}{\delta\omega_0}$ hinzu.

\renewcommand{\figurename}{Abb.}
\renewcommand{\tablename}{Tab.}

\section{Fehlerabschätzungen}\label{kap:fehlerermittlung}
\begin{figure}
		\begin{center}
			\subfloat[Phasenakkumulation]{
				\includegraphics[width=0.48\textwidth]{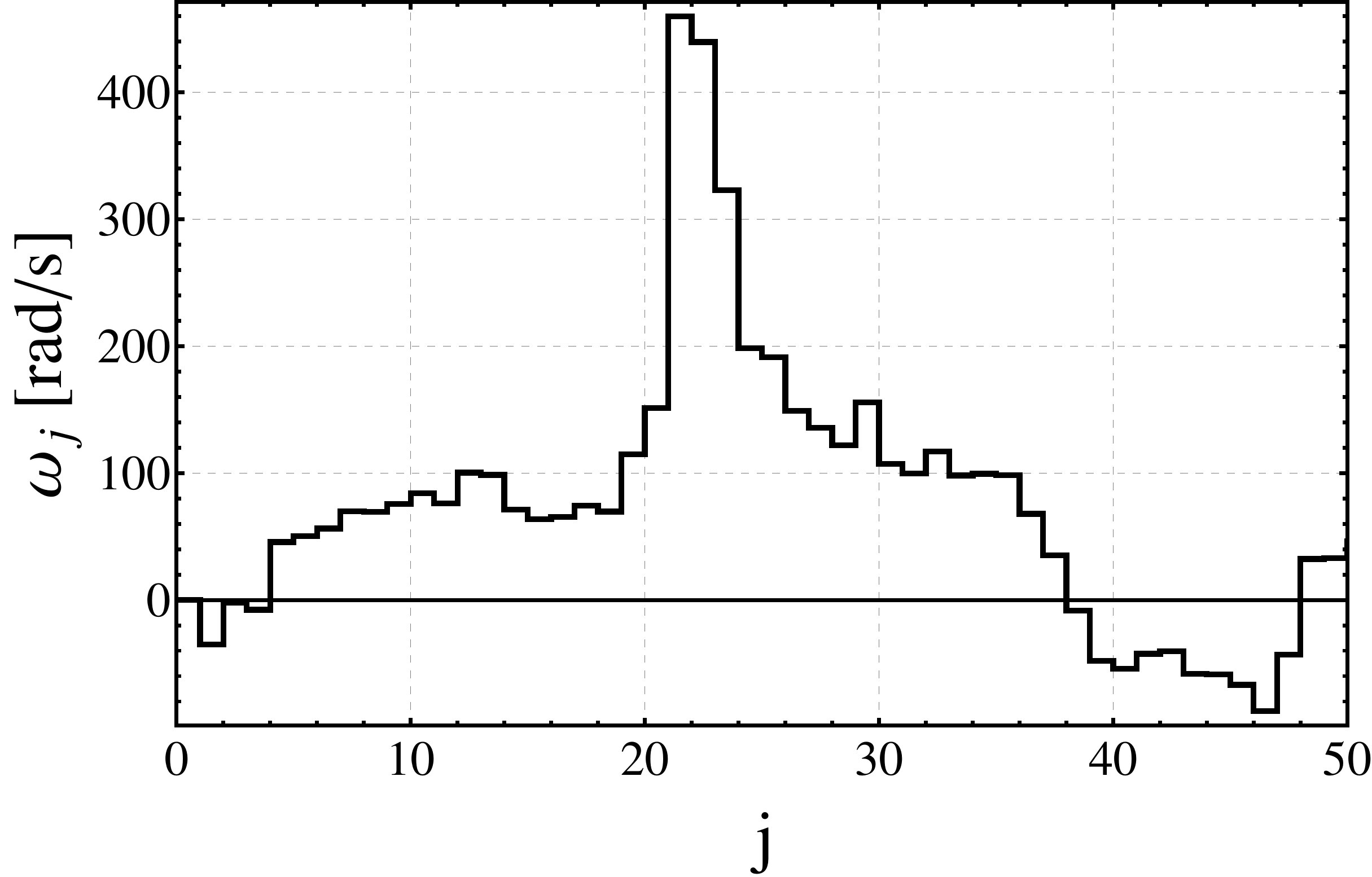}\label{fig:phase-accumulation-large}
			}
			\subfloat[Fehler in Phasenakkumulation]{
				\includegraphics[width=0.48\textwidth]{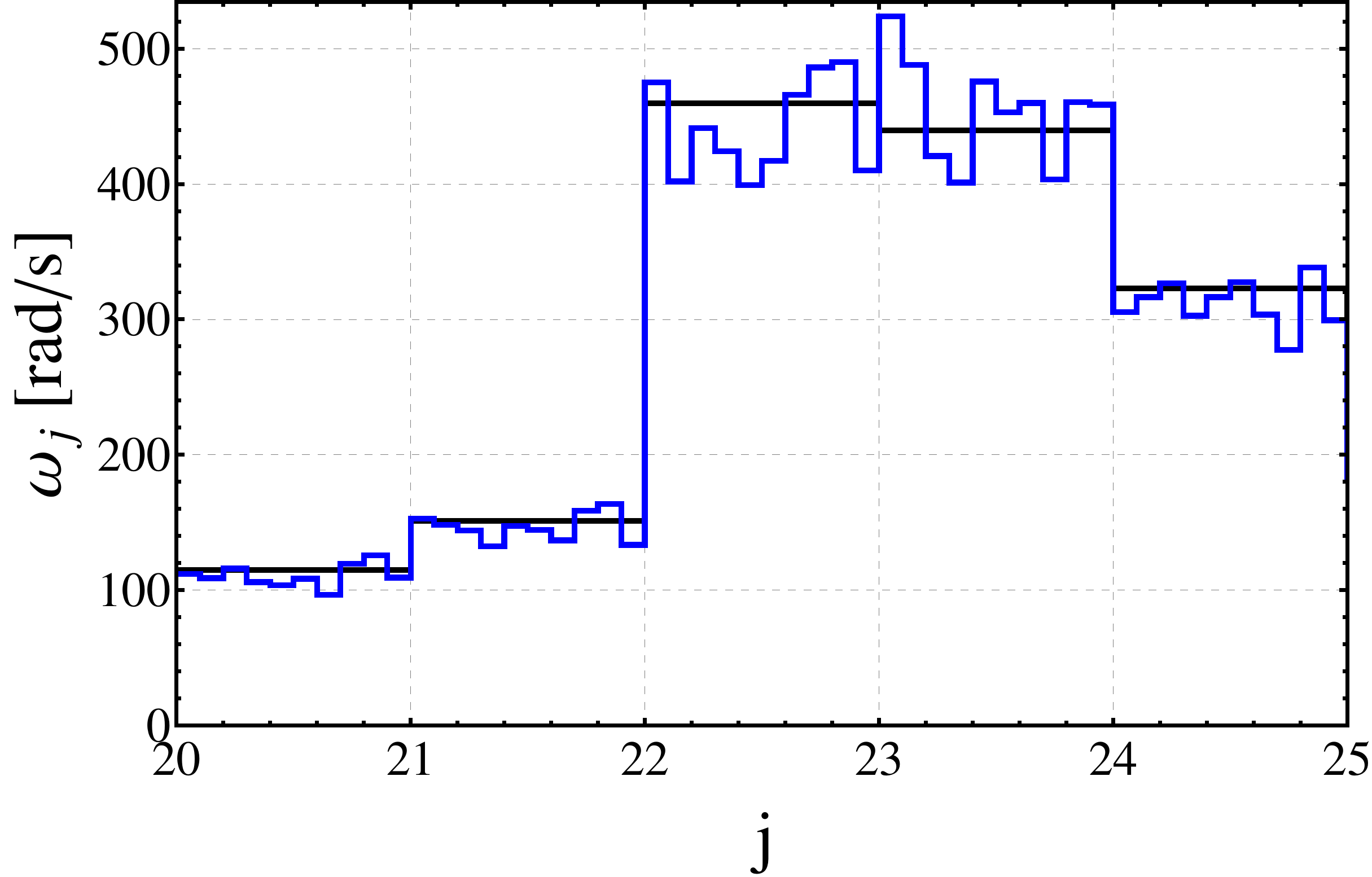}\label{fig:phase-accumulation-zoom}
			}	
		\end{center}		
		\caption{Diskreter Verlauf des erfahrenen Offresonanzfeldes einer Trajektorie. Die Phase zur Zeit $t$ ergibt sich aus dem Integral über den Frequenzverlauf. In Abb. \ref{fig:phase-accumulation-zoom} ist schematisch der zu einigen Schritten gehörende Verlauf von höher aufgelösten Mikrotrajektorien gezeigt. Die über die Mikrotrajektorien akkumulierte Phase weicht von der des grob aufgelösten Zeitschritts ab.}
		\label{fig:phase-accumulation}
	\end{figure}
\subsection{Phasenfehler pro Zeitschritt}
\begin{figure}
		\begin{center}		
			\subfloat[Radiale Verteilung der Fehler $\Delta\omega$ bei den Phaseninkrementen. Die Streuung um den $1/r^3$-Verlauf folgt aus dem zufälligen Verlauf der Mikrotrajektorien. Für Bereiche mit stärkerem Gradienten nimmt die Streuung daher zu.]{
				\includegraphics[width=0.45\textwidth]{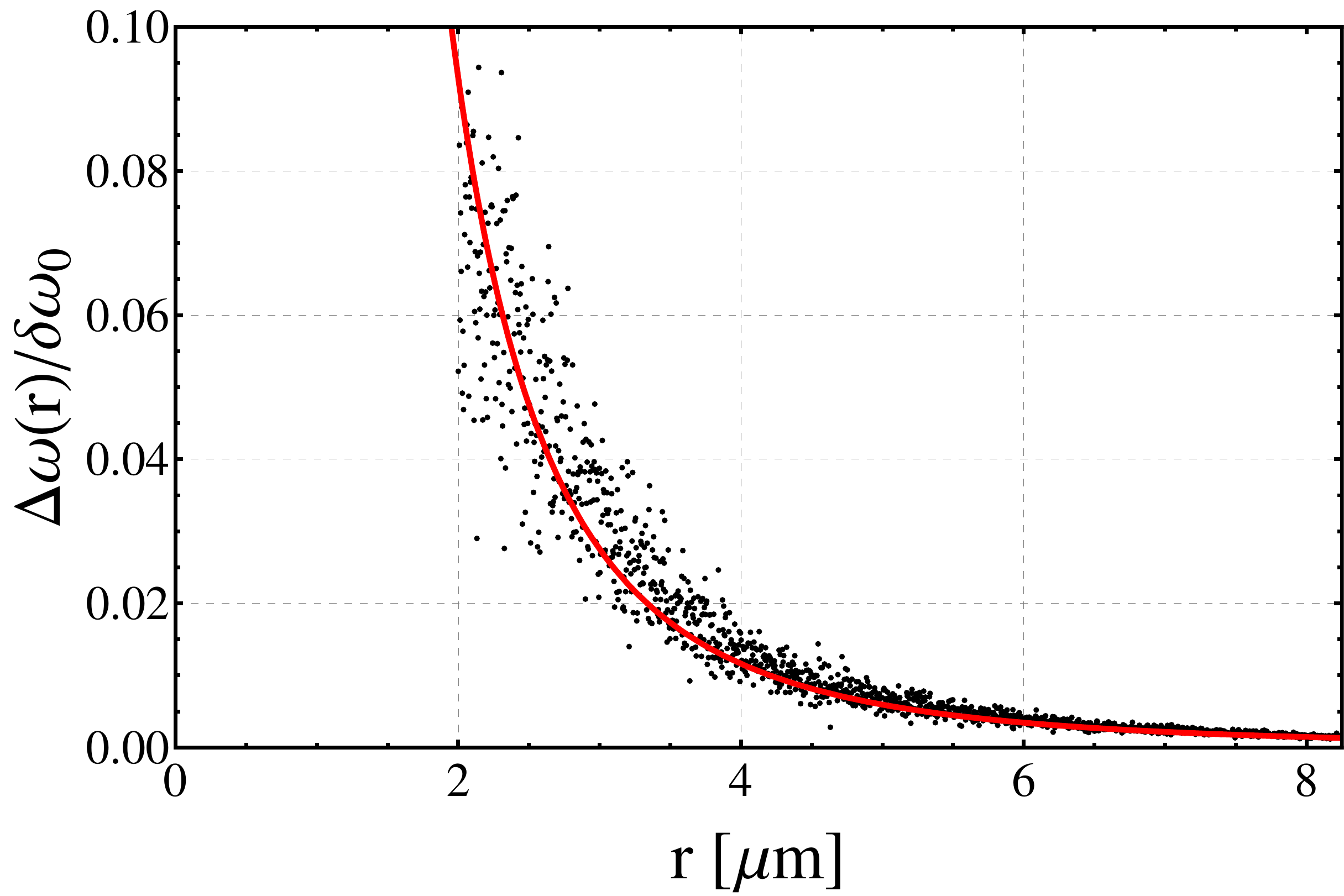}\label{fig:error-radius-dependency}
			}		
			\subfloat[Für größere Schrittweiten $\Delta t$ nimmt auch der Fehler $\Delta\omega$ bei den Phaseninkrementen zu. Für $\sigma\approx0.4\mum$ folgt ein relativer Fehler von $\omega(\vec{r})$ von ca. $10\%$.]{
				\includegraphics[width=0.45\textwidth]{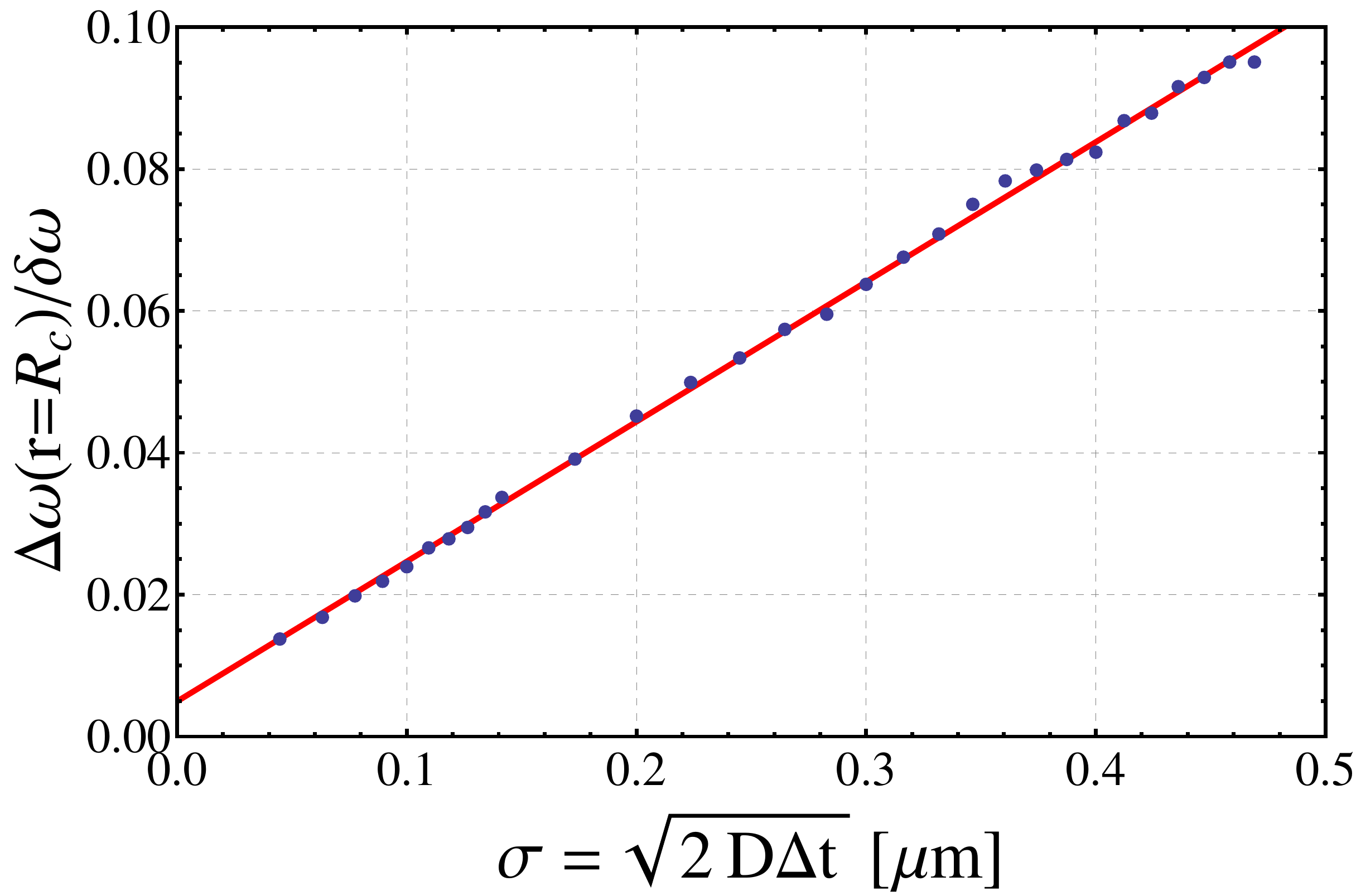}\label{fig:error-sigma-dependency}
			}
			\\
			\subfloat[Die Räumliche Verteilung der Fehler lässt keine Rückschlüsse auf eine mögliche Winkelabhängigkeit von $\Delta\omega$ zu.]{
				\includegraphics[width=0.62\textwidth]{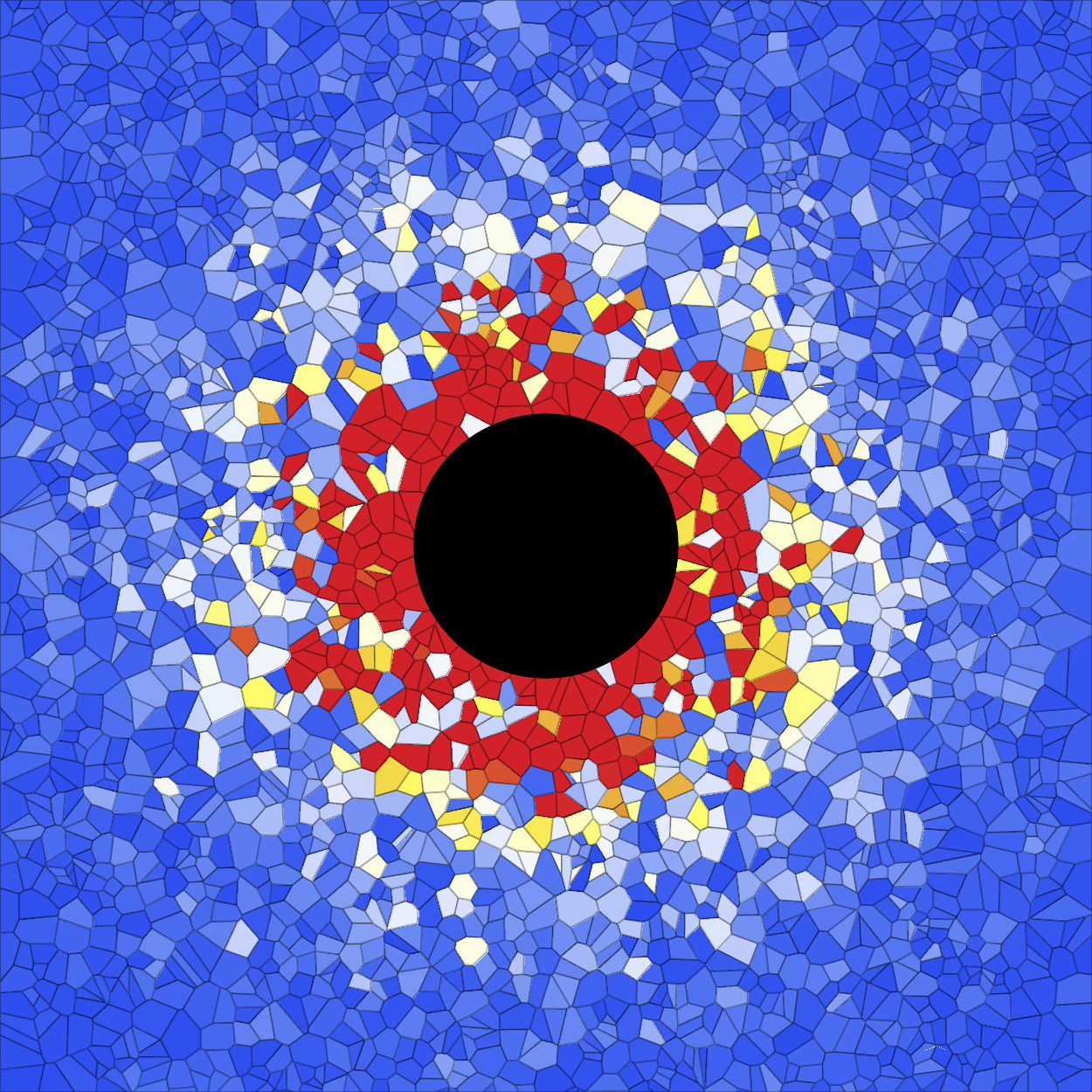}\label{fig:error-map}
			}
		\end{center}		
		\caption{Die Simulationsparameter waren $R_c\!=\!2\mum$, $\eta\!=\!5.9\%$, $\omega_0\!=\!936\radps$, $\sigma=0.447214\mum$ (entspricht $D=1\mum^2/\ms$ und $\Delta t=0.1\ms$), die Auflösung des Interpolationsgitters betrug $\Delta x=0.02\mum$. 	Zur Fehlerbestimmung wurden $100000$ zufällige Schritte erzeugt. Für die Ermittlung der radialen Abhängigkeit wurde über Kreisringe der Dicke $dr\!=\!4.125\mbox{nm}$ gemittelt. Der maximale relative Fehler von $\omega_j$ direkt am Rand der Kapillaren beträgt etwa $10\%$. Daraus ergibt sich ein maximaler Phasenfehler pro Zeitschritt von ca. $\delta\omega_0\cdot\Delta t\cdot10\%\approx0.01\pi$.}
		\label{fig:error-map-analysis}
	\end{figure}

In Gl. (\ref{eq:phase-accumulation}) gehen mit jedem Zeitschritt $j$ Fehler im Phaseninkrement
	\begin{align}
		\Delta t\omega_j=\Delta t(\overline{\omega_j}\pm\Delta\omega_j) = \Delta t (\overline{\omega(\vec{r})}+\Delta\omega(\vec{r}))
		\label{eq:phase-accumulation-error}
	\end{align}
einer Trajektorie ein. Diese sind zum Einen durch die zeitliche Diskretisierung des Random Walks (siehe Abb. \ref{fig:phase-accumulation-zoom}), zum Andern durch die räumliche Diskretisierung des Off\-resonanz\-feldes auf dem Interpolationsgitters verursacht. Man kann jedoch erreichen, dass der Fehler durch die Interpolation vernachlässigbar klein gegenüber dem der zeitlichen Diskretisierung ist. Dazu muss nur die Auflösung des Interpolationsgitters höher als die mittlere zurückgelegte Schrittlänge $\sigma$ sein. Dies ist in zwei Dimensionen und für "kleinere" Simulationsboxen leicht möglich. Es wird daher im Folgenden nur auf das Zustandekommen des ersten Effekts eingegangen. Ziel ist es eine Obergrenze für $\Delta t$ abzuschätzen und so durch geeignete Anpassung der Simulationsparameter die Rechenzeit erheblich zu verkürzen.\newline	
Zur Abschätzung werden zunächst kurze Trajektorien bei sehr kurzen mittleren Schrittweiten $\sigma'\ll\sigma$ erzeugt und deren Frequenzverlauf integriert (für die meisten durchgeführten Simulationen gilt $\sigma\geq0.1\mum$ und $\sigma'=0.005\mum$). Diese Trajektorien entsprächen bei normaler Schrittweite einem einzelnen Schritt $\sigma$ (vgl. Abb. \ref{fig:phase-accumulation-zoom}). Dadurch werden die mikroskopischen Bewegungen während eines einzelnen Zeitschritts $\Delta t$ berücksichtigt. Die Strecke von Start- und Endpunkt dieser Mikrotrajektorie wird dann als einzelner Zeitschritt mit Länge $\Delta t$ interpretiert und das mittlere Phaseninkrement $\overline{\omega(\vec{r})}=\omega(\frac{1}{2}(\vec{r}_{\mbox{\tiny start}}+ \vec{r}_{\mbox{\tiny end}}))$ berechnet. Ein Vergleich der Summe der mikroskopischen Phaseninkremente mit $\overline{\omega(\vec{r})}$ liefert eine Abschätzung für $\Delta\omega(\vec{r})$. Diese Abschätzung wird immer genauer je feiner die Mikrotrajektorie aufgelöst wird.\newline
Die Abhängigkeit von $\Delta\omega$ von $\vec{r}$ erklärt sich wie folgt: Ist das Offresonanzfeld $\omega(\vec{r})$ in der Umgebung von $\vec{r}$ wenig gekrümmt, so erwartet man kaum Abweichungen, da die numerische Integration über $\omega(\vec{r})$ gut funktioniert, bei starken Gradienten hingegen kann die Integration die akkumulierte Phase deutlich über- bzw. unterschätzen. Die einzelnen Fehler $\Delta\omega_j$ sind also primär abhängig vom Ort und der verwendeten Schrittweite $\sigma$.\newline
Abb. \ref{fig:error-radius-dependency} zeigt die radiale Verteilung von $\Delta\omega(\vec{r})$ für eine Simulation mit linearer Interpolation des Feldes ($\Delta x=0.02\mum$). In Abb. \ref{fig:error-sigma-dependency} ist die Zunahme des maximalen Fehlers $\Delta \omega_{\mbox{\tiny max}}=\Delta\omega(r=R_c)$ mit zunehmender mittlerer Schrittweite, bzw. steigender Diffusion bei gleich bleibendem $\Delta t$ gezeigt. Eine Winkelabhängigkeit der Fehler ist aus Abb. \ref{fig:error-map} nicht erkennbar.
Aus Gl. \ref{eq:single-capillary-field} folgt für den Betrag des Gradienten des Offresonanzfeldes
	\begin{align}
		\left|\vec{\nabla}\omega(\vec{r})\right|=\sqrt{\left(\frac{\partial\omega(\vec{r})}{\partial x}\right)^2+\left(\frac{\partial\omega(\vec{r})}{\partial y}\right)^2}=2\delta\omega_0{R_c}^2 \frac{1}{r^3}.
		\label{eq:kapillarfeld-gradient}
	\end{align}
Dies stimmt gut mit dem ermittelten Verlauf in Abb. \ref{fig:error-radius-dependency} überein. Der Fehler ist also hauptsächlich vom Betrag des Gradienten abhängig. Aus Abb. \ref{fig:error-sigma-dependency} ist auch eine lineare Abhängigkeit von $\Delta\omega_{\mbox{\tiny max}}$ von $\sigma=\sqrt{2D\Delta t}$ erkennbar. Man findet folglich insgesamt
	\begin{equation}
		\Delta\omega_{\mbox{\tiny max}}\  \widetilde{\propto}\  \delta\omega_0 \frac{R_c^2}{r^3}  \sigma .
		\label{eq:error-bla}
	\end{equation}
Bei stärkeren Offresonanzen und höheren Diffusionskonstanten müssen also kleinere Zeitschritte gewählt werden.\newline

\subsection{Fehlerfortpflanzung des Phasenfehlers pro Zeitschritt}
Aus Gl. (\ref{eq:phase-accumulation}) und Gl. (\ref{eq:phase-accumulation-error}) folgt für den Phasenfehler einer Trajektorie zum Zeitpunkt $t$
	\begin{align}
		\Delta\Phi(t)\leq\left\{	\begin{array}{cl}
													\mbox{A:} & 	\displaystyle\sqrt{m}\;\Delta t\;\Delta\omega_{\mbox{\tiny max}} 				\\
													\mbox{B:} & 	\displaystyle m		    \;\Delta t\;\Delta\omega_{\mbox{\tiny max}} 
								\end{array}\right. \ \ \mbox{mit}\ \ m=\left\lfloor \frac{t}{\Delta t}\right\rfloor.
		\label{eq:phasenfehler}
	\end{align}
Dabei gilt näherungsweise die optimistische Abschätzung A, da die $\Delta\omega_j$ weitestgehend zufällig verteilt sind. Im schlimmsten Fall (B) sind alle $\Delta\Phi_j$ gleich, solche Trajektorien sind jedoch äußerst unwahrscheinlich und fallen daher bei einer späteren Mittelung über alle Trajektorien kaum ins Gewicht. Wegen der Periodizität von $\cos$ und $\sin$ in Gl. (\ref{eq:magnetisierung}) sind kleine relative Fehler in der Offresonanz nicht ausreichend. Da für $\Delta\Phi (t)=\pi$ bereits jegliche Phaseninformation verloren ist folgt für die absoluten Fehler $\Delta \omega_{\mbox{\tiny max}}$ die Bedingung
	\begin{align}
		\renewcommand\arraystretch{1.5}
		\Delta\omega_{\mbox{\tiny max}} < \left\{	\begin{array}{cl}
											 \mbox{A:} & 	\displaystyle\frac{\pi}{\sqrt{m}\Delta t}	 					\\
											 \mbox{B:} & 	\displaystyle\frac{\pi}{ m	    \Delta t}			
								\end{array}\right..								
		\label{eq:max_phasenfehler}
	\end{align}
Für den (komplexen) Fehler der Magnetisierung $\Delta M(t)$ ergibt sich nach dem gaußschen Fehlerfortpflanzungsgesetz und Gl. (\ref{eq:magnetisierung})
	\begin{align}
		\left[\Delta M(t)\right]^2	&= \sum_{n=0}^{N-1}{\left[\left(\frac{\partial M(t)}{\partial\Phi_n(t)}\right)^2 \left(\Delta\Phi(t)\right)^2\right]}		\notag	\\	
									&= \frac{\Delta \Phi (t)^2}{N} \sum_{n=0}^{N-1}{\left[-\sin(\Phi_n(t))+i\cos(\Phi_n(t))\right]^2}.
		\label{eq:magnetisierungs_fehler}
	\end{align}
Wegen
	\begin{align}
		\sum_{n=0}^{N}{z_n z_n^{*}} \geq (\sum_{n=0}^{N}{z_n})(\sum_{n=0}^{N}{z_n})^{*}\quad\mbox{und}\quad\left|-\sin(\varphi)+i\cos(\varphi)\right|=1
		\label{eq:cauchy_schwarz}
	\end{align}
lässt sich eine Obergrenze für den Fehler der Magnetisierung zu
	\begin{align}
		|\Delta M(t)| \leq \Delta\omega_{\mbox{\tiny max}} \Delta t \sqrt{\frac{m}{N}}\quad\mbox{Fall A}
		\label{eq:abs-magnetisation-error}
	\end{align}
abschätzen.
Damit ist es nun möglich die Zeitschrittweite $\Delta t$ so zu wählen, dass zum Einen die Phaseninformationen im Simulationsintervall erhalten bleiben, zum Anderen aber nicht unnötig Rechenzeit durch eine zu feine Zeitauflösung verschwendet wird. Vergleiche einer analytischen Lösung für das Krogh-Modell mit Simulationsdaten (siehe Kap. \ref{kap:mathieu-analytic-vs-simulation}) sowie der Vergleich von zeitlich sehr hoch aufgelösten mit niedriger aufgelösten Simulationen zeigen das in Gl. (\ref{eq:max_phasenfehler}) von Fall (A) ausgegangen werden kann. Der Random Walk erweist sich somit trotz der lokalen Korrelationen der $\Delta\omega(\vec{r})$ auf einer Trajektorie als äußerst robust bzgl. Fehleranfälligkeit. Für Diffusionskonstanten $D\leq2\mum^2/\ms$ und Offresonanzen $\dom_0\leq1000\radps$ führt die meist verwendete Schrittweite $\Delta t=0.1\ms$ zu $\Delta\omega_{\mbox{\tiny max}}\leq150\radps$. Mit Gl. (\ref{eq:phasenfehler}) folgt für $m=10000$ (bzw. $t=1s$) der maximale Phasenfehler $\Delta\Phi(1s)\approx\pi/2$. Obwohl dieser Fehler bereits sehr hoch erscheint ist die Genauigkeit der Simulation deutlich höher als angenommen (vgl. Abb. \ref{fig:analytic-vs-simulation}). Dies liegt u.a. daran, dass die Trajektorien einen Großteil der Zeit in großem Abstand zu den Kapillaren verlaufen, und somit der mittlere Phasenfehler pro Zeitschritt deutlich unter $\Delta t\Delta\omega_{\mbox{\tiny max}}$ liegt.

\subsection{Statistische Fehler}
Im Zuge der Mittelung über die $N$ Trajektorien in Gl. \ref{eq:magnetisierung} wird während der Simulation zu\-sätz\-lich zu $M(t)$ auch die Standardabweichung $\Delta  M_{T,sim}(t)$ von $\Real(M(t))$ und $\Imag(M(t))$ bestimmt. Um dabei nicht alle Trajektorien im Speicher behalten zu müssen wird der Verschiebungssatz
	\begin{equation}
		\sum_{i=1}^N \left(x_i - \bar{x}\right)^2 = \left( \sum_{i=1}^N x_i^2 \right) - N \bar{x}^2= \left( \sum_{i=1}^N x_i^2 \right) - \frac{1}{N}\left(\sum_{i=1}^N x_i\right)^2
	\label{eq:verschiebungssatz}
	\end{equation}
angewandt. Vergleiche von Simulationsdaten mit verschiedenen $N$ mit der analytischen Lösung des Krogh-Modells zeigen sowohl die erwartete Abnahme von $\sigma(t)$ mit $\sqrt{N}$ (bis hin zu $N=2048000$), als auch eine gute Übereinstimmung von Simulation und Theorie. Für $N\approx50000$ überwiegt $\sigma(t)$ zudem deutlich den aus Gl. \ref{eq:abs-magnetisation-error} folgenden Fehler.
	
\subsection{Analytische Lösung vs. Simulation}\label{kap:mathieu-analytic-vs-simulation}
\begin{figure}		
		\begin{center}\includegraphics[width=0.75\textwidth]{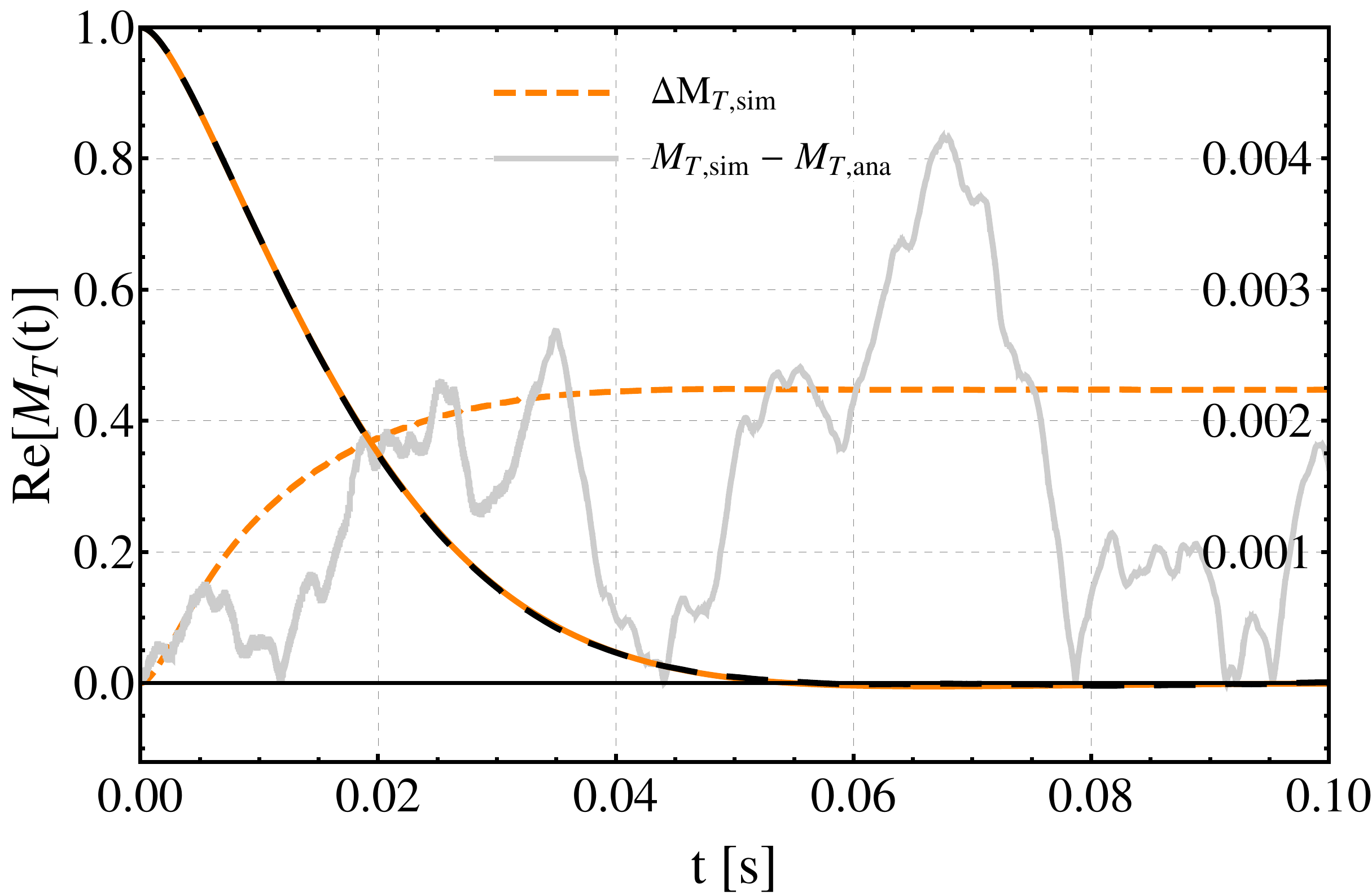}\end{center}
		\caption{Vergleich der analytischen Lösung nach \cite{ZienerPHDThesis} mit Simulationsdaten (linke Skala). Nur bei Betrachtung der Differenz $M_{T,sim} - M_{T,ana}$ ist ein Unterschied im Verlauf erkennbar (rechte Skala). Dieser Unterschied liegt etwa in der Größenordnung der statistischen Fehler $\Delta M_{T,sim}$ der Simulation (rechte Skala). ($\delta\omega_0=936\mathsf{rad/s}$, $D=1\mum^2/\ms$, $R_c=2\mum$, $\eta=5.33\%$, $N=100000$)}
		\label{fig:analytic-vs-simulation}
	\end{figure}

Mittels einer Implementierung der Matrixgleichung aus Kap. \ref{kap:mathieu} in MATHEMATICA\textsuperscript{\textregistered} wurde der Winkelanteil des Separationsansatzes gelöst (siehe Anhang \ref{appendix:mathieu}). Die Eigenwerte $\lambda_{nm}$ (siehe \cite{ZienerPHDThesis}) der radialen Eigenfunktionen wurden numerisch bestimmt. Mit dem Spektrum der Eigenwerte $a_{2m}$ und $\lambda_{nm}$ wurde dann der analytische Signalverlauf des Krogh-Modells für verschiedene Parameterkombinationen $\dom_0$, $D$, $R_c$ und $\eta$ berechnet. Für hohe $\dom_0$ und $R_c$, bzw. den Grenzfall $D\,\rightarrow\,0$ steigt wegen rapide zunehmendem $q$ auch die Anzahl der zu berücksichtigenden Eigenwerte schnell an. Abb. \ref{fig:analytic-vs-simulation} zeigt eine exemplarische Gegenüberstellung von Simulationsdaten mit der aus der analytischen Lösung gewonnenen Reihendarstellung. Für die gewählten Parameter ergibt sich $q=3.745 \mathsf{i}$. Vom Eigenwertspektrum wurden nur die ersten vier Eigenwerte verwendet.

\chapter{Ergebnisse}
\renewcommand{\figurename}{Abb.}
\renewcommand{\tablename}{Tab.}

\section{Mathieu-Funktionen}\label{kap:mathieu}
\begin{figure}
		\begin{center}
		\subfloat[$m=0$, $r=0$]{
			\includegraphics[width=0.75\columnwidth]{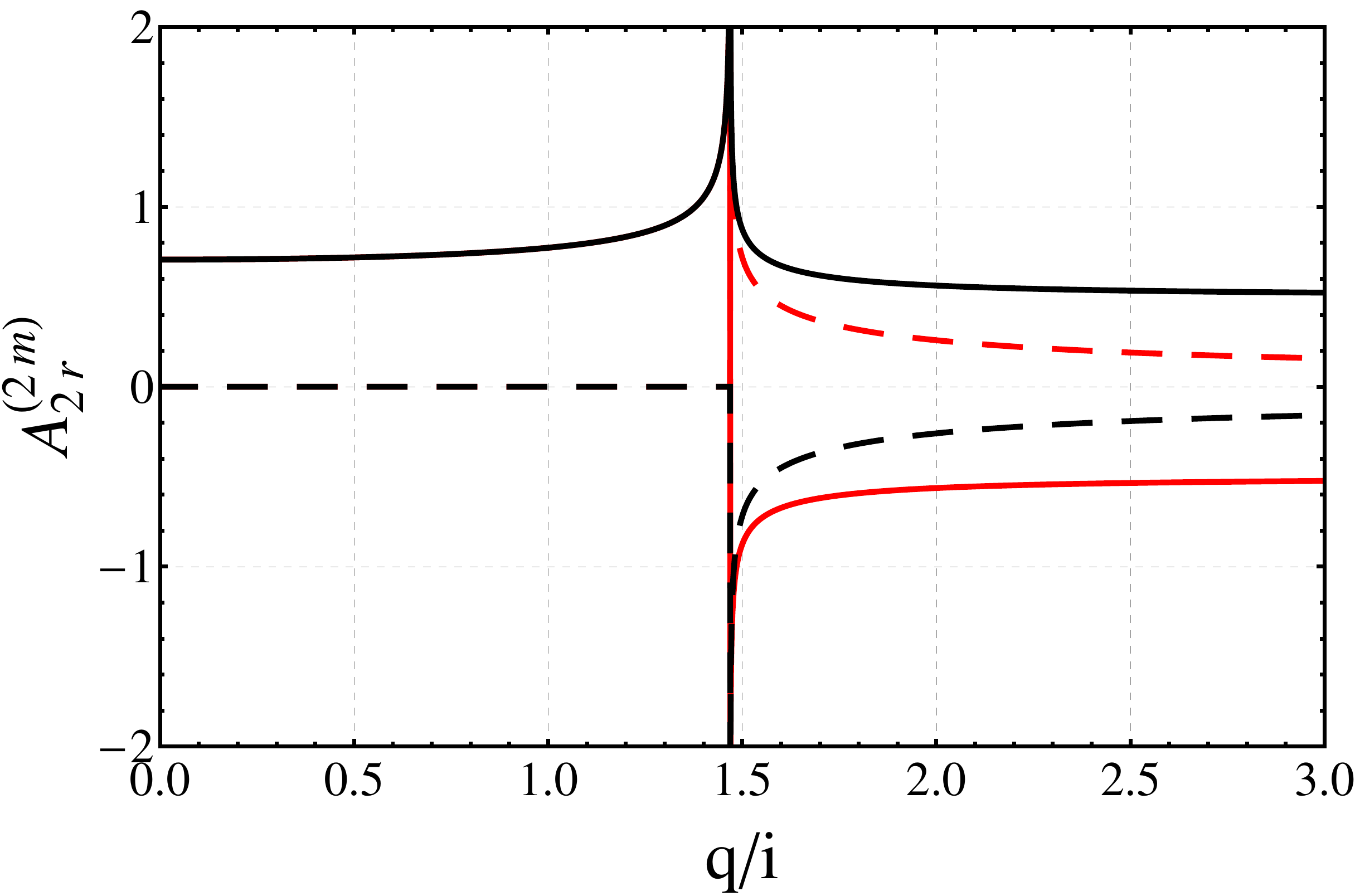}\label{fig:eigenvector-inversion-A00}
		}
		\\
		\subfloat[$m=4$, $r=4$]{
			\includegraphics[width=0.75\columnwidth]{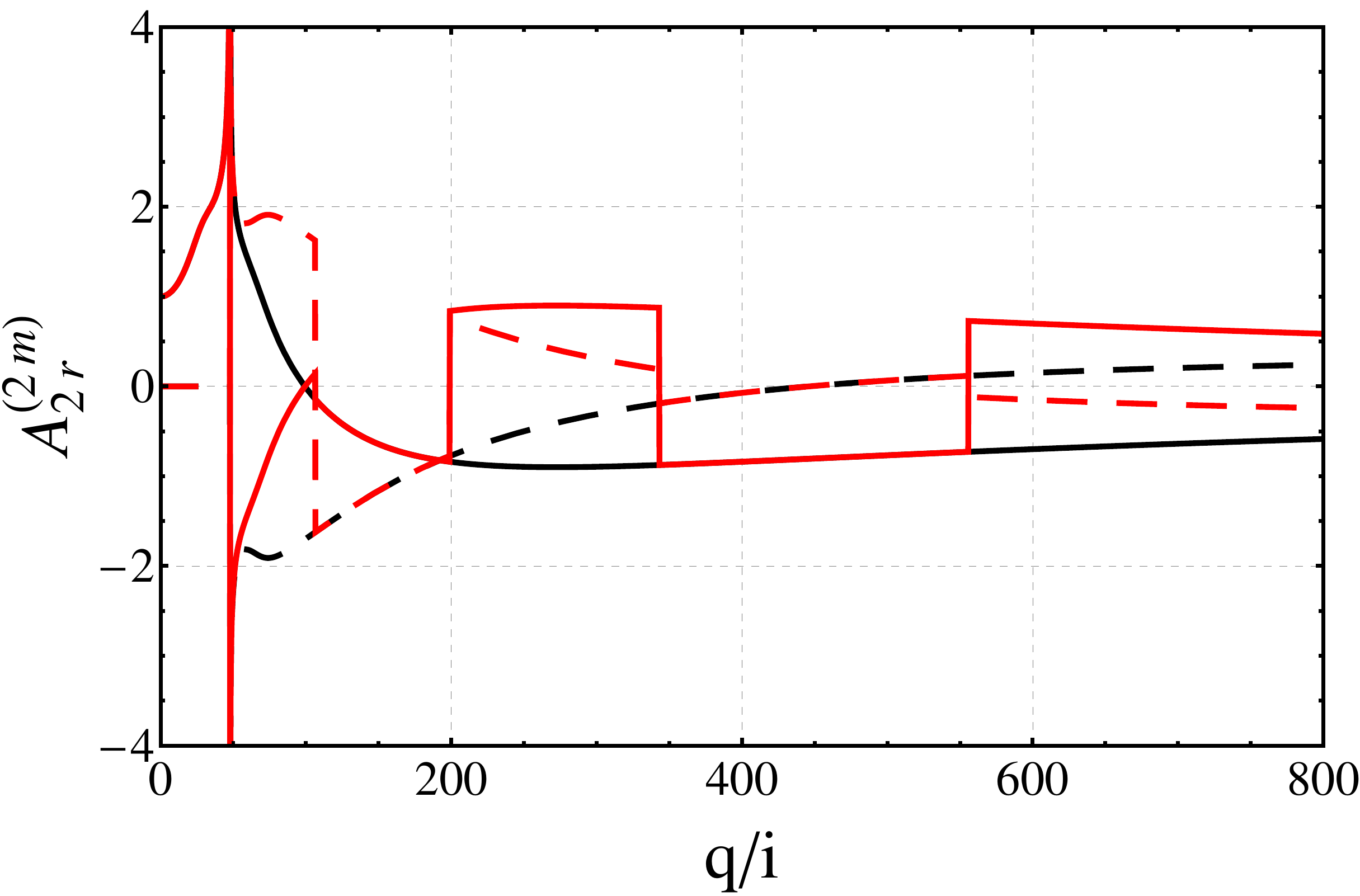}\label{fig:eigenvector-inversion-A1010}
		}
		\end{center}
		\caption{Das verwendete numerische Verfahren zum Berechnen der Eigenwerte (siehe Anhang \ref{appendix:mathieu}) konvergiert für bestimmte $q$ in die negativen Eigenvektoren $-\vec{A}'^{(2m)}$. Dadurch entstehen für einige Komponenten $r$ Sprünge zwischen $-A'^{(2m)}_{2r}$ und $+A'^{(2m)}_{2r}$ (rot). Auch $A_0^{(0)}$ zeigt ohne Korrektur für $q\,\rightarrow\,\infty$ falsches asymptotisches Verhalten (rot). Mit der Richtungsanpassung nach Gl. (\ref{eq:fourier-coefficients-sign}) die Komponenten stetig in $q$ (schwarz).\newline
		durchgezogen: $\Real(A'^{(2m)}_{2r})$, gestrichelt: $\Imag(A'^{(2m)}_{2r})$}
		\label{fig:eigenvector-inversion}
	\end{figure}

Für reflektive Randbedingungen an der Kapillare und dem äußeren Rand des Versorgungs\-zylinders kann die Bloch-Torrey-Gleichung (\ref{eq:bloch-torrey}) für das Krogh-Modell analytisch gelöst werden \cite{ZienerPHDThesis}. Durch Separation erhält man dabei für den Winkelanteil die Mathieu'sche Differentialgleichung
	\begin{align}
		\frac{\mathsf{d}^2\Phi_m(\varphi)}{\mathsf{d}\varphi^2}+\left[a_m-2q\cos{(2\varphi)}\right]\Phi_m(\varphi)=0
		\quad\mbox{mit}\quad q=\frac{\mathsf{i}R_c^2}{2 D}\delta\omega_0.
		\label{eq:mathieu-dgl} 
	\end{align}
Dabei ist $a_m$ eine von dem Parameter $q$ abhängige Separationskonstante. Im Gegensatz zu gut dokumentierten Mathieu-Differentialgleichungen mit reellem Parameter $q$ folgt aus der Bloch-Torrey-Gleichung allerdings ein rein imaginäres $q$. Vorhandene Softwarelösungen (z.B. MATHEMATICA\textsuperscript{\textregistered} und Matlab\textsuperscript{\textregistered}) stellen zwar für reelle $q$ Funktionen zur Verfügung, für den eher exotischen Fall rein imaginärer Parameter $q$ gibt es aber entweder keine oder nur fehlerhafte Implementierungen. Im Folgenden soll daher kurz auf die Eigenschaften der Mathieu-Funktionen eingegangen werden, welche eine korrekte Implementierung von Mathieu-Funktionen mit rein imaginärem Parameter $q$ ermöglicht. Die eigentliche Implementierung mittels MATHEMATICA\textsuperscript{\textregistered} findet sich im Anhang \ref{appendix:mathieu}. Eine ausführliche Beschreibung weiterer Eigenschaften der Mathieu-Funktionen findet sich z.B. in Kapitel 20 in \cite{AbramowitzStegun} oder in \cite{Mechel}.\newline
Wegen der Anfangsbedingungen und der Periodizität des Feldes kommen für die Lösung der Bloch-Torrey-Gleichung im Krogh-Modells nur die geraden, $\pi$-periodischen Mathieu-Funktionen $\Phi_m(\varphi)=\mathsf{ce}_{2m}(\varphi)$ in Frage. Diese sind orthonormal und lassen sich in Fourier-Reihen entwickeln:
	\begin{align}
		\int_{0}^{2\pi}\mathsf{d}\varphi \mathsf{ce}_{2m}(\varphi)\mathsf{ce}_{2m'}(\varphi)=\pi\delta_{m m'}\label{eq:mathieu-orthonormal}\\
		\mathsf{ce}_{2m}(\varphi)=\sum_{r=0}^{\infty}A_{2r}^{(2m)}\cos{(2 r \varphi)}.\label{eq:mathieu-fourrier-series}
	\end{align}
Setzt man Gl. (\ref{eq:mathieu-fourrier-series}) in Gl. (\ref{eq:mathieu-dgl}) ein, so erhält man eine Rekursionsformel für die $A_{2r}^{(2m)}$ \cite{AbramowitzStegun}
	\begin{align}
		\begin{array}{rcrrcrrcl}
					   & &			&		   		& & q & A_2^{(2m)}  		&=& a_{2m} A_0^{(2m)}		\\
		2 q A_0^{(2m)} &+& 4		& A_2^{(2m)}    &+& q & A_4^{(2m)}  		&=& a_{2m} A_2^{(2m)}		\\
	 q A_{2r-2}^{(2m)} &+& (2r)^2	& A_{2r}^{(2m)} &+& q & A_{2r+2}^{(2m)}		&=& a_{2m} A_{2r}^{(2m)}\quad\mbox{für}\quad r \geq 2 .   		\\
		\end{array}
	\end{align}
Dieses Gleichungssystem lässt sich (analog zu \cite{ChaosCador2002}) auch in Matrixform (Gl. (\ref{eq:mathieu-matrix-equation}) und Gl. (\ref{eq:mathieu-matrix-equation-short})) darstellen.
	\begin{align}
		\begin{pmatrix}
		 0         & \sqrt{2} q & 0			&  		&  		 &			&\\
		 \sqrt{2}q & 4          & q			& 0		&  		 &			&\\
		 0         & q			& 16		&\ddots	& 0 	 &			&\\
				   & 0			& \ddots	&\ddots	& q		 & 0		&\\
				   & 			& 0			& q		& (2r)^2 & \ddots	&\\
				   &  			& 			& 0		& \ddots & \ddots	&\\
		\end{pmatrix} \times\begin{pmatrix} \sqrt{2}A_0 \\ A_2 \\ A_4 \\ \vdots \\ A_{2r} \\ \vdots \\ \end{pmatrix}
		& =a_{2m}\begin{pmatrix} \sqrt{2}A_0 \\ A_2 \\ A_4 \\ \vdots \\ A_{2r} \\ \vdots \\ \end{pmatrix}\label{eq:mathieu-matrix-equation}
	\end{align}
	\begin{align}
		\widehat{T} \times \vec{A}^{(2m)} = a_{2m}  \vec{A}^{(2m)} \label{eq:mathieu-matrix-equation-short}
	\end{align}
Bei einer praktischen Anwendung dieser Matrixgleichung bricht man die Rekursion nach $k$ Schritten ab, löst also numerisch das Eigenwertproblem einer Submatrix $\widehat{T}'(k)$ von $\widehat{T}$ der Größe $k \times k$ und findet so genäherte Eigenwerte $a_{2m}'(k)$ und zugehörige Eigenvektoren ${\vec{A}}^{'(2m)}(k)$. In der Arbeit \cite{Ikebe1996599} wird die Konvergenz der Eigenwerte für $k\,\rightarrow\,\infty$ gezeigt. Zusätzlich liefert sie eine Abschätzung für die Fehler $a_{2m}-a_{2m}'(k)$ in Abhängigkeit von $q$ und $k$:
	\begin{align}
		a_{2m} - a_{2m}'(k) \approx q {A'}_{2k}^{(2m)}(k)\cdot{A'}_{2k+2}^{(2m)}(k+1)\label{eq:mathieu-eigenvalue-error-estimation} 
	\end{align}
Schätzt man die Größenordnung von $q$ aus Gl. (\ref{eq:mathieu-dgl}) ab, so ergibt sich für realistische Gewebeparameter $q\approx1\mathsf{i}$ ($D\approx1\mum^2/\ms$, $\delta\omega_0\approx100\radps$ und $R_c\approx4\mum$).\newline
Nach Gl. (\ref{eq:mathieu-eigenvalue-error-estimation}) sind die Fehler der Eigenwerte $a_{2m}'(k)$ (mit $m \leq 15$) für  $q$ bis zu $q \leq 250\mathsf{i}$ und einer Matrixgröße $k=25$ verschwindend gering. Für $k=25$ kann man also bis weit in den Bereich der statischen Dephasierung (z.B. bis zu $D\approx0.01\mum^2/\ms$) oder bis hin zu sehr hohen Offresonanzen ($\delta\omega_0\approx20000\radps$) bzw. zu sehr großen Radien ($R_c\approx50\mum$) Rechnungen mit minimalen Fehlern durchführen.\newline
Eine weitere Möglichkeit um festzustellen, ob die Matrixgröße $k$ eine ausreichend hohe Genauigkeit liefert, bietet die Parseval-Relation (Gl. (68) in \cite{ZienerPHDThesis}): 
	\begin{align}
	\sum_{m=0}^{k} \left[{A'}_0^{(2m)}(k)\right]^2=\frac{1}{2}.\label{eq:fourrier-coefficients-parselval}
	\end{align}
Für kleine $q$ ($q\approx1\mathsf{i}$) kann die Summe sogar schon für $m=4$ abgebrochen werden. Da die ${A'}_0^{(2m)}$ ab einem bestimmten Schwellwert $q_l$ des imaginären Parameters $q$ i.A. komplexwertig sind, sollte auf jeden Fall bis zu einem geradzahligen Index $k$ summiert werden. Nur dann können sich wegen Gl. (87) aus \cite{ZienerPHDThesis}
	\begin{align}
		A_{2r}^{(4l+2)}=(-1)^r A_{2r}^{(4l)*}\quad\mbox{für}\quad q > q_l
		\label{eq:mathieu-complex-conjugated-A4l}
	\end{align}
die imaginären Anteile der Summanden in Gl. (\ref{eq:fourrier-coefficients-parselval}) gegenseitig aufheben. Eine Summation bis zum Index $m=k$ ist allerdings nicht sinnvoll, da für $k\rightarrow m$ die ${A'}_0^{(2m)}$ wegen der endlichen Matrixgröße stark fehlerbehaftet sind. Zuletzt kann noch die Orthonormalitätsrelation (\ref{eq:mathieu-orthonormal}) überprüft werden.\newline
Sind die Eigenvektoren $\vec{A}^{(2m)}$ ermittelt, müssen sie noch nach Gl. (\ref{eq:mathieu-orthonormal}) normiert werden.\newline
Die numerischen Verfahren zur Lösung des Gleichungssystems (\ref{eq:mathieu-matrix-equation-short}) können außerdem für verschiedene $q$ auch in negative Eigenvektoren konvergieren. Diese Problematik wird in Abb. \ref{fig:eigenvector-inversion} dargestellt. Nach \cite{Mechel} ergeben sich für große $q$ die $A_{2r}^{(2m)}$ zu
	\begin{align}
		A_0^{(2m)}\approx2^{-m-1/4}\frac{\sqrt{(2m)!}}{m!}(\pi\sqrt{q})^{-1/4}.
		\label{eq:asymptodic-fourier-coefficients}
	\end{align}
Damit folgt für die Vorzeichen von Real- und Imaginärteil der $A_{0}^{(2m)}$ bei rein imaginärem, großem $q$
	\begin{align}
		\sgn\left(\Real\left(A_0^{(2m)}\right)\right)=+1\quad\mbox{und}\quad\sgn\left(\Imag\left(A_0^{(2m)}\right)\right)=-1.
		\label{eq:fourier-coefficients-sign}
	\end{align}
Für kleine $q$ lässt sich Gl. (97) aus \cite{ZienerPHDThesis} verwenden um Vorzeichen von Real- und Imaginär\-teil zu finden:
	\begin{align}
		A_0^{(2m)}\propto \mathsf{i}^m\quad\mbox{für}\quad q<p_{\left\lfloor m/2\right\rfloor}
		\label{eq:mathieu-fourier-coefficients-sign-2}
	\end{align}
Der Realteil von $A_0^{(0)}$ weist also an der Polstelle bei $q_0\approx1.5\mathsf{i}$ (siehe Abb. \ref{fig:eigenvector-inversion-A00}) keinen Vorzeichenwechsel auf und ist für alle $q$ positiv. Sollte das numerisch gefundene $A_0^{(0)}$ dies nicht erfüllen, so wird das Vorzeichen des kompletten Eigenvektors, also die Vorzeichen aller $A_{2r}^{(0)}$, invertiert (siehe Abb. \ref{fig:eigenvector-inversion}). Wegen Gl. (\ref{eq:mathieu-complex-conjugated-A4l}) sind auch die Vorzeichen für $m=1$ bekannt und der zugehörige Eigenvektor kann wenn nötig invertiert werden. Mit Hilfe von Gl. (\ref{eq:fourier-coefficients-sign}) und (\ref{eq:mathieu-fourier-coefficients-sign-2}) kann analog zu $m=0$ für jedes $m$ auf richtige Vorzeichen geprüft werden. Wegen des Vektorcharakters der $A_{2r}^{(2m)}$ reicht immer bereits die Kenntnis des richtigen Vorzeichens von $A_0^{(2m)}$ aus um den kompletten Vektor entsprechend zu korrigieren.\newline
Mit den MATHEMATICA\textsuperscript{\textregistered} Funktionen in Anhang \ref{appendix:mathieu} wird das von MATHEMATICA\textsuperscript{\textregistered} bereitgestellte Standardverfahren zur Lösung von Eigenwertproblemen genutzt. Dieses berücksichtigt jedoch nicht die speziellen oben beschriebenen Anforderungen an die Eigenwerte und Eigenvektoren. Die gefundenen Eigenwerte und Eigenvektoren müssen daher im Nachhinein den richtigen Mathieu-Funktionen zugeordnet werden. Die Eigenvektoren werden außerdem normiert und um ein möglicherweise falsches Vorzeichen korrigiert. Um schließlich eine Näherung der eigentlichen Mathieu-Funktion zu erhalten werden die numerisch gefundenen Eigenvektoren in Gl. (\ref{eq:mathieu-fourrier-series}) eingesetzt werden. Dabei ist auf die Gewichtung $\sqrt{2}$ in der $A_0^{(2m)}$ in Gl. (\ref{eq:mathieu-matrix-equation}) zu achten.\newline
Mit der Darstellung der Mathieu-Funktionen als Fourier-Reihe nach Gl. (\ref{eq:mathieu-fourrier-series}) kann nun der Winkelanteil $\Phi(\varphi)$ aus dem Separationsansatz in \cite{ZienerPHDThesis} berechnet werden. Für die komplexen Bessel-Funktionen, die für den Radialteil der Separation benötigt werden, stellt MATHEMATICA\textsuperscript{\textregistered} bereits eine korrekten Implementierung zur Verfügung. Zeit-, Winkel- und Radialteil der Separation können jetzt für beliebige Parameter des Krogh-Modells berechnet werden, der räumliche und zeitliche Verlauf der transversalen Magnetisierung $m_T(\vec{r},t)$ ist vollständig bestimmt.

\renewcommand{\figurename}{Abb.}
\renewcommand{\tablename}{Tab.}

\section{Frequenzverteilungen von Quadratischem und Hexagonalem Gitter}
In den folgenden Kapiteln sollen die Unterschiede und Gemeinsamkeiten zwischen den Frequenzverteilungen von Krogh-Modell und hexagonalem bzw. quadratischem Gitter dargestellt werden. Das quadratische Gitter wurde mit dem Hintergedanken auf eventuell andere Anwendungsbereiche außerhalb der Biologie bzw. Medizin berechnet. Es stellt außerdem neben dem hexagonalen Gitter die zweite universelle zweidimensionale Gitter\-anordnung dar.\newline
Zur Berechnung der Frequenzverteilungen wurde Gl. (\ref{eq:offresonance-distribution}) in diskretisierter Form auf die Interpolationsgitter aus Kap. \ref{kap:feldberechnung} angewandt. Mit $\Delta x$ als Auflösung des Interpolationsgitters und $a\cdot b$ als Volumen der Einheitszelle ergeben sich $N$ Flächenelemente
	\begin{align}
 		N=\left\lceil \frac{a}{\Delta x}\right\rceil\cdot\left\lceil\frac{b}{\Delta x}\right\rceil,
	\end{align}
wobei die beiden Brüche jeweils auf die nächste ganze Zahl aufgerundet werden. Für jedes der Flächenelemente wurde mit dem in Kap. \ref{kap:recursive-collision-detection} vorgestellten Suchbaum geprüft, ob der Mittelpunkt des Flächenelements im Inneren einer Kapillare liegt. Dies führt dazu, dass das Histogramm der Offresonanzen nur aus ca. $N'=N(1-\eta)$ Stichproben aufgebaut wird.\newline

\subsection{Quadrat-Gitter}
\begin{figure}
		\begin{center}
		\subfloat[$\alpha=0^\circ$]{
			\includegraphics[width=0.45\columnwidth]{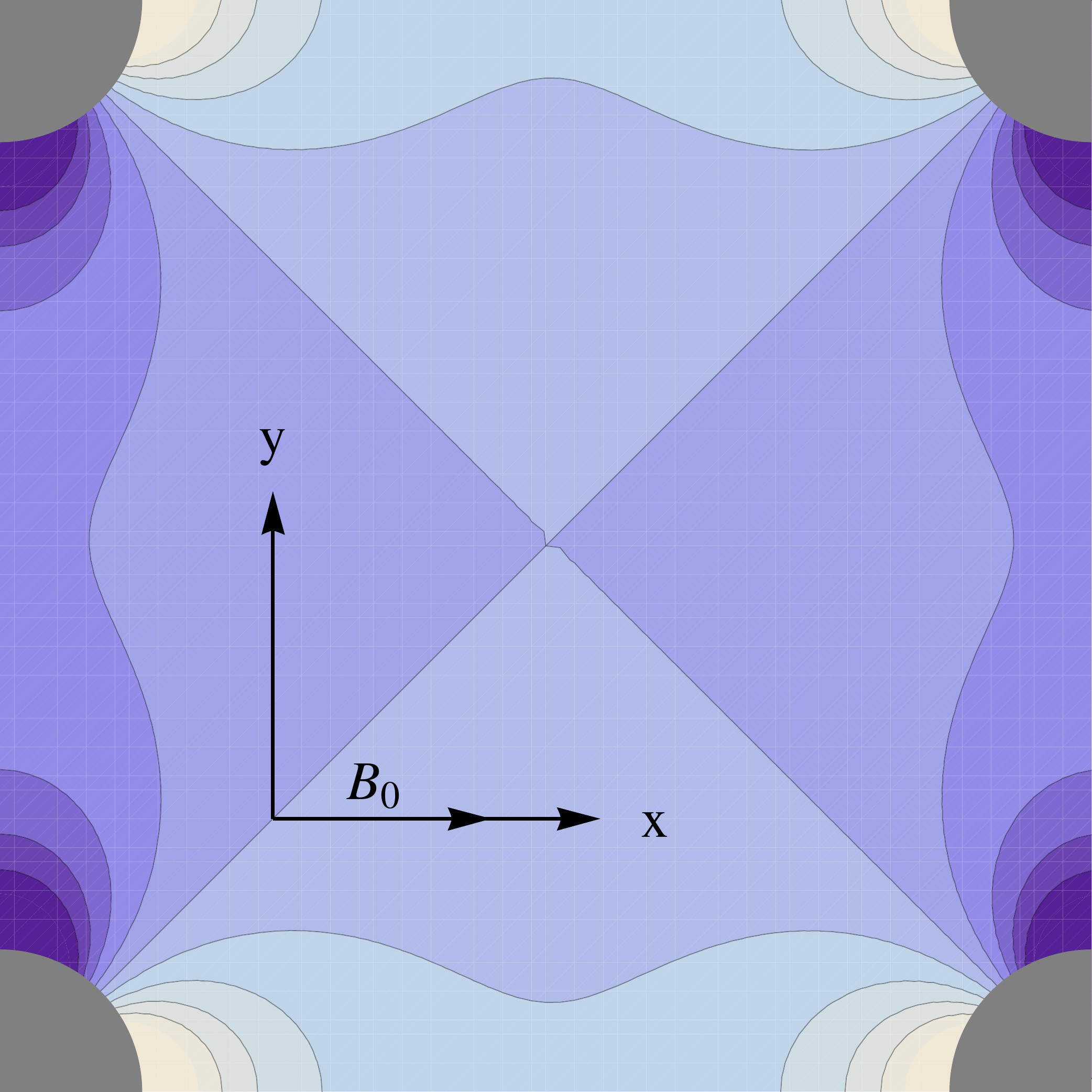}\label{fig:square-contour}
		}
		\subfloat[$\alpha=30^\circ$]{
			\includegraphics[width=0.45\columnwidth]{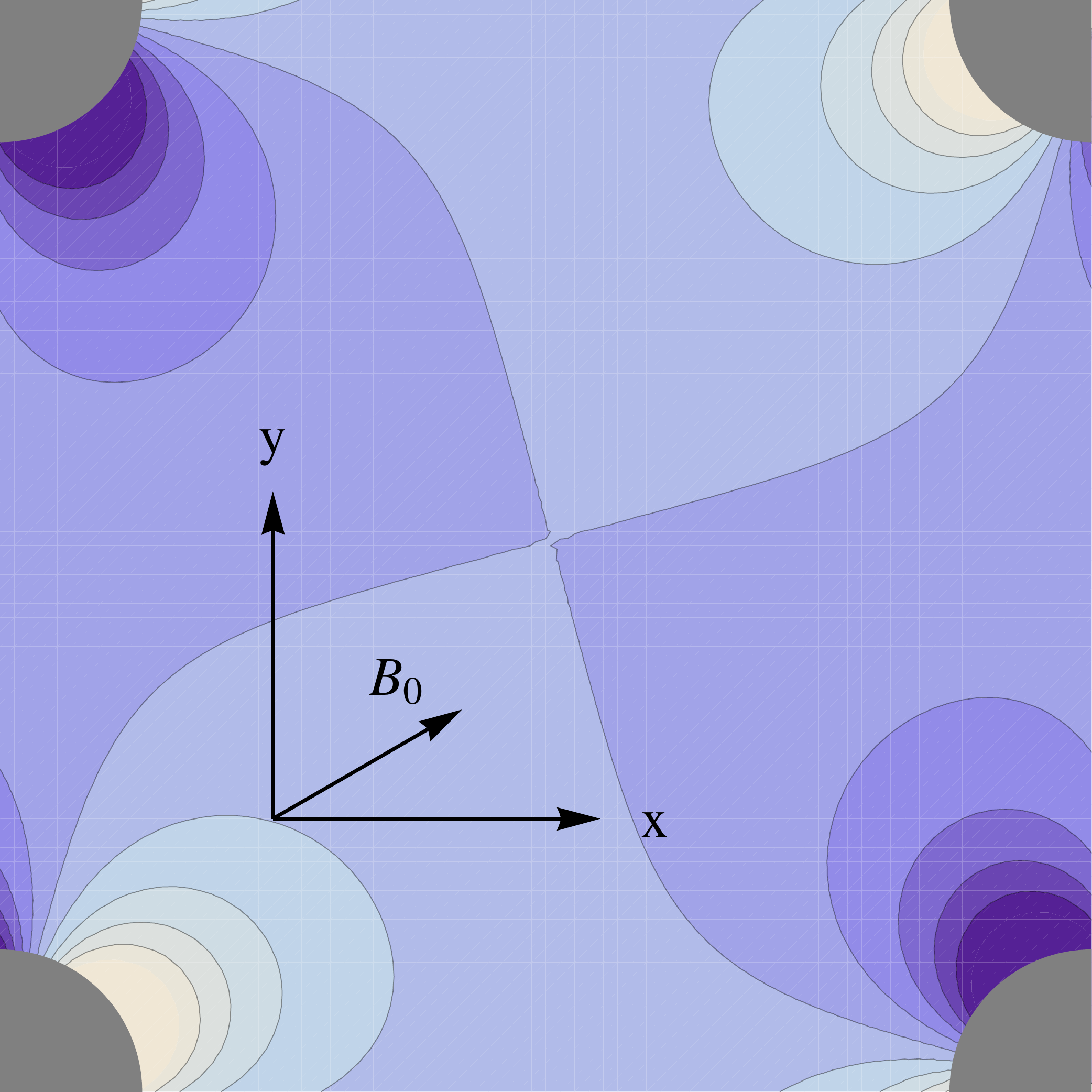}\label{fig:square-contour-30deg}
		}
		\end{center}
		\caption{Insgesamt beinhaltet eine Elementarzelle bzw. Simulationsbox des quadratischen Gitters nur eine Kapillare. Das Feld ist von der Verkippung zwischen Magnetfeld und Gittervektor abhängig.}
		\label{fig:square-contours}
	\end{figure}

\begin{figure}
		\begin{center}
		\subfloat[$\alpha=0^\circ$]{
			\includegraphics[width=0.49\columnwidth]{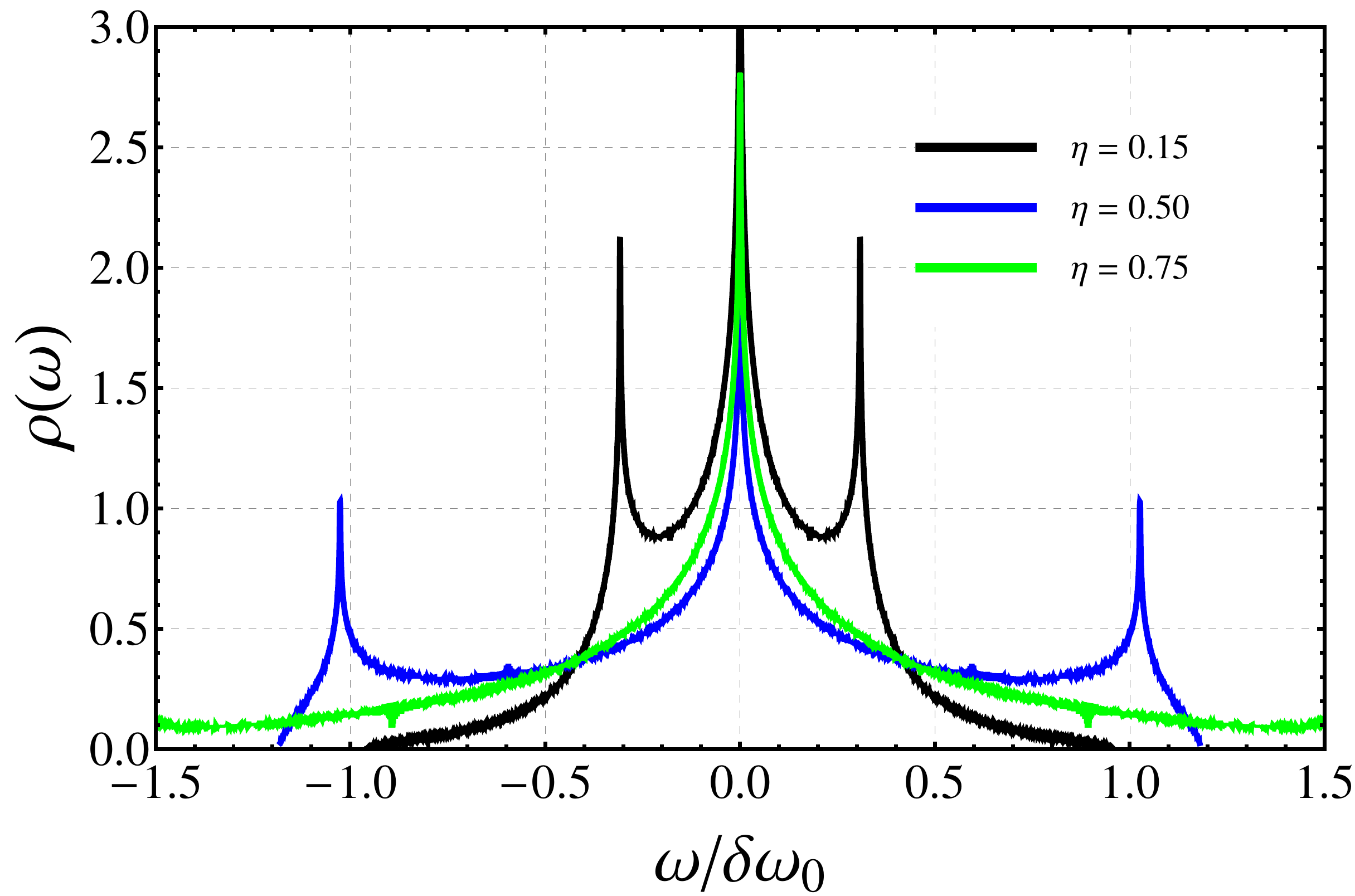}\label{fig:square-eta}
		}		
		\subfloat[$\eta\approx6.3\%$]{
			\includegraphics[width=0.49\columnwidth]{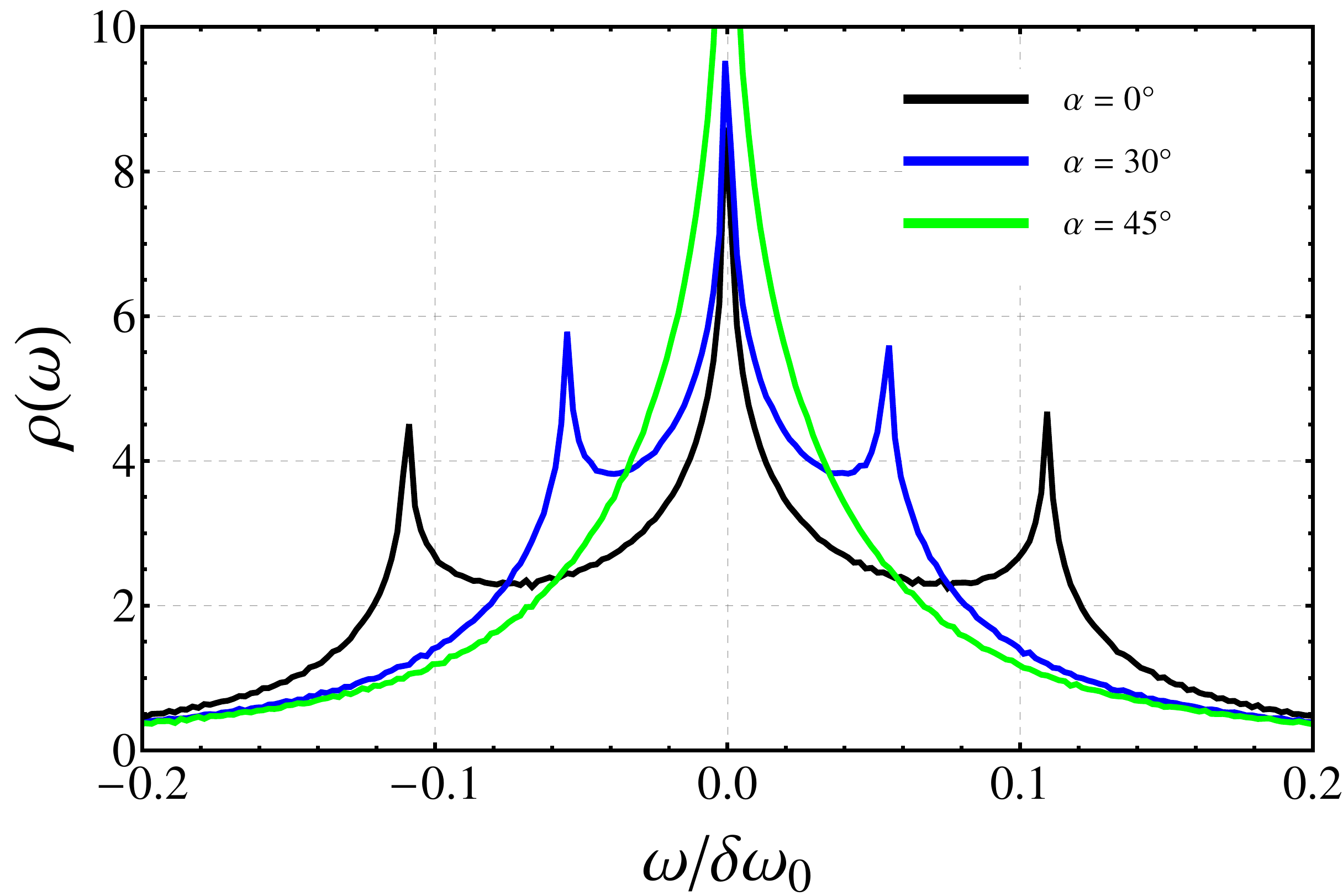}\label{fig:square-alpha}
		}
		\\
		\subfloat[$\alpha=0^\circ$]{
			\includegraphics[width=0.49\columnwidth]{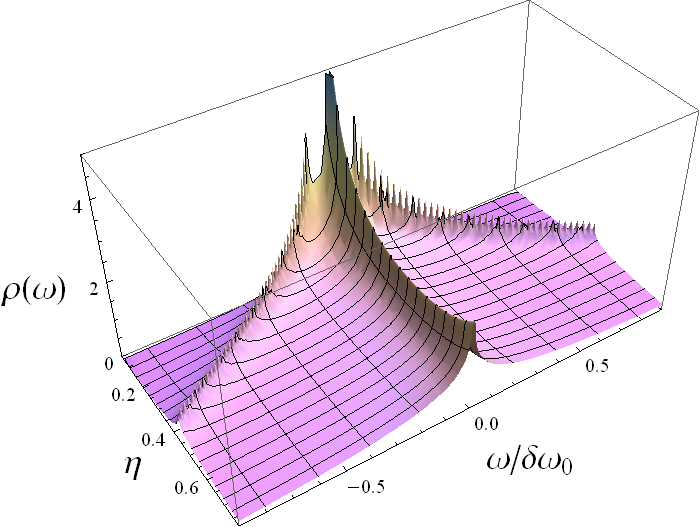}\label{fig:square-eta-3D}
		}		
		\subfloat[$\eta\approx6.3\%$]{
			\includegraphics[width=0.49\columnwidth]{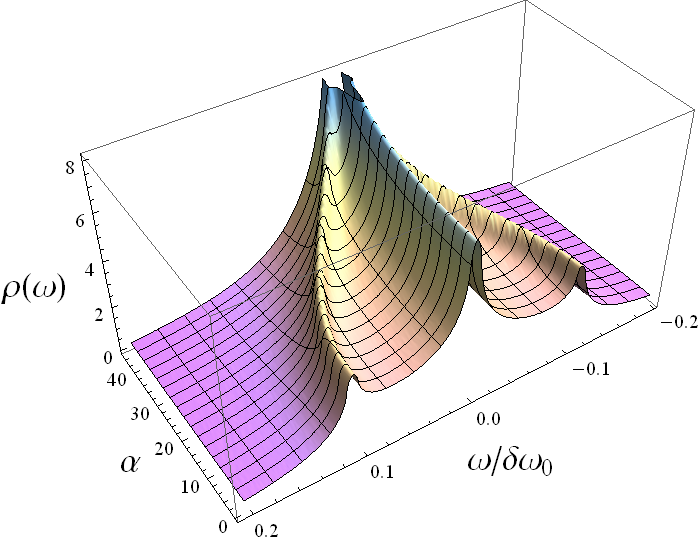}\label{fig:square-alpha-3D}
		}
		\\
		\subfloat[$\alpha=0^\circ$]{
			\includegraphics[width=0.49\columnwidth]{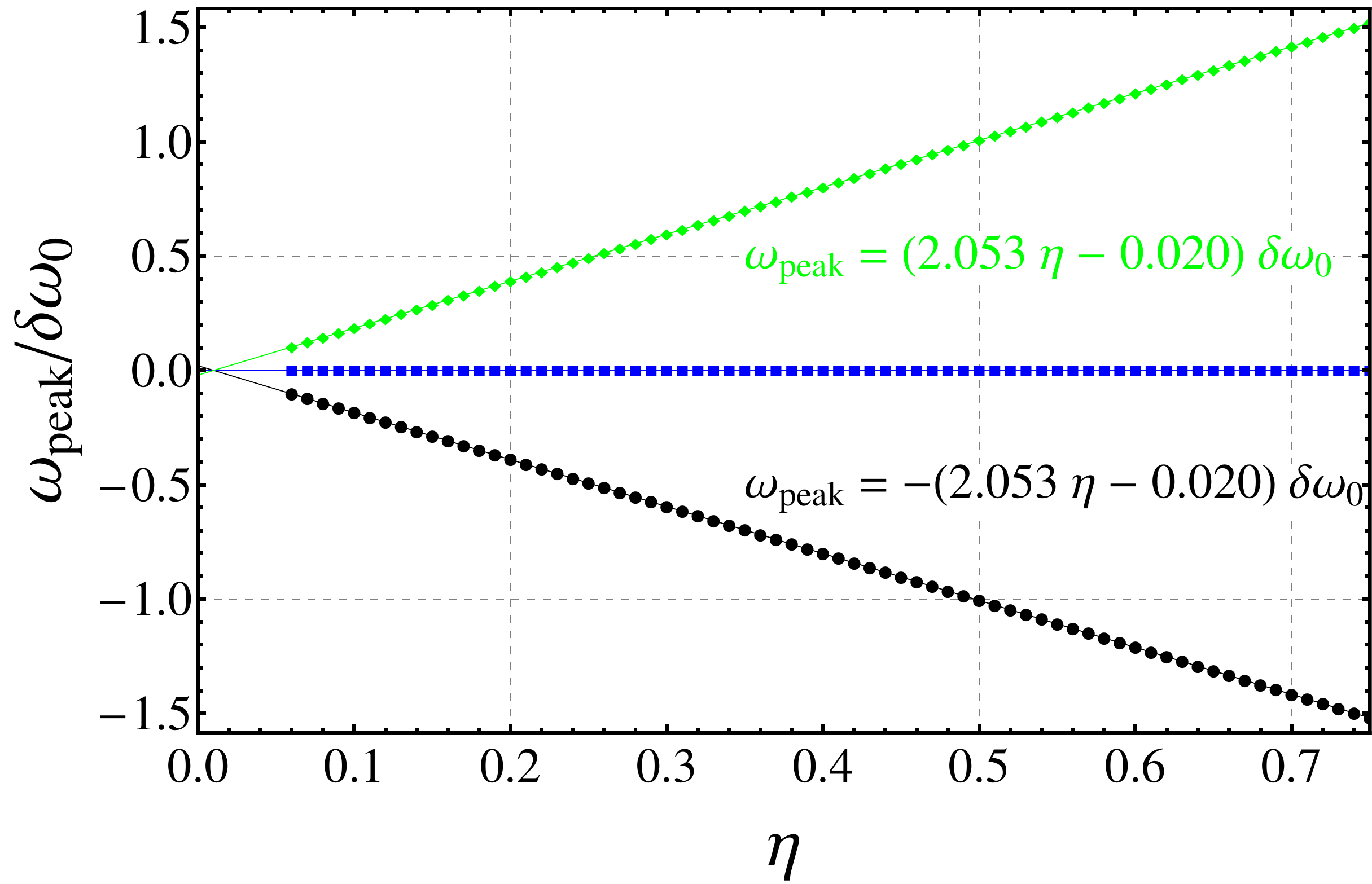}\label{fig:square-eta-peaks}
		}		
		\subfloat[$\eta\approx6.3\%$]{
			\includegraphics[width=0.49\columnwidth]{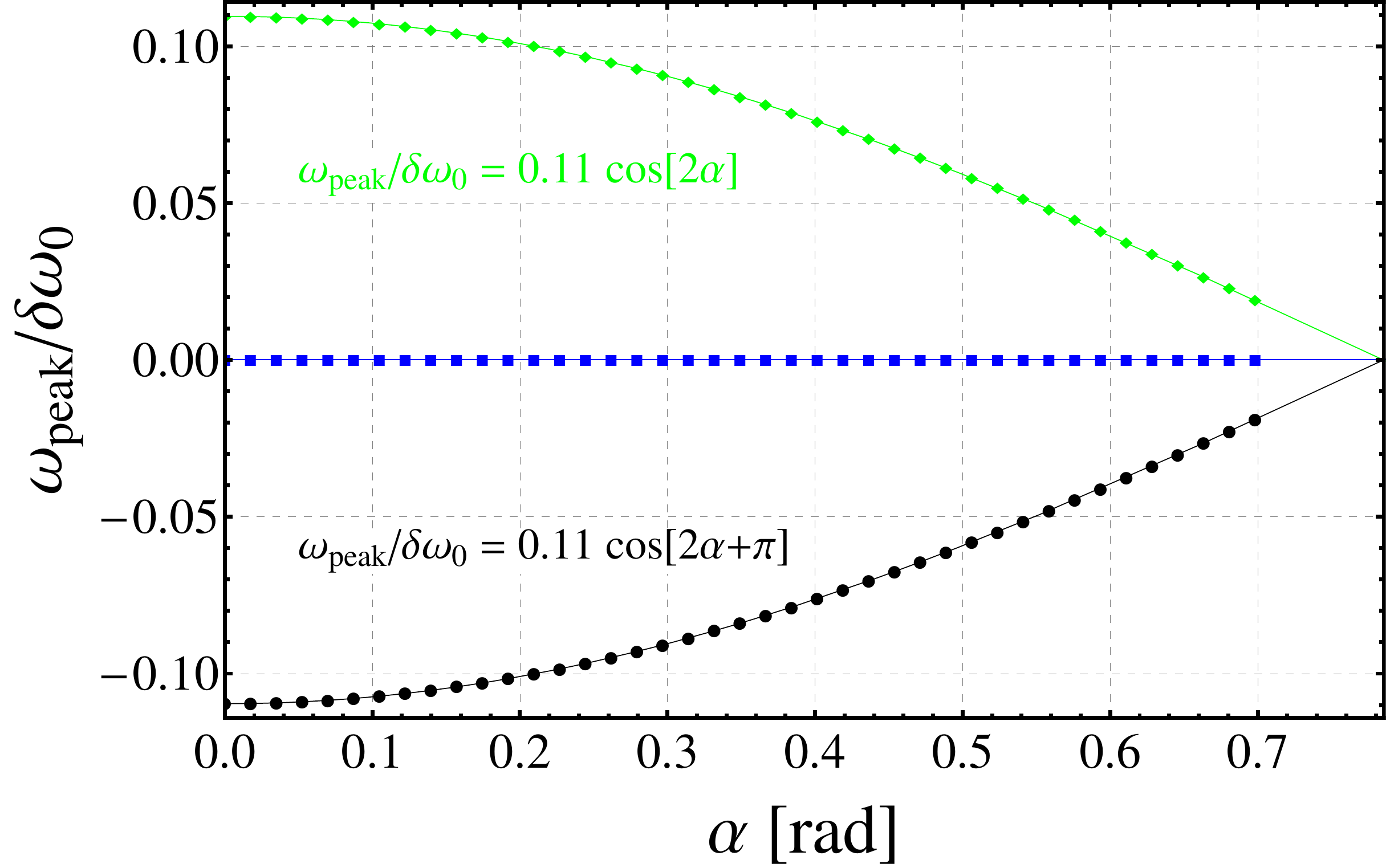}\label{fig:square-alpha-peaks}
		}
		\end{center}
		\caption{Die Frequenzverteilung $\rho(\omega)$ des quadratischen Gitters unter Variation von $\alpha$ und $\eta$. Bei konstantem $\alpha$ wandern die Peaks für steigendes $\eta$ linear nach außen. Bei konstantem $\eta$ wandern sie für $\alpha\rightarrow45^\circ$ nach innen. Das Rauschen in Abb. \ref{fig:square-eta} und in den Peaks von Abb. \ref{fig:square-eta-3D} wird durch die Diskretisierung des Feldes verursacht.}
		\label{fig:square-offres-distributions}
	\end{figure}

Die Felder des quadratischen Gitters wurden zunächst in einer Elementarzelle der Größe $15.355\mum\times15.355\mum$ mit einer Auflösung von $\Delta x=0.02\mum$ berechnet. Die Kapillarradien wurden so gewählt, dass sich $\eta=0.05$ ergibt. Dies führt zu $N'\approx560000$. Der Winkel $\alpha$ wurde von $\alpha=0^\circ$ bis zu $\alpha=45^\circ$ in $1^\circ$-Schritten erhöht. Wegen der Spiegelsymmetrie der Frequenzverteilung bzgl. der $\omega$-Achse ist die $\alpha$-Abhängigkeit so vollständig abgedeckt. Abb. \ref{fig:square-contours} zeigt exemplarisch den Unterschied der Felder für $\alpha=0^\circ$ und $\alpha=30^\circ$ in einer Elementarzelle.\newline
Ausgehend von $\eta=0.05$ können dann über die in Kap. \ref{kap:feldberechnung} erläuterte Skalierung auch Offresonanzverteilungen für $\eta$ bis hin zu $\eta_{\mbox{\tiny max}}=\frac{R_0^2\pi}{(2R_0)^2}=\frac{\pi}{4}\approx 78\%$ berechnet werden. Abb. \ref{fig:square-offres-distributions} zeigt eine Übersicht der Eigenschaften von $\rho(\eta,\alpha)$.\newline
Für zunehmendes $\eta$ wandern die Peaks von $\rho(\omega)$ wie im Krogh-Modell nach außen. Während jedoch im Krogh-Modell $\omega_{\mbox{\tiny peak}}=\eta\dom_0$ gilt, ergibt sich für das quadratische Gitter nach Abb. \ref{fig:square-eta-peaks} $\omega_{\mbox{\tiny peak}}\approx2\eta\dom_0$. Durch die Überlagerung der Felder der einzelnen Kapillaren liegt auch die maximale Offresonanz immer über $\dom_0$. Für $\eta_{\mbox{\small max}}$ ergibt sich  nach Abb. \ref{fig:square-eta-peaks} die maximale Offresonanz zu $\dom_{\mbox{\tiny max}}\approx\frac{\pi}{2}\dom_0$. Für Drehungen des Gitters um $\alpha$ ergibt sich ebenfalls ein einfacher funktionaler Zusammenhang. Insgesamt gilt
\begin{align}
		\begin{array}{rcrcl}
			\omega_{\mbox{\tiny peak}}(\eta) 	&=&\pm(2.053\eta-0.020)\dom_0 	& \mbox{für} & \alpha=0^\circ\\
			\omega_{\mbox{\tiny peak}}(\alpha)	&=&0.11\cos\left(2(\alpha+\frac{\pi}{4})\pm\frac{\pi}{2}\right)\dom_0 & \mbox{für} & \eta\approx6.3\%.\\
		\end{array}
		 \label{eq:static-square-functions}
\end{align}
Ob sich Steigung bzw. Amplitude aus Gl. (\ref{eq:static-square-functions}) für $\alpha\neq0^\circ$ und anderes $\eta$ ändern oder nicht, wurde nicht überprüft. Die Fit-Fehler von Steigung und Amplitude liegen weit unter $1\%$.
	
\subsection{Hexagonales Gitter}
\begin{figure}
		\begin{center}
		\subfloat[$\alpha=0^\circ$]{
			\includegraphics[width=0.45\columnwidth]{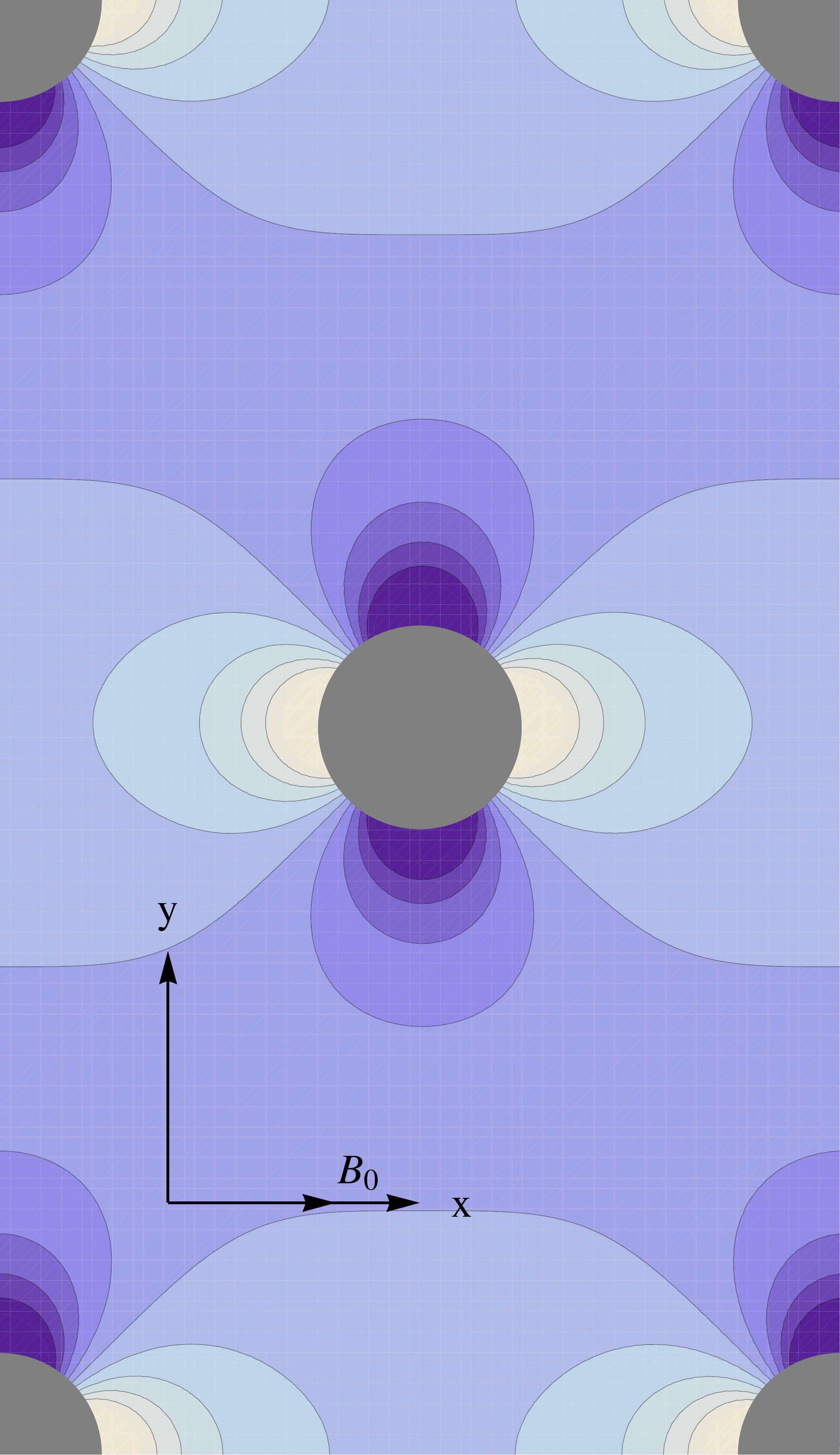}\label{fig:hex-contour}
		}
		\subfloat[$\alpha=10^\circ$]{
			\includegraphics[width=0.45\columnwidth]{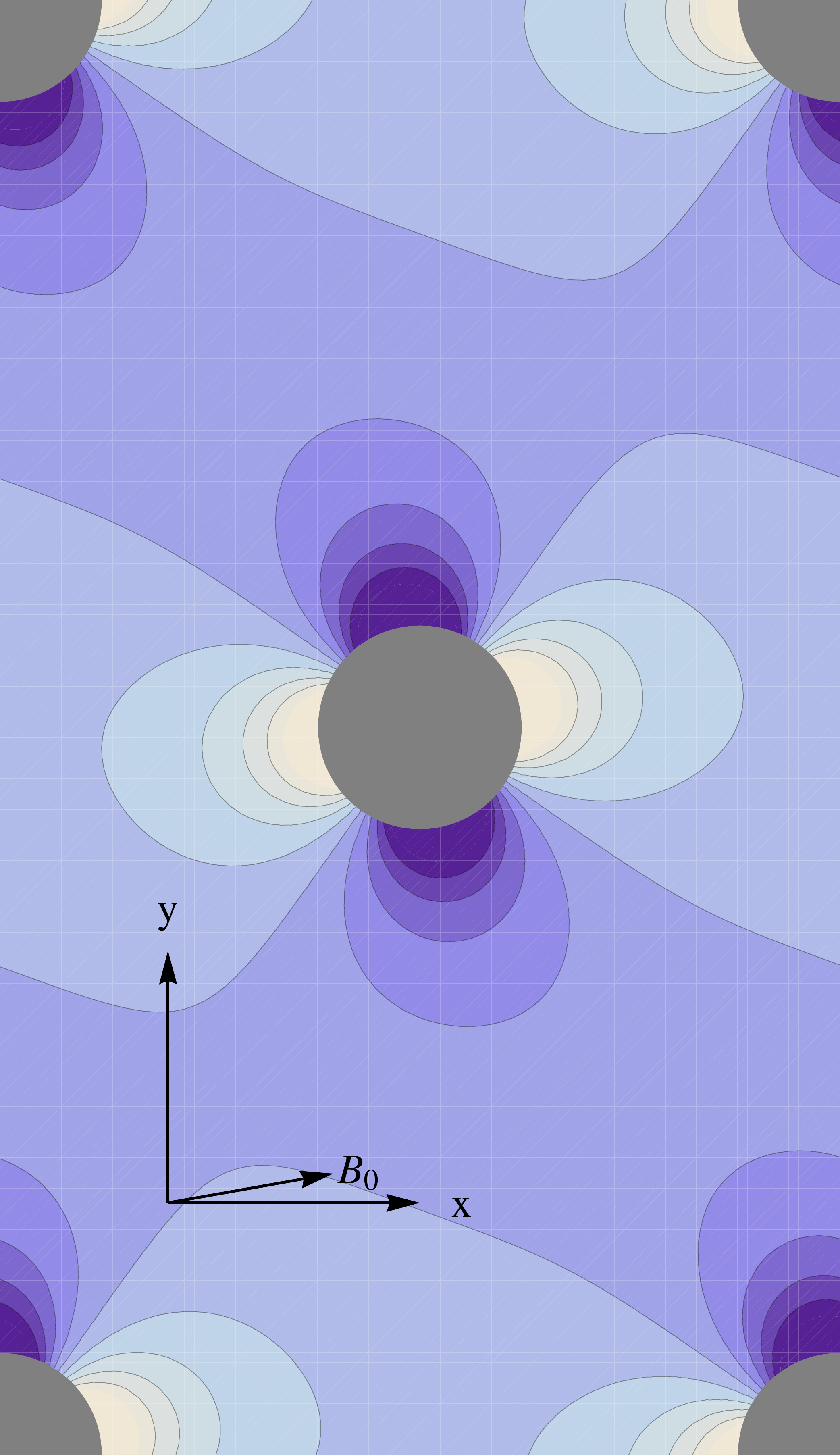}\label{fig:hex-contour-10deg}
		}
		\end{center}
		\caption{Feldverteilung eines regelmäßigen hexagonalen Kapillargitters. Da das Gitter nicht mehr rotationssymmetrisch ist wird der Winkel $\alpha$ benötigt um die Orientierung des Gitters bzgl. des Magnetfeldes zu beschreiben.}
		\label{fig:hex-contours}
	\end{figure}

\begin{figure}
		\begin{center}
		\subfloat[$\alpha=0^\circ$]{
			\includegraphics[width=0.49\columnwidth]{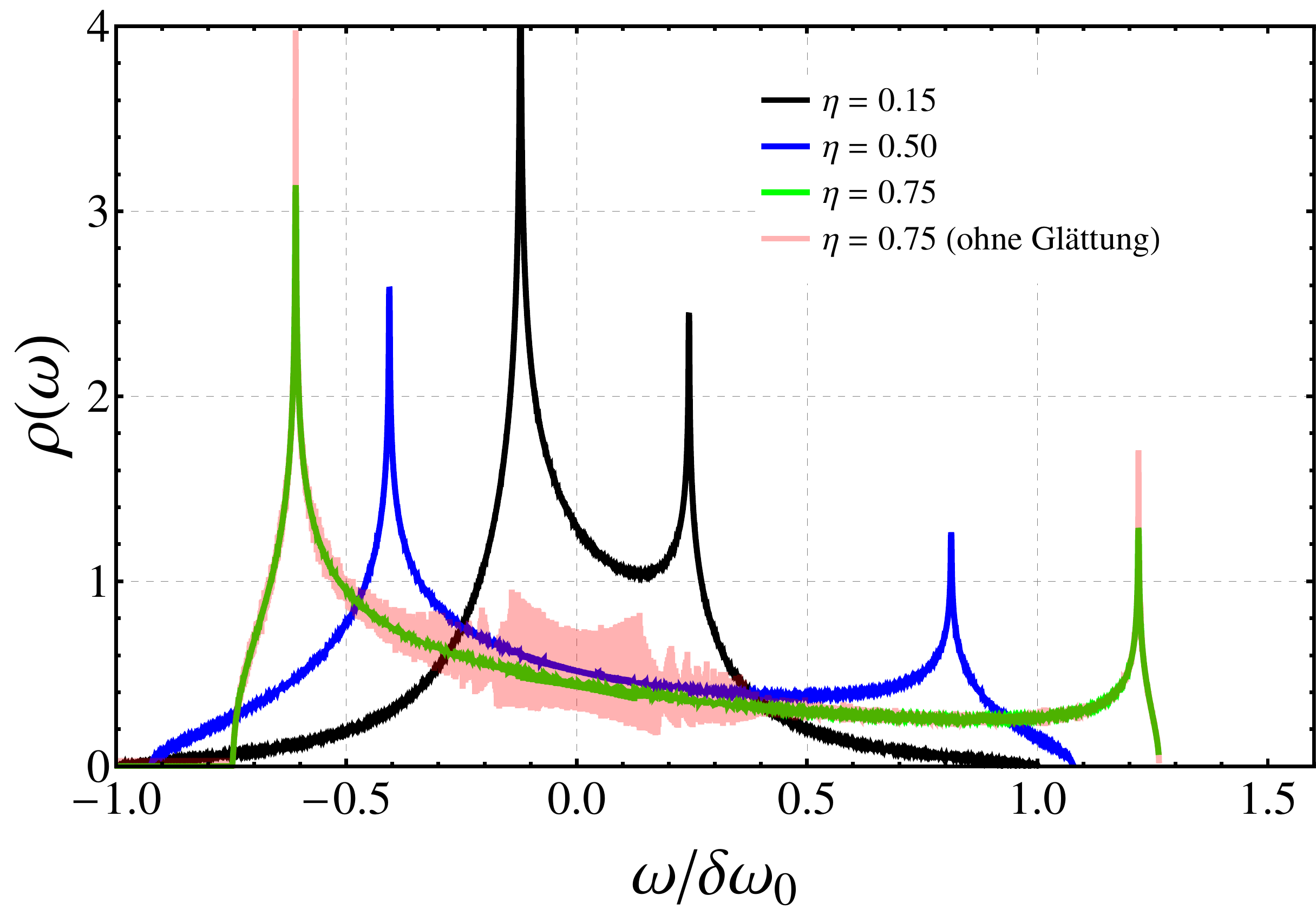}\label{fig:hex-eta}
		}		
		\subfloat[$\eta=10\%$]{
			\includegraphics[width=0.49\columnwidth]{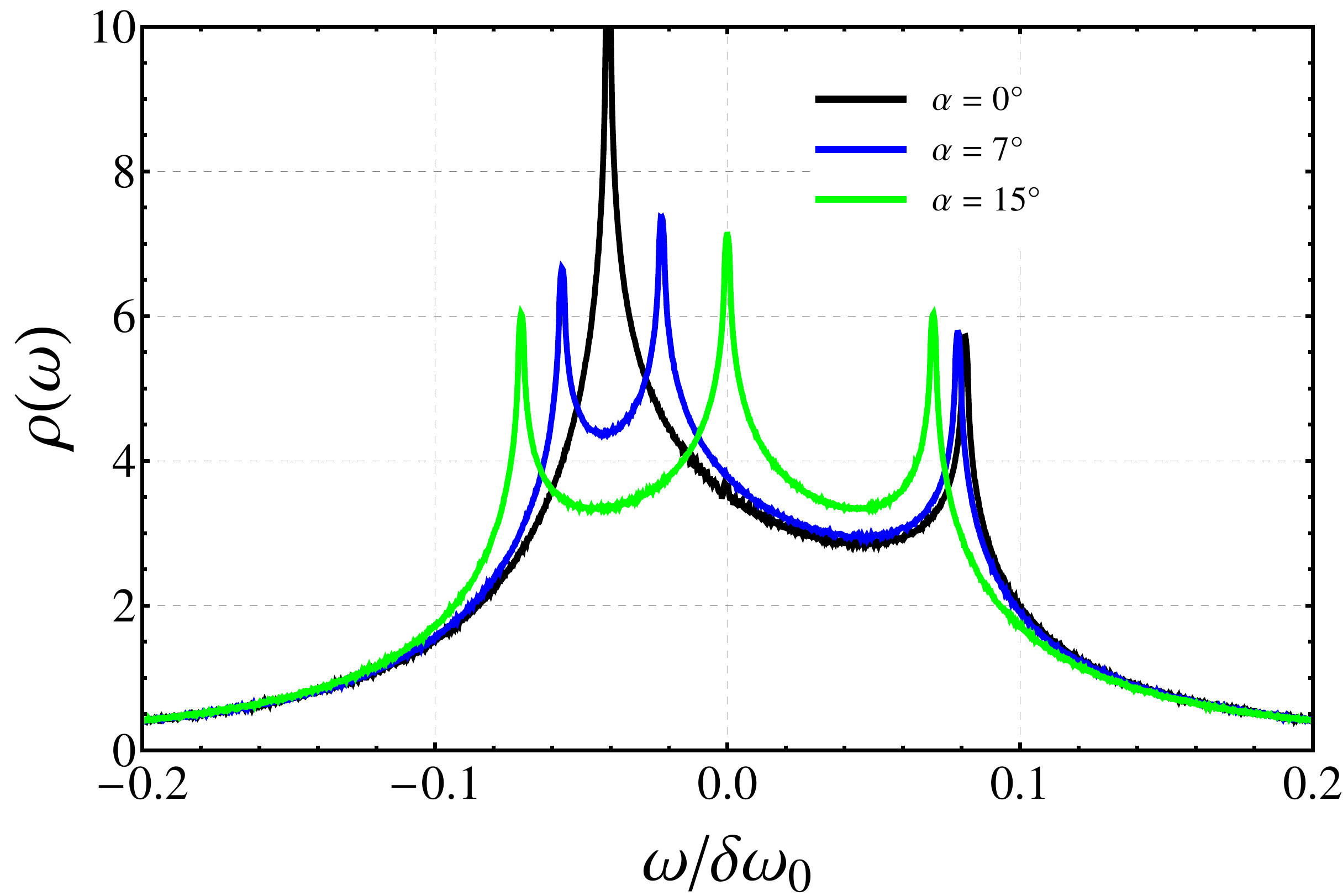}\label{fig:hex-alpha}
		}
		\\
		\subfloat[$\alpha=0^\circ$]{
			\includegraphics[width=0.49\columnwidth]{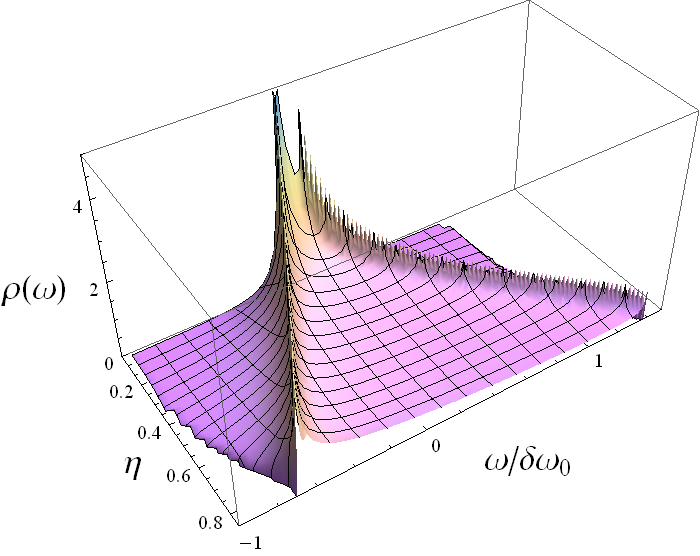}\label{fig:hex-eta-3D}
		}		
		\subfloat[$\eta=10\%$]{
			\includegraphics[width=0.49\columnwidth]{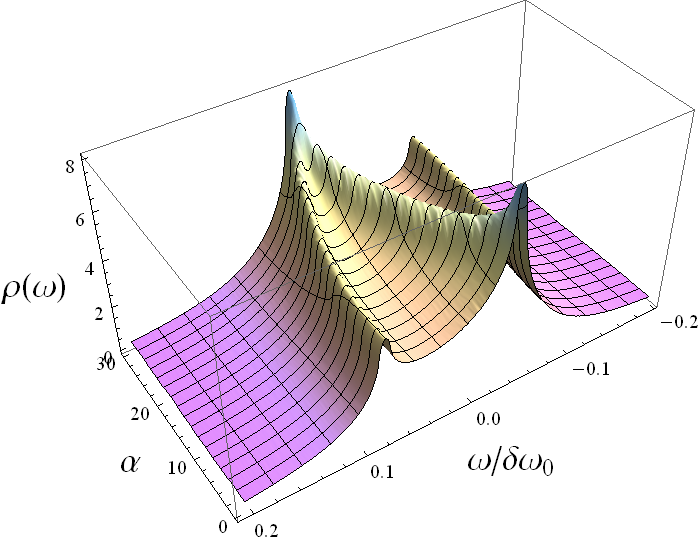}\label{fig:hex-alpha-3D}
		}
		\\
		\subfloat[$\alpha=0^\circ$]{
			\includegraphics[width=0.49\columnwidth]{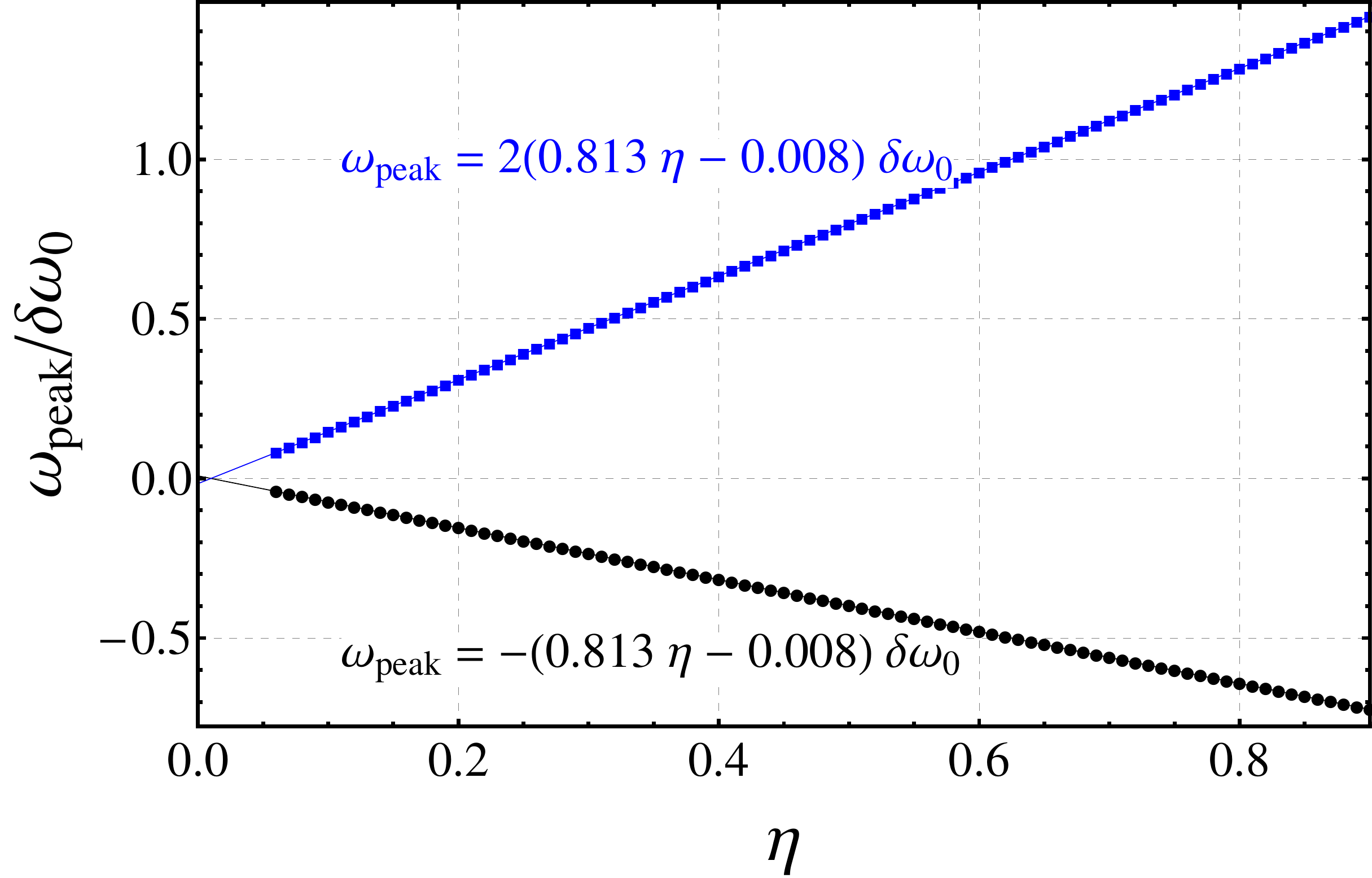}\label{fig:hex-eta-peaks}
		}		
		\subfloat[$\eta=10\%$]{
			\includegraphics[width=0.49\columnwidth]{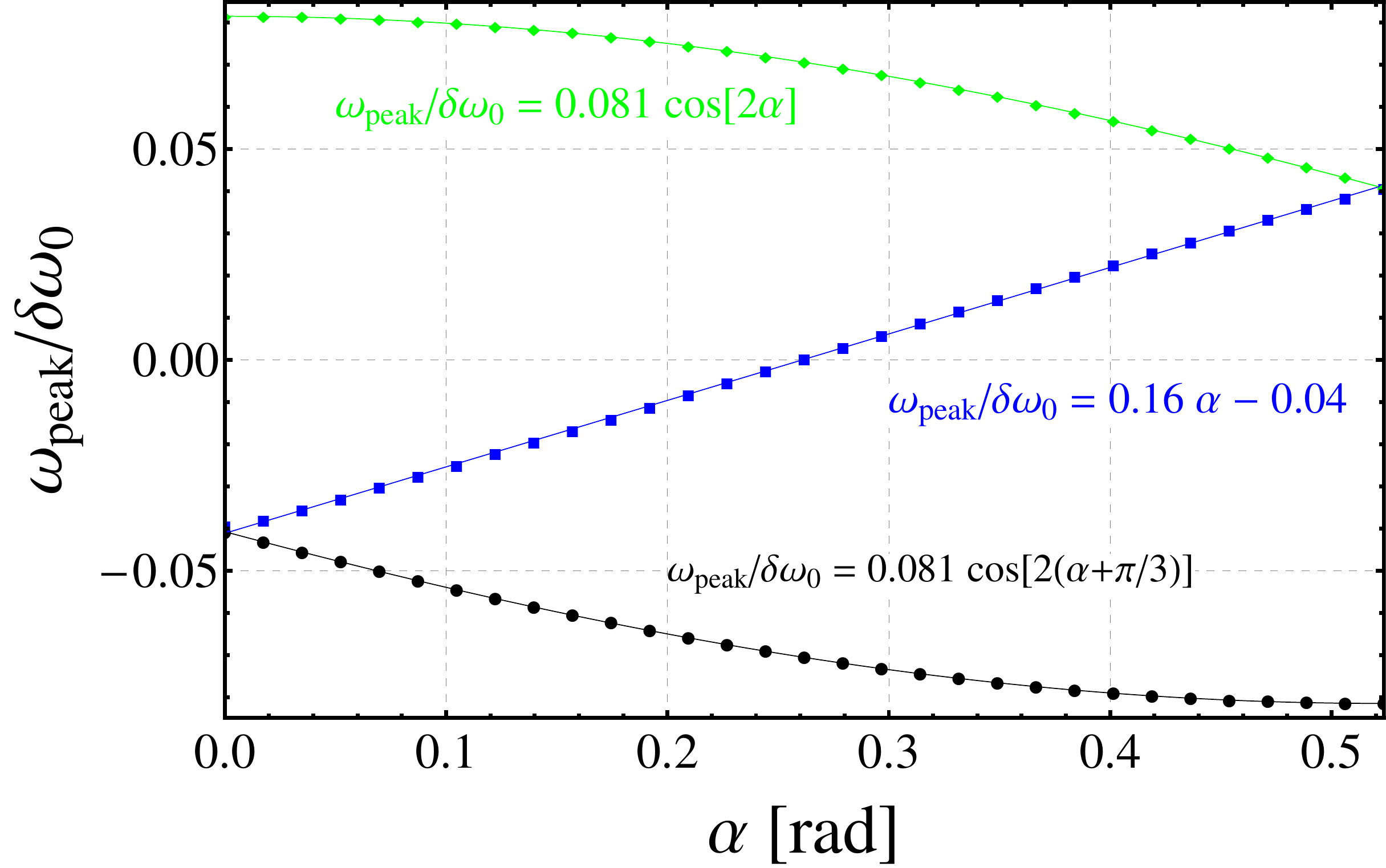}\label{fig:hex-alpha-peaks}
		}
		\end{center}
		\caption{Die Frequenzverteilung $\rho(\omega)$ des hexagonalen Gitters unter Variation von $\alpha$ und $\eta$. Auch hier kann das durch die Diskretisierung bedingte Rauschen nicht vollständig durch Glättung unterdrückt werden.}
		\label{fig:hex-offres-distributions}
	\end{figure}

Für das hexagonale Gitter wurden die Felder für $\alpha=0^\circ$ bis $\alpha=30^\circ$ ebenfalls in $1^\circ$ Schritten berechnet. Abb. \ref{fig:hex-contours} zeigt die Konfiguration einer Einheitszelle. Die Auflösung des Feldes betrug für  $\alpha\neq0^\circ$ wie beim quadratischen Gitter $0.02\mum$. Für $\alpha=0^\circ$ wurde das Feld zusätzlich mit einer deutlich höheren Auflösung von $0.005\mum$ berechnet. Das maximale $\eta$ des hexagonalen Gitters ist gegeben durch 
	\begin{align}
		\eta_{\mbox{\tiny max}}=\frac{2R_c^2\pi}{(2R_c)^2\sqrt{3}}=\frac{\pi}{2\sqrt{3}}\approx 90\%.
	\end{align}
Abb. \ref{fig:hex-offres-distributions} zeigt die Übersicht für das hexagonale Gitter. Wie schon beim quadratischen Gitter findet man einfache Zusammenhänge für die Positionen der Peaks in Abhängigkeit von $\eta$
	\begin{align}
			\begin{array}{rclcl}
				\omega_{\mbox{\tiny peak1}}(\eta) 			&=& 2(0.813\eta-0.008) \dom_0		&\mbox{für} & \alpha=0^\circ	\\
				\omega_{\mbox{\tiny peak2}}(\eta) 			&=& -(0.813\eta-0.008) \dom_0		&\mbox{für} & \alpha=0^\circ.	\\														
			\end{array}				
			\label{eq:static-hex-functions-eta}
	\end{align}
Während es für $\alpha=0^\circ$ nur zwei Peaks gibt, spaltet für $\alpha\neq0^\circ$ einer der Peaks auf und wandert mit steigendem $\alpha$ in Richtung des anderen Peaks um bei $\alpha=30^\circ$ mit diesem zu verschmelzen. Es ergibt sich
	\begin{align}
			\begin{array}{rclcl}
				\omega_{\mbox{\tiny peak1}}(\alpha)	&=& 0.081\dom_0\cos\left(2\alpha\right)  						& \mbox{für} & \eta=10\%		\\
				\omega_{\mbox{\tiny peak2}}(\alpha)	&=& (0.16\alpha-0.04)\dom_0							  					& \mbox{für} & \eta=10\%		\\
				\omega_{\mbox{\tiny peak3}}(\alpha)	&=& 0.081\cos\left(2(\alpha+\frac{\pi}{3})\right) 	& \mbox{für} & \eta=10\%. 		\\													
			\end{array}				
			\label{eq:static-hex-functions-alpha}
	\end{align}
Durch Spiegelung an der $y$-Achse lässt sich die Frequenzverteilung $\rho(\alpha=30^\circ)$ in $\rho(\alpha=0^\circ)$ überführen.

\subsection{Diskretisierungsfehler}
Durch die Diskretisierung von Gl. (\ref{eq:offresonance-distribution}) könenn in Frequenzverteilungen, v.a. im niedrigen Frequenzbereich, Artefakte auftreten (siehe Abb. \ref{fig:hex-eta}). Hauptursache für die Artefakte ist die Regelmäßigkeit der Abtastung des Feldes auf dem Interpolationsgitter aus Kap. \ref{kap:feldberechnung}. Für die Abbildungen wurden daher alle Daten durch die Bildung eines gleitenden Durchschnitts über zehn benachbarte Frequenzen geglättet. Für die später durchgeführten Simulationen mit dynamischer Dephasierung spielen die Artefakte hingegen keine Rolle, da bei dem Random-Walk die regelmäßige Abtastung durch Interpolation für beliebige Punkte ersetzt wird.\newline
In Abb. \ref{fig:square-eta-peaks} und Abb. \ref{fig:hex-eta-peaks} bildet der Verlauf der Peakpositionen keine perfekte Ursprungsgerade. Dies deutet darauf hin, dass in beiden Fällen die Summation in Gl. (\ref{eq:cyclic-offresonance-field}) noch nicht vollständig konvergiert ist. Für mehr Summanden ergäben sich systematisch höhere Off\-resonanzen, die Peaks von $\rho(\omega)$ würden weiter nach außen wandern und die Achsenabschnitte der Geraden verschwinden. Ein Fehler von ca. $2\%$ in der Stärke der Offresonanzen ist jedoch bzgl. der anderen Simulationsfehler vernachlässigbar. 

\renewcommand{\figurename}{Abb.}
\renewcommand{\tablename}{Tab.}

\section{2D1CP}\label{kap:2D1CP-Analysis}
\subsection{Feldgrö\ss e und Kapillaranzahl}
\begin{figure}
		\begin{center}\includegraphics[width=0.75\columnwidth]{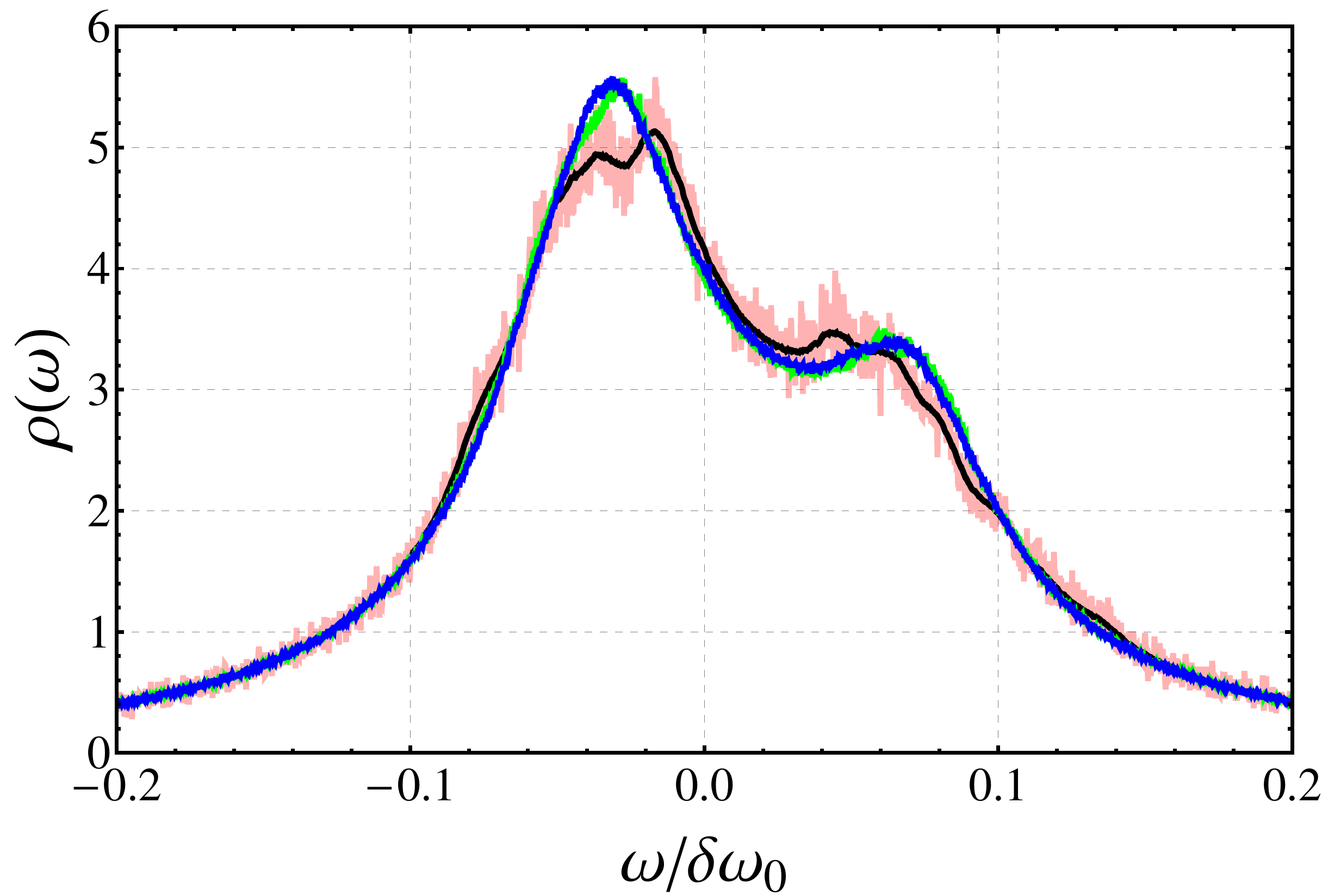}\end{center}
		\caption{Durch eine größere Simulationsbox mit mehr beinhalteten Kapillaren sinkt der Einfluss der spezifischen Kapillaranordnung auf $\rho(\omega)$. Für $32$ Kapillaren (pink) wurde  $\rho(\omega)$ geglättet(schwarz) um einen besseren Vergleich mit einer $200$ (grün) bzw. $1250$ (blau) Kapillaren enthaltenden Konfiguration zu ermöglichen. Die grüne und die blaue Kurve sind direkt vergleichbar, da für die 200er Simulationsbox eine erhöhte Feldauflösung gewählt wurde. Deutlich ist in den Peaks für $32$ Kapillaren eine Unterstuktur zu erkennen, die durch die individuelle Anordnung der Kapillaren verursacht wird. Bereits für $200$ Kapillaren ist diese durch die verbesserte Statistik praktisch vollständig verschwunden. Für die Konfigurationen galt $\eta=0.05$. Die Standardabweichung für die Positionierung der Kapillaren betrug $\sigma=1\mum$.}		
		\label{fig:capillary-count}		
	\end{figure}

Wie im vorangegangenen Kapitel wurden zunächst die Frequenzverteilungen für verschiedene 2D1CP-Konfigurationen nach Gl. (\ref{eq:offresonance-distribution}) bestimmt. Dabei tritt das Problem auf, dass $\rho(\omega)$ stark von der charakteristischen Anordnung (d.h. dem Mikrozustand) des Plasmas abhängen kann. Um dies zu umgehen müssen ausreichend viele Kapillaren in der Simulationsbox enthalten sein.\newline
Um den Einfluss der Anzahl der Kapillaren in einer Simulationsbox auf die Frequenz\-verteilung abschätzen zu können, wurden Konfigurationen mit bis zu $1250$ Kapillaren berechnet. Die Positionen der Kapillaren wurden um die regulären hexagonalen Gitterpunkte nach einer Normalverteilung verschoben. Um mehr Kapillaren zu berücksichtigen, muss die Simulationsbox größer werden. Bei gleich bleibender Feldauflösung nimmt also die Anzahl an Stützpunkten, aus welchen das Histogramm berechnet, zu. Dadurch wird $\rho(\omega)$ mit zunehmender Kapillarzahl immer glatter (Abb. \ref{fig:capillary-count}). Zudem ist zu erkennen wie die Abhängigkeit von einer spezifischen Anordnung verschwindet.\newline
Wenn nicht anders erwähnt werden im Folgenden bei allen unregelmäßigen Konfigurationen $200$ Kapillaren verwendet, was einen Kompromiss aus Rechenaufwand und Genauigkeit darstellt. Ausgehend von $\ICD\approx25\mum$ führt dies zu einer Feldgröße von ca. $250\mum\times430\mum$ liegt also etwa in der Größenordnung eines Voxels. Für noch niedrigere Kapillardichten ($\ICD>25\mum$) steigt der Einfluss der individuellen Kapillarverteilung bis schließlich bei einer Kapillare pro Voxel wieder eine stark charakteristische Frequenzverteilung (ähnlich wie im hexagonalen/quadratischen Gitter bzw. Krogh-Modell) erreicht wird. Für ein Voxel mit nur einer (großen) Kapillare wurden bereits in \cite{Kennan94} und \cite{Sedlacik07} ausführliche Untersuchungen durchgeführt.

\subsection{2D1CP mit nicht uniformen Radien}
\begin{figure}
		\begin{center}
		\subfloat[Radien konstant, Anordnung gestört]{
			\includegraphics[width=0.75\columnwidth]{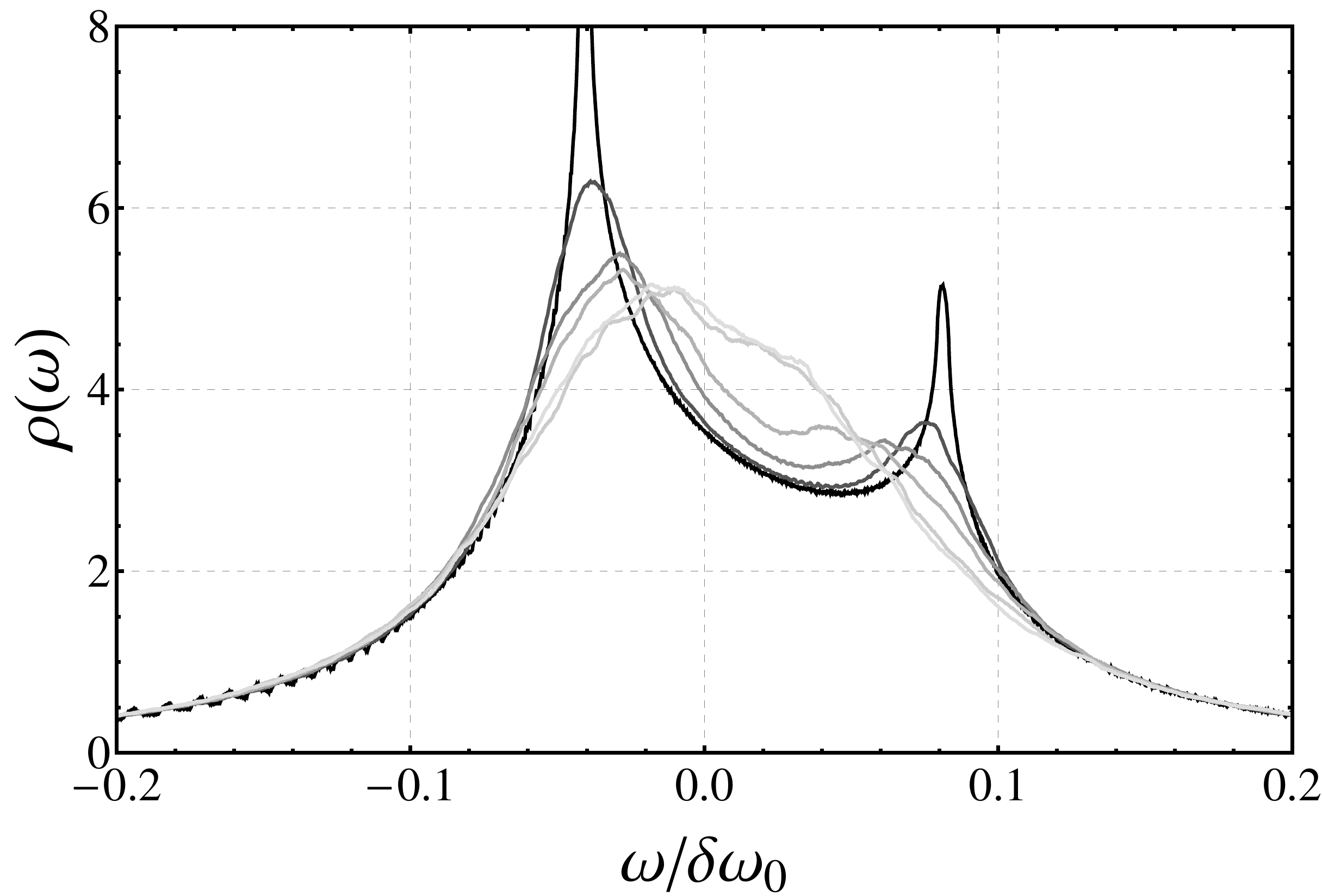}\label{fig:location-distribution}
		}
		\\
		\subfloat[Radien normalverteilt, Anordnung regelmäßig]{
			\includegraphics[width=0.75\columnwidth]{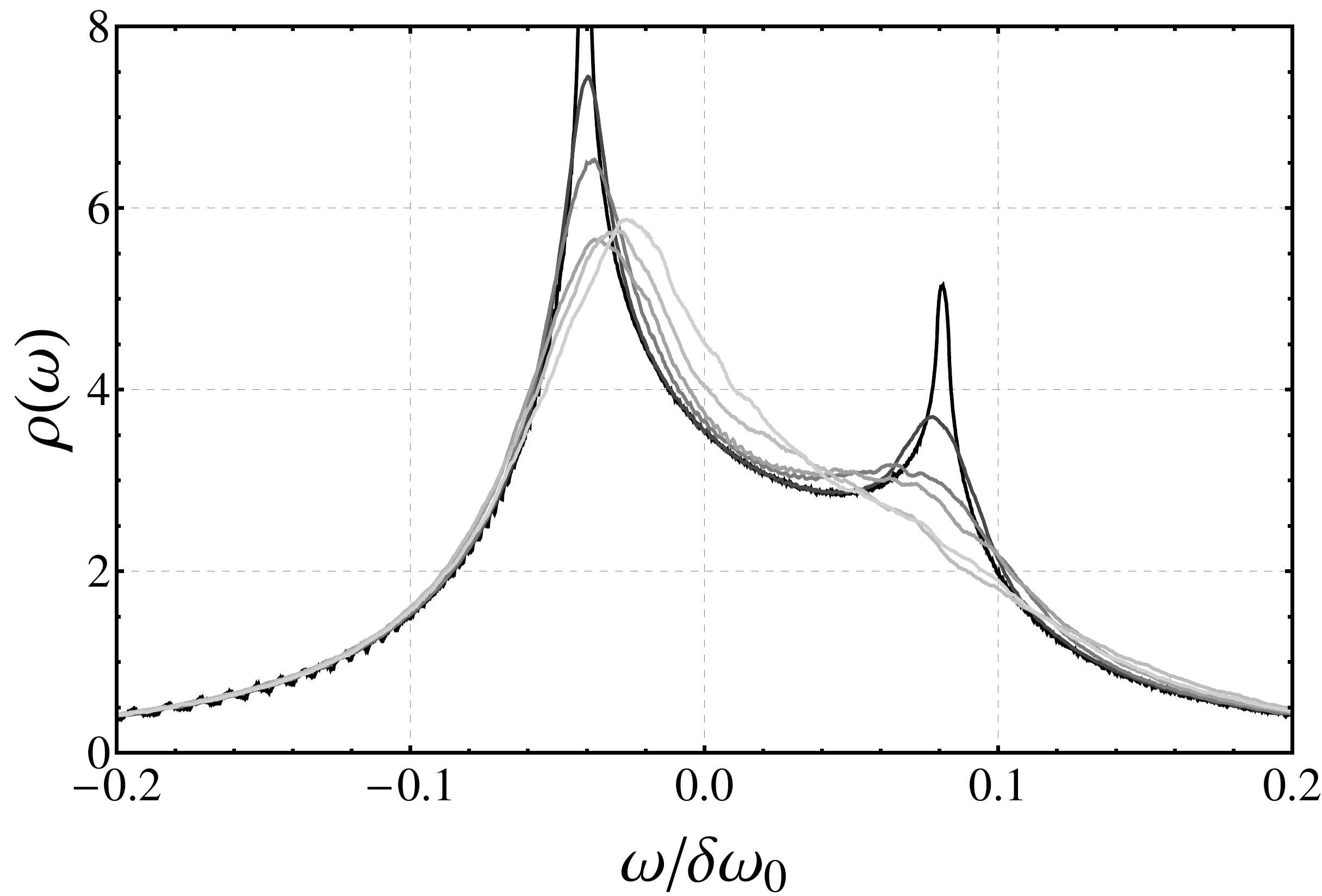}\label{fig:radial-distribution}
		}
		\end{center}
		\caption{Abb. \ref{fig:location-distribution} zeigt die Frequenzverteilung von einem hexagonalem Gitter, bei welchem die Gitterpunkte zufällig um die regulären Gitterpunkte verschoben sind (Standardabweichung $\sigma=0$ bis $2.5\mum$ mit $\Delta\sigma=0.5\mum$ von schwarz nach grau). In Abb. \ref{fig:radial-distribution} sind die Kapillaren regelmäßig angeordnet, die Radien jedoch normalverteilt ($\mu=1.94\mum$ und $\sigma=0$ bis $0.5\mum$ mit $\Delta\sigma=0.1\mum$ von schwarz nach grau). 
In beiden Fällen zeigt sich ein ähnlicher Effekt: Die Peaks wandern Richtung Mitte und verschmieren. Offensichtlich ist jedoch die asymmetrische Form der Frequenzverteilung robuster gegenüber der Variation der Radien. Vergleicht man $\sigma/\ICD=2.5\mum/16.5\mum\approx15\%$ aus Abb. \ref{fig:location-distribution} mit $\sigma/R_c=0.5\mum/1.94\mum\approx25\%$ aus Abb. \ref{fig:radial-distribution} so bleibt bei der Variation der Radien die Asymmetrie weitgehend erhalten.}
		\label{fig:lattice-disturbances}
	\end{figure}

\begin{figure}
		\begin{center}\includegraphics[width=0.75\columnwidth]{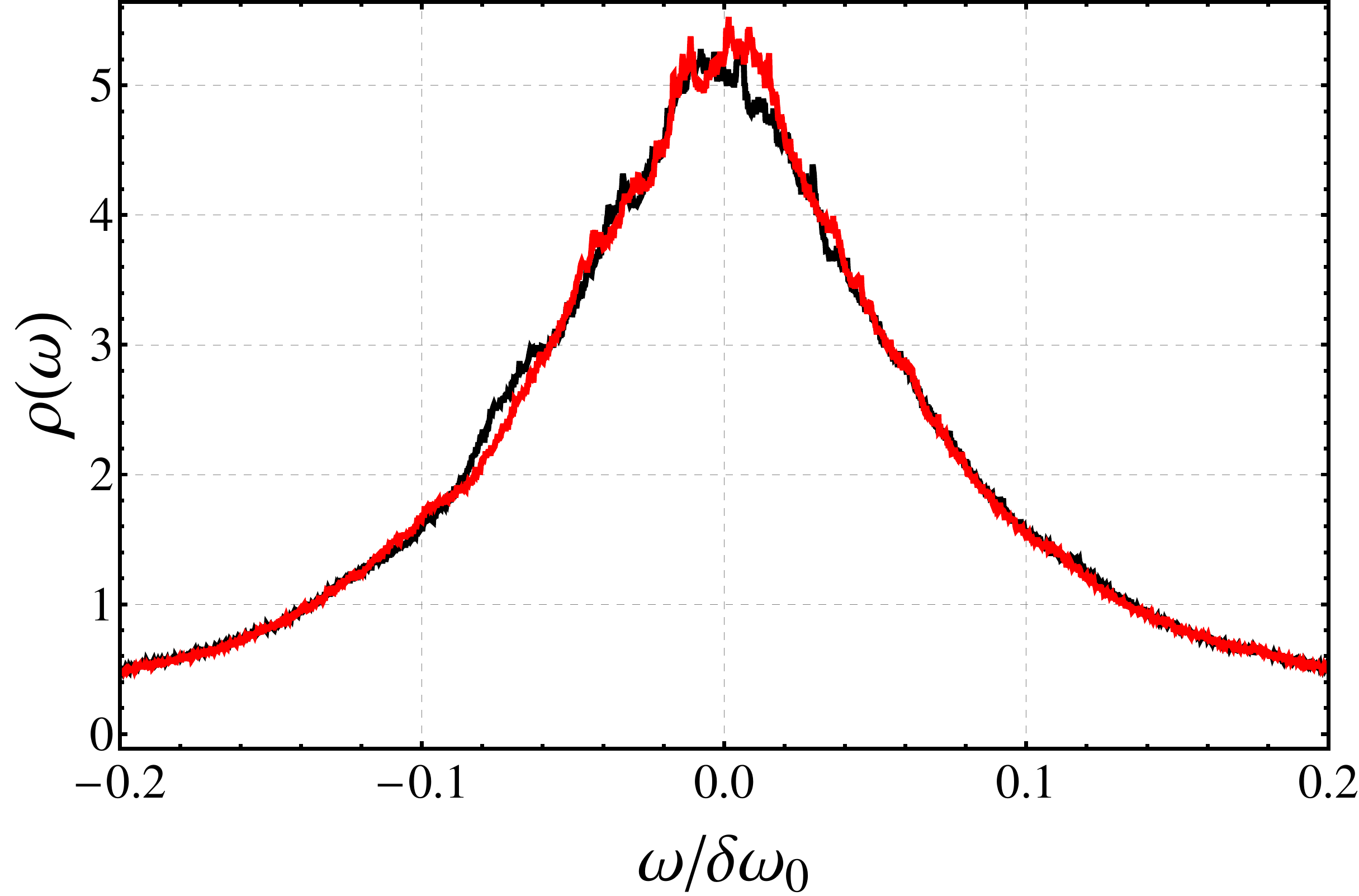}\end{center}
		\caption{Ab einer gewissen Temperatur hat eine zusätzliche Variation der Radien praktisch keinen Einfluss mehr auf die Frequenzverteilungen der Plasma-Konfigurationen. $\Gamma=4$ (entspricht nach \cite{Karch2006} etwa realem Gewebe), $\eta\approx6\%$, $\ICD=16.5\mum$ (schwarz: Radien konstant, rot: Radien normalverteilt)}		
		\label{fig:2D1CP-with-radius-variation}		
	\end{figure}

Um zu prüfen welchen Einfluss die Uniformität der Radien auf die statische Frequenzverteilung des 2D1CP hat, wird zunächst wieder von einer regelmäßigen hexagonalen Anordnung ausgegangen. Erzeugt man solche Konfigurationen, wobei die einzelnen Radien einer Normalverteilung folgen, so beginnen die charakteristischen Peaks der Frequenzverteilung mit zunehmender Breite der Radienverteilung zu "zerfließen" und driften dabei leicht Richtung Koordinatenursprung (Abb. \ref{fig:lattice-disturbances}). Dies ähnelt dem Effekt welchen auch die zufällige Anordnungen der Kapillaren aus Abb. \ref{fig:capillary-count} und Abb. \ref{fig:location-distribution} zeigt.\newline
Aufbauend auf dieser Tatsache lässt sich argumentieren, dass bei hinreichend starker Störung durch eine zufällige Positionierung, also bei den 2D1CP-Konfigurationen mit niedrigem $\Gamma$, der Einfluss unregelmäßiger Radien weitgehend vernachlässigt werden kann (siehe Abb. \ref{fig:2D1CP-with-radius-variation}). Für kleine RBV und nicht zu große Schwankungen der $R_c$ sollte sich diese Argumentation auch auf dynamische ($D>0$) Relaxationsprozesse übertragen lassen. Im weiteren Verlauf gilt für die betrachteten 2D1CP-Verteilungen daher immer $R_{c,i}=R_c=\mbox{const.}$.

\subsection{Abhängigkeit der statischen Frequenzverteilungen von der Plasmatemperatur}
\begin{figure}
		\begin{center}
		\subfloat[$\Gamma=2$(rot), $5$(schwarz), $140$(blau), $250$(grün) und $800$(orange))]{
			\includegraphics[width=0.75\columnwidth]{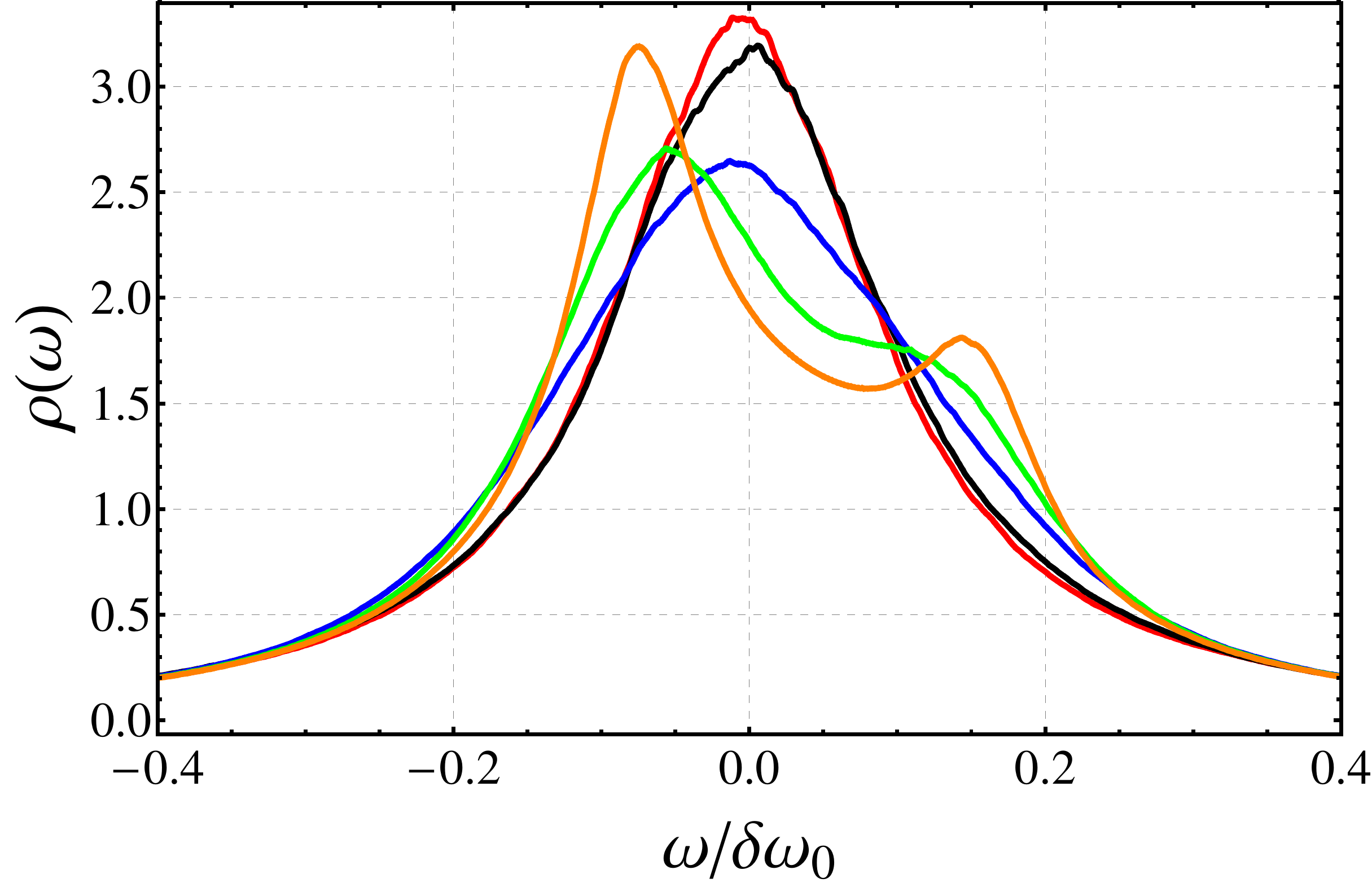}\label{fig:2D1CP-gamma-frequency-2D}
		}
		\\
		\subfloat[]{
			\includegraphics[width=0.75\columnwidth]{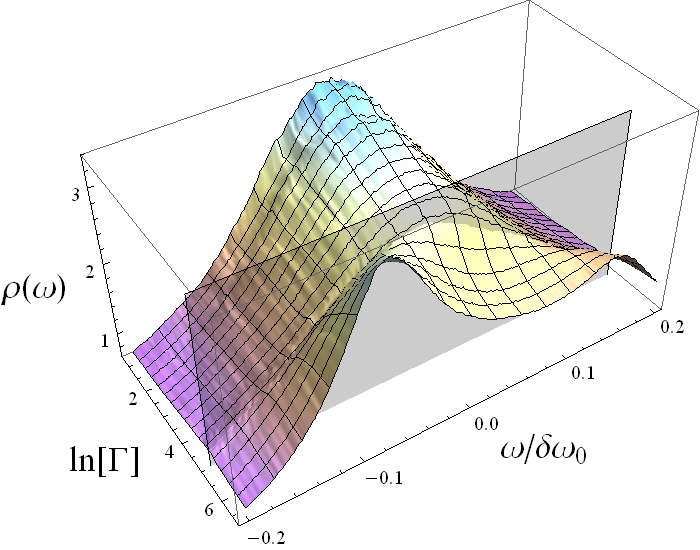}\label{fig:2D1CP-gamma-frequency-surface}
		}
		\end{center}
		\caption{Mit abnehmendem $\Gamma$ (abnehmender Ordnung) laufen die zwei Peaks die Frequenz\-verteilung in einen einzelnen zentralen lorentz-förmigen Peak. Bei $\Gamma\approx140$ kristallisiert das Plasma (graue Ebene in Abb. \ref{fig:2D1CP-gamma-frequency-surface}), oberhalb dieses Wertes ist noch die Ähnlichkeit mit der Frequenzverteilung des hexagonalen Gitters erkennbar. $\eta=10\%$, $\ICD=16.5\mum$}
		\label{fig:2D1CP-gamma-frequency-dependency}
	\end{figure}

\begin{figure}
		\begin{center}
		\subfloat[Sobald das Plasma kristallisiert ($\Gamma\approx140$) macht es einen deutlichen Unterschied, ob der Realteil oder der Absolutbetrag zur Auswertung verwendet wird. Das Signal ist dann nicht mehr rein reell und die Frequenzverteilung ist kein Lorentz-Profil mehr. Für kleine $\Gamma$ ist das Frequenzprofil weitestgehend symmetrisch und lorentz-förmig. Daher unterscheiden sich die verschiedenen Auswertemethoden hier kaum.]{
			\includegraphics[width=0.76\columnwidth]{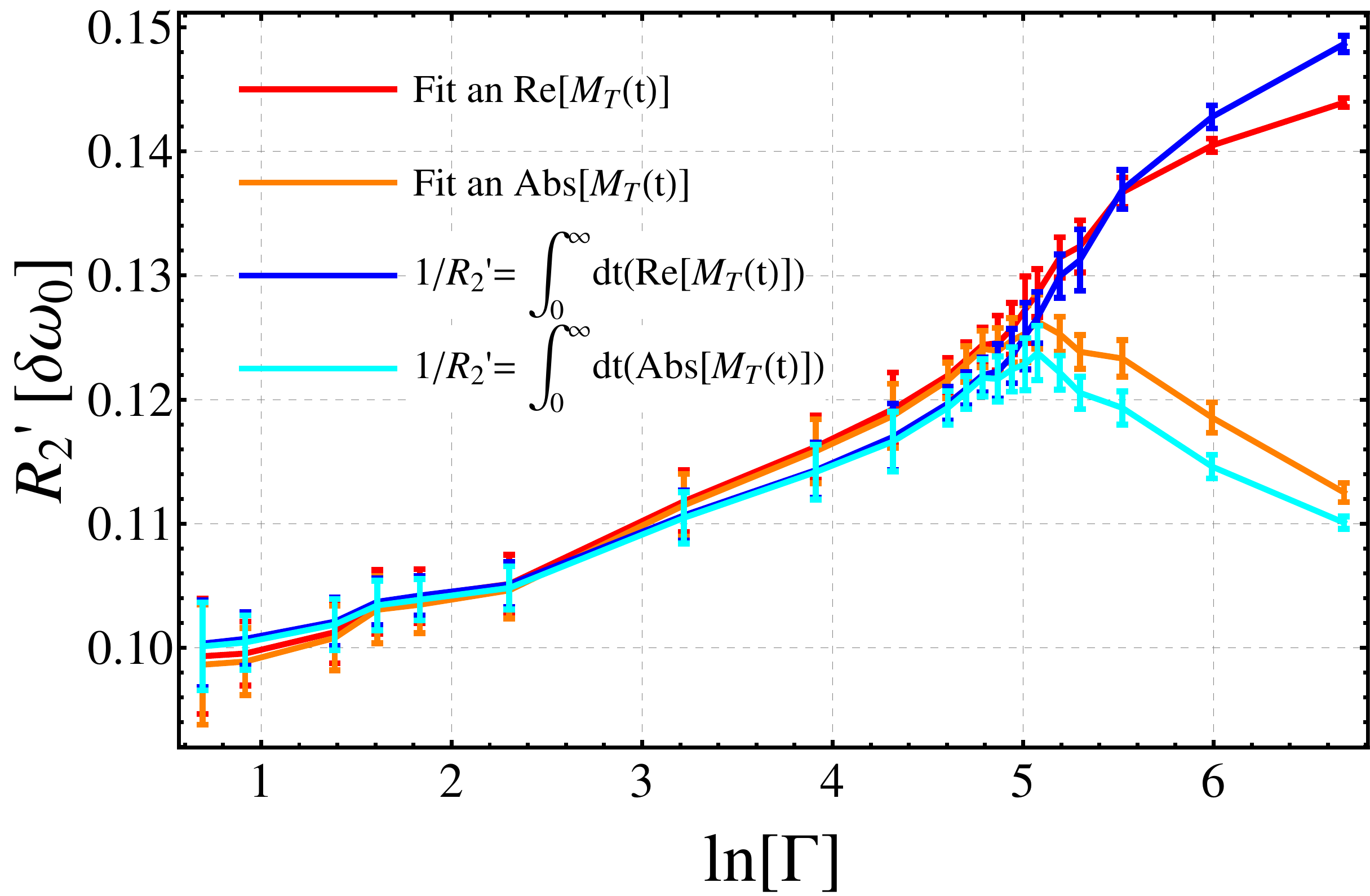}\label{fig:2D1CP-gamma-R2}
		}\\
		\subfloat[Für Integration bzw. Fit an den Absolutbetrag ergibt sich für $\Gamma\approx140$ ein Maximum der Relaxationszeiten. Wegen der für $\Gamma>140$ auftretenden Oszillationen wird $R_2'$ dann wieder kleiner.]{
			\includegraphics[width=0.76\columnwidth]{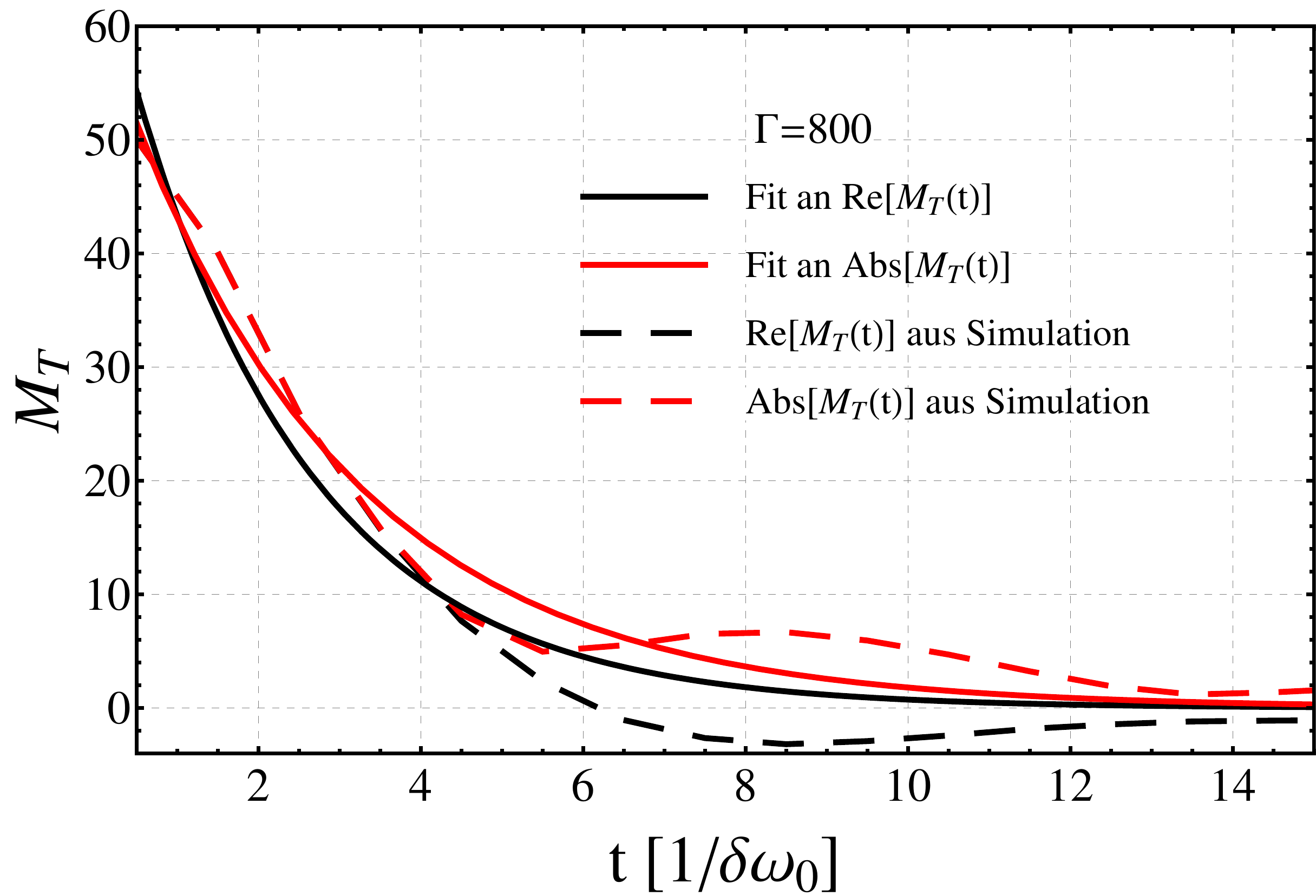}\label{fig:relaxation-rate-maximum}
		}
		\end{center}
		\caption{Abhängigkeit der Relaxationszeit im Static-Dephasing-Regime von der Plasmatemperatur. Aus Abb. \ref{fig:2D1CP-gamma-R2} lässt sich der Unterschied von $R_2'(\Gamma=2)$ und $R_2'(\Gamma=5)$ zu ca. $0.005\delta\omega_0$ bestimmen. Für jedes $\Gamma$ wurden zehn verschiedene Konfigurationen erstellt und Mittelwert und Standardabweichung der $R_2'$ bestimmt.($\eta=0.1$, $\ICD=16.5\mum$)}		
		\label{fig:2D1CP-gamma-R2-dependency}
	\end{figure}

\begin{figure}
		\begin{center}\includegraphics[width=0.8\columnwidth]{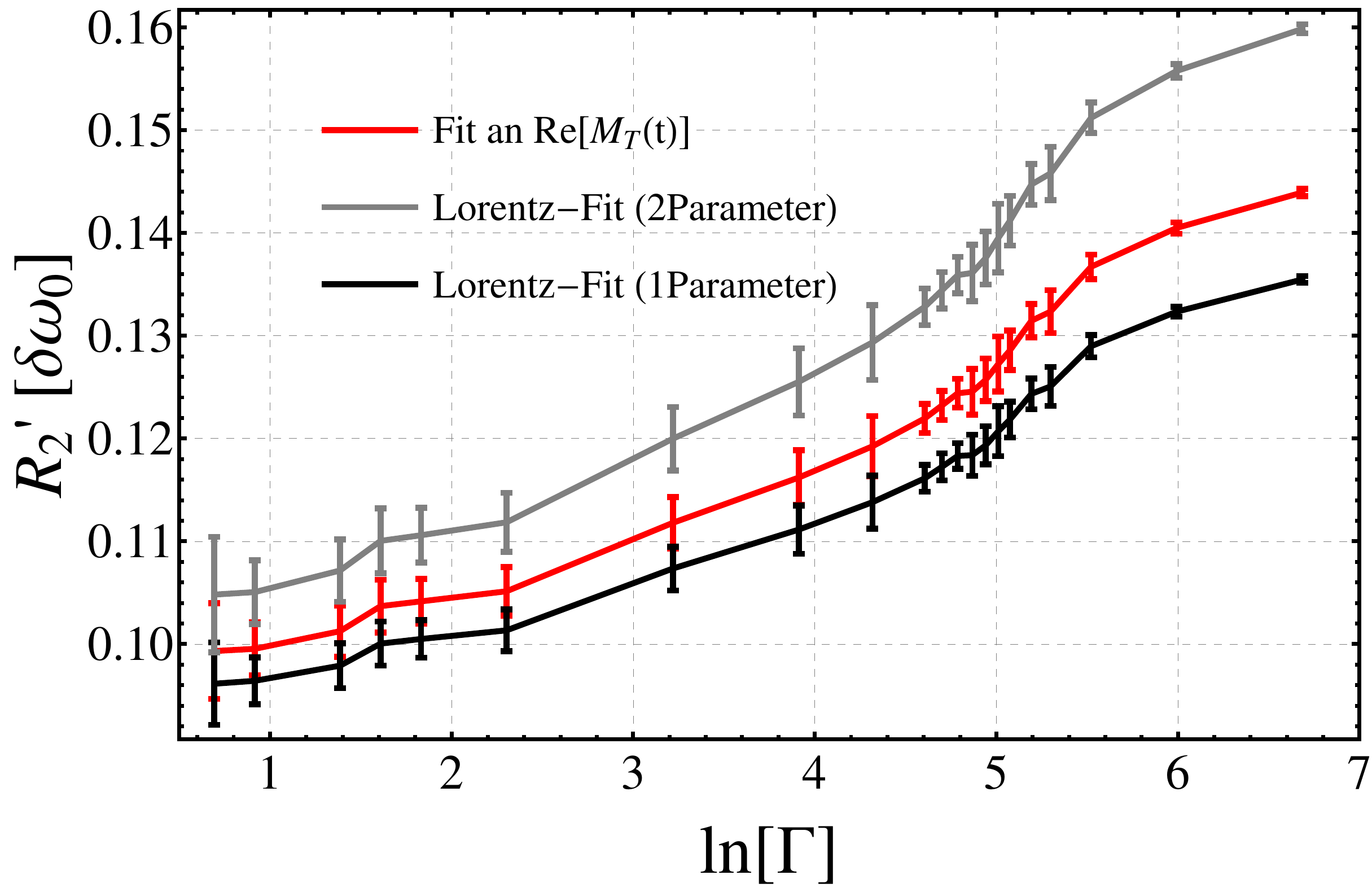}\end{center}
		\caption{Dass der Versuch einen nicht monoexponentiellen Verlauf durch eine einzige Relaxations\-zeit zu klassifizieren, zu Problemen führt, wird besonders beim Fitten an ein Lorentz-Profil im Frequenzraum deutlich: Je nach angenommener Fit-Funktion (festes oder freies $A$ nach Gl. (\ref{eq:lorentz-profile})) ergeben sich deutlich unterschiedliche Relaxationsraten. Die Relaxationszeiten aus dem Lorentz-Fit geben grob das Verhalten von Integration bzw. Fit an den Realteil der Magnetisierung wieder.($\eta=0.1$, $\ICD=16.5\mum$)}		
		\label{fig:2D1CP-log-gamma-R2}
	\end{figure}

Wie in Kap. \ref{kap:introduction} erwähnt bestand ein Hauptziel dieser Arbeit darin zu prüfen, ob aus der $T2$- bzw. $T2^*$-Zeit Rückschlüsse auf die mikroskopische Anordnung der Kapillaren gezogen werden können. Außerdem wurde in \ref{kap:2D1CP-Theory} erläutert, wie der Parameter $\Gamma$ einer solchen Konfiguration als Charakterisierungsmerkmal des Gewebes verwendet werden kann. Es stellt sich also die Frage, ob aus einer gemessenen Relaxationszeit $T_2^*$ ein aussagekräftiger Wert für $\Gamma$ ermittelt werden kann. Um diese Frage zu beantworten wurden zunächst nur die statischen Felder untersucht und keine Signalverläufe simuliert. Zeigt sich bereits in den statischen Verteilungen nur eine verschwindende Abhängigkeit von $\Gamma$, so würde diese durch die Diffusion, die prinzipiell die zur spezifischen Konfiguration gehörende Frequenzverteilung "verschmiert", nur weiter abgeschwächt werden.\newline
Abb. \ref{fig:2D1CP-gamma-frequency-dependency} zeigt die Abhängigkeit der statischen Frequenzverteilung von $\Gamma$. Für verschiedene $\Gamma$ zwischen $2$ und $800$ wurden dafür jeweils 10 Plasmakonfigurationen (mit je 200 Kapillaren) erzeugt und über diese gemittelt. Da für die hohen Werte von $\Gamma$ der in Kap. \ref{kap:2D1CP-Theory} beschriebene Metropolis Algorithmus nur sehr langsam (bzw. nur mit sehr hohem Rechenaufwand) in ein Gitter kristallisiert, wurde das Plasma nicht langsam gekühlt, sondern ausgehend vom exakten hexagonalen Gitter erhitzt.\newline
Für die gemittelten statischen Frequenzverteilungen wurden dann die zugehörigen $R_2'$-Raten bestimmt. Folgende Auswertungen wurden durchgeführt:
\begin{itemize}
	\item Lorentz-Fit im Frequenzraum mit dem freien Parameter $R_2'$ bzw. $\gamma$ bei festem $A=\frac{1}{\pi}$ und $\mu=0$ nach Gl. (\ref{eq:lorentz-profile}). Für $A=\frac{1}{\pi}$ erfüllt der Lorentz-Peak die gleiche Normierung wie die Frequenzverteilung.
	\item Lorentz-Fit im Frequenzraum mit zwei freien Parametern $A$ und $R_2'$ bei festem $\mu=0$ nach Gl. (\ref{eq:lorentz-profile}).
	\item Monoexponentieller Fit an den Realteil des FID.
	\item Monoexponentieller Fit an den Absolutbetrag des FID.
	\item Anwendung der Mean-Relaxation-Time-Approximation auf den Realteil des FID: $T_2^*=\int_0^\infty\mathsf{d}t \mathsf{Re}(M_T(t))$
	\item Anwendung der Mean-Relaxation-Time-Approximation auf den Absolutbetrag des FID: $T_2^*=\int_0^\infty\mathsf{d}t \mathsf{Abs}(M_T(t))$
\end{itemize}
Bei den letzten vier Fällen folgt der FID aus der Fourier-Transformation von $\rho(\omega)$. Da $\rho(\omega)$ nicht vollkommen symmetrisch und auch kein perfekter Lorentz-Peak ist (siehe Kap. \ref{kap:offresonance-distribution}), gilt $\mathsf{Im}(M_T(t))\neq0$. Der monoexponentielle Fit und die Integration können also sowohl für den Absolutbetrag $\mathsf{Abs}(M_T(t))$, oder unter der Annahme einer zumindest näherungsweise symmetrischen Frequenzverteilung und damit weitestgehend reeller Magnetisierung, für $\mathsf{Re}(M_T(t))$ durchgeführt werden.\newline
In Abb. \ref{fig:2D1CP-gamma-R2-dependency} ist der Verlauf der Relaxationsraten in Abhängigkeit von $\Gamma$ aufgetragen. Für den Absolutbetrag von $M_T(t)$ ergibt sich für $\Gamma\approx140$ ein Maximum in der Relaxationsrate, während bei Betrachtung des Realteils die Relaxationsrate stetig weiter ansteigt. Da die Offresonanzverteilung nicht mehr symmetrisch ist, beginnt der FID für $\Gamma>140$ zu oszillieren. Diese Oszillationen können wegen der Absolutwertbildung besser mit kleineren Relaxationsraten angenähert werden (siehe Abb. \ref{fig:relaxation-rate-maximum}).\newline
Während sich für $\Gamma<140$ nach Abb. \ref{fig:2D1CP-gamma-R2} mehr oder weniger die gleichen Relaxationsraten ergeben, liefert die Auswertung im Frequenzraum durchweg zwei deutlich unterschiedliche Verläufe (Abb. \ref{fig:2D1CP-log-gamma-R2}). Für zwei freie Fitparameter des Lorentz-Profils ergeben sich systematisch niedrigere, für den Fit mit nur einem freien Parameter höhere Relaxationsraten.\newline
In Tab. \ref{tab:2D1CP-Relaxations} sind die aus Abb. \ref{fig:2D1CP-gamma-R2-dependency} folgenden Relaxationszeiten für verschiedene Offresonanzstärken bei zwei nach \cite{Karch2006} realistischen $\Gamma$s, sowie für $\Gamma=800$ (entspricht in etwa dem hexagonalen Gitter) aufgelistet.
	
\begin{table}
		\begin{center}
		\begin{tabular}[ht]{lc|c|c|c|c|c|c|c|c} 		
& 
$\displaystyle\frac{R_{2,sim}}{\dom_0}$	&
\multicolumn{2}{c}{$\displaystyle\frac{T_{2,sim}^*(1.5T)}{\ms}$} &
\multicolumn{2}{|c}{$\displaystyle\frac{T_{2,sim}^*(7T)}{\ms}$} &
\multicolumn{2}{|c}{$\displaystyle\frac{T_2^*(1.5T)}{\ms}$}	&
\multicolumn{2}{|c}{$\displaystyle\frac{T_2^*(7T)}{\ms}$}	\\[0.5ex] 
		  \hline
$\dom_0 [\frac{\mbox{\small rad}}{\mbox{\small s}}]$  & & 100 & 200 & 500 & 1000 & 100 & 200 & 500 & 1000			\\[0.5ex] 
		  \hline 
		  \hline
	$\Gamma=2$			& 0.0996								& 100.4	&	50.2	&	20.1	&	10.0	&	36.4	&	26.7	&	14.8	&	8.5		\\
	$\Gamma=5$			& 0.1035								&	96.6	&	48.3	&	19.3	&	9.7		& 35.9	& 26.1	&	14.4	& 8.3		\\
	$\Gamma=800$		& 0.1463								& 68.3	&	34.2	&	13.7	& 6.8		&	31.1	&	21.4	& 11.0	&	6.1		\\
			\hline
	$\Gamma=4$ 		  &	-											& 395 &	111	& 25	& 	10 &	50 & 39	 & 17	 & 8				\\
	$\Gamma=\infty$ &	-											&	502	& 126	& 21	&	 	 7 &	51 & 38	 & 15  & 6				\\
		\end{tabular}
		\end{center}
		\caption{Relaxationszeiten für statische (oben) und dynamische (unten) Dephasierung. Im Statischen wurde $R_2'$ für $\Gamma=2$ und $\Gamma=5$ über die vier verschiedenen Auswertemethoden in Abb. \ref{fig:2D1CP-gamma-R2} gemittelt, für $\Gamma=800$ wurde nur über die Auswertung der Realteile gemittelt. Für $\dom_0$ wurden näherungsweise die Werte aus Tab. \ref{tab:offres-influence} verwendet. Nach \cite{ZienerPHDThesis} wurde $T_2=57\ms$ als typischer Wert für das Myokard angenommen. Die Relaxationszeiten mit Diffusion folgen aus dem Fit an $\mathsf{Re}(M_T(t))$ in Abb. \ref{fig:2D1CP-dynamic-dephasing}. Für dynamisch und statisch gilt $\eta=10\%$, für die dynamisch Dephasierung wurde $\ICD=17\mum$ (d.h. $R_c=2.82\mum$) und $D=1\mum^2/\ms$ gewählt. Weder im statisch noch im dynamisch dephasierenden Regime hat der Ordnungsgrad des Plasmas nennenswerten Einfluss auf die Relaxationszeiten.}
		\label{tab:2D1CP-Relaxations}
	\end{table}

\subsection{Dynamisch Dephasierendes Regime}
\begin{figure}
		\begin{center}
		\subfloat[$B=1.5T$]{
			\includegraphics[width=0.8\columnwidth]{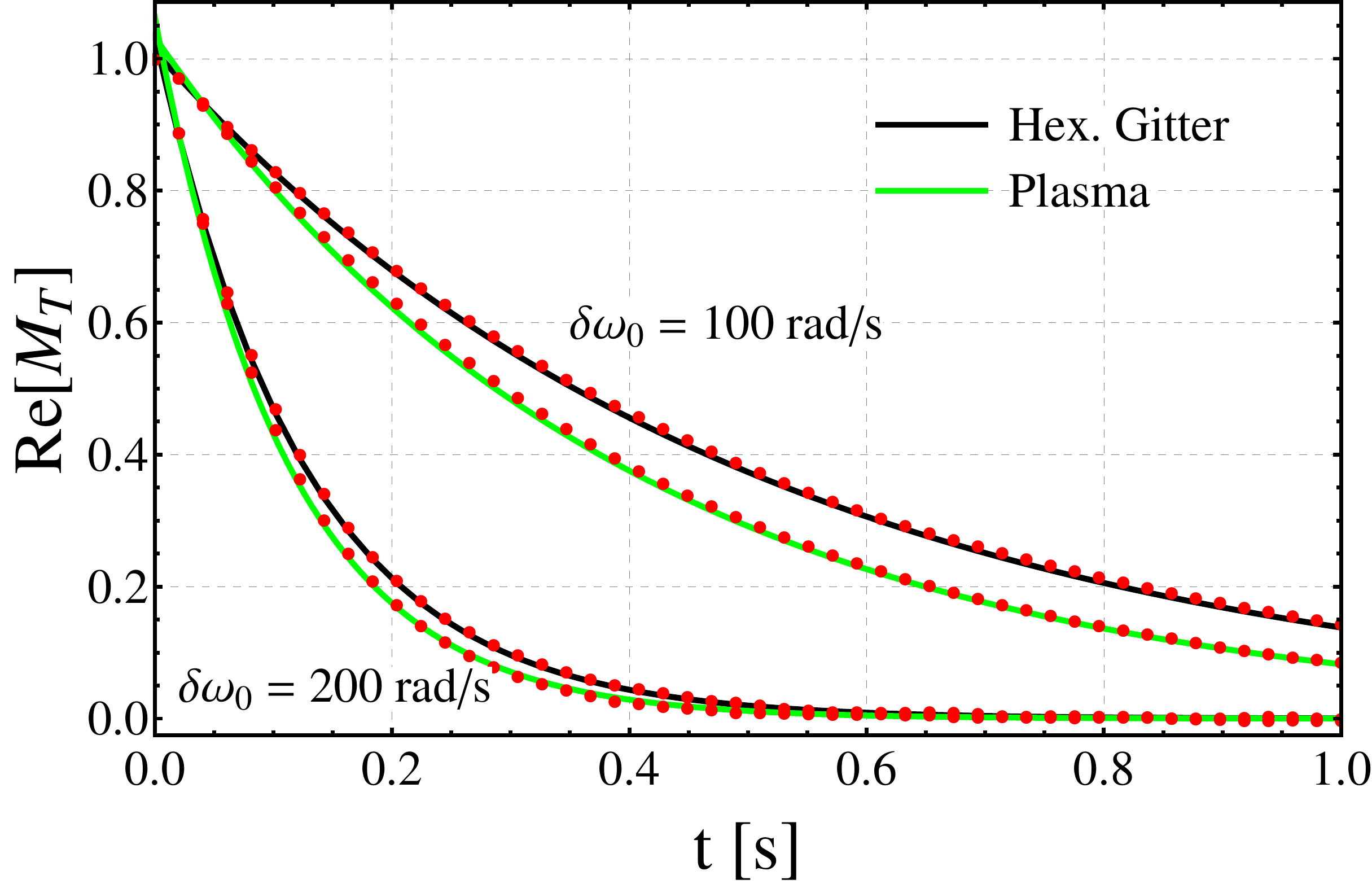}\label{fig:2D1CP-dynamic-fid-1.5T}
		}
		\\
		\subfloat[$B=7T$]{
			\includegraphics[width=0.8\columnwidth]{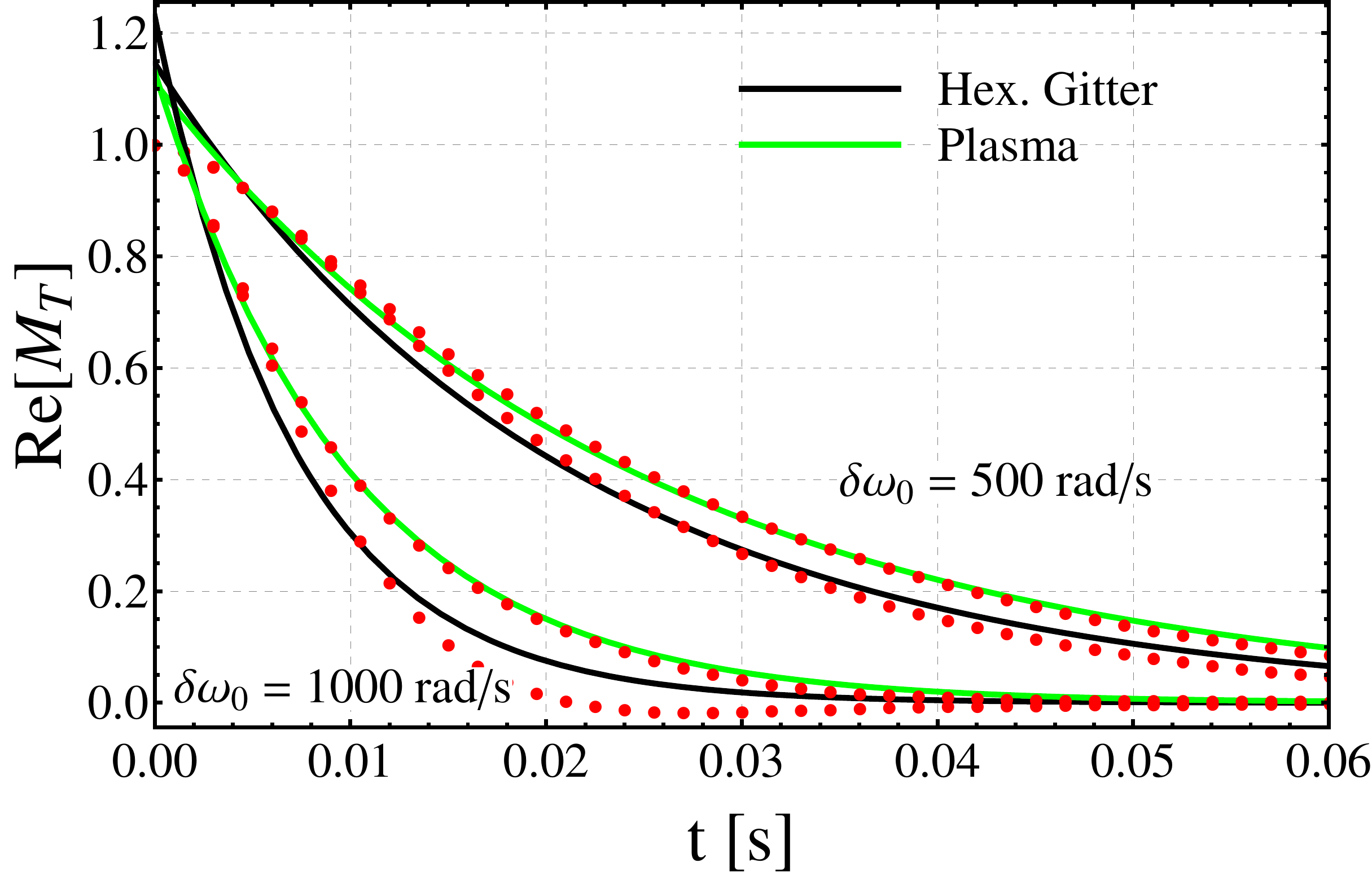}\label{fig:2D1CP-dynamic-fid-7T}
		}
		\end{center}
		\caption{Simulationsdaten (gepunktet) und Fit des FID bei dynamischer Dephasierung. Während für niedrige Offresonanzen das Signal der Plasma Verteilung schneller abklingt, liefert für hohe Offresonanzen das hexagonale Gitter höhere Relaxationsraten. Zu beachten ist, dass der FID-Verlauf aus der Simulation noch keine intrinsische Relaxation berücksichtigt. Der Imaginärteil von $M_T$ ist in allen Fällen kleiner als $0.1$. $\ICD=17\mum$, $\eta=10\%$, $D=1\mum^2/\ms$, $R_c=2.82\mum$}		
		\label{fig:2D1CP-dynamic-dephasing}
	\end{figure}

Betrachtet man ein dynamisch dephasierendes Regime, skalieren die Relaxationsraten $R_2'$ nicht mehr wie bei der statischen Dephasierung linear mit den Offresonanzen. Für $\Gamma=4$ und die hexagonale Geometrie (entspricht $\Gamma=\infty$) wurden daher auch Simulationen mit aktiver Diffusion bei verschiedenen Offresonanzstärken durchgeführt. Die Relaxationszeiten wurden aus einem monoexponentiellen Fit an den Realteil der FIDs bestimmt (siehe Abb. \ref{fig:2D1CP-dynamic-dephasing}) und sind somit direkt mit den Relaxationszeiten der statischen Dephasierung vergleichbar (siehe Tab. \ref{tab:2D1CP-Relaxations}). Auffällig ist vor allem, dass unter berücksichtigung der Diffusion für höhere Offresonanzen das hexagonale Gitter, für niedrige Offresonanzen aber das Plasma höhere Relaxationsraten liefert. Untersucht man das Frequenzspektrum des FID der dynamischen Dephasierung im hexagonalen Gitter, so stellt man fest, dass dieses zwar näherungsweise lorentz-förmig ist, der Mittelwert $\mu$ der Verteilung bei starken Offresonanzen jedoch leicht von null abweicht. Bei den Plasma-Konfigurationen verschwindet der Mittelwert der Verteilung unabhängig von der Offresonanzstärke und der Diffusion, allein durch die Statistik über die zufällige Anordnung der Kapillaren.
	\newline
	
\subsection{Fazit}
Die verschiedenen Relaxationszeiten welche sich aus den unterschiedlichen Fit-Methoden ergeben zeigen deutlich die Problematik die sich ergibt, wenn man versucht den FID als monoexponetiellen Verlauf zu nähern. Um gemessene Relaxationszeiten mit der Simulation vergleichen zu können, muss man sich daher zuvor in beiden Fällen für das gleiche Auswerteverfahren entscheiden. Da in der Simulation alle Größen direkt greifbar sind, kann diese Wahl an die Gegebenheiten einer Messung angepasst werden.\newline
Der systematische Unterschied in den Relaxationszeiten für die Auswertung nach den zwei verschiedenen Lorentz-Profilen lässt sich wie folgt begründen. Lässt man $A$ als Freiheitsgrad zu, so kann in der Frequenzraummitte die Offresonanzverteilung besser durch das Lorentz-Profil angenähert werden, die Relaxationszeit kann besser an die niederfrequenten Anteile des Zerfalls angepasst werden. Für festes $A$ ist diese Anpassung nicht möglich. Auch ist zu bedenken, dass die Offresonanzverteilung nach außen hin begrenzt ist, für $|\omega|>\omega_{max}$ gilt offensichtlich $\rho(\omega)=0$. Der FID kann also im Static-Dephasing niemals ein perfekter monoexponentieller Zerfall sein, da dafür beliebig hohe Offresonanzen existieren müssten. Eine bessere Anpassung an die niederfrequenten Anteile scheint also sinnvoll.\newline
Bezogen auf die medizinische Anwendung zeichnet sich aus Tab. \ref{tab:2D1CP-Relaxations}, abgesehen vom praktisch verschwindend geringen Einfluss von $\Gamma$ auf die Relaxationszeiten, ein weiteres Problem ab. Selbst wenn die Abhängigkeit $R_2^*(\Gamma)$ stärker ausgeprägt wäre und sich in der Realität messtechnisch erfassen ließe, so bleibt die Ungewissheit über die Stärke der Offresonanzen. Offensichtlich hängt das Relaxationsverhalten weit kritischer von $\dom_0$ ab als von $\Gamma$. Da $\dom_0$ jedoch in der Praxis mit großen Ungewissheiten behaftet ist, lässt sich einer gemessenen Relaxationszeit ein riesiger Bereich des Parameters $\Gamma$ zuordnen. Nach Tab. \ref{tab:2D1CP-Relaxations} kann in dynamischer wie statischer Dephasierung daher selbst zwischen extremer Ordnung ($\Gamma=800$) und extremer Unordnung ($\Gamma=2$) nicht mehr unterschieden werden. Zusätzlich hängt $R_2^*$ natürlich auch von der Diffusion im Gewebe ab, die für das Myokard bisher als exakt $1\mum^2/\ms$ angenommen wurde, tatsächlich jedoch ebenfalls nur auf einen gewissen Bereich eingegrenzt ist. Beide Probleme werden auch im nächsten Kapitel weiter untersucht.\newline
In möglichen anderen Anwendungsbereichen, die viele der bisherigen Modellannahmen besser widerspiegeln, lässt sich eventuell auch im dynamisch dephasierenden Regime noch auf die Größenordnung von $\Gamma$ schließen, wenn die Diffusion und die Offresonanzen genau bekannt sind.

\section{Interkapillare Abstände}
	\begin{figure}
		\begin{center}\includegraphics[width=\textwidth]{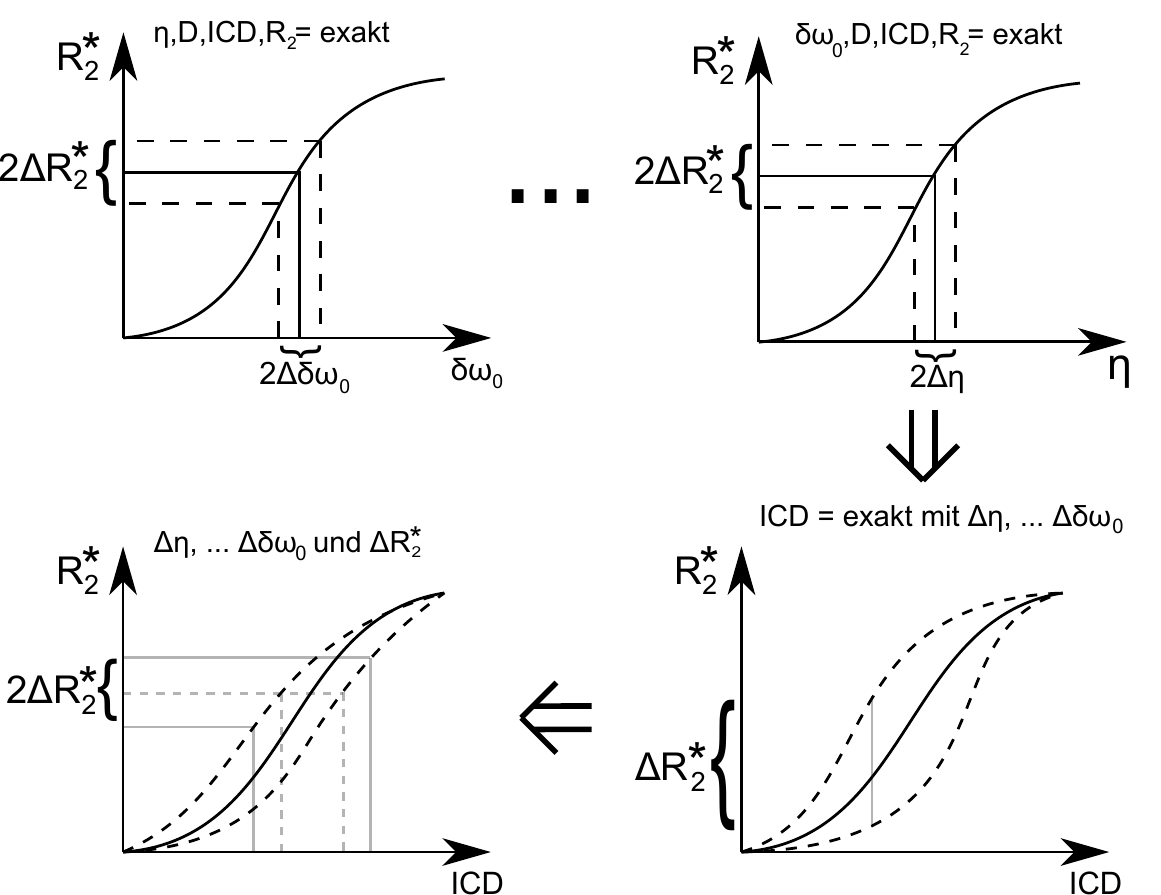}\end{center}
		\caption{Mit den Simulationen können die funktionalen Abhängigkeiten der Relaxationsrate von den einzelnen Parametern $\eta$, $\dom_0$ $\ICD$ und $D$ sehr genau wiedergegeben werden. Die einzelnen Abhängigkeiten geben jedoch nur entsprechende Schnitte im vierdimensionalen Parameterraum wieder, da für jede Abhängigkeit die anderen Parameter konstant gehalten werden müssen. In der Realität lassen sich die Parameter $\eta$, $\dom_0$ und $D$ nur innerhalb ihres Fehlerintervalls angeben. Für jede Ungewissheit der eingehenden Größen folgt daher eine Bandbreite der möglichen Relaxationsraten (oben). Insgesamt definieren die verschiedenen Fehlerintervalle $\Delta\eta$, $\Delta\dom_0$ und $\Delta D$ für jeden $\ICD$ ein dreidimensionales Teilvolumen des vierdimensionalen Parameterraums. Die Schwankungen der Relaxationsrate innerhalb dieses Volumens entspricht der Bandbreite der Relaxationsrate die sich jedem $\ICD$ Wert zuordnen lässt (rechts unten). Wenn man diese Zuordnung umkehrt können einer Relaxationsrate eine Reihe möglicher $\ICD$s zugewiesen werden (links unten, grau gestrichelt). Die Relaxationsrate (z.B. aus einer echten Messung) ist jedoch ebenfalls mit Fehlern behaftet. Die Bandbreite möglicher $\ICD$ Werte wird dadurch noch größer (links unten, grau).}
		\label{fig:auswertung}
	\end{figure}
\begin{figure}
		\begin{center}\includegraphics[height=0.8\textheight]{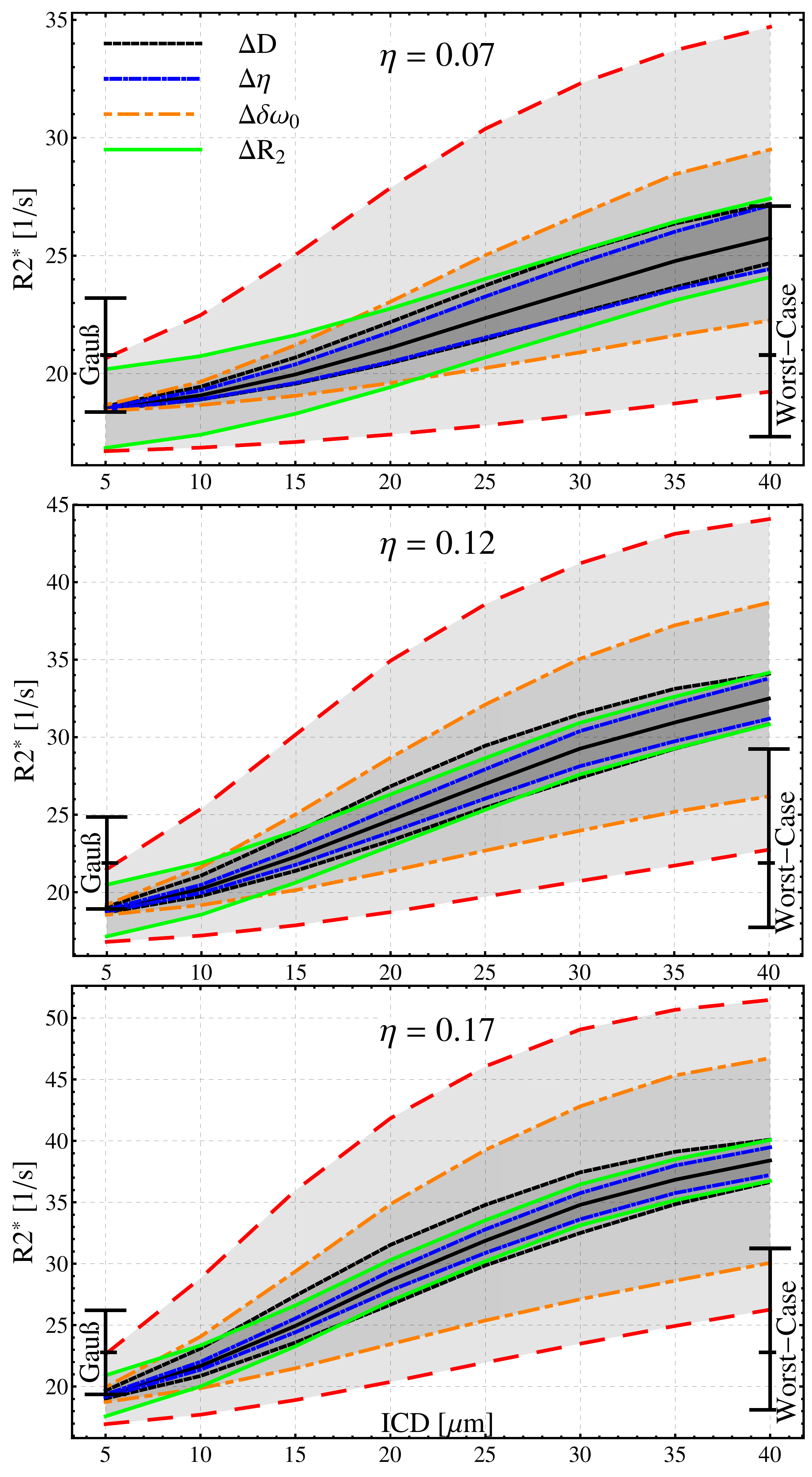}\end{center}
		\caption{Die mittlere Linie zeigt den Verlauf der Relaxationszeit (Gl. (\ref{eq:r2s-funktion})) in Abhängigkeit vom $\ICD$ bei 1.5T ($\overline{\dom_0}=150\radps$), $R_2=18.3s^{-1}$ und $D=1.5\mum^2/\ms$. Für konstantes $\eta$ nehmen dabei für steigenden $\ICD$ auch die Kapillarradien $R_c$ zu. Die umfassenden Linien bzw. schattierten Bereiche geben die durch die einzelnen Größen verursachten maximalen Abweichungen wieder (z.B. $R_2^*\left(\overline{\eta}, \ICD, \overline{\dom_0}, \overline{D}+\Delta D,\overline{R_2}\right)$ und $R_2^*\left(\overline{\eta}, \ICD, \overline{\dom_0}, \overline{D}-\Delta D,\overline{R_2}\right)$, siehe auch Abb. \ref{fig:auswertung}). Die äußerste Umfassung gibt den Worst-Case, d.h. die schlimmst mögliche Kombination der einzelnen Fehler an (Gl. (\ref{eq:r2s-worst-case-approximation})).}
		\label{fig:icd-analysis-1.5T}
	\end{figure}
\begin{figure}
		\begin{center}\includegraphics[height=0.8\textheight]{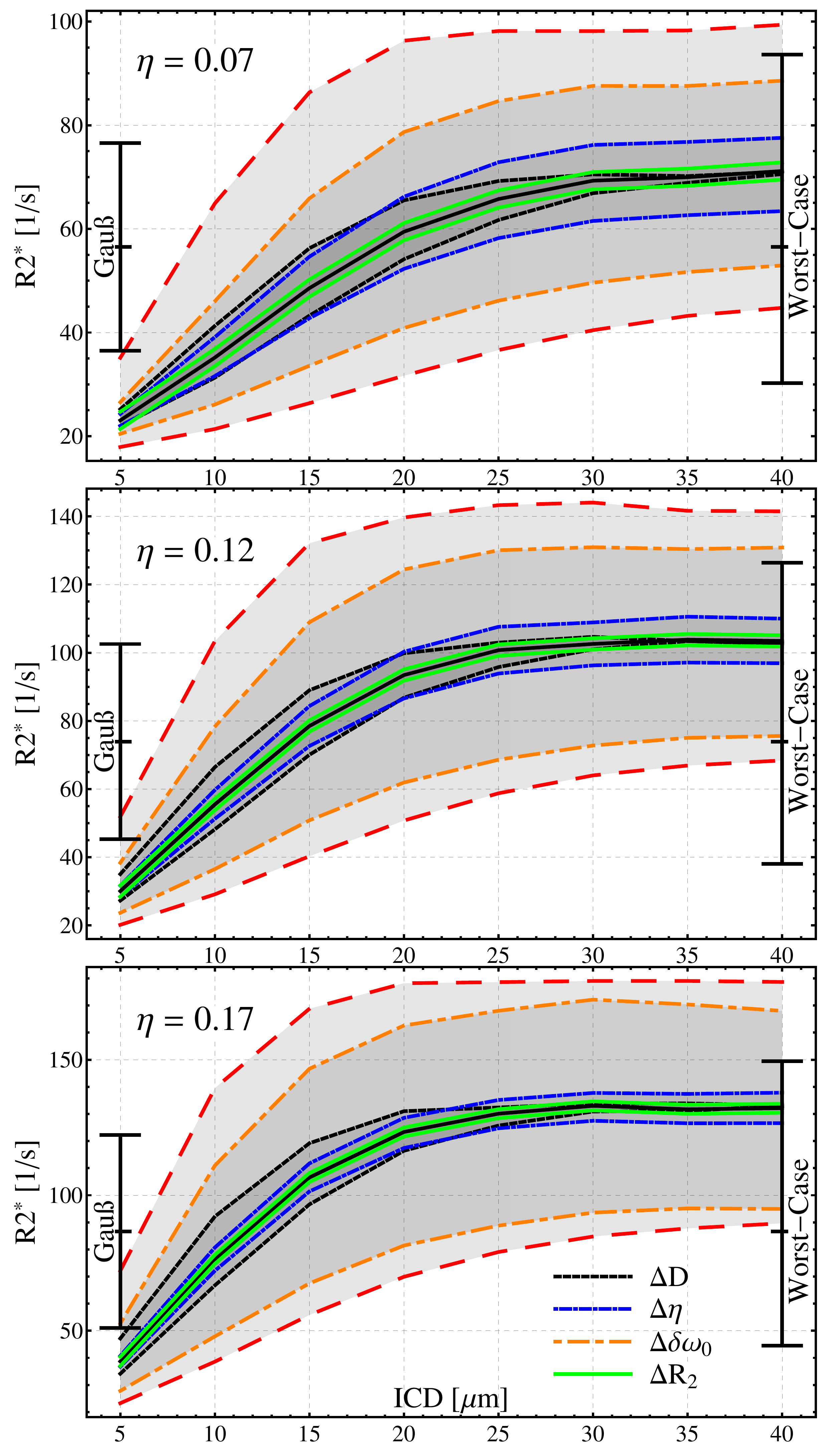}\end{center}
		\caption{Relaxationszeit (Gl. (\ref{eq:r2s-funktion})) in Abhängigkeit vom $\ICD$ bei 7T ($\overline{\dom_0}=750\radps$), $R_2=18.3s^{-1}$ und $D=1.5\mum^2/\ms$. Für konstantes $\eta$ nehmen dabei für steigenden $\ICD$ auch die Kapillarradien $R_c$ zu. Bei einer Zuordnung eines $\ICD$ Bereichs zu einer Relaxationsrate müssen auch die Fehler der Relaxationsrate berücksichtigt werden (siehe graue Linien in Abb. \ref{fig:auswertung} links unten). Dazu wurden die Fehler der Relaxationsrate nach gaußscher und Worst-Case-Abschätzung (siehe Kap. \ref{kap:relaxationsrate-realistisch}) ebenfalls markiert.}
		\label{fig:icd-analysis-7T}
	\end{figure}

In den vorangegangenen Kapiteln wurde primär der Unterschied im Relaxationsverhalten bedingt durch den Ordnungsparameter $\Gamma$ und $\dom_0$ untersucht. Die weiteren in das Modell einfließenden Parameter $R_c$, $\eta$, $\ICD$ und $D$ wurden jedoch konstant gehalten. Im Folgenden soll nun auch der Einfluss der restlichen Parameter auf die Relaxation untersucht werden. Eine konsequente Verwendung von Relaxationsraten an Stelle der Relaxationszeiten vermeidet dabei Schreib- und Rechenaufwand. Da außerdem $R_c$, $\eta$ und $\ICD$ über die in Tab. \ref{tab:model-parameters} gegebene Funktion verknüpft sind, ergibt sich die Relaxationsrate nach Gl. (\ref{eq:r2s-relax-rate-approximation}) für die verschiedenen Modellgeometrien als Funktion der folgenden Parameter
	\begin{align}
		R_2^*=R_2^*\left(\eta,\ICD,\dom_0,D,R_2\right)\approx R_{2,sim}^*\left(\eta,\ICD,\dom_0,D\right)+ R_2.
		\label{eq:r2s-funktion}
	\end{align}
Der vierdimensionale Parameterraum dieser Funktion wurde im Rahmen dieser Arbeit zunächst nur für $D=1.0\mum^2/\ms$ untersucht. Für die Parameter $\eta=0.04$ bis $\eta=0.20$ und $\ICD=5\mum$ bis $\ICD=40\mum$ wurden Schrittweiten $0.01$ für $\eta$ und $1\mum$ für den $\ICD$ verwendet. Als Offresonanzen wurden die Werte $\dom_0=\{50,100,150, 200, 250,300, 500, 750, 1000\}\radps$ verwendet. Für den Ordnungsgrad wurde $\Gamma=4$ (entspricht nach \cite{Karch2006} in etwa realem Gewebe) und $\Gamma=\infty$ für das hexagonale Gitter simuliert.\newline
Für einen einzigen Wert der Diffusionskonstante führt dies bereits zu $17\cdot46\cdot9\cdot2\approx14000$ voneinander unabhängigen Simulationen. Bei einer Rechenzeit von ca. $3$h pro Simulation (ein Kern, Intel Xeon $2000$GHz) führt das zu einer Gesamtrechenzeit von mehr als $1500$ Tagen auf einer CPU. Durch Verwendung verschiedener Rechner mit mehreren Kernen konnten die Simulationen innerhalb einer Zeitspanne von ca. zwei Monaten durchgeführt werden.\newline
Für höhere Diffusionskonstanten $D=1.5\mum^2/\ms$ und $D=2.0\mum^2/\ms$ wurden daher die anderen Parameter auf eine kleinere Auswahl beschränkt. In dem so definierten Volumen des Parameterraums wurde Gl. \ref{eq:r2s-funktion} dann durch lineare Interpolation zwischen den Stützpunkten approximiert. Für alle Parameterkombinationen wurde außerdem unter Ausnutzung des entwickelten Algorithmus für die Mathieu-Funktionen (Kap. \ref{kap:mathieu}) die aus der analytischen Lösung nach \cite{ZienerPHDThesis}, Gl. (154), folgenden Relaxationszeiten für das Krogh-Modell berechnet.\newline
Die Simulation liefert für vorgegebenen Eingabewerte $\eta$, $\dom_0$, $\ICD$ und $D$ genau eine Relaxationsrate ohne Fehler. Um die Bandbreite der zu diesen Relaxationsraten möglichen $\ICD$ Werte zu bestimmen, muss jedoch berücksichtigt werden dass in der Realität die Gewebeparameter nur innerhalb ihrer Fehlergrenzen bekannt sind (siehe auch Abb. \ref{fig:auswertung}). Es wird daher davon ausgegangen, dass in Gl. \ref{eq:r2s-funktion} folgenden Größen fehlerbehaftet sind
	\begin{align}		
		\begin{array}{rcrcl}
			\eta				&=&\overline{\eta} 				&+&\Delta\eta 				\\
			\dom_0			&=&\overline{\dom_0}			&+&\Delta\dom_0				\\
			D						&=&\overline{D} 					&+&\Delta D						\\
			R_{2,real}	&=&\overline{R_{2,real}} 	&+&\Delta R_{2,real}.	\\
		\end{array}	
		\label{eq:params-with-error}
	\end{align}
Daraus folgt für jeden als exakt angenommenen $\ICD$ mit Gl. (\ref{eq:r2s-funktion}) ein Fehler $\Delta R_{2,sim}^*$. 
Als Mittelwerte und Fehler werden für die weitere Auswertung angenommen
	\begin{align}		
		\begin{array}{rclccc}
		\eta 				&=&(7\pm1)\% 											&	\mbox{bzw.}&(12\pm1)\% 				&	\mbox{bzw.}\quad(17\pm1)\% 					\\
		\dom_0			&=&(150\pm50)\frac{\mathsf{rad}}{\mathsf{s}}							&	\mbox{bzw.}&(750\pm250)\frac{\mathsf{rad}}{\mathsf{s}}	&															\\
		D						&=&(1.5\pm0.5)\frac{\mum^2}{\ms}					&																		&																		\\
		T_{2,real}	&=&(55\pm5)\ms										&																		&																		\\
		R_{2,real}	&=&(18.3\pm1.7)\frac{1}{\mathsf{s}}			&																		&																		\\
		\end{array}
		\label{eq:error-containing-parameters}
	\end{align}
Die $\dom_0$ sind wie im letzten Kapitel an Tab. \ref{tab:offres-influence}, also $1.5T$ und $7T$ angelehnt. Die $\eta$-Werte decken in etwa den in \cite{Waller00} ermittelten Bereich ab. Auch die Diffusionskonstante und $R_2$ decken den üblicherweise in der Literatur verwendeten Bereich ab (siehe z.B. \cite{ZienerPHDThesis} oder \cite{Bauer99}).\newline
Verläufe von Gl. (\ref{eq:r2s-funktion}) für die verschiedenen Offresonanzen und RBVs sind in Abb. \ref{fig:icd-analysis-1.5T} und Abb. \ref{fig:icd-analysis-7T} aufgetragen. Für konstantes $\eta$ nehmen dabei für steigenden $\ICD$ auch die Kapillarradien $R_c$ zu. Für jede in die Funktion eingehende Größe wurde außerdem die durch sie verursachte maximale Abweichung sowie die durch alle Fehler zusammen verursachte maximale Abweichung eingetragen (siehe auch Abb. \ref{fig:auswertung}). Vor allem bei hohen Feldstärken bzw. höherem RBV ist der Fehler der Offresonanzen dominant.
	
\subsection{Realistische Relaxationsraten für reales Gewebe}\label{kap:relaxationsrate-realistisch}
Um eine realistische Relaxationsrate für reales Gewebe abzuschätzen muss zunächst von einem bekannten realistischen $\ICD$ ausgegangen werden. Nach \cite{Bauer99} gilt $R_c\approx(2.5\pm0.5)\mum$. Daraus folgt mit $\eta$ aus Gl. (\ref{eq:error-containing-parameters}) und Tab. \ref{tab:model-parameters} für den mittleren $\ICD$
\begin{align}
	\ICD=\frac{\sqrt{2\pi}}{2\cdot3^{1/4}}\left(\frac{(\overline{R_c}+\Delta R_c)}{\sqrt{\overline{\eta}-\Delta\eta}}+\frac{(\overline{R_c}-\Delta Rc)}{\sqrt{\overline{\eta}+\Delta\eta}}\right).
	\label{eq:mean-icd}
\end{align}
Für stärkere Offresonanzen und höhere RBV ergibt sich eine schnellere Relaxation, höhere Diffusion hingegen senkt die Relaxationsrate. Minimale und maximale Relaxationsraten ergeben sich also im schlimmsten Fall aus Gl. (\ref{eq:r2s-funktion}), (\ref{eq:error-containing-parameters}) und (\ref{eq:mean-icd}) zu
\begin{align}
	\begin{array}{rcl}
		R_{2,min}^*&=&R_2^*(\overline{\eta}-\Delta\eta,\ICD,\overline{\dom_0}-\Delta\dom_0,\overline{D}+\Delta D) + \overline{R_{2,real}}	- \Delta R_{2,real} \\
		\overline{R_2^*}&=&R_2^*(\overline{\eta}					 ,\ICD,\overline{\dom_0}						 ,\overline{D}) 					+ \overline{R_{2,real}} 						\\
		R_{2,max}^*&=&R_2^*(\overline{\eta}+\Delta\eta,\ICD,\overline{\dom_0}+\Delta\dom_0,\overline{D}-\Delta D) + \overline{R_{2,real}}	+ \Delta R_{2,real}. \\
	\end{array}
	\label{eq:r2s-worst-case-approximation}
\end{align}
Tab. \ref{tab:2D1CP-Relaxations-icd} listet für das Plasma-Modell ($\Gamma=4$) die so abgeschätzten Relaxationsraten bzw. Relaxationszeiten auf. Ausgehend von voneinander unabhängigen Fehlern kann wie in Kap. \ref{kap:fehlerermittlung} auch die gaußsche Fehler\-fort\-pflanzung verwendet werden
	\begin{align}
		\left(\Delta R_{2,gauss}^*\right)^2 = \left(\frac{\partial R_2^*}{\partial\dom_0}\right)^2\Delta\dom_0^2
																				+ \left(\frac{\partial R_2^*}{\partial D}\right)^2\Delta D^2
																				+ \left(\frac{\partial R_2^*}{\partial \eta}\right)^2\Delta\eta^2
																				+ \left(\frac{\partial R_2^*}{\partial R_2}\right)^2\Delta R_2^2.
		\label{eq:r2s-gauss-approximation}
	\end{align}
Die daraus folgenden Fehler sind ebenfalls in Tabelle \ref{tab:2D1CP-Relaxations-icd} zu finden. In beiden Fällen folgt der Großteil der Fehler wie bereits im vorangegangenen Kapitel aus der Ungewissheit über die Offresonanzen. Die aus Gl. (\ref{eq:params-with-error}) und (\ref{eq:r2s-funktion}) folgende Ungewissheit $\Delta R_{2,sim}^*$ sollte nicht mit einem aus einer tatsächlichen $R_2^*$-Messung folgenden Messfehler $\Delta R_{2,real}^*$ verwechselt werden. Bei einer $T_2^*$-Messung ist der Messfehler durch den Messprozess und das Messgerät bestimmt und wird in der Größenordnung von einigen Millisekunden liegen. Das oben bestimmte $\Delta R_{2,sim}^*$ wird ausschließlich durch die in Gl. (\ref{eq:params-with-error}) aufgelisteten und somit in die Simulation eingehenden Ungewissheiten verursacht und kann sich deutlich von $\Delta R_{2,real}^*$ unterscheiden.\newline
In Abb. \ref{fig:icd-analysis-1.5T} und Abb. \ref{fig:icd-analysis-7T} sind beide Abschätzungen eingetragen. Die Worst-Case-Abschätzung ist identisch mit der maximalen Aufspaltung der Relaxationsraten für den mittleren $\ICD$ (siehe Tab. \ref{tab:2D1CP-Relaxations-icd}).

	\begin{table}
		\renewcommand\arraystretch{1}
		\begin{center}
		\begin{tabular}[ht]{l|c|c|cccc|ccc} 	

		$\dom_0$ & $\eta$ & $\ICD$	&	$R_{2,min}^*$ & $\overline{R_2^*}$ &	$R_{2,max}^*$ &$\Delta R_{2, gauss}^*$ & $T_{2,min}^*$ & $T_{2,avg}^*$ &	$T_{2,max}^*$ \\
		\hline
		\hline
		\multirow{3}{0.9cm}{\begin{sideways}\parbox{2.5cm}{$\displaystyle(150\pm50)\frac{\mathsf{rad}}{\mathsf{s}}$}\end{sideways}}
						& $ (7\pm1)\%$	&	 18.4					&		17.3				& 20.7					& 26.9					& 	2.4									&	37.1(43.2)	& 48.3 				& 57.8(54.7)\\[1.5ex]
						& $(12\pm1)\%$	&	 13.9					&		17.7				&	21.8					& 29.1					& 	3.0									& 34.4(40.3)	&	45.8				& 56.5(53.1)\\[1.5ex]
						& $(17\pm1)\%$	&	 11.6					&		18.1				& 22.7					& 31.1					&		3.5									& 32.1(38.2)	&	44.1				& 55.4(52.1)\\[1.5ex]
		\hline
		\multirow{3}{0.9cm}{\begin{sideways}\parbox{2.8cm}{$\displaystyle(750\pm250)\frac{\mathsf{rad}}{\mathsf{s}}$}\end{sideways}}
						& $ (7\pm1)\%$	&	 18.4					&		29.9				& 56.4					& 94.4					& 	 9.9								&	10.6(15.1)	& 17.7				& 33.4(21.5)\\[1.5ex]
						& $(12\pm1)\%$	&	 13.9					&		37.7				&	73.9					& 127.7					&		12.6								&	 7.8(11.5)	&	13.5				& 26.5(16.3)\\[1.5ex]
						& $(17\pm1)\%$	&	 11.6					&		44.2				&	87.3					& 152.6					& 	15.1								&	 6.6( 9.7)	&	11.4				& 22.6(13.8)\\[1.5ex]
		\end{tabular}
		\end{center}
		\caption{Aus Simulationen bestimmte Relaxationsraten (in $s^{-1}$) und Relaxationszeiten (in $\ms$) für $\Gamma=4$.	Für die gaußsche Fehlerabschätzung sind die Relaxationszeiten in Klammern angegeben. Die $\ICD$ Werte sind in $\mum$ gegeben.}
		\label{tab:2D1CP-Relaxations-icd}
	\end{table}

\subsection{Rückschlussmöglichkeit auf ICD}\label{kap:icd-ruecklschluss}
Möchte man aus den Relaxationszeiten auf den $\ICD$ rückschließen, muss dazu die Zuordnung aus Gl. (\ref{eq:r2s-funktion}) nach $\ICD$ aufgelöst, also die Funktion
	\begin{align}
		\ICD=\ICD\left(R_2^*,\eta,\dom_0,D,R_{2,real}\right)
		\label{eq:icd-funktion}
	\end{align}
aufgestellt werden. Da u.a. $\eta$ in diese Gleichung einfließt, muss das $RBV$ der zu Grunde liegenden Kapillarstruktur bekannt sein. Diese kann z.B. wie in \cite{Waller00} durch $T_1$-Messungen ermittelt werden.\newline
Bei Auflösen von Gl. (\ref{eq:r2s-funktion}) nach Gl. (\ref{eq:icd-funktion}) führt der diskret abgetastete Parameterraum zu Problemen. Im Allgemeinen ist auch nicht gegeben, dass überhaupt eine eindeutige Umkehrfunktion existiert. Um von den Relaxationszeiten auf den ICD zu schließen wird Gl. (\ref{eq:icd-funktion}) daher mittels Abb. \ref{fig:icd-analysis-1.5T} und Abb. \ref{fig:icd-analysis-7T} graphisch ausgewertet. Dazu sind in den Abbildungen auch die im vorangegangenen Kapitel ermittelten realistischen Relaxationsraten und deren Fehlerbereiche markiert. Die Fehler der anderen in Gl. (\ref{eq:icd-funktion}) eingehenden Größen sind bereits durch die verschiedenen Bänder markiert. Für fast alle Kombinationen aus $\eta$ und $\dom_0$ lassen sich nach der Worst-Case-Abschätzung  dem Relaxationsbereich $\ICD$-Werte aus der kompletten Bandbreite der untersuchten $\ICD$s zuordnen. Auch mit der optimistischeren gaußschen Abschätzung der Relaxationsraten (linker Balken in Abb. \ref{fig:icd-analysis-1.5T} und \ref{fig:icd-analysis-7T}) lässt sich, wegen der großen Ungewissheit über die Offresonanzen, dieser Bereich kaum eingrenzen.\newline
Es zeigt sich jedoch, dass mit der im nächsten Kapitel eingeführten Näherung für dreidimensionale Voxel die Offresonanzen genauer eingegrenzt werden können. Für eine Auswertung der Abhängigkeit der Relaxationsraten vom $\ICD$ bei konstanten Radien und entsprechend variablem RBV sei daher auf später verwiesen. Dort wird zudem auch die Worst-Case Abschätzung nach Gl. (\ref{eq:r2s-worst-case-approximation}) durch eine gaußsche Abschätzung nach Gl. (\ref{eq:r2s-gauss-approximation}) ersetzt.

\subsection{Von 2D nach 3D}\label{kap:2D-to-3D}
	\begin{figure}
		\begin{center}\includegraphics[width=0.7\textwidth]{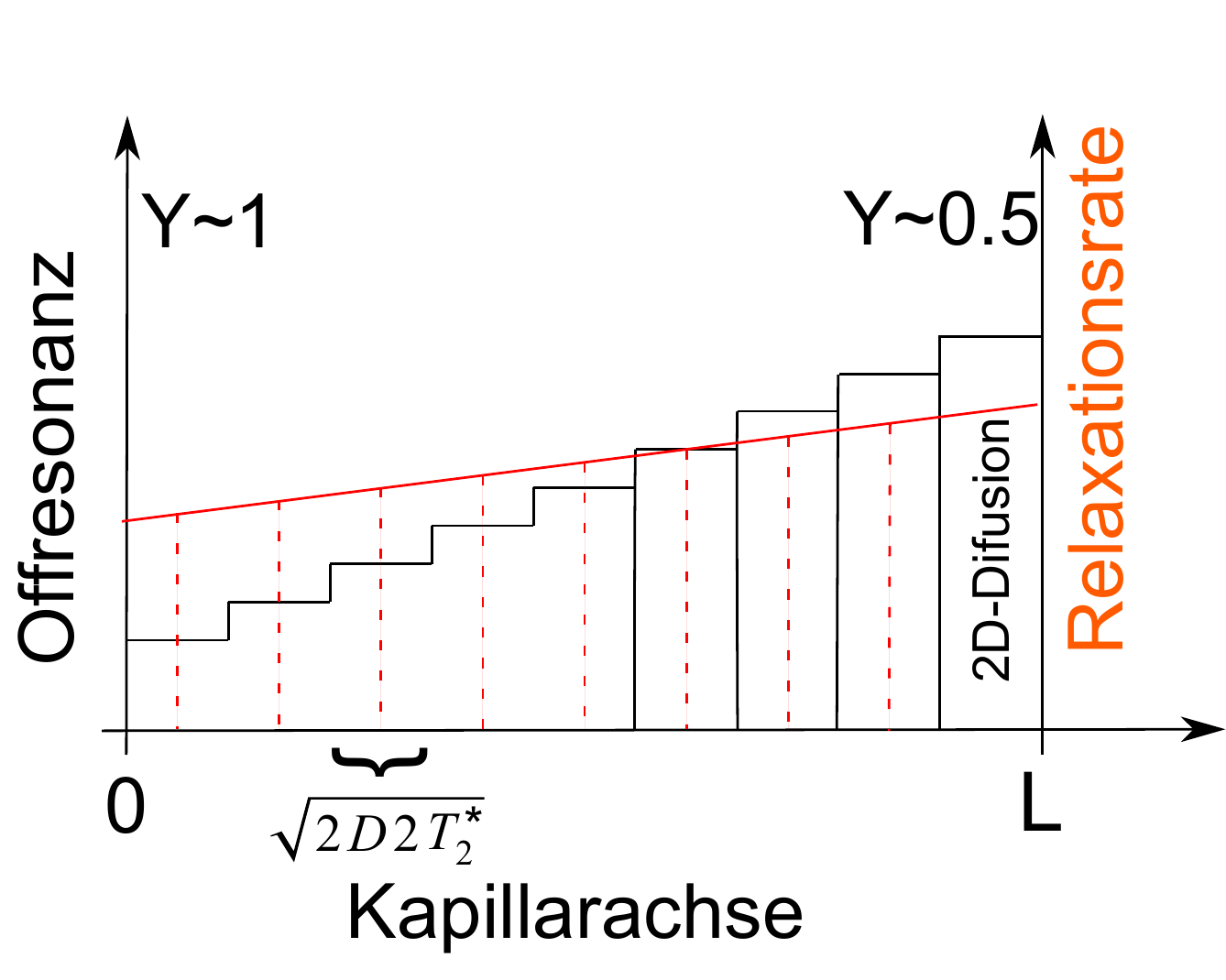}\end{center}
		\caption{Während der Relaxation können die Protonen nur über einen kleinen Bruchteil $\sqrt{2D\cdot2T_2^*}$ der Kapillarlänge diffundieren. Obwohl die Offresonanzstärke entlang der Kapillare variiert, da der Oxygenierungsgrad des Blutes $Y$ in Richtung des venösen Endes abnimmt, wird in diesem Bereich $\dom_0$ als konstant angenommen. Als grobe Näherung wird eine lineare Abnahme der Offresonanzen und der Relaxationszeiten entlang der Kapillarachse angenommen. Die Transversalmagnetisierung des dreidimensionalen Voxels kann dann näherungsweise berechnet werden, indem man über die Magnetisierungen der einzelnen Scheiben mittelt.}
		\label{fig:2D-to-3D}
	\end{figure}
	\begin{figure}
		\begin{center}\includegraphics[width=0.7\textwidth]{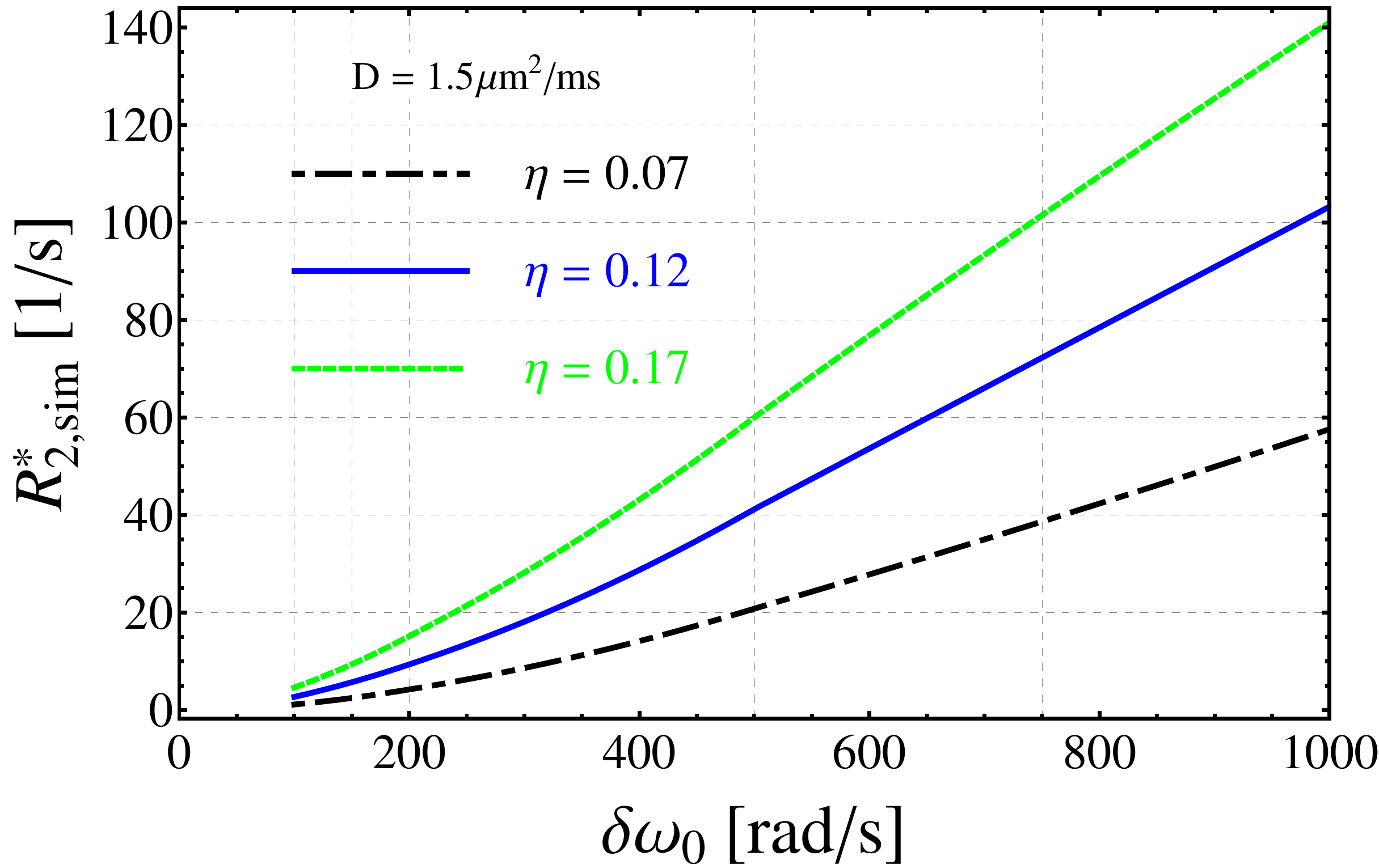}\end{center}
		\caption{Trägt man die Relaxationsrate gegen die Offresonanz auf, so ergibt sich über weite Bereiche ein näherungsweise linearer Zusammenhang. Für den $\ICD$ wurden entsprechend Tab. \ref{tab:2D1CP-Relaxations-icd} bzw. Gl. (\ref{eq:mean-icd}) realistische Werte gewählt. Die vertikalen Gitterlinien markieren die durch die Simulation erfassten Schnitte des Parameterraums senkrecht zur Offresonanzachse.}
		\label{fig:offresonance-linearity}
	\end{figure}
\begin{figure}
		\begin{center}
		\subfloat[Die Relaxationsrate des dreidimensionalen Voxels nach Gl. (\ref{eq:magnetization-3D-integrated-monoexp-approx}) ist fast identisch mit der mittleren Relaxationsrate der zweidimensionalen Schichten.]{
			\includegraphics[width=0.75\columnwidth]{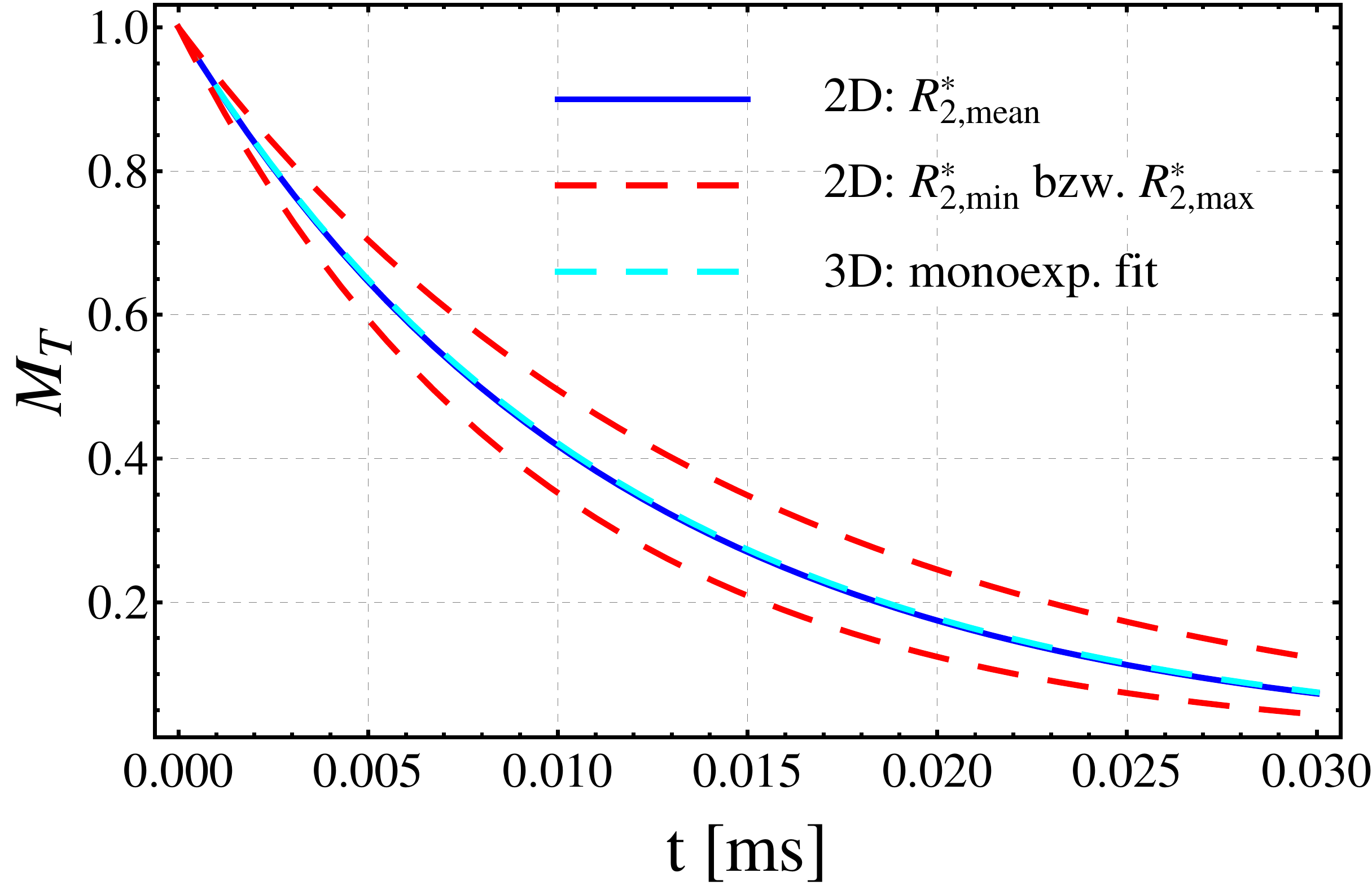}\label{fig:2D-to-3D-exp-decay}
		}\\
		\subfloat[Erst im Langzeitverhalten zeigen sich Unterschiede zwischen der diskreten Mittelung nach Gl. (\ref{eq:magnetization-3D-approx-linear}), der kontinuierlichen Mittelung nach Gl. (\ref{eq:magnetization-3D-integrated}) und dem monoexponentiell genäherten Verlauf nach Gl. (\ref{eq:magnetization-3D-integrated-monoexp-approx}).]{
			\includegraphics[width=0.75\columnwidth]{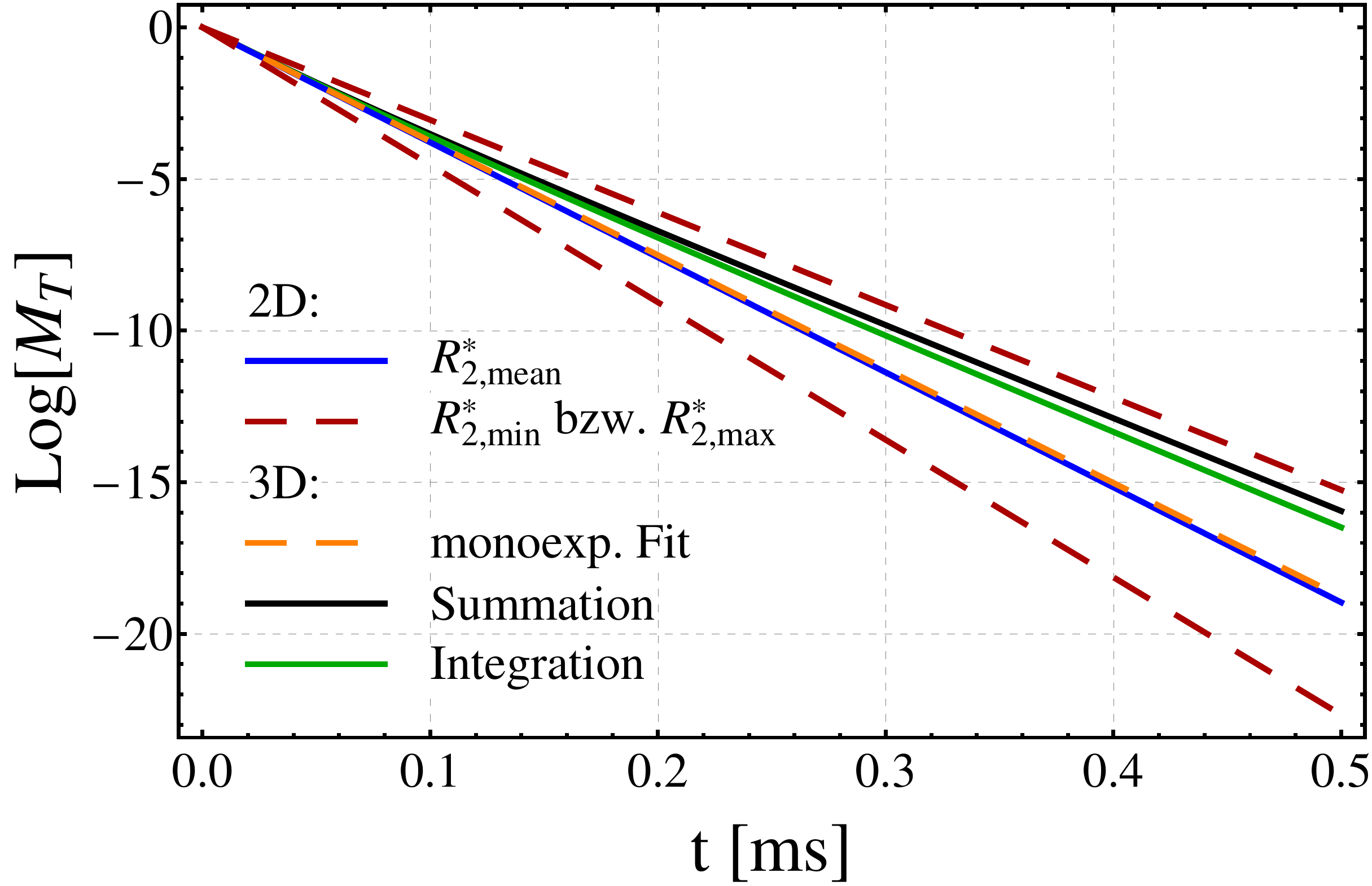}\label{fig:2D-to-3D-exp-decaylog}
		}
		\end{center}
		\caption{Zweidimensionaler und gemittelter dreidimensionaler Zerfall der Transversalmagnetisierung für $D=1.5\mum^2/\ms$, $\eta=0.17$ und $\dom_0=750\radps$. Die verschiedenen Re\-la\-xations\-raten ergeben sich zu $R_{2,min}^*=70.3\mathsf{s}^{-1}$, $R_{2,max}^*=104.4\mathsf{s}^{-1}$, $R_{2,mean}^*=87.3\mathsf{s}^{-1}$ und $R_{2,3D}^*=86.5\mathsf{s}^{-1}$. Die nach Gl. (\ref{eq:r2s-worst-case-approximation}) berechnete Relaxationsrate $\overline{R_2^*}$ beträgt $86.2s^{-1}$. Die durch den Oxygenierungsgrad verursachten Fehler in der Relaxationszeit einer zweidimensionalen Scheibe werden also durch eine Mittelung über die Scheiben weitestgehend aufgehoben. Für die intrinsische Relaxation wurde $R_2=18.3s^{-1}$ angenommen.}
		\label{fig:2D-to-3D-decays}
	\end{figure}
\begin{figure}
		\begin{center}
		\subfloat[Die relative Abweichung zwischen $R_{2,mean}^*$ aus Gl. (\ref{eq:2D-to-3D-mean-relaxation}) und $R_{2,3D}^*$ aus Gl. (\ref{eq:magnetization-3D-integrated-monoexp-approx}) hängt von $R_{2,mean}^*$ selbst und der Differenz $R_{2,max}^*-R_{2,min}^*$ ab.]{
			\includegraphics[width=0.75\columnwidth]{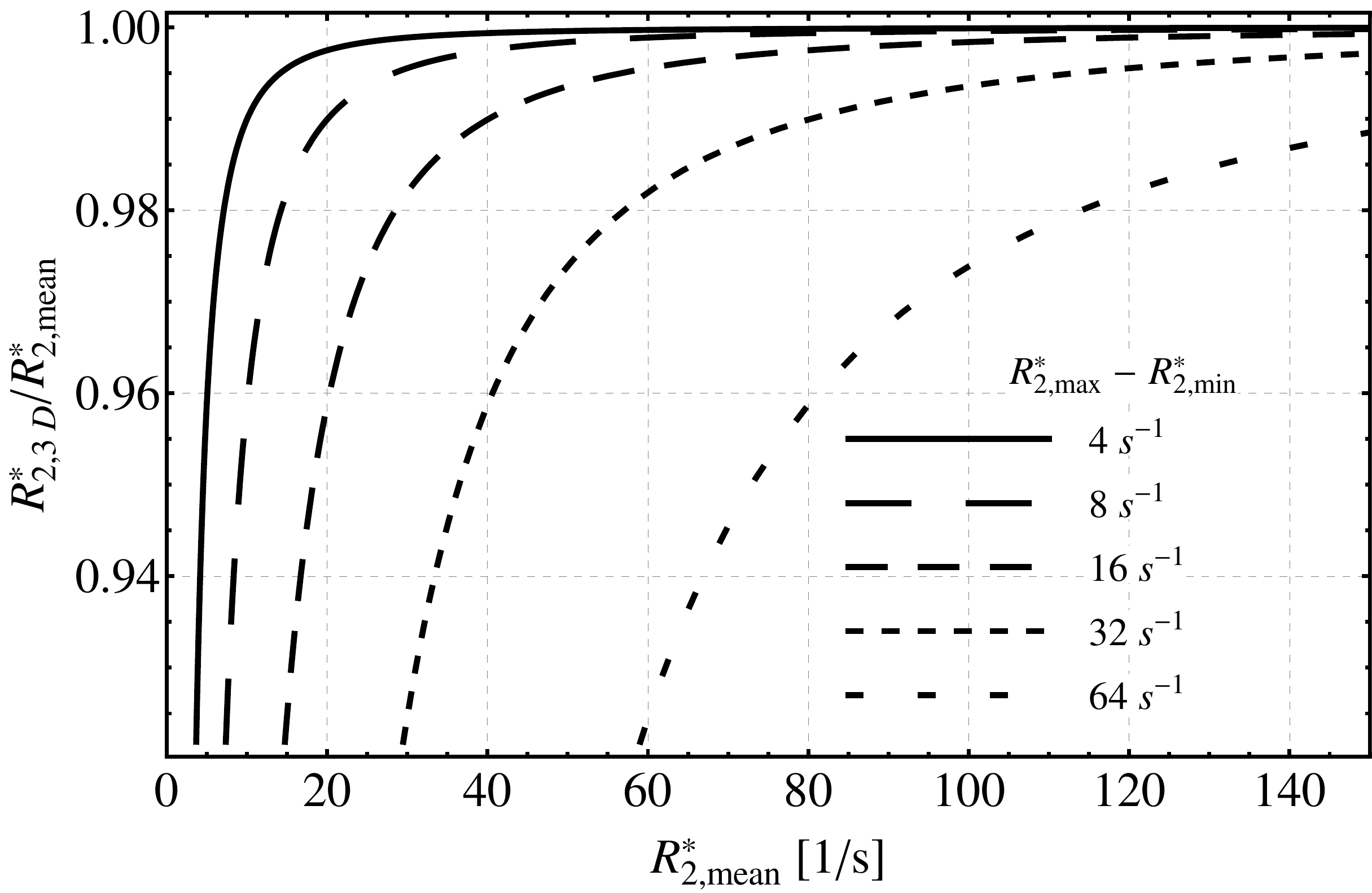}\label{fig:2D-to-3D-relaxation-errors}
		}\\
		\subfloat[]{
			\includegraphics[width=0.75\columnwidth]{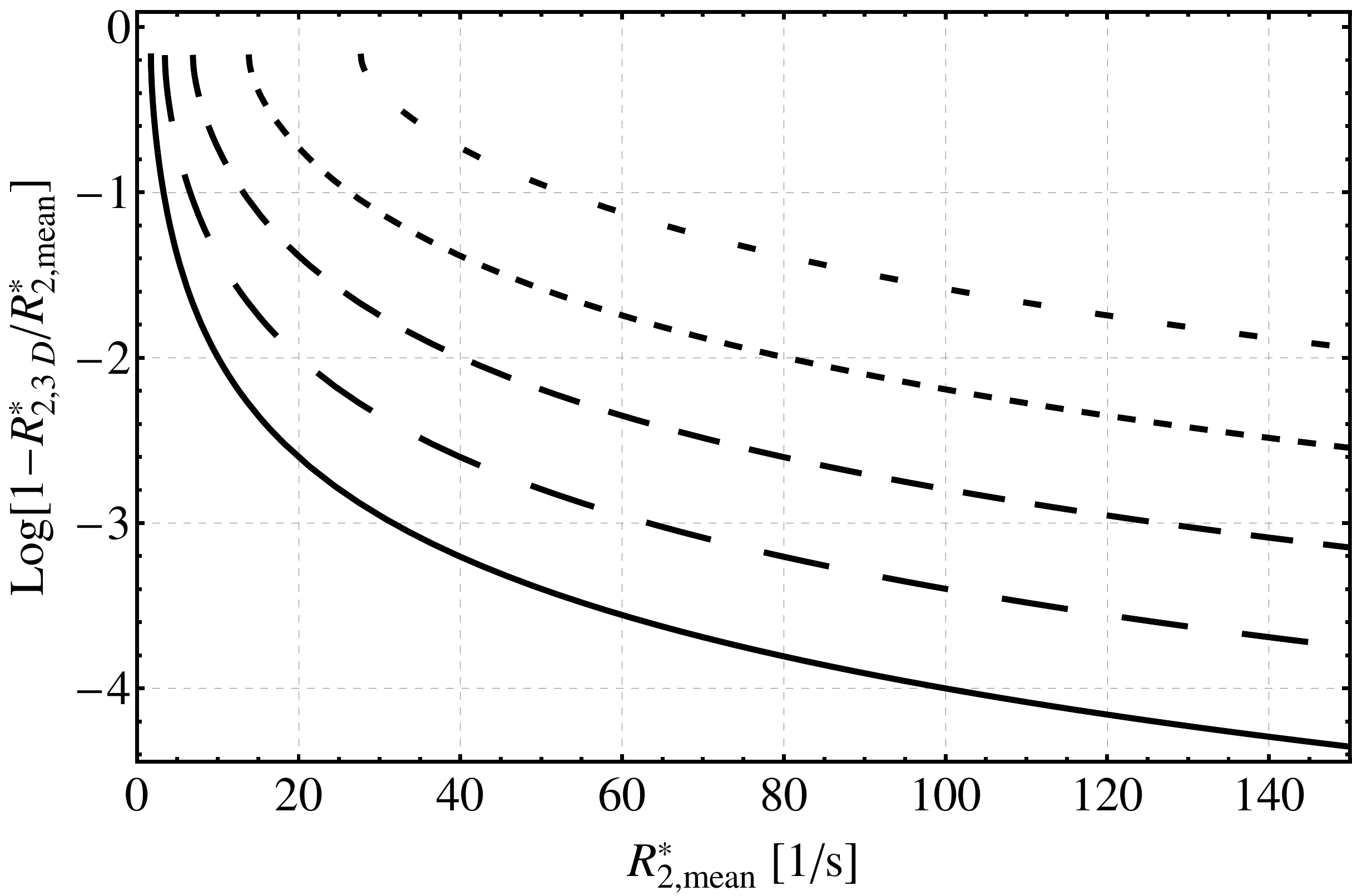}	\label{fig:2D-to-3D-relaxation-errors-log}
		}
		\end{center}
		\caption{Obwohl $R_{2,max}^*-R_{2,min}^*$  hier eigentlich nur die durch den Oxygenierungsgrad verursachte Aufspaltung der Relaxationsraten angibt, können die zu $\Delta\dom_0$ gehörenden Bänder aus Abb. \ref{fig:icd-analysis-1.5T} und Abb. \ref{fig:icd-analysis-7T} als grobe Vergleichsbasis herangezogen werden. Dort sind jedoch auch noch die anderenen in Tab. \ref{tab:offres-influence} aufgelisteten Fehler in $\Delta\dom_0$ enthalten, weshalb die Aufspaltung zwischen $R_{2,max}^*$ und $R_{2,min}^*$ überschätzt wird. Für 1.5T liegt die Aufpaltung unter $15\mathsf{s}^{-1}$, für 7T bei unterhalb von $80\mathsf{s}^{-1}$. Mit $R_{2,mean}^*\approx \overline{R_2^*}$ aus Tab. \ref{tab:2D1CP-Relaxations-icd} folgt aus Abb. \ref{fig:2D-to-3D-relaxation-errors} für 1.5T ein Unterschied zwischen $R_{2,mean}^*$ und $R_{2,3D}^*$ von weniger als $5\%$. Für 7T wird die Abweichung stärker.}
		\label{fig:2D-to-3D-relaxation-error}
	\end{figure}
\begin{figure}
		\begin{center}\includegraphics[height=0.83\textheight]{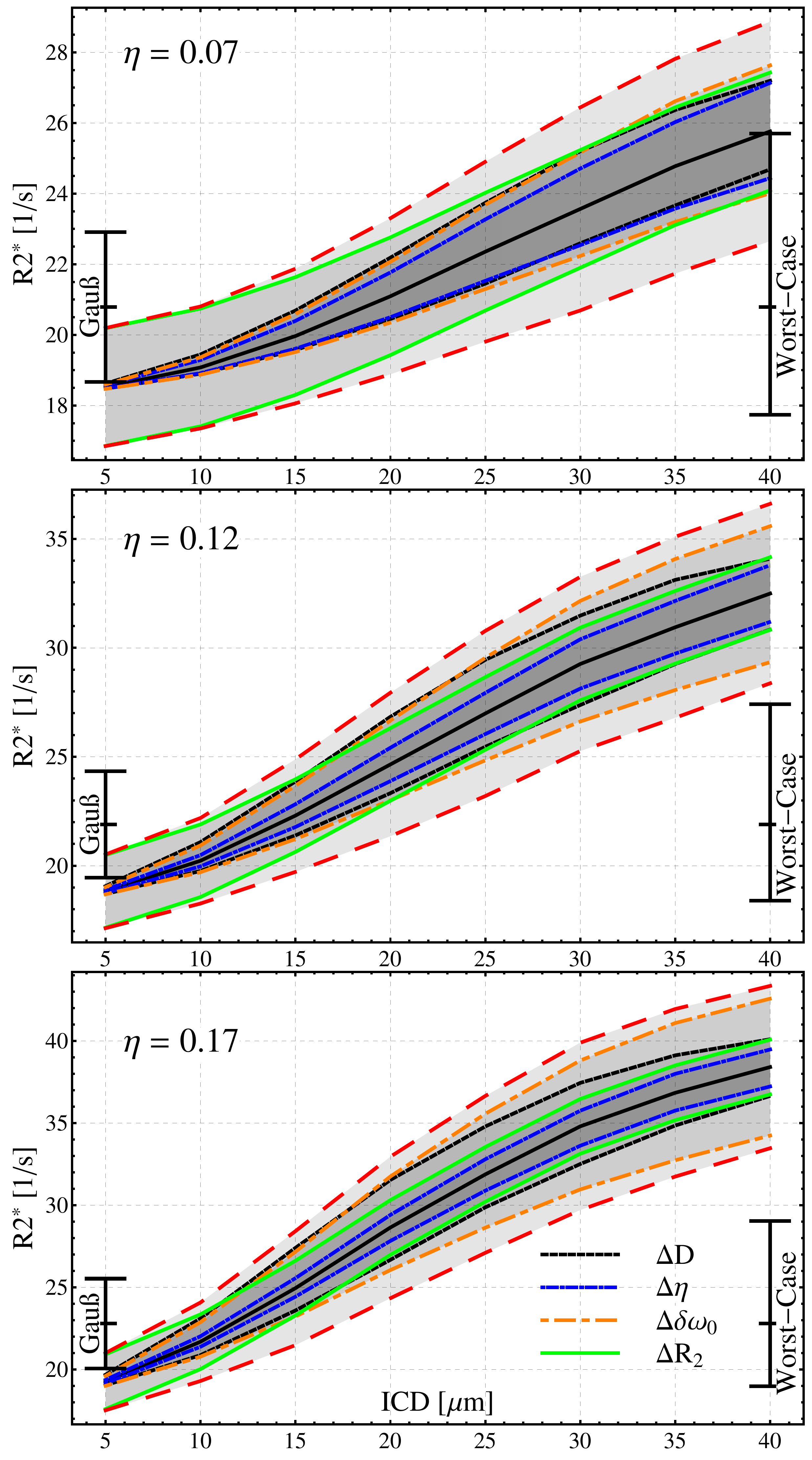}\end{center}
		\caption{Verlauf der Relaxationszeit in Abhängigkeit vom $\ICD$ bei 1.5T, $R_2=18.3s^{-1}$ und $D=1.5\mum^2/\ms$ bei konstantem $\eta$ (steigender $\ICD$ $\Rightarrow$ steigende $R_c$). Die umfassenden Linien bzw. schattierten Bereiche geben die durch die einzelnen Größen verursachten maximalen Abweichungen wieder. Für den Fehler der Offresonanzen wurde unter Vernachlässigung der Oxygenierung $\Delta\dom_0=25\radps$ angenommen. Das äußerste Band fasst alle Fehlerquellen in einer gaußschen Näherung zusammen.}
		\label{fig:icd-analysis-1.5T-offres-correction}
	\end{figure}
\begin{figure}
		\begin{center}\includegraphics[height=0.83\textheight]{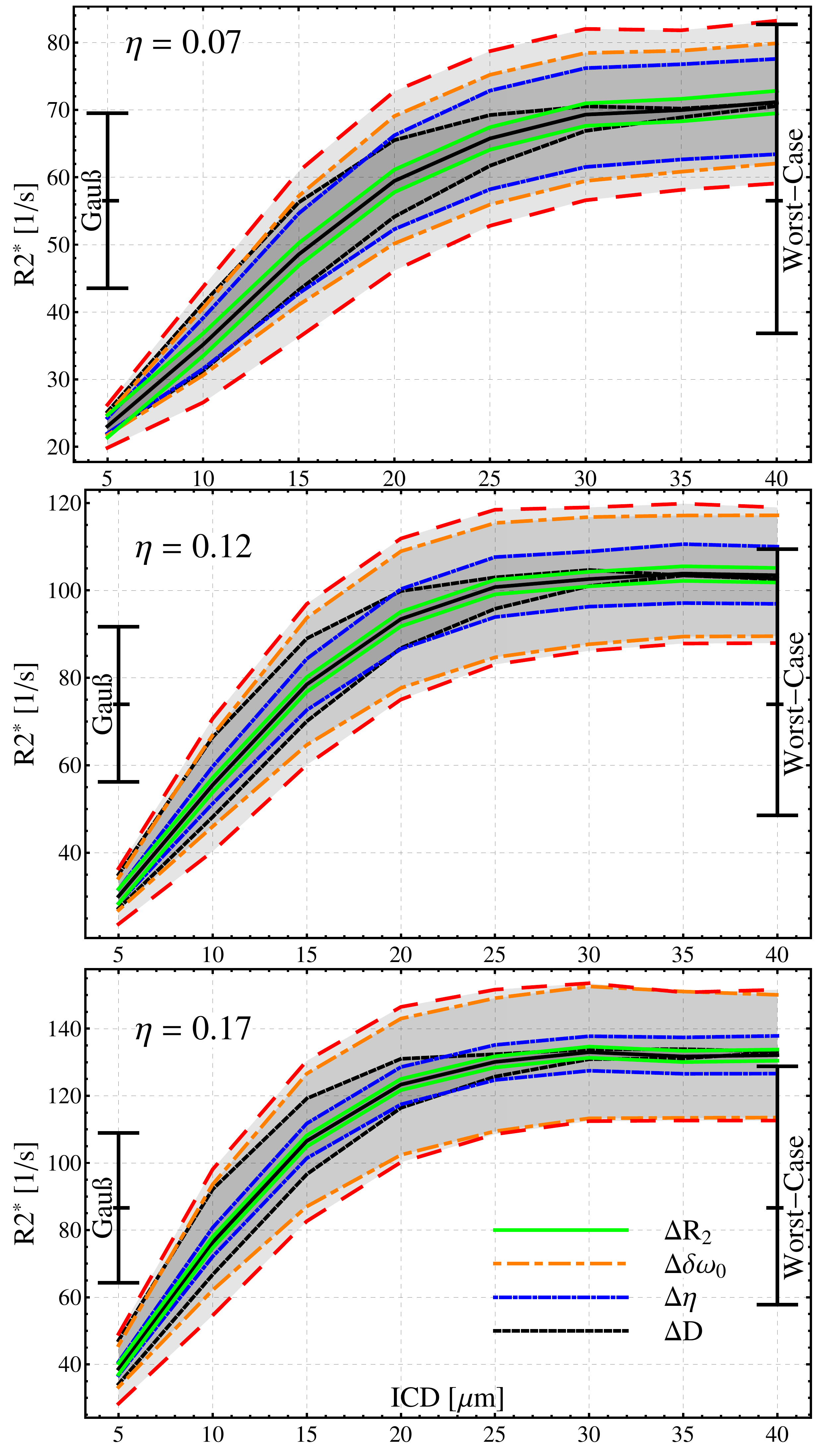}\end{center}
		\caption{Verlauf der Relaxationszeit in Abhängigkeit vom $\ICD$ bei 7T, $R_2=18.3s^{-1}$ und $D=1.5\mum^2/\ms$ bei konstantem $\eta$ (steigender $\ICD$ $\Rightarrow$ steigende $R_c$). Durch die deutlich optimistischere Fehlerabschätzung lässt sich nun der zu einer Relaxationszeit gehörende Bereich möglicher $\ICD$ Werte genauer eingrenzen. Für den Fehler der Offresonanzen wurde unter Vernachlässigung der Oxygenierung $\Delta\dom_0=125\radps$ angenommen.}
		\label{fig:icd-analysis-7T-offres-correction}
	\end{figure}
\begin{figure}
		\begin{center}\includegraphics[height=0.83\textheight]{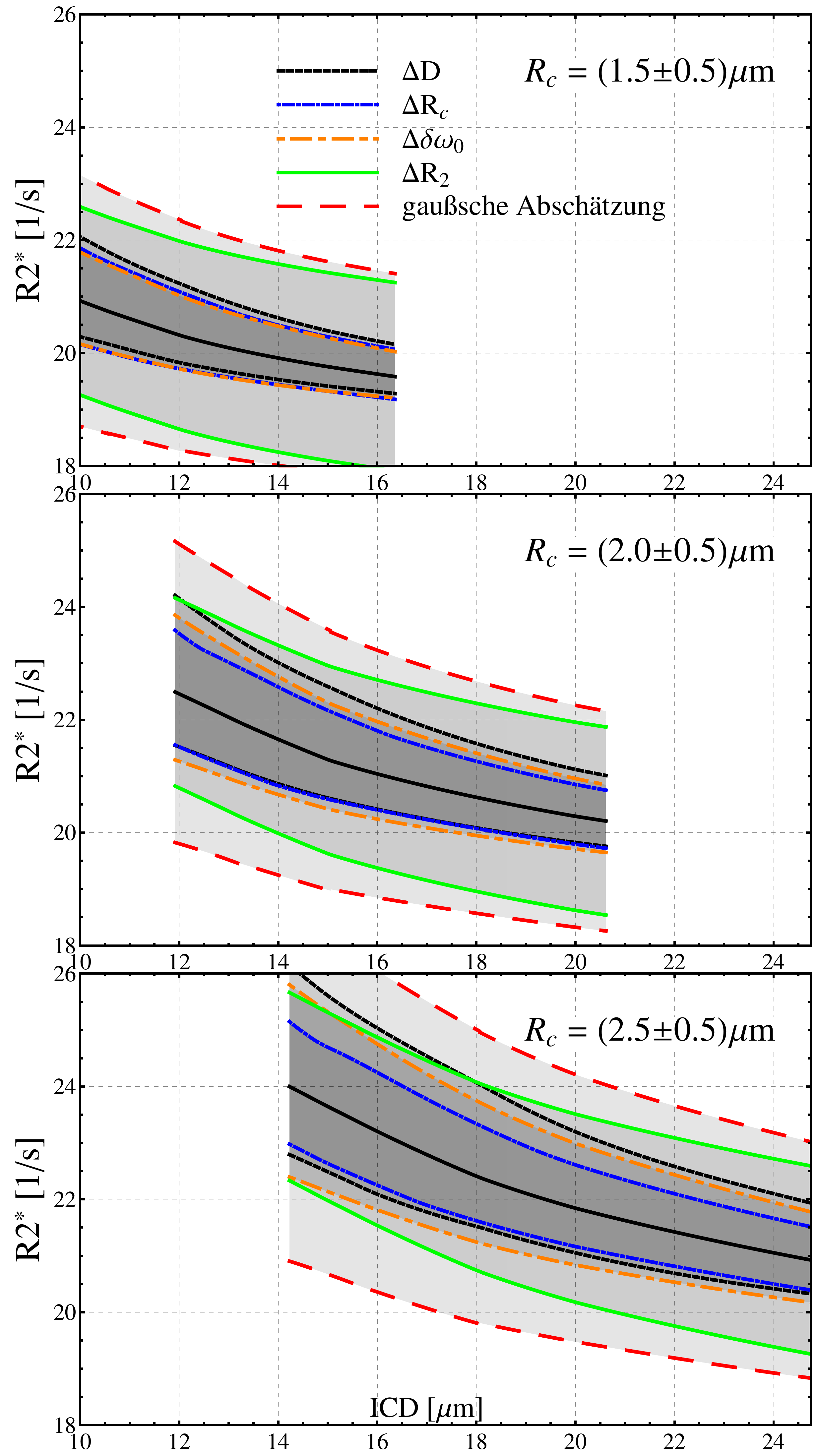}\end{center}
		\caption{Verlauf der Relaxationszeit in Abhängigkeit vom $\ICD$ bei 1.5T, $R_2=18.3s^{-1}$ und $D=1.5\mum^2/\ms$ bei konstantem $R_c$ (steigender $\ICD$ $\Rightarrow$ sinkendes RBV). Die umfassenden Linien bzw. schattierten Bereiche geben die durch die einzelnen Größen verursachten maximalen Abweichungen wieder. Für den Fehler der Offresonanzen wurde unter Vernachlässigung der Oxygenierung $\Delta\dom_0=25\radps$ angenommen. Das äußerste Band fasst alle Fehlerquellen in einer gaußschen Näherung zusammen.}
		\label{fig:icd-analysis-1.5T-offres-correction-const-radius}
	\end{figure}
\begin{figure}
		\begin{center}\includegraphics[height=0.83\textheight]{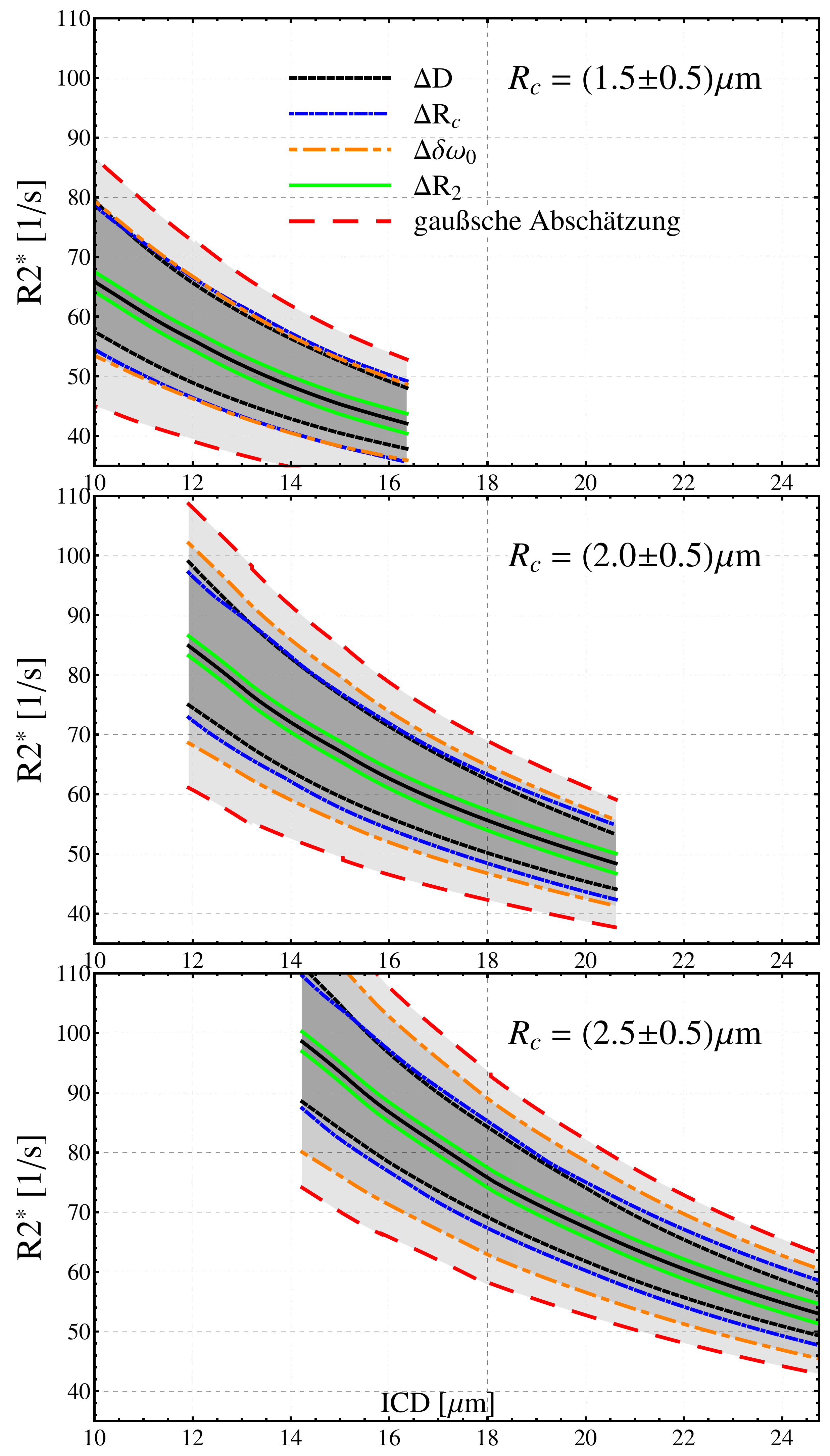}\end{center}
		\caption{Verlauf der Relaxationszeit in Abhängigkeit vom $\ICD$ bei 7T, $R_2=18.3s^{-1}$ und $D=1.5\mum^2/\ms$ bei konstantem $R_c$ (steigender $\ICD$ $\Rightarrow$ sinkendes RBV). Durch die deutlich optimistischere Fehlerabschätzung lässt sich nun der zu einer Relaxationszeit gehörende Bereich möglicher $\ICD$ Werte genauer eingrenzen. Für den Fehler der Offresonanzen wurde unter Vernachlässigung der Oxygenierung $\Delta\dom_0=125\radps$ angenommen.}
		\label{fig:icd-analysis-7T-offres-correction-const-radius}
	\end{figure}

Wie aus den Abb. \ref{fig:icd-analysis-1.5T} und \ref{fig:icd-analysis-7T} zu erkennen ist, macht die Ungewissheit $\Delta\dom_0$ (orangene Linie) meist den Hauptteil der Bandbreite möglicher Relaxationsraten aus. Nach Tab. \ref{tab:offres-influence} folgt der große Fehler $\Delta\dom_0$ hauptsächlich aus dem Oxygenierungsgrad $Y$. Zwi\-schen arteriellem und venösem Ende geben die Kapillaren einen Großteil des mitgeführten Sauerstoffs ans Gewebe ab \cite{Duncker08}, die zweidimensionalen Eigenschaften des verwendeten Modells verhindern also eine scharfe Eingrenzung von $Y$. Die bisher gefundenen Relaxations\-raten setzen nämlich eine konstante Offresonanz $\dom_0$ entlang der gesamten Kapillarachse voraus.\newline
Möchte man das zweidimensionale Modell besser an die realen Gegebenheiten anpassen, so muss man berücksichtigen, dass das Signal eines dreidimensionalen Voxels aus vielen unterschiedlichen zweidimensionalen "Scheiben" mit jeweils eigenen Offresonanzstärken zusammengesetzt wird. Diese Scheiben haben die Dicke $\overline{d}=\sqrt{2D\cdot 2 T_2^*}$, also die Stecke die ein Magnetisierungspaket im Mittel während der Relaxation zurücklegt. Ausgehend von $T_2^*<T_2\approx50\ms$ und $D\approx1.5\mum^2/\ms$ ergibt sich $\overline{d}\approx17\mum$. Bei ca. $100\mum$ Kapillarlänge muss also von etwa fünf bis zehn verschiedenen Scheiben ausgegangen werden. Diese Dicke ist zwar über die Relaxationszeit $T_2^*$ indirekt von den Offresonanzen abhängig, die intrinsische Relaxation mit $T_2$ gibt jedoch eine maximale Dicke vor (Abb. \ref{fig:2D-to-3D}).\newline
Für die über ein dreidimensionales Volumen gemittelte Magnetisierung ergibt sich also
	\begin{align}
		M_{T,3D}(t)\approx\frac{1}{N}\sum_{j=1}^N{\exp(-R_{2,j}^*\cdot t)}.
		\label{eq:magnetization-3D-approx}
	\end{align}
Wobei der Index $j$ über die verschiedenen Scheiben läuft und $R_{2,j}^*$ die aus der Simulation folgenden Relaxationszeiten für die entsprechenden Offresonanzen sind. Nimmt man als Näherung einen linearen Zusammenhang der Offresonanzen entlang der Kapillarachse an, so folgt mit Abb. \ref{fig:offresonance-linearity} in guter Näherung auch ein entsprechend linearer Verlauf der Relaxationsraten entlang der Kapillarachse. Damit ergibt sich
	\begin{align}
		M_{T,3D}(t)&\approx\frac{1}{L/\Delta l}\sum_{j=0}^{L/\Delta l}{\exp(-R_2^*(j\Delta l)\cdot t)} \quad\mbox{mit}\label{eq:magnetization-3D-approx-linear}\\
		R_2^*(l)   &=      \frac{l}{L}(R_{2,max}^*-R_{2,min}^*)+R_{2,min}^*.																					\label{eq:magnetization-3D-approx-relaxations-linear}	
	\end{align}
$\Delta l$ als Scheibendicke sollte etwa in der Größenordnung von $\overline{d}$ liegen und die Gesamtlänge $L$ ganzzahlig teilen. $R_{2,max}^*$ und  $R_{2,min}^*$ bezeichnen die Relaxationsraten am arteriellen bzw. venösen Ende der Kapillaren, $R_{2,mean}^*$ den Mittelwert aus $R_{2,min}^*$ und  $R_{2,max}^*$. Der Faktor $L^{-1}(R_{2,max}^*-R_{2,min}^*)+R_{2,min}^*$ in Gl. (\ref{eq:magnetization-3D-approx-relaxations-linear}) entspricht in der linearen Näherung der Ableitung $\partial R_2^*/\partial l$.\newline
Für niedrige Diffusionskonstanten, kurzes $T_2$ oder sehr lange Kapillaren gehen immer mehr Summanden in Gl. (\ref{eq:magnetization-3D-approx-linear}) ein. Die Summe kann dann in ein Integral überführt werden
	\begin{align}
				M_{T,3D}(t)	&=\frac{1}{L}\int_0^L{\mathsf{d}l \exp(-R_2^*(l)\cdot t)}\notag\\[0.3ex]
								&=\frac{\exp(-R_{2,min}^* \cdot t)-\exp(-R_{2,max}^*\cdot t)}{t\cdot(R_{2,max}^*-R_{2,min}^*)}.
		\label{eq:magnetization-3D-integrated}
	\end{align}
Nähert man Gl. (\ref{eq:magnetization-3D-integrated}) wiederum durch einen monoexponentiellen Zerfall mit der Relaxations\-rate $R_{2,3D}^*$ ergibt sich eine minimale Abweichung
	\begin{align}
				\mathsf{min}||\frac{\exp(-R_{2,min}^* \cdot t)-\exp(-R_{2,max}^*\cdot t)}{t\cdot(R_{2,max}^*-R_{2,min}^*)}-\exp(-R_{2,3D}^*\cdot t)||\notag\\
				\mbox{für}\quad R_{2,3D}^*=\frac{1}{6}\left(R_{2,max}^*+R_{2,min}^*+\sqrt{R_{2,max}^{*2}+14 R_{2,max}^* R_{2,min}^*+R_{2,min}^{*2}}\right).
		\label{eq:magnetization-3D-integrated-monoexp-approx}
	\end{align}
Der Verlauf von $R_{2,3D}^*$ in Abhängigkeit von $R_{2,max}^*$ und $R_{2,min}^*$ ist in Abb. \ref{fig:2D-to-3D-relaxation-error} aufgetragen. Wie zu sehen ist, gibt es für hohe Relaxationsraten $R_{2,mean}^*$ auch bei großen Abständen zwischen $R_{2,min}^*$ und $R_{2,max}^*$ nur einen geringen Unterschied zu $R_{2,3D}^*$.
Die Relaxationsrate $R_2$ lässt sich sowohl bei Summation als auch bei Integration als multiplikativer Faktor abspalten.\newline
Mit den Annahmen aus Tab. \ref{tab:offres-influence} ist etwa die Hälfte des Offresonanzfehlers durch die Oxygenierung verursacht.  $R_{2,min}^*$ und $R_{2,max}^*$ folgt mit Gl. \ref{eq:r2s-funktion} für gegebene andere Parameter daher zu
	\begin{align}
		R_{2,min}^*\approx &\left(\eta,\ICD,\dom_0-0.5\Delta\dom_0,D\right)+ R_{2,real}\notag\\[1ex]
		R_{2,max}^*\approx &\left(\eta,\ICD,\dom_0+0.5\Delta\dom_0,D\right)+ R_{2,real}\notag\\[1ex]
		R_{2,mean}^*=&\frac{1}{2}(R_{2,min}^* +	R_{2,max}^*)\approx \overline{R_2^*}.		\label{eq:2D-to-3D-mean-relaxation}
	\end{align}
In Abb. \ref{fig:2D-to-3D-exp-decay} und Abb. \ref{fig:2D-to-3D-exp-decaylog} ist der aus Gl. (\ref{eq:magnetization-3D-approx-linear}) und Gl. (\ref{eq:magnetization-3D-integrated}) folgende Verlauf $M_{T,3D}(t)$ aufgetragen. Der Vergleich mit der Relaxationsrate der mittleren Scheibe zeigt, dass der aus dem dreidimensionalen Voxel folgende multiexponentielle Zerfall sehr gut durch die mitt\-le\-re zweidimensionale Relaxationsrate wiedergegeben wird. Die Fehler der Off\-resonanzen durch den Oxygenierungsgrad haben also in einem dreidimensionalen Voxel nur geringen Einfluss auf die Relaxationsrate. Unter der Annahme es gelte $R_{2,mean}^*\approx R_{2,3D}^*$ kann man in Tab. \ref{tab:offres-influence} also den durch die Oxygenierung $Y$ bedingten Fehler der Offresonanzen vernachlässigen. Ersetzt man zudem die Worst-Case-Abschätzung aus Gl. (\ref{eq:r2s-worst-case-approximation}) ebenfalls durch eine gaußsche Abschätzung, so schrumpfen die Fehlerbereiche aus Abb. \ref{fig:icd-analysis-1.5T} und Abb. \ref{fig:icd-analysis-7T} deutlich zusammen.\newline
Wie Abb. \ref{fig:icd-analysis-1.5T-offres-correction} und Abb. \ref{fig:icd-analysis-7T-offres-correction} zu entnehmen ist, lässt sich bei konstantem $\eta$ so der Bereich möglicher $\ICD$-Werte zwar deutlich besser, aber immer noch nicht sehr genau eingrenzen. Abb. \ref{fig:icd-analysis-1.5T-offres-correction-const-radius} und Abb. \ref{fig:icd-analysis-7T-offres-correction-const-radius} liefern die Relaxationsraten für konstante Kapillarradien und somit sinkendem RBV für steigenden $\ICD$. Da beim Abtasten des Parameterraums das RBV nur einem realistischen Bereich zwischen $\eta=0.05$ und $\eta=0.20$ variiert wurde, kann für die außgewählten $R_c$ nicht über die komplette $\ICD$-Achse geplottet werden, da sich für $\eta$ nach Tab. \ref{tab:model-parameters} Werte außerhalb des angenommenen physiologischen Bereichs ergeben würden.\newline
Unter Berücksichtigung des linearen Gradienten in den Offresonanzen entlang der Kapillarachse muss auch die Gültigkeit des zweidimensionalen Modells neu betrachtet werden. Geht man von einer Kapillarlänge $L\approx100\mum$ aus, so ergibt sich entlang der Kapillare Gradient der Offresonanzen von ca. $50\,\mathsf{rad}/(\mathsf{s}\cdot\mum)$ bei $1.5T$ und ca. $250\,\mathsf{rad}/(\mathsf{s}\cdot\mum)$ bei $7T$. Bei einer Scheibendicke $\overline{d}\approx17\mum$ folgt also innerhalb einer Scheibe noch ein Unterschied von ca. $10\radps$ bei $1.5T$ bzw. $60\radps$ bei $7T$. Zwar sinkt die Relaxationszeit $T_2^*$ mit steigenden Off\-resonanzen, wodurch auch die Scheibendicke kleiner gewählt werden kann, trotzdem ist davon auszugehen, dass mit steigender Feldstärke das zweidimensionale Modell an Genauigkeit einbüßt. Die Diffusion entlang des linearen Offresonanzgradienten kann dann nicht mehr vernachlässigt werden.\newline
Sollte der Oxygenierungsgrad nicht linear entlang der Kapillarachse abnehmen, so muss analog zu Abb. \ref{fig:2D-to-3D} geprüft werden wie stark die linearen Off\-resonanz\-gradienten innerhalb einzelner Scheiben ausfallen. Sind die Scheiben noch dick genug, bzw. der Off\-resonaz\-gradient vernachlässigbar klein, kann die Abhängigkeit der Relaxationsrate von der Scheibenposition $R_2^*(l)$ aus Gl. (\ref{eq:magnetization-3D-approx-relaxations-linear}) entsprechend angepasst werden. Je nach funktionalem Zusammenhang von $R_2^*(l)$ sind Gl. (\ref{eq:magnetization-3D-integrated}) und Gl. (\ref{eq:magnetization-3D-integrated-monoexp-approx}) dann nicht mehr analytisch lösbar.
	
\subsection{Vergleich der Relaxation in Krogh-Modell, Plasma-Verteilung, hexagonalem Gitter und Strong-Collision Näherung}
\begin{figure}
		\begin{center}\includegraphics[width=0.7\textwidth]{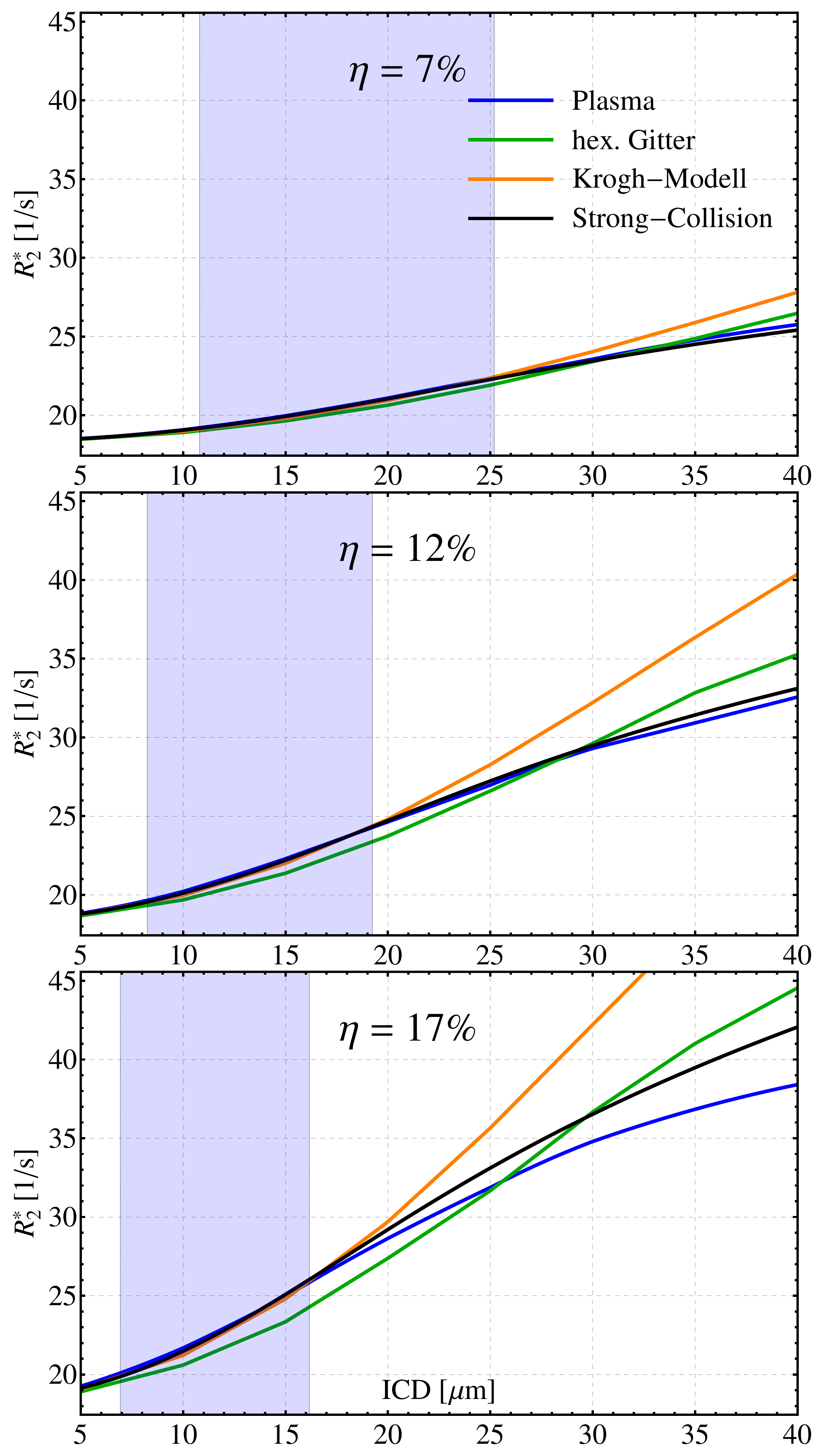}\end{center}
		\caption{Relaxationsrate der verschiedenen untersuchten Modellgeometrien für $D=1.5\mum^2/ms$, $\dom_0=150 \mathsf{rad}/\mathsf{s}$, $R_2=18.3s^{-1}$ konstantem $\eta$ (steigender $\ICD$ $\Rightarrow$ steigende $R_c$). Für $R_c\approx2.5\mum$ ist der nach Gl. (\ref{eq:mean-icd}) folgende $\ICD$-Bereich blau markiert.}
		\label{fig:icd-comparison-offres-150}
	\end{figure}

\begin{figure}
		\begin{center}\includegraphics[width=0.7\textwidth]{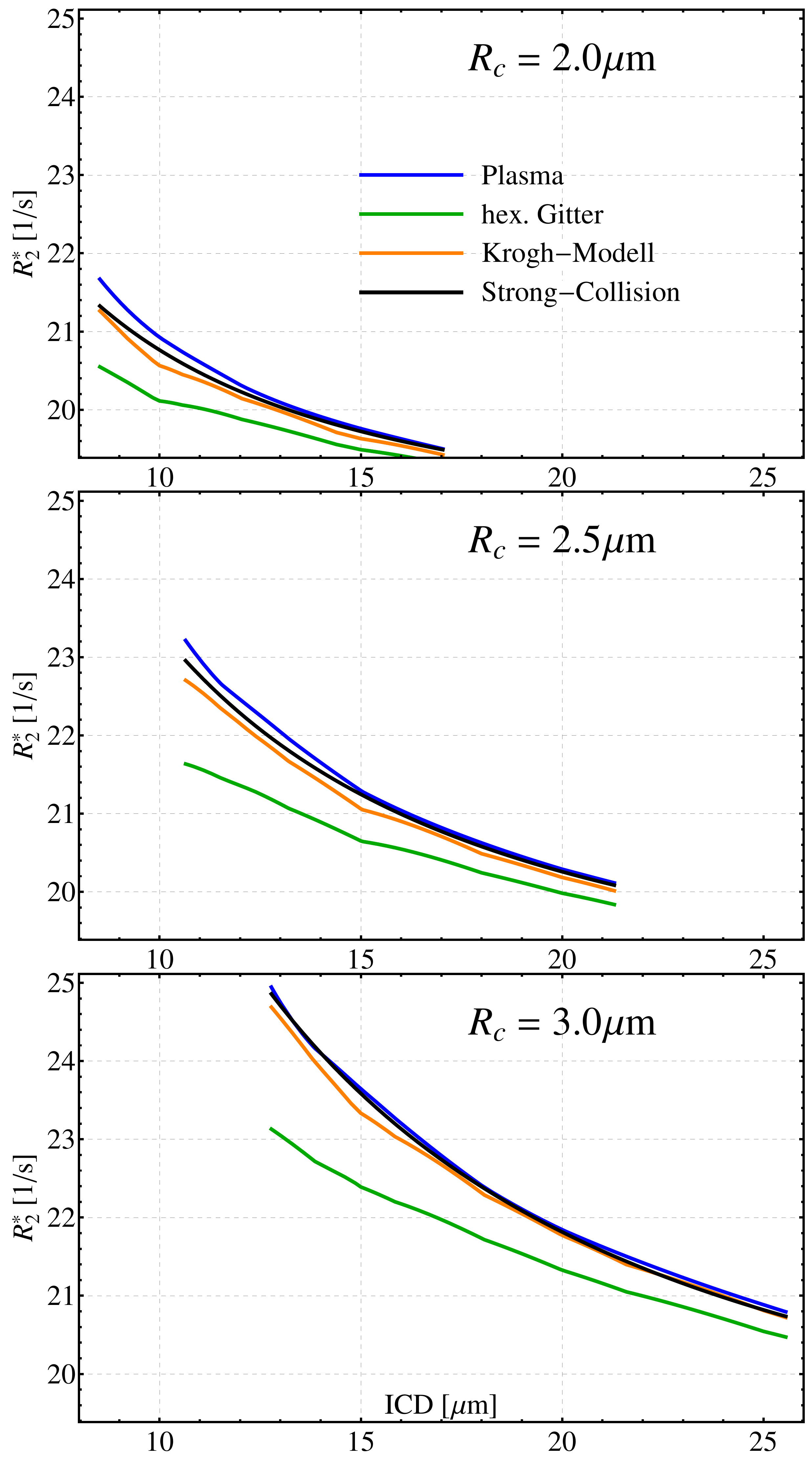}\end{center}
		\caption{Relaxationsraten der verschiedenen Modelle für $D=1.5\mum^2/\ms$, $\dom_0=150\radps$ und $R_2=18.3s^{-1}$ bei konstanten $R_c$ (steigender $\ICD$ $\Rightarrow$ sinkendes RBV). Da bei den Simulationen der Parameter $\eta$ als Freiheitsgrad verwendet und nur im Bereich $0.05$ bis $0.20$ untersucht wurde, kann für konstante $R_c$ nach  Tab. \ref{tab:model-parameters} nicht über den kompletten $\ICD$ Bereich interpoliert werden.}
		\label{fig:icd-comparison-radii-const-150}
	\end{figure}

\begin{figure}
		\begin{center}\includegraphics[width=0.7\textwidth]{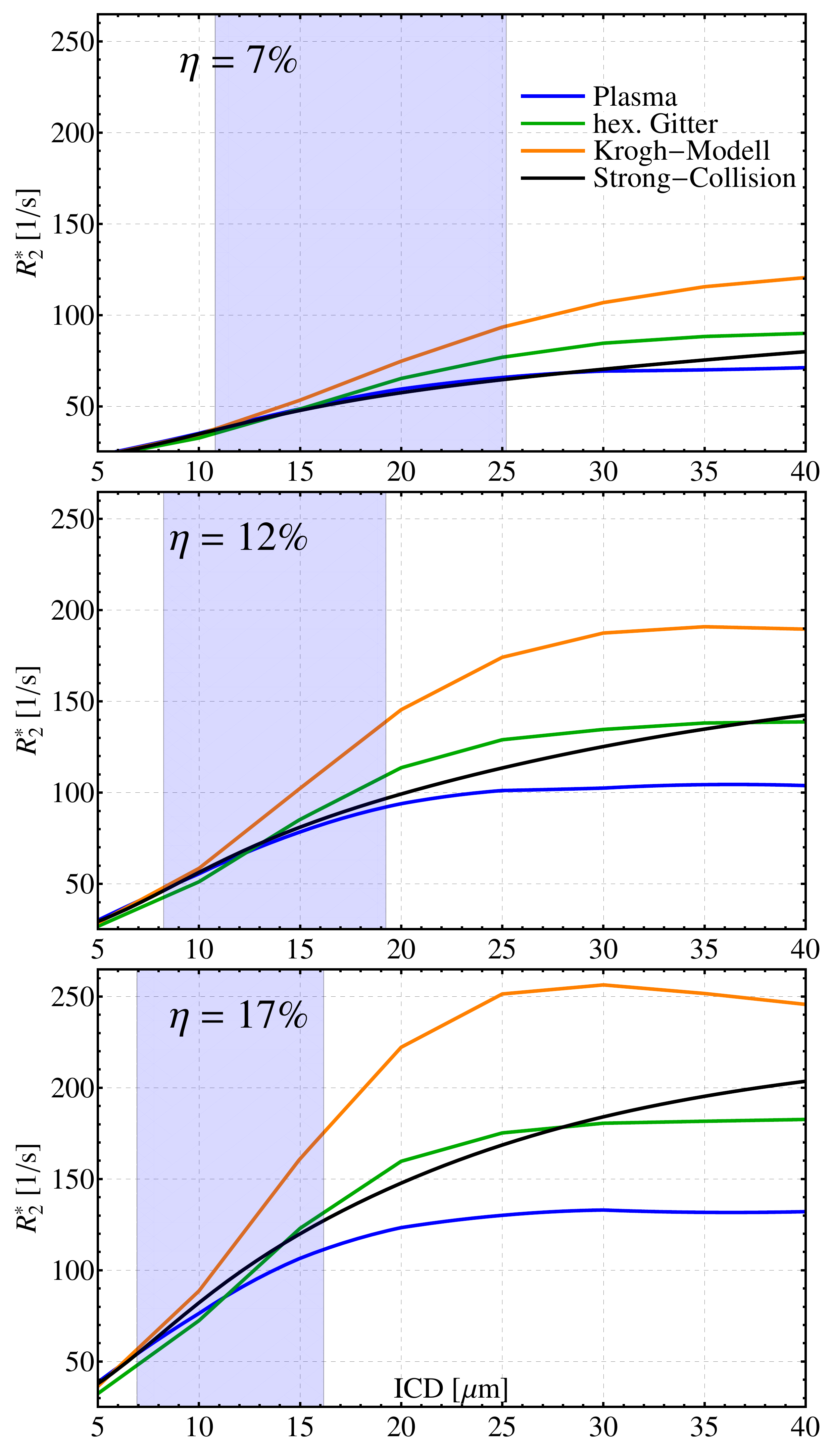}\end{center}
		\caption{Relaxationsrate der verschiedenen untersuchten Modellgeometrien für $D=1.5\mum^2/ms$, $\dom_0=750\mathsf{rad}/\mathsf{s}$, $R_2=18.3s^{-1}$ und konstantem $\eta$ (steigender $\ICD$ $\Rightarrow$ steigende $R_c$). Für $R_c\approx2.5\mum$ ist der nach Gl. (\ref{eq:mean-icd}) folgende $\ICD$-Bereich blau markiert.}
		\label{fig:icd-comparison-offres-750}
	\end{figure}

\begin{figure}
		\begin{center}\includegraphics[width=0.7\textwidth]{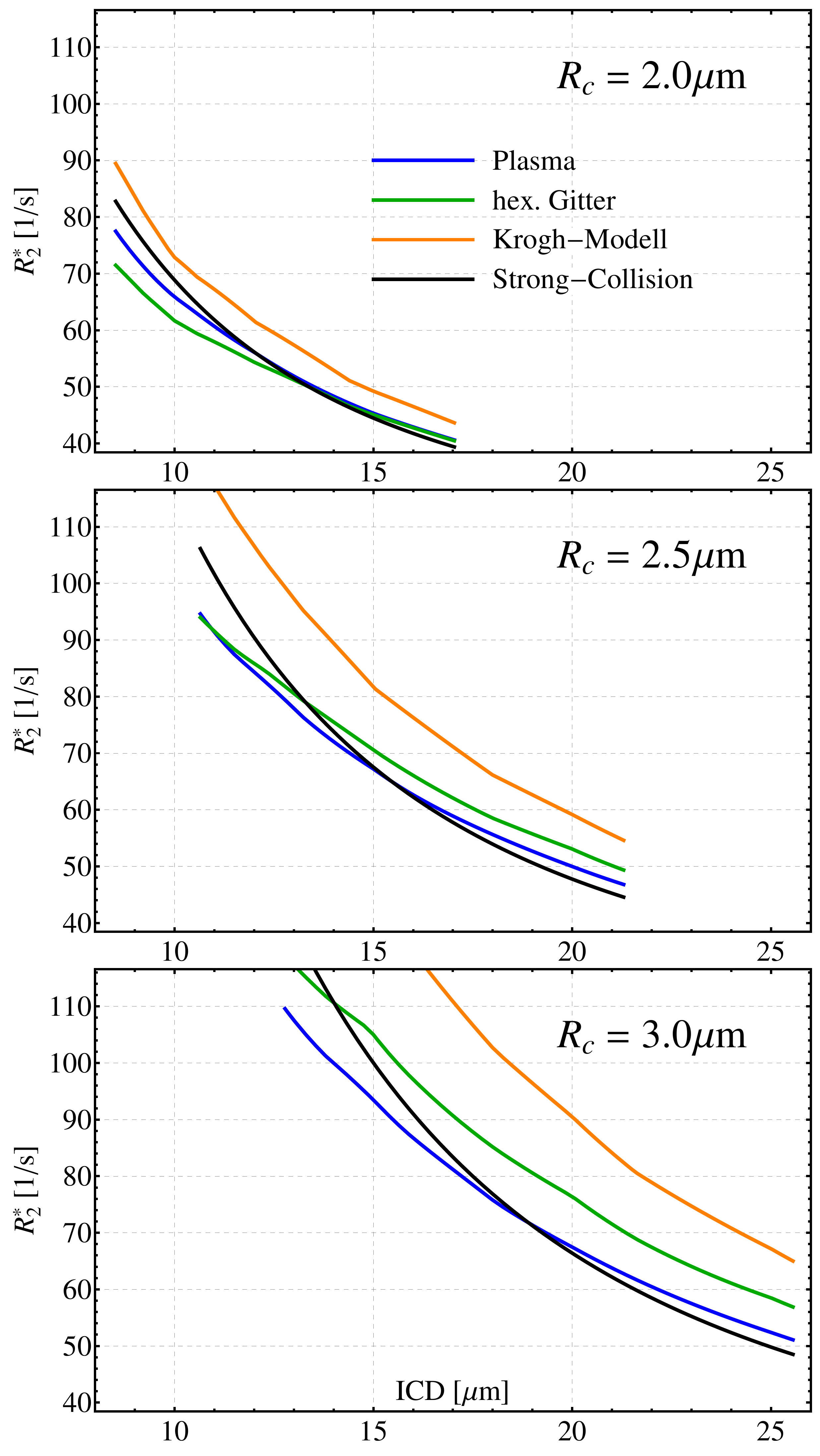}\end{center}
		\caption{Relaxationsraten der verschiedenen Modelle für $D=1.5\mum^2/\ms$, $\dom_0=750\radps$ und $R_2=18.3s^{-1}$ bei konstanten $R_c$ (steigender $\ICD$ $\Rightarrow$ sinkendes RBV).}
		\label{fig:icd-comparison-radii-const-750}
	\end{figure}

\begin{figure}
		\begin{center}\includegraphics[width=0.67\textwidth]{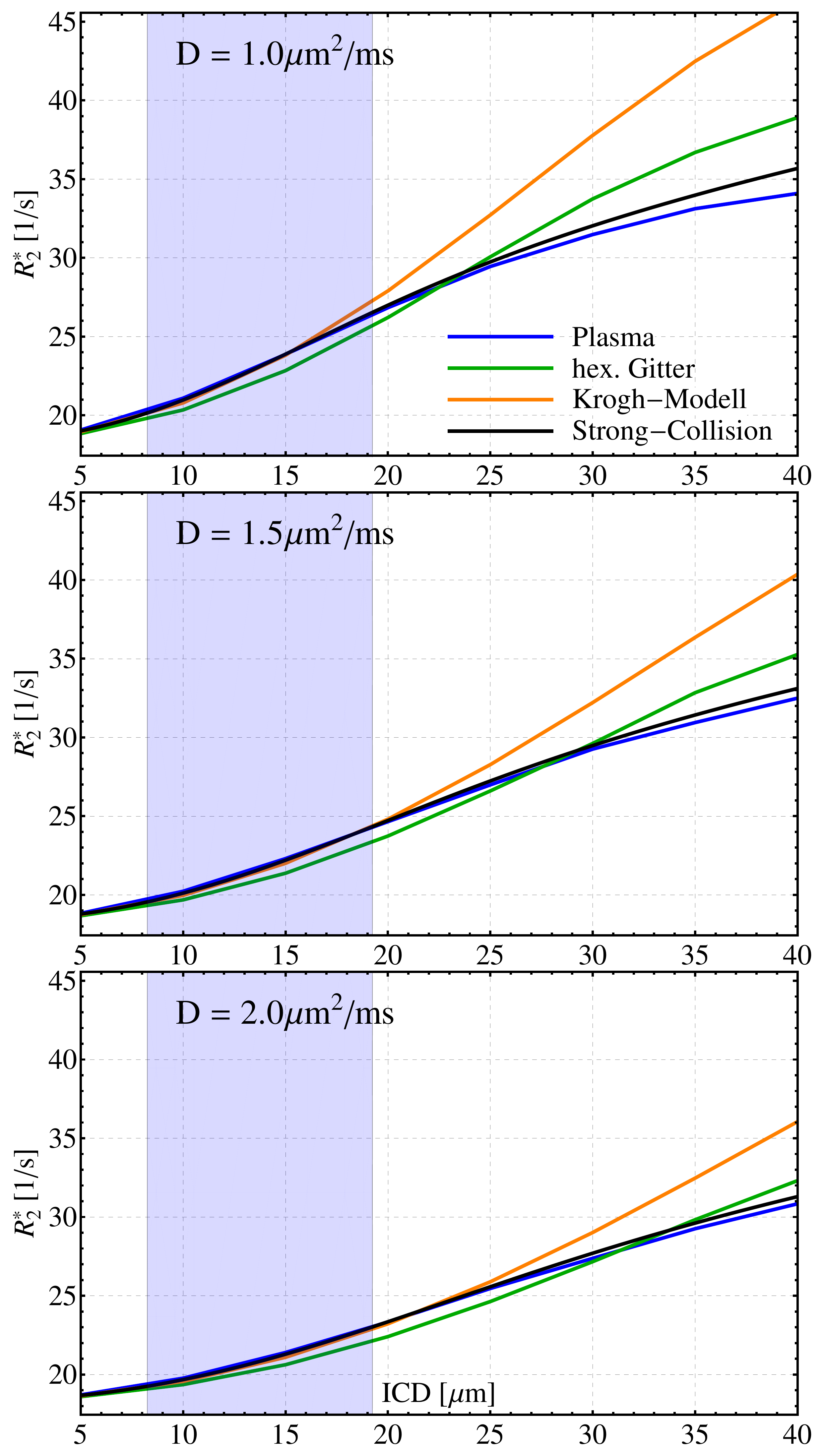}\end{center}
		\caption{Relaxationsraten der verschiedenen Modelle für $\dom_0=150\radps$ und $R_2=18.3s^{-1}$ bei konstantem $\eta=0.12$ (steigender $\ICD$ $\Rightarrow$ steigende $R_c$). Bei niedrigen Feldstärken bestehen zwischen den verschiedenen Modellen im klinisch relevanten $\ICD$-Bereich (schattiert) nur geringe Unterschiede in den vorausgesagten Relaxationsraten. Für niedrige $\ICD$ muss auch Diffusionsrate ausreichend klein sein, um bereits die Aufspaltung in die verschiedenen Static-Dephasing-Grenzfälle erkennen zu können.}
		\label{fig:icd-comparison-konvergence-diffusion-150}
	\end{figure}

\begin{figure}
		\begin{center}\includegraphics[width=0.67\textwidth]{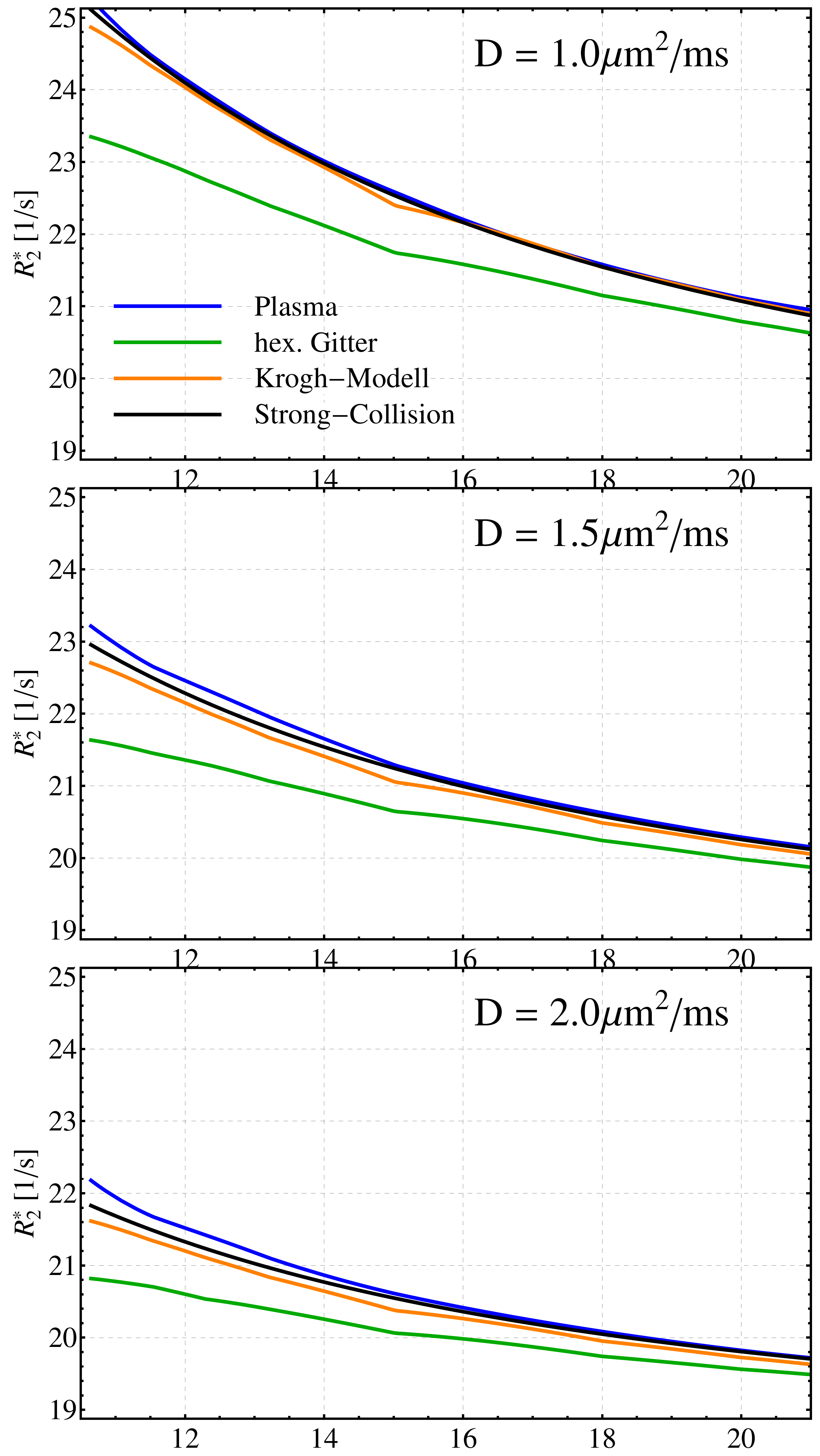}\end{center}
		\caption{Relaxationsraten der verschiedenen Modelle für $\dom_0=150\radps$ und $R_2=18.3s^{-1}$ bei konstanten $R_c=2.5\mum$ (steigender $\ICD$ $\Rightarrow$ sinkendes RBV).}
		\label{fig:icd-comparison-konvergence-diffusion-150-const-radii}
	\end{figure}

\begin{figure}
		\begin{center}\includegraphics[width=0.7\textwidth]{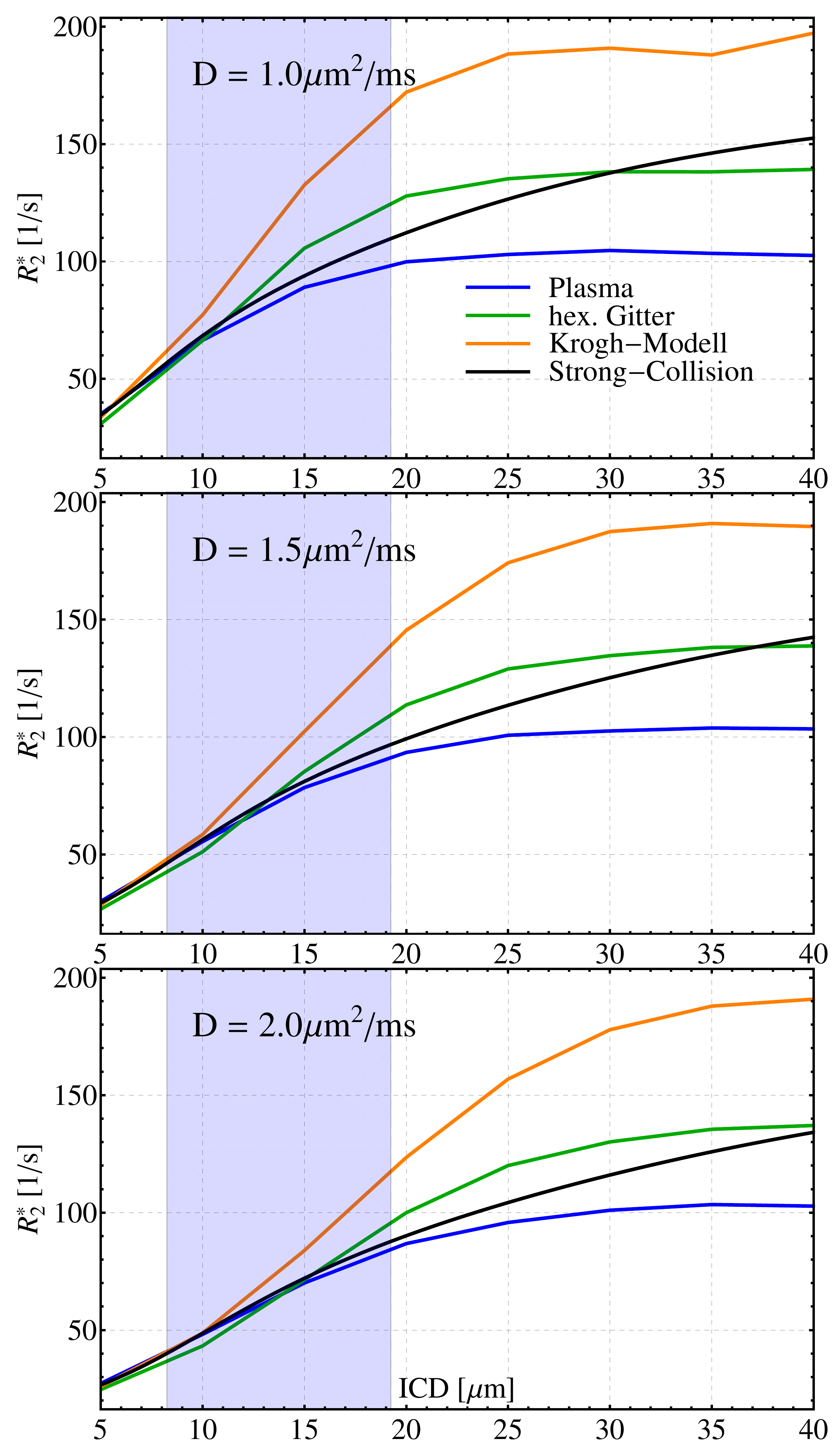}\end{center}
		\caption{Relaxationsraten der verschiedenen Modelle für $\dom_0=750\radps$ und $R_2=18.3s^{-1}$ bei konstantem $\eta=0.12$ (steigender $\ICD$ $\Rightarrow$ steigende $R_c$). Mit zunehmender Diffusionsrate konvergieren die verschiedenen Modelle immer langsamer, d.h. erst für einen höheren $\ICD$, in ihre maximale Relaxationsrate $R_2'$. Bereits im klinisch relevanten $\ICD$-Bereich (schattiert) weichen die vom Krogh-Modell vorhergesagten Relaxationsraten deutlich von denen des Plasma-Modells ab.}
		\label{fig:icd-comparison-konvergence-diffusion-750}
	\end{figure}

\begin{figure}
		\begin{center}\includegraphics[width=0.67\textwidth]{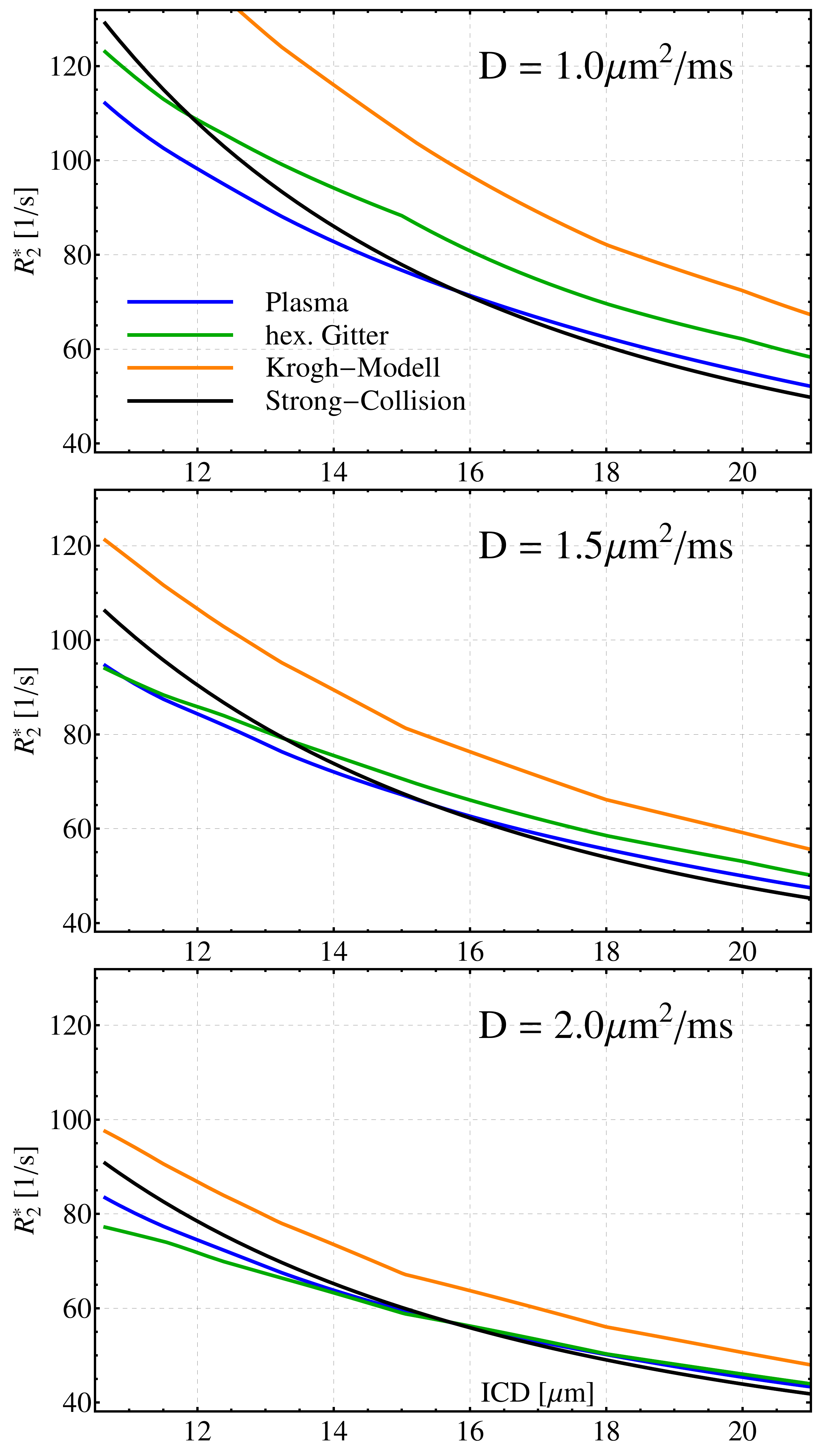}\end{center}
		\caption{Relaxationsraten der verschiedenen Modelle $\dom_0=750\radps$ und $R_2=18.3s^{-1}$ bei konstanten $R_c=2.5\mum$ (steigender $\ICD$ $\Rightarrow$ sinkendes RBV).}
		\label{fig:icd-comparison-konvergence-diffusion-750-const-radii}
	\end{figure}

In Abb. \ref{fig:icd-comparison-offres-150} und Abb. \ref{fig:icd-comparison-offres-750} ist der Verlauf der Relaxationsrate in Abhängigkeit vom $\ICD$ für die verschiedenen Modellgeometrien bei konstantem $D$, konstantem $\dom_0$ für konstante $\eta$ gezeigt. Die Abbildungen \ref{fig:icd-comparison-konvergence-diffusion-150} und \ref{fig:icd-comparison-konvergence-diffusion-750} zeigen den Einfluss der Diffusionskonstanten auf die Relaxationszeiten. Zusätzlich zur analytischen Lösung des Krogh-Modells wurde auch die Näherungslösung des Krogh-Modells durch den Strong-Collision Ansatz berechnet \cite{BauerPRL99}. Die Relaxationsraten des Krogh-Modells folgen aus Gl. (154) in \cite{ZienerPHDThesis}.\newline
Mit den abgetasteten Punkten im Parameterraum lässt sich auch die Abhängigkeit der Relaxationsrate vom $\ICD$ bei konstanten Kapillarradien zeigen. Während in Abb. \ref{fig:icd-comparison-offres-150} und Abb. \ref{fig:icd-comparison-offres-750} der Parameter $\eta$ konstant gehalten wurde, die Kapillarradien also mit steigendem $\ICD$ zunehmen, zeigen Abb. \ref{fig:icd-comparison-radii-const-150}, Abb. \ref{fig:icd-comparison-radii-const-750}, Abb. \ref{fig:icd-comparison-konvergence-diffusion-150-const-radii} und Abb. \ref{fig:icd-comparison-konvergence-diffusion-750-const-radii} die Abhängigkeit der Relaxationsraten für konstante $R_c$, also sinkendem $\eta$ für steigenden $\ICD$. 
 


	

\subsection{Kontrastmittelanwendung}
\begin{figure}
		\begin{center}\includegraphics[width=0.65\textwidth]{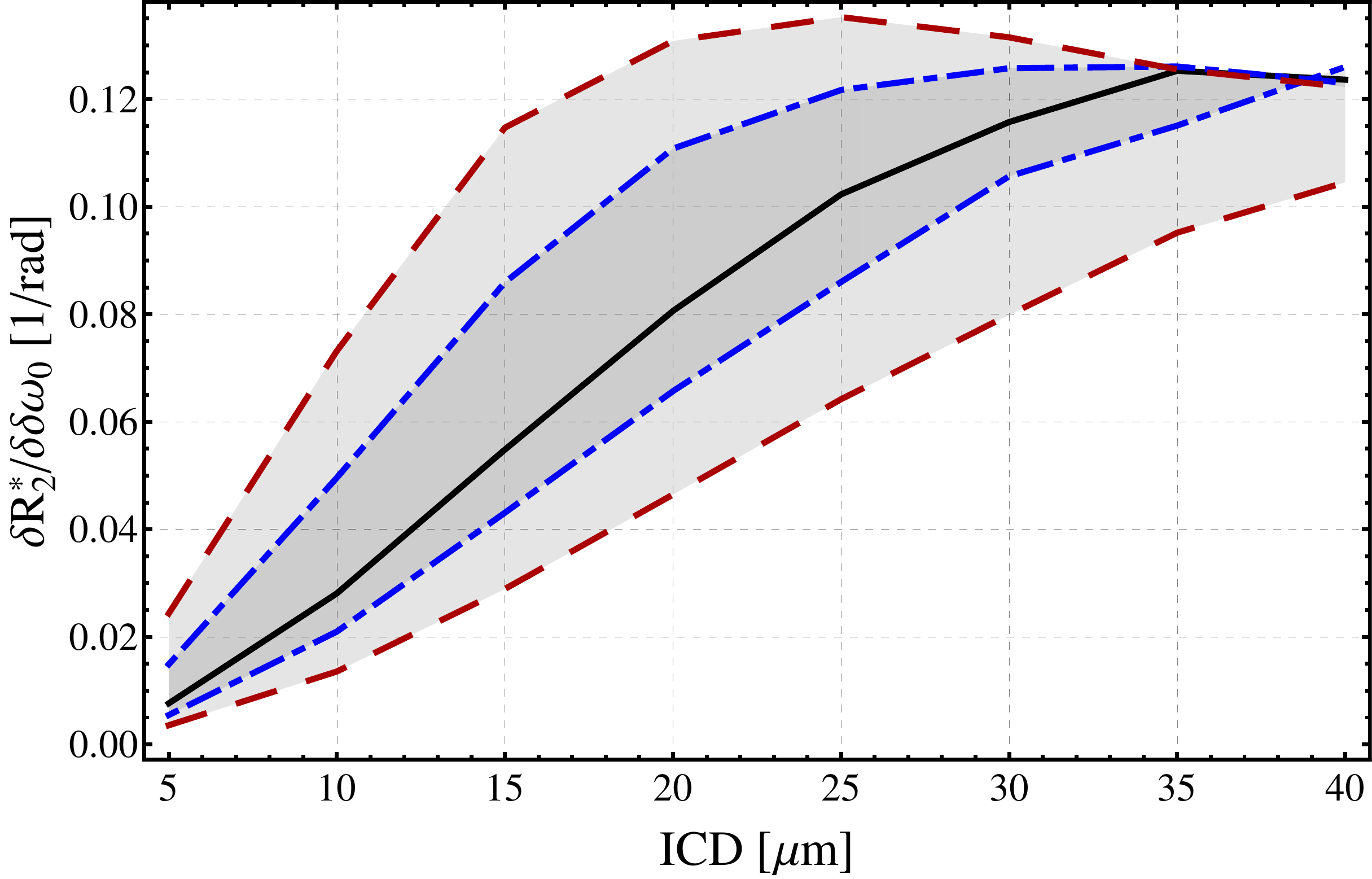}\end{center}
		\caption{Quotient aus Relaxationszeitänderung und Offresonanzverschiebung durch Kontrastmittelgabe für ein konstantes $\eta$ (steigender $\ICD$ $\Rightarrow$ steigende $R_c$). Die Ableitung $\partial R_2^*/\partial\dom_0$ ist auch weiterhin eine Funktion von $\ICD$, $\dom_0$ und den anderen Parametern (hier: $D=1.5\mum^2/\ms$, $\dom_0=150\radps$ und $\eta=0.12$). Der schattierte Bereich gibt die entsprechende Bandbreite der aus dem Fehler folgenden möglichen Differenzenquotienten an. Für den inneren Bereich wurde nur $\Delta\dom_0=50\radps$ berücksichtigt und die Fehler der anderen eingehenden Größen vernachlässigt. Bei dem äußeren Bereich wurden auch die Fehler $\Delta D=0.5\mum^2/\ms$ und $\Delta\eta=0.01$ in einer Worst-Case-Abschätzung berücksichtigt. Für $R_c\approx2.5\mum$ ist ein Wert von $\delta R_2/\delta\dom_0\approx0.05$ zu erwarten. Nach der Definition von $R_2^*$ aus Gl. (\ref{eq:r2s-funktion}) verschwindet die Abhängigkeit von $R_2$ beim differenzieren nach $\dom_0$. Die Kontrastmittel können aber auch Einfluss auf die intrinsische Relaxation $R_2^i$ haben was hier nicht berücksichtigt wurde.}
		\label{fig:differenzenquotient}
	\end{figure}

Wie bereits in Kap. \ref{kap:offres-strength} beschrieben kann durch die Verabreichung von intravaskularen Kontrastmitteln der Suszeptibilitätsunterschied $\Delta\chi$ zwischen Kapillare und Gewebe, und somit auch die Stärke der Offresonanzen variiert werden. Im Gegensatz zur schlecht messbaren tatsächlichen Offresonanzstärke $\dom_0$ kann die Verschiebung der Offresonanzen (im Folgenden $\delta\dom_0$, nicht zu verwechseln mit $\Delta\dom_0$) durch die Kontrastmittel über die verabreichte Kontrastmittelkonzentration abgeschätzt werden. Der Unterschied der Relaxationsrate vor und nach Kontrastmittelvergabe (im Folgenden $\delta R_2^*$) und die Verschiebung der Offresonanzen stellen dann evtl. bessere Rückschlussmöglichkeiten auf den $\ICD$ zur Verfügung, indem das Problem der großen $\Delta\dom_0$ teilweise umgangen wird. Im Folgenden wird wie nach \cite{Kennan94} von einer Zunahme von $\Delta\chi$ um ca. $30-40\%$ nach Kontrastmittelvergabe ausgegangen. Dies entspricht etwa einer Konzentration von 1mM des Kontrastmittels Gd-DTPA \cite{Kennan94}. Die Offresonanzen verschieben sich dadurch um ca. $\delta\dom_0\approx50\radps$ von $\dom_0=150\radps$ auf ca. $200\radps$ bei 1.5T und um $\delta\dom_0\approx250\radps$ von $\dom_0=750\radps$ auf ca. $1000\radps$ bei 7T.\newline
Sind $\delta R_2^*$ und $\delta\dom_0$ bekannt, so kann analog zur Kap. \ref{kap:icd-ruecklschluss} dem Quotienten aus $\delta R_2^*$ und $\delta\dom_0$ ein $\ICD$-Bereich zugeordnet werden. In den Bereichen des Parameterraums, in denen dieser Differenzenquotient nur schwach von $\dom_0$ abhängt, ist dann der durch $\Delta\dom_0$ verursachte Fehler deutlich geringer. Abb. \ref{fig:differenzenquotient} zeigt eine Auftragung von $\delta R_2^*/\delta\dom_0$ gegen den $\ICD$ exemplarisch für ein konstantes $\eta$.\newline
Während bei $\dom_0=150\radps$ in Abb. \ref{fig:differenzenquotient} ein weitgehend linearer Zusammenhang zum $\ICD$ besteht, konvergiert der Differenzenquotient für hohe Offresonanzen genau wie die Relaxationsrate gegen einen Grenzwert. Je früher diese Konvergenz eintritt, desto schwerer wird es aus den gemessenen Relaxationsraten und der Offresonanzverschiebung den $\ICD$ einzugrenzen.

\subsection{Fazit}
Die Möglichkeit aus einer $R_2^*$-Messung auf den $\ICD$ Rückschlüsse ziehen zu können sinkt und steigt mit der Kenntnis der anderen in Gl. (\ref{eq:r2s-funktion}) eingehenden Parameter. Lassen sich Diffusionskonstante, RBV und Offresonanzen noch deutlich genauer eingrenzen als in den vorangegangenen Kapiteln angenommen, so kann mit dem Nachschlagewerk über den durchsuchten Parameterraum mit einer gemessenen Relaxationszeit direkt ein Intervall möglicher $\ICD$-Werte bestimmt werden.\newline
Viele der beschriebenen Eigenschaften lassen sich gut über die statischen Frequenzverteilungen der entsprechenden Geometrien begründen. Im Krogh-Modell und hexagonalen Gitter führen die zwei Peaks in der statischen Frequenzverteilung, auch noch unter Einfluss der Diffusion, zu einem breiteren Frequenzspektrum des FID, und somit zu höheren Relaxationsraten. Bei der Plasmaverteilung hingegen ist bereits die statische Frequenzverteilung deutlich schmaler, was durchweg zu niedrigeren Relaxationsraten führt.\newline
Mit zunehmenden Kapillarabständen konvergieren die Relaxationsraten bei konstantem $\eta$ gegen den Static-Dephasing Grenz\-wert $R_2'$. Die Bedingung $\dom_0 R_c^2/D\gg1$ ist hier gut erfüllt. Wie vorhergesagt, tritt je nach Offresonanz und RBV, diese Konvergenz unterschiedlich schnell ein. Während für $\dom_0=750\radps$ in Abb. \ref{fig:icd-comparison-offres-750} die Konvergenz für alle $\eta$ gut zu sehen ist, ist sie in Abb. \ref{fig:icd-comparison-offres-150} nur für $\eta=0.17$ angedeutet. In diesem Konvergenzbereich ist es prinzipiell schwierig, den funktionellen Zusammenhang aus Gl. (\ref{eq:r2s-funktion}) nach Gl. (\ref{eq:icd-funktion}) aufzulösen. Die Diffusion kann die Dephasierung der Magnetisierung nicht mehr kompensieren, da dafür die Protonen über zu große Strecken hinweg wandern müssten, um signifikante Unterschiede im Offresonanzfeld spüren zu können.  Es zeigt sich auch, dass sowohl die in \cite{Magma05} abgeschätzte Relaxationsrate $2\dom_0\eta/(1+\eta)$ für das Krogh-Model als auch der Static-Dephasing Grenzwert des hexagonalen Gitters deutlich größer ausfallen als die des Plasmas. Dies folgt direkt aus den statischen Frequenzverteilungen.\newline
Durch Spin-Echo-Experimente kann im Static-Dephasing Grenzfall ein Großteil der Magnetisierung im Echo refokussiert werden. Wenn die intrinsische Relaxation nicht zu schnell ist, hängt im Konvergenzbereich möglicherweise die T2-Zeit signifikant von den Kapillarabständen ab. Bedingung dafür ist, dass Protonen zwischen den Nahfeldern verschiedener Kapillaren wechseln können, bevor die Magnetisierung weitgehend durch die intrinsische Relaxation ausgelöscht ist. Bei hinreichend hoher Diffusion bzw. langsamer intrinsischer Relaxation sind also Spin-Echo-Experimente vielversprechender als der normale FID (siehe z.B. \cite{JMR2010}).\newline
Bei konstant gehaltenem $\eta$ ergeben sich für kleine $\ICD$ und niedrige Offresonanzen im realistischen $\ICD$-Bereich nur kleine Unterschiede zwischen den vier Modell\-geo\-metrien (Abb. \ref{fig:icd-comparison-konvergence-diffusion-150}). In diesem Bereich besteht zwischen $\ICD$ und Relaxationsrate in guter Näherung ein linearer Zusammenhang. Und der Verlauf des FID ist hauptsächlich durch das Nahfeld der Kapillaren beeinflusst. Bei höheren Offresonanzen können sich die verschiedenen Modelle jedoch bereits im niedrigen $\ICD$-Bereich deutlich voneinander unterscheiden (Abb. \ref{fig:icd-comparison-konvergence-diffusion-750}). Vor allem für kleines $\eta$ und niedrige Offresonanzen gibt gibt die Strong-Collision Näherung das Verhalten des Plasma-Modells am besten wieder.\newline
Analysiert man das Relaxationsverhalten für konstante $R_c$ (steigender $\ICD$ $\Rightarrow$ sinkendes RBV) ergibt sich ein völlig anderer Grenzfall. Da das RBV für $\ICD\rightarrow\infty$ verschwindet, konvergiert die Relaxationsrate gegen ihren intrinsischen Anteil. Die intrinsische Relaxationsrate ist unabhängig vom Modell und die (absolute) Aufspaltung zwischen den verschiedenen Modellen nimmt daher mit steigendem $\ICD$ ab.\newline
Auch unter Annahme konstanter Kapillarradien ist eine Zuordnung eines $\ICD$-Bereichs zur Relaxationsrate $R_2^*$ schwierig und kritisch von der Kenntnis der anderen Parameter abhängig (Abb. \ref{fig:icd-analysis-1.5T-offres-correction-const-radius}  und Abb. \ref{fig:icd-analysis-7T-offres-correction-const-radius}).\newline
Wie im letzten Kapitel angedeutet, bietet sich abhängig von der Feldstärke, evtl. eine Vorher-Nachher-Messung der Relaxationszeit bzgl. einer Kontrastmittelvergabe, als Möglichkeit, doch noch Informationen über den $\ICD$ zu gewinnen an. Die möglicherweise mit einem Kontrastmittel verbundene beschleunigte intrinsische Relaxation $R_2^i$ wurde nicht berücksichtigt, da wie in \cite{Bauer99} von einer undurchlässigen Kapillarwand ausgegangen wird. Die beschleunigte intrinsische Relaxation bleibt somit auf das Innere der Kapillare begrenzt und kann wegen der niedrigen $\eta$ vernachlässigt werden.\newline

\chapter{Diskussion}
Bei der numerischen Implementierung der Bloch-Torrey-Gleichung über den Random-Walk Ansatz wurde viel Wert auf einfache Erweiterbarkeit, z.B. auf differenziertere Randbedingungen oder für Spin-Echo Experimente und gute Performanz bzw. effiziente Parallelisierung gelegt. So ist eine umfangreiche Programmbibliothek entstanden, die zunächst einer sehr gründlichen Fehlerabschätzung unterzogen wurde. Auch die korrekte Implementierung der in die Simulation einfließenden Modellannahmen, z.B. die reflektiven Randbedingungen an den Kapillaren, wurden getestet.\newline
Durch Entwicklung einer korrekten Näherung der komplexen Mathieu-Funktionen als Reihendarstellung konnte dann anhand der analytischen Lösung der Bloch-Torrey-Gleichung nach \cite{ZienerPHDThesis} die numerische Implementierung anhand des Krogh-Modells verifiziert werden.\newline
Ausgehend von den statischen Frequenzverteilungen verschiedener Kapillaranordnungen wurden dann speziell das hexagonale Gitter und das besser an reales Gewebe angepasste 2D1CP-Modell nach \cite{Karch2006} einer genaueren Analyse unterzogen. Dazu wurde zunächst der Metropolisalgorithmus auf das 2D1CP angewandt um Kapillarkonfigurationen für verschiedene $\Gamma$ zu erzeugen. Unter Verwendung verschiedener Computeserver wurde dann über mehrere Monate hinweg ein großer Bereich des durch die Freiheitsgrade der Simulation aufgespannten Parameterraums abgetastet. Für Feldstärken von 1.5T bzw. 7T wurden so eine Karte der Relaxationsraten für die verschiedenen Dimensionen des Parameterraums (Offresonanzstärke $\dom_0$, Regional Blood Volume $\eta$, Kapillarabstände $\ICD$ und Diffusionskonstante $D$) erstellt.\newline
Die aus den Simulationen folgenden Daten wurden unter Berücksichtigung verschiedener Aspekte, wie z.B. der zweidimensionalen Näherung oder einer möglichen Anwendung von intravaskularen Kontrastmitteln, einer ausführlichen Analyse unterzogen. Der Fokus lag dabei auf der Prüfung ob und in welchem Rahmen eine Relaxationszeit Informationen über die Anordnung der Kapillaren, speziell den Kapillarabständen, liefert.\newline
Mit einem Vergleich der Simulationsdaten mit den analytisch berechneten Relaxationsraten des Krogh-Modells, kann der mögliche Anwendungsbereich des Krogh-Modells eingegrenzt werden. Zwar geben für dicht angeordnete Kapillaren das Krogh-Modell und seine Näherung durch die Strong-Collision das gleiche Verhalten wie das hexagonale Gitter und die 2D1CP-Konfigurationen wieder, bereits für realistische Kapillarabstände weicht jedoch das Krogh-Modell deutlich von den Simulationsdaten ab. Da die 2D1CP-Anordnungen das dem realen Gewebe am nächsten kommende hier behandelte Modell ist, sollte es daher auch für den direkten Anwendungsbezug verwendet werden.\newline
Es zeigt sich, dass die Möglichkeit aus einer Relaxationsrate Rückschlüsse auf die Mikrostruktur innerhalb eines Voxels ziehen zu können äußerst kritisch von der Kenntnis der Diffusionsrate, der Offresonanzen und dem RBV abhängt. Sind nur grobe Schätzwerte für diese drei Parameter gegeben, so kann mit einer Relaxationsrate nicht einmal der mittlere Kapillarabstand $\ICD$ sinnvoll eingegrenzt werden. Lassen sich die Parameter hingegen mit hoher Genauigkeit angeben, so lässt sich mit dem erstellten Nachschlagewerk für die Relaxationsraten direkt ein $\ICD$-Bereich eingrenzen.\newline
Gibt man einen maximalen Fehler für den $\ICD$ vor der nicht überschritten werden soll, so kann mit den erzeugten Datensätzen auch die mindestens notwendige Genauigkeit der anderen Parameter abgeschätzt werden. Aufgrund der Vierdimensionalität des Parameterraums hängen die erforderlichen Mindestgenauigkeiten jedoch gegenseitig voneinander ab, weswegen eine allgemeine Auswertung in diesem Rahmen nicht praktikabel, für konkrete Fälle aber prinzipiell möglich wäre.\newline
Die Zuordnung einer Relaxationsrate zu einem Ordnungsgrad $\Gamma$ bleibt jedoch problematisch. Selbst bei einer sehr optimistischen Fehlerabschätzung könnte bestenfalls zwischen einer sehr regelmäßigen hexagonalen Anordnung ($\Gamma>140$) und einer relativ zufälligen Verteilung ($\Gamma\approx4$) unterschieden werden. Eine Klassifizierung realer Gewebe über ihren Ordnungsgrad wie in \cite{Karch2006} ist praktisch unmöglich.\newline
Als weiterer Ausblick erscheint vor allem die Simulation von Spin-Echo Experimenten und eine Erweiterung der Random-Walk Implementierung auf drei Dimensionen vielversprechend.\newline
Bei Spin-Echos werden deutlich größere Zeitskalen betrachtet als beim FID. Während für hohe $\ICD$ der FID als statische Dephasierung genähert werden kann und die Relaxationsrate $R_2^*$ gegen $R_2'$ konvergiert, nimmt für steigenden $\ICD$ der Einfluss langreichweitiger Diffusion auf die Transversalmagnetisierung zu. Bei einer Simulation von Spin-Echo Experimenten müssen aber zwei Punkte berücksichtigt werden. Zum Einen werden in den Simulationen mehr Zeitschritte benötigt um den längeren Zeitbereich zu erfassen, zum Anderen muss eine der Modellannahmen neu geprüft werden: Wie in \cite{Bauer99} bzw. \cite{Donahue94} beschrieben ist auf den relativ kurzen Zeitskalen des FID der Austausch zwischen intra- und extravaskularer Magnetisierung vernachlässigbar. In der Zeitskala der Spin-Echos kann dieser Austausch jedoch relevant werden, da z.B. eine schnellere intrinsische Dephasierung im Inneren der Kapillare, durch den Protonenaustausch auch auf die extravaskulare Magnetisierung übertragen wird. Außerdem kommt bei SE Simulationen die Inter-Echo-Zeit $T_E$ zu den anderen Freiheitsgraden der Simulation hinzu. Eine Erweiterung der Simulation auf dreidimensionale Geometrien würde zum Einen einige der kritischen Modellannahmen der zweidimensionalen Diffusion umgehen, zum Anderen auch neue Anwendungsbereiche wie z.B. die Untersuchung von kugelförmigen Offresonanzquellen (z.B. in \cite{Ziener07} oder \cite{Ziener201038}) eröffnen.\newline
Da mittlerweile eine deutlich höhere Rechenleistung und mehr Arbeitsspeicher zur Verfügung stehen, sind Spin-Echo und 3D-Simulationen nun auch in einem vertretbaren Zeitrahmen realisierbar.


\appendix
\part*{Anhang} 
\renewcommand{\appendixtocname}{Anhang}
\renewcommand{\figurename}{Abb.}
\renewcommand{\tablename}{Tab.}

\chapter{Mathieu Matrix Implementierung in MATHEMATICA}\label{appendix:mathieu}
\lstset{basicstyle=\ttfamily,tabsize=2,numbers=left,numberstyle=\footnotesize,stepnumber=1,frame=single}
Im Folgenden bezeichnet \texttt{MatrixDimension} die globale zu setzende Matrixgröße $k$. \texttt{Digits} gibt an wieviele Dezimalstellen von MATHEMATICA\textsuperscript{\textregistered} in internen Berechnungen zu verwenden sind. \texttt{Eps} ist die angenommene numerische Genauigkeit der Ergebnisse.\newline
Mit der Funktion \texttt{MathieuMatrix[q]} wird die Rekursionsmatrix aus Gl. (\ref{eq:mathieu-matrix-equation}) für ein spezifisches $q$ initialisiert.
\begin{lstlisting}
MathieuMatrix[q_]:=Module[{matrix},
	matrix=Table[Table[
		If[Abs[i-j]==1,q,If[i==j,(2i)^2,0]],
		{i,0,MatrixDimension-1}],{j,0,MatrixDimension-1}];
	matrix[[2,1]]=Sqrt[2]q;
	matrix[[1,2]]=Sqrt[2]q;
	matrix
];
\end{lstlisting}
Die Funktion \texttt{SortEV[v1,v2]} gibt an ob der Eigenwert \texttt{v1} zu einem kleineren $m$ gehört als \texttt{v2} und stellt somit auch im komplexen eine Ordnungsrelation der Eigenwerte zur Verfügung.
\begin{lstlisting}
SortEV[v1_,v2_]:=If[Abs[Re[v1]-Re[v2]]<Eps,
	Im[v1]<Im[v2],
	Re[v1]<Re[v2]
];
\end{lstlisting}
Die Funktion \texttt{A[q]} löst das Eigenwertproblem der zu $q$ gehörigen Matrix mit dem Befehl \texttt{Eigensystem[]}.
\texttt{Eigensystem[]} hat als ersten Rückgabewert eine Liste mit Eigenwerten (\texttt{vals}). Der zweite Rückgabewert ist eine Liste der Eigenvektoren (\texttt{vecs}) also eine $k\times k$ Matrix. Da die Rückgabewerte von \texttt{Eigensystem[]} unsortiert sind, wird in Zeile 3 zunächst die nötige Permutation der Eigenvektoren ermittelt. Dies geschieht über die zu den Eigenvektoren gehörigen Eigenwerte mit Hilfe der Ordnungsrelation \texttt{SortEV[v1,v2]}. Zeile 4 sortiert dann die erste Dimension der Eigenvektormatrix entsprechend der Permutation um. In der For-Schleife werden die Eigenvektoren normiert und falls nötig um $180^\circ$ gedreht, um numerisch bedingte Sprünge zwischen positiven und negativen Eigenvektoren zu korrigieren. Am Schluss wird noch die $\sqrt{2}$ Normierung der ersten Komponente rückgängig gemacht. \texttt{A[q][[m+1,r+1]]} liefert dann den Fourierkoeffizienten $A_{2r}^{(2m)}$.
\begin{lstlisting}
A[q_]:=Module[{vals,vecs,norm,permutation,flip,m},
	{vals,vecs} = Eigensystem[N[MathieuMatrix[q],Digits]];
	permutation = Ordering[vals,All,SortEV[#1,#2]&];
	vecs        = vecs[[permutation]];
	For[m=0,m<MatrixDimension,m++,
		norm        = Total[vecs[[m+1]]^2];
		vecs[[m+1]] = vecs[[m+1]]/Sqrt[norm];
		If[Mod[m,2]==0,
			(*I^m is Real*)flip=Sign[Re[vecs[[m+1,1]]]]*Sign[Re[I^m]],
			(*I^m is Imag*)flip=Sign[Im[vecs[[m+1,1]]]]*Sign[Im[I^m]]
		];
		vecs[[m+1]]=vecs[[m+1]]*flip;
	];
	vecs[[All,1]]/=Sqrt[2];
	vecs
];
\end{lstlisting}
Die Funktion \texttt{a[q]} gibt einen korrekt sortierten Vektor mit den Eigenwerten der zu $q$ gehörigen Matrix zurück.
\begin{lstlisting}
a[q_]:=Sort[Eigenvalues[N[MathieuMatrix[q],Digits]],SortEV[#1,#2]&];
\end{lstlisting}
Die Funktion \texttt{Ce2m[q,m]} gibt ein Funktionsobjekt für die zu $q$ gehörige $2m$-te Mathieu-Funktion zurück.
\begin{lstlisting}
Ce2m[q_,m_]:=Module[{coeff=A[q]},
	Function[{phi},Sum[coeff[[m+1,i]]*Cos[2(i-1)phi]],{i,1,MatrixDimension}]]
];
\end{lstlisting}
Verwendet werden kann das Funktionsobjekt z.B. wie folgt:
\begin{lstlisting}
ce2=Ce2m[4I,1];
Plot[Re[ce2[phi]],{phi,0,2Pi}]
\end{lstlisting}


\bibliographystyle{alpha}
\bibliography{dipl_bib}
\clearpage

\noindent \LARGE Erklärung\newline\newline
\normalsize Hiermit versichere ich, dass ich die vorliegende Arbeit selbstständig verfasst, und keine anderen als die angegebenen Quellen und Hilfsmittel verwendet habe.\newline\newline
Würzburg, 28.07.2011\newline\newline\newline\newline
Martin Rückl
\end{document}